\documentclass[12pt,a4paper]{book}

\usepackage[french]{babel}

\usepackage{graphicx}

\def\baselinestretch{1.3}

\renewcommand{\SS}{\scriptscriptstyle}
\newcommand{\p}{\makebox[0pt][l]{/}p}

\begin{document}

\pagenumbering{roman}

\begin{titlepage}

\begin{center}

\vspace{-4cm}

\hspace*{9cm} {\Large \bf UCL-IPT-00-12}

\vspace{7mm}

Universit\'e catholique de Louvain\\
Facult\'e des Sciences,
D\'epartement de Physique\\
Institut de Physique Th\'eorique

\vspace{3cm}

{\LARGE \bf Propagation et oscillations\\
\vspace{3mm} en th\'eorie des champs}

\vspace{2cm}

Dissertation pr\'esent\'ee par Michael Beuthe\\
 en vue de l'obtention du grade de Docteur en Sciences\\
Promoteur: Professeur Jacques Weyers\\
Louvain-la-Neuve, le 4 septembre 2000

\vspace{2cm}

{\bf Abstract:}
\end{center}
After a review of the problems associated with the conventional treatment of particle
oscillations, an oscillation formula is derived within the framework of quantum field
theory. The oscillating particle is represented by its propagator and the initial
and final states by wave packets. It is obviously relativistic from the start and
moreover applies both to stable (neutrinos) and unstable particles (K and B mesons,
unstable neutrinos). CPLEAR and DAFNE experiments are studied as examples, with
special attention directed to CP violation. The problems resulting from equal
energies/momentum/velocities prescriptions are analyzed and solved. Oscillations of
associated particles are found to be nonexistent. The relativistic generalization
of the Wigner-Weisskopf equation is also derived.

\phantom{blabla}
\vspace*{10cm}

\end{titlepage}%

\newpage

{\large \bf Remerciements}

\vspace{0.5cm}

Je voudrais tout d'abord remercier mon promoteur, M. Jacques Weyers,
pour m'avoir accord\'e sa confiance pendant toutes ces ann\'ees,
ainsi que de m'avoir permis d'aborder des sujets vari\'es dont cette dissertation
n'est qu'un reflet.

J'aimerais exprimer ma reconnaissance envers M. Jean Pestieau,
qui m'a propos\'e plusieurs collaborations, dont l'une est \`a l'origine
de ce travail; cette th\`ese n'existerait pas sans lui.
Je remercie \'egalement MM. Gabriel L\'opez Castro et Ricardo Gonzalez Felipe
qui ont particip\'e \`a ces collaborations. Non seulement ils m'ont
beaucoup appris mais il a \'et\'e un vrai plaisir de travailler avec eux.

Merci aussi aux lecteurs de cette th\`ese, qui ont accept\'e de faire partie
du jury, MM. les professeurs Luis Alvarez-Gaum\'e, Jean-Pierre Antoine,
Jean-Marie Fr\`ere et Jean-Marc G\'erard. Je remercie ce dernier autant pour
les nombreuses discussions sur la relativit\'e g\'en\'erale que sur la
ph\'enom\'enologie des particules \'el\'ementaires.

Merci enfin \`a mon jury officieux mais impitoyable, Tilio Rivoldini et bien s\^ur Jeanne De Jaegher.

\frontmatter
\tableofcontents

\setlength{\topmargin}{0cm}
\setlength{\textheight}{21cm}

\mainmatter

\addcontentsline{toc}{chapter}{Introduction}
\pagestyle{myheadings}\markboth{INTRODUCTION}{INTRODUCTION}

\chapter*{Introduction}

Bien que la th\'eorie des champs serve principalement au calcul de processus
physiques microscopiques, elle permet aussi l'\'etude de ph\'enom\`enes
macroscopiques.
Par macroscopique j'entends une distance ou un temps tels que leur inverse
est beaucoup plus grand en unit\'es naturelles ($\hbar = 1, \, c=1$) que
l'\'energie typique impliqu\'ee dans le processus.
Le ph\'enom\`ene le plus fondamental de ce type est la propagation
d'une particule entre deux points de l'espace-temps, s\'epar\'es par une
distance et/ou un temps macroscopiques.

En physique classique, la probabilit\'e de d\'etecter une particule
stable est cons\-tante au cours du temps, par d\'efinition m\^eme de la
stabilit\'e,
tandis que la pro\-ba\-bi\-li\-t\'e de d\'etecter une particule instable d\'ecro\^{\i}t
au cours du temps comme une exponentielle: \mbox{${\cal P}(t) = e^{-t/\tau}$},
o\`u $\tau$ est le temps de vie moyen.
Ce comportement d\'ecouvert par Rutherford est d\'eriv\'e en physique classique
bien que la d\'esint\'egration soit un ph\'enom\`ene quantique.
En physique quantique, cette image fonctionnant merveilleusement
bien du point de vue exp\'erimental est modifi\'ee sous deux aspects.

En premier lieu, la brique de base de la physique quantique n'est plus la
pro\-ba\-bi\-li\-t\'e de d\'etection mais l'amplitude de probabilit\'e, dont la norme
au carr\'e est en correspondance avec la probabilit\'e classique de d\'etection.
Cette amplitude suit une \'evolution d\'eterministe. La m\'ecanique quantique
nous dit que l'amplitude de d\'etection d'une particule stable de masse $m$
oscille dans le rep\`ere au repos de la particule comme une exponentielle:
\mbox{${\cal A}(t) \sim \exp (-imt)$}. La probabilit\'e de d\'etection d'une
particule stable libre reste donc constante au cours du temps. 
La r\'ealit\'e de ce concept quantique se manifeste le plus clairement \`a travers
l'interf\'erence obser\-vable entre plusieurs amplitudes, dont des exemples
v\'erifi\'es exp\'erimentalement sont l'~exp\'e\-rience \`a deux trous pour des
\'electrons,
l'effet Aharonov-Bohm et les oscillations dans le syst\`eme $K^0\overline{K^0}$.

En deuxi\`eme lieu, les principes fondamentaux de la th\'eorie quantique interdisent
une forme exactement exponentielle pour la loi de d\'esint\'egration.
La raison en est que les produits de d\'esint\'egration peuvent se recombiner
pour reformer l'\'etat initial \cite{fonda}. On peut le voir en divisant
l'intervalle de temps $t$ en deux sous-intervalles $t_1$ et $t_2$.
L'amplitude de probabilit\'e doit satisfaire \`a l'\'equation
$$
  {\cal A}(t_1+t_2) = {\cal A}(t_1) \, {\cal A}(t_2)
  + <\!\Psi_{inst} \,| e^{-iHt_2} |\,\Psi_{des}(t_1) \!> \, ,
$$
o\`u $\Psi_{inst}$ symbolise l'\'etat instable et $\Psi_{des}(t_1)$
repr\'esente l'\'etat des produits de d\'esint\'e\-gration au temps $t_1$.
En ne gardant que le premier terme du membre de droite de cette \'equation,
on obtient la formule classique de d\'esint\'egration exponentielle.
Le second terme, purement quantique, modi\-fie ce comportement:
les produits de d\'esint\'egration peuvent retourner \`a
l'\'etat initial\footnote{ D'autres d\'emonstrations de l'existence de corrections non exponentielles
existent, tant en m\'ecanique quantique \cite{fonda,khalfin,sakurai}, qu'en
th\'eorie des champs \cite{schwinger,jacob,brown}.}.
Bien que l'existence th\'eorique de ces corrections \`a la loi de Rutherford
soit connue depuis longtemps, elle n'a pas
pu \^etre test\'ee exp\'erimentale\-ment jusqu'\`a pr\'esent \cite{greenland}.
L'ordre de grandeur de ces corrections peut \^etre estim\'e en m\'ecanique
quantique, mais il vaut mieux recourir tout de suite \`a la th\'eorie des
champs, puisqu'elle constitue le seul cadre th\'eorique coh\'erent pour la
description des particules instables. Dans ce formalisme, les contributions
exponentielles sont en correspondance avec les p\^oles complexes du propagateur
de la particule \cite{brown,peierls}, et l'amplitude de propagation
macroscopique oscille dans le rep\`ere au repos de la particule comme une
exponentielle: \mbox{${\cal A}(t) \sim \exp (-imt - \Gamma t/2)$}.
La partie imaginaire de l'exposant s'interpr\`ete comme le produit de la masse
de la particule et du temps tandis que la partie r\'eelle de l'exposant s'interpr\`ete
comme le produit de l'inverse du temps de vie moyen par le temps.
Les contributions non exponentielles proviennent de l'existence de seuils de
production d'\'etats \`a plusieurs particules ainsi que par des facteurs de
seuil figurant dans les amplitudes de production et de d\'etection de la
particule se propageant \cite{schwinger,jacob}. On peut v\'erifier dans ce
cadre que ces corrections sont n\'egligeables pour toute propagation
macroscopique. Nous y reviendrons.

Dans ce travail, nous allons nous int\'eresser principalement \`a
{\it l'oscillation de particules}.
Si les particules en question sont quasiment d\'eg\'en\'er\'ees en masse
et m\'elang\'ees par une interaction, des effets d'interf\'erence
appara\^{\i}tront sous la forme d'oscillations spatiales de la
probabilit\'e de d\'etection des particules. Ces effets d\'ependent
directement de la diff\'erence de masse entre les \'etats se propageant. Cette
diff\'erence doit \^etre tr\`es petite pour qu'une oscillation spatiale
macroscopique soit observable. Il faut donc v\'erifier que ce qui a pu \^etre
n\'eglig\'e \`a l'\'echelle de la masse dans l'analyse de la propagation d'une
seule particule peut encore \^etre n\'eglig\'e \`a l'\'echelle de la
diff\'erence de masse entre des \'etats quasiment d\'eg\'en\'er\'es. Dans le
cas du syst\`eme $K^0\overline{K^0}$, la diff\'erence de masse causant
l'oscillation est du m\^eme ordre que les corrections habituellement
n\'eglig\'ees en $\Gamma/m$:
$\Delta m/m \approx \Gamma_{\SS S}/m \approx 10^{-14}$!
D'autres probl\`emes se posent \`a propos de l'interface classique/quantique.
En effet, bien que les oscillations dans la probabilit\'e de d\'etection soient
de nature purement quantique, leur observation n'est possible que si les
particules sont bien localis\'ees dans l'espace, comme des particules
classiques. Nous d\'evelopperons ces sujets ult\'erieurement.

Pourquoi les oscillations de particules suscitent-elles tant d'attention?
Deux motivations principales sous-tendent cet int\'er\^et:
d'une part, l'\'etude de la violation CP dans les syst\`emes oscillants
$K^0\overline{K^0}$ et $B^0\overline{B^0}$, et d'autre part
la mesure des masses des neutrinos au moyen de leurs oscillations.

Bien que la pr\'ediction \cite{pais} et l'observation \cite{lande}
des oscillations des kaons datent de plus de quarante ans,
le ph\'enom\`ene voit son int\'er\^et sans cesse renouvel\'e.
La premi\`ere raison est que le syst\`eme des kaons neutres fut
jusqu'\`a l'ann\'ee pass\'ee le seul endroit o\`u une violation CP
se manifestait.
L'observation \cite{christenson} de cette vio\-la\-tion dans les oscillations
des kaons, dite {\it violation CP indirecte}, est maintenant doubl\'ee par
l'observation \cite{burkhardt} d'une violation CP dans leurs
d\'esint\'egrations, dite {\it violation CP directe}.
Si la violation indirecte est vue comme une perturbation d'ordre
$\epsilon$ d'un syst\`eme $K^0\overline{K^0}$ respectant CP,
les exp\'eriences montrent que $\epsilon \sim 10^{-3}$ et que la
violation CP directe est de l'ordre de $\epsilon^2$. Il est donc important de
disposer d'une des\-cription coh\'erente du syst\`eme \`a l'ordre
${\cal O}(\epsilon^2)$. Ce n'est pourtant pas le cas du formalisme
couramment utilis\'e dans la litt\'erature,
qui est l'approximation de Wigner-Weisskopf \cite{wigner}.
Dans cette approche, l'\'evolution temporelle des \'etats instables
ob\'eit \`a une \'equation effective de type Schr\"odinger
dont l'hamiltonien non hermitien permet les d\'esint\'egrations.
Par cons\'equent, les matrices de diagonalisation de l'hamiltonien
sont en g\'en\'eral non unitaires, les \'etats correspondants ne sont
pas orthogonaux et leur normalisation ne peut se faire sans
ambigu\"{\i}t\'e. Ce probl\`eme est souvent n\'eglig\'e, car, dans la
limite o\`u la sym\'etrie CP est respect\'ee, la matrice de
diagonalisation est unitaire\footnote{Les sym\'etries
CPT et CP imposent toutes les deux l'\'egalit\'e des \'el\'ements diagonaux
mais la sym\'etrie CP impose en plus l'\'egalit\'e des \'el\'ements non
diagonaux de l'hamiltonien \cite{lipshutz};
ces contraintes permettent une diagonalisation d'une matrice
non hermitienne par une matrice unitaire. En toute g\'en\'eralit\'e, la
condition n\'ecessaire et suffisante pour diagonaliser une matrice par
une transformation unitaire est que la matrice soit normale, c'est-\`a-dire
qu'elle commute avec son hermitienne conjugu\'ee.}.
Il en r\'esulte que la non-orthogonalit\'e des \'etats propres
de masse est de l'ordre de $\epsilon$ et que l'ambigu\"{\i}t\'e de
normalisation est encore plus petite, de l'ordre de $\epsilon^2$.
C'est pourquoi elle est g\'en\'eralement n\'eglig\'ee pour la description
de la violation CP indirecte. On a vu ci-dessus qu'il ne peut plus en
\^etre question dans l'\'etude de la violation CP directe qui est du
m\^eme ordre de grandeur.

Une seconde raison pour s'int\'eresser aux oscillations de kaons est la
transposition imm\'ediate du formalisme appropri\'e aux kaons \`a
d'autres syst\`emes m\'eson/anti-m\'eson. Le plus prometteur est
le $B^0\overline{B^0}$, dont les oscillations ne sont observ\'ees que
depuis 1987 \cite{albrecht1} tandis qu'une premi\`ere mesure de la
violation CP dans ce syst\`eme a \'et\'e effectu\'ee r\'ecemment
\cite{affolder}. Bien que le
formalisme de Wigner-Weisskopf n'ait pas \'et\'e jusqu'\`a pr\'esent
en contradiction avec les donn\'ees exp\'erimentales, il n'est pas
certain que cette description poss\`ede le m\^eme degr\'e de validit\'e
pour mod\'eliser la violation CP du syst\`eme $B^0\overline{B^0}$, o\`u les
corrections relativistes sont beaucoup plus importantes.

Une derni\`ere raison d'\'etudier ces syst\`emes de m\'eson/antim\'eson
est l'\'etude exp\'erimentale
de corr\'elations quantiques macroscopiques (effet EPR) de deux kaons
ou m\'esons {\it B} oscillant simultan\'ement \cite{lipkin1}.

L'autre motivation principale de s'int\'eresser aux oscillations
spatio-tem\-po\-rel\-les de la probabilit\'e de d\'etection concerne les
neutrinos.
Historiquement, les premi\`eres donn\'ees exp\'erimentales sugg\'erant des
neutrinos massifs proviennent de l'anomalie du flux des neutrinos solaires,
qui est bien inf\'erieur aux pr\'edictions des mod\`eles solaires les plus
sophistiqu\'es \cite{davis,bahcall,bilenky}.
En 1968, Bruno Pontecorvo attribue le d\'eficit en neutrinos
\'electroniques \`a des oscillations entre neutrinos de diff\'erents nombres
leptoniques \cite{pontecorvo}.
Plus tard, des solutions recourant \`a des neutrinos instables
sont aussi propos\'ees \cite{petcov}.
Plus r\'ecemment, on assiste \`a un retour en force du concept de neutrinos
massifs en raison de l'ob\-ser\-va\-tion d'une anomalie angulaire dans le
flux des neutrinos atmosph\'eriques \cite{fukuda}.
Les deux mod\`eles les plus
simples r\'esolvent ce probl\`eme en attribuant une masse aux neutrinos.
Le premier mod\`ele explique l'anomalie par une oscillation de neutrinos
\cite{learned} tandis que le deuxi\`eme recourt \`a un neutrino instable
\cite{barger}.
Une derni\`ere raison pour soup\c{c}onner une masse aux
neutrinos provient de l'exp\'erience LSND \cite{lsnd}.
Quelle que soit l'explication de ces diff\'erentes anomalies, il para\^{\i}t
difficile d'\'eviter le
recours \`a des neutrinos massifs, ce qui n'est de toute fa\c{c}on
interdit par aucun principe fondamental. Les anomalies cit\'ees
impliquent des masses pour les neutrinos extr\^emement faibles par
rapport aux \'energies typiques des processus concern\'es et donc
des oscillations macroscopiques sur de tr\`es grandes distances.
D\`es lors, il s'av\`ere crucial de disposer d'une description
coh\'erente de la propagation de neutrinos massifs, puisque les
effets d'oscillations r\'esultent de la compensation presque totale
d'\'energies-impulsions des diff\'erents \'etats oscillants,
en raison de la petitesse des diff\'erences de masses par rapport \`a
l'\'energie de la par\-ti\-cule.
Notons enfin que la masse des neutrinos rend possible
une violation CP dans le secteur leptonique.
Etant donn\'e que l'explication de l'anomalie des neutrinos
atmosph\'eriques par une d\'esint\'egration reste toujours possible,
il me para\^{\i}t int\'eressant d'\'etablir un formalisme o\`u les
deux ph\'enom\`enes d'oscillation et de d\'esint\'egration
coexistent. 
Cet objectif nous ram\`ene au syst\`eme $K^0\overline{K^0}$, o\`u la
d\'esint\'egration et l'oscillation vont de pair. Dans ce travail,
les ana\-lyses des propagations des kaons et des neutrinos seront
regroup\'ees dans un seul formalisme d\'eriv\'e \`a partir des
principes de la th\'eorie des champs.

Dans le premier chapitre, le concept d'oscillation est d\'efini en
m\'ecanique quantique et le traitement traditionnel de ces oscillations
est pr\'esent\'e. Les difficult\'es soulev\'ees par cette approche
sont pass\'ees en revue, avec pour conclusion que seul un calcul en th\'eorie
des champs peut clarifier ces probl\`emes.
Dans le deuxi\`eme chapitre, un mod\`ele simplifi\'e de propagation
d'une particule est \'etabli. Il nous conduit \`a l'\'etude de la
transform\'ee de Fourier du propagateur, \`a l'obtention de la traditionnelle
loi d'\'evolution en exponentielle ainsi que des corrections non
exponentielles n\'egligeables pour des propagations macroscopiques.
Le troisi\`eme chapitre est consacr\'e \`a l'application de ce formalisme \`a la
propagation de particules en m\'elange: apr\`es diagonalisation de
leur propagateur, la probabilit\'e de propagation et d'oscillation de
ces particules est calcul\'ee dans le mod\`ele simplifi\'e du deuxi\`eme
chapitre.
On constate alors que ce mod\`ele ne suffit pas \`a r\'epondre \`a toutes
les interrogations initiales, bien qu'il \'eclaircisse d\'ej\`a une s\'erie
de points d\'elicats.
Dans le chapitre sui\-vant, un mod\`ele de propagation plus r\'ealiste est
d\'evelopp\'e, toujours dans le cadre de la th\'eorie des champs.
Les \'etats entrants et sortants y sont mod\'elis\'es par des paquets d'ondes.
Le calcul de la probabilit\'e de d\'etection d'une particule instable fournit
la r\'eponse attendue. Les corrections non exponentielles sont aussi
recalcul\'ees et restent n\'egligeables. Il ne reste plus qu'\`a appliquer ce
mod\`ele aux oscillations de particules, ce qui est l'objet du cinqui\`eme chapitre.
Avant de d\'eriver la probabilit\'e d'oscillation, on v\'erifie que les
corrections non exponentielles propres aux m\'elanges de particules sont
n\'egligeables. Une probabilit\'e d'oscillation d\'ependant uniquement de
la distance est ensuite obtenue et analys\'ee terme par terme. Les diff\'erents
facteurs y figurant repr\'esentent la d\'ecroissance exponentielle due \`a
l'instabilit\'e de la particule, l'oscillation due au m\'elange, la
disparition de l'oscillation \`a grande distance due \`a la d\'ecoh\'erence
ainsi que les conditions d'observabilit\'e des oscillations dues aux conditions
de production et de d\'etection.
Le chapitre se termine par une comparaison de nos r\'esultats avec les
diff\'erentes approches existantes.
Au sixi\`eme chapitre, on \'etablit des prescriptions de calcul simplifiant
l'application de nos formules \`a des cas concrets. Ensuite, certaines quantit\'es
observables des exp\'eriences CPLEAR et DA$\Phi$NE sont calcul\'ees dans notre
mod\`ele. Enfin, le travail se termine par une ana\-lyse ph\'enom\'enologique
rigoureuse de la violation CP dans le syst\`eme m\'eson/antim\'eson.

La nouveaut\'e de ce travaille r\'eside en l'application de la th\'eorie des champs
\`a la description du m\'elange de particules instables, dans le but de traiter de
fa\c{c}on unifi\'ee les cas des m\'esons K et B et des neutrinos, avec l'espoir d'en
tirer des le\c{c}ons int\'eressantes pour chaque cas particulier.
Comme nouveau r\'esultat, nous obtenons une formule relativiste et directement
applicable aux exp\'eriences, puisque ne d\'ependant que de la distance de propagation
et de l'observation des \'etats initiaux et finaux. Le recours \`a des concepts
classiques ext\'erieurs au formalisme est \'evit\'e. Le traitement de la particule
oscillante comme \'etat interm\'ediaire permet de montrer l'inanit\'e des questions
suivantes, infiniment ressass\'ees dans de nombreux articles: quelle est
l'\'energie-impulsion de la parti\-cule oscillante?, comment d\'efinir des \'etats
physiques non orthogonaux?, comment d\'efinir un \'etat propre de saveur?
(notre r\'eponse \`a cette derni\`ere question n'est pas originale).
Plusieurs controverses concernant la longueur d'oscillation et l'oscillation \'eventuelle
des particules associ\'ees \`a la production de la particule oscillante trouvent enfin une
solution claire dans notre formalisme, qui fournit sans ambigu\"{\i}t\'e la
g\'en\'eralisation relativiste de la formule de Wigner-Weisskopf pour un m\'elange de
particules.

Les conditions d'observabilit\'e des oscillations (n\'ecessit\'e d'une incertitude sur
l'impulsion de la particule oscillante, existence d'une longueur de coh\'erence) sont
\'etudi\'ees pour la premi\`ere fois dans le cas des particules instables.
Comme sous-produit de notre analyse, nous avons obtenu une estimation, nouvelle en
th\'eorie des champs, des corrections non exponentielles \`a la propagation d'un m\'elange
de parti\-cules.
Comme nouvelles applications, nous avons trait\'e rigoureusement la ph\'eno\-m\'e\-no\-logie de
la violation CP \`a partir de nos formules. Le cas d'une oscillation
double (\mbox{$\phi \!\rightarrow\! K\overline{K}$}) est aussi \'etudi\'e pour la premi\`ere fois en
th\'eorie des champs.

\pagestyle{headings}

\chapter{Oscillations en m\'ecanique quantique}
\label{oscillationsMQ}

\section{Oscillations temporelles en m\'ecanique\\ quantique}
\label{oscillationtempMQ}

L'\'etude de l'oscillation de la probabilit\'e de d\'etection pour un
m\'elange de particules se fait habituellement dans le cadre de la m\'ecanique
quantique.
Nous allons travailler avec des particules scalaires instables. Le spin des
neutrinos est n\'eglig\'e dans ce traitement. On verra \`a la section
\ref{oscifermion} que le spin peut \^etre n\'eglig\'e si les masses des
particules m\'elang\'ees sont petites par rapport \`a leurs impulsions. 
Avant de d\'efinir la notion d'oscillation,
il faut discuter du m\'elange de particules, qui est
une des conditions sine qua non d'apparition des oscillations.

\subsection{Le m\'elange de particules}

Le concept de m\'elange d'\'etats rel\`eve des principes fondamentaux de la
th\'eorie quantique.
Le principe de la mesure implique que la mesure d'une observable
projette l'\'etat quantique initial sur un vecteur propre de l'op\'erateur
associ\'e \`a cet observable.
Il est donc n\'ecessaire de choisir une base d'\'etats physiques constitu\'ee de
vecteurs propres de l'op\'erateur en question.
Par cons\'equent, le choix de mesurer une caract\'eristique particuli\`ere
d'un \'etat quantique d\'etermine automatiquement le choix de la base des
\'etats physiques. Les diff\'erentes bases associ\'ees aux diff\'erents
op\'erateurs sont reli\'ees par des transformations lin\'eaires; un \'etat
d\'evelopp\'e dans une base donn\'ee peut \^etre vu comme un {\it m\'elange}
des \'etats propres de la base en question.
Un exemple bien connu de m\'elange est donn\'e par l'exp\'erience de
Stern-Gerlach: si l'\'etat initial a son spin up selon l'axe $x$, la mesure de
son spin selon l'axe $y$ nous oblige \`a consid\'erer cet \'etat initial dans la base
de l'op\'erateur mesurant le spin selon l'axe $y$, c'est-\`a-dire que l'\'etat
initial est d\'ecompos\'e en un m\'elange d'un \'etat ayant son spin up selon
$y$ et d'un \'etat ayant son spin down selon $y$.
Un autre exemple est celui du syst\`eme $K^0\overline{K^0}$ o\`u l'on n\'eglige la
violation CP: un $K^0$ initial
se d\'esint\'egrant en \'etats propres sous CP (deux ou trois pions),
l'identification des produits de d\'esint\'egration \'equivaut \`a une mesure
de la CP-parit\'e. Le $K^0$ initial doit donc \^etre d\'ecompos\'e en un
m\'elange d'\'etats de CP-parit\'e positive et d'\'etats de CP-parit\'e
n\'egative.

Le concept d'{\it oscillation} pr\'esuppose, d'une part, un m\'elange de
particules de masses dif\-f\'e\-ren\-tes:
il doit \^etre en principe possible de mesurer les
caract\'eristiques des particules dans deux bases diff\'erentes, dont une
est constitu\'ee d'\'etats propres de masse.
D'autre part, la pr\'esence d'une
oscillation dans la probabilit\'e de d\'etection interdit qu'une
mesure soit effectu\'ee dans la base d'\'etats propres de masse.
Pourquoi cette base intervient-elle? La raison en est qu'elle se distingue de toutes
les autres bases par le fait qu'elle est la base d'\'etats propres de
l'op\'erateur d'\'evolution dans le temps $\hat H$. Elle intervient donc comme
base interm\'ediaire dans tout processus o\`u une \'evolution temporelle est
pr\'esente, sans qu'il soit n\'ecessaire d'effectuer une mesure de la masse.
Les diff\'erents \'etats propres de masse \'evoluent diff\'eremment,
mais tant qu'une mesure ne les distingue pas, le principe quantique de
superposition lin\'eaire dicte que l'amplitude totale
du processus est la somme des amplitudes des processus indistinguables.
Dans ces circonstances, une oscillation dans la probabilit\'e de d\'etection de
la particule appara\^{\i}tra en raison de l'interf\'erence entre les
amplitudes partielles correspondant aux diff\'erents \'etats propres de masse.
En conclusion, la probabilit\'e de d\'etection d'une particule se propageant
pr\'esentera des {\it oscillations} spatio-temporelles {\it si} l'\'etat
initial de la particule et son \'etat final ne sont pas des \'etats propres de
masse {\it et} s'il est impossible de d\'eterminer quel \'etat propre de masse
a contribu\'e comme \'etat interm\'ediaire (ce qui implique que les
diff\'erences de masse doivent \^etre petites).

Pr\'ecisons les deux bases pertinentes au ph\'enom\`ene d'oscillation.
L'hamiltonien doit pouvoir \^etre subdivis\'e en un hamiltonien de propagation
$H_{propag}$ d\'ecrivant la propagation libre des particules et un hamiltonien
d'interaction $H_{int}$ d\'ecrivant les interactions produisant ces particules
et \'eventuellement leur d\'etection.
Ces deux parties de l'hamiltonien pourront \^etre distingu\'ees s'il existe
un nombre quantique, appel\'e {\it saveur}, pr\'eserv\'e par $H_{int}$ mais
viol\'e par $H_{propag}$.
La base appropri\'ee \`a la production de
particules est constitu\'ee de vecteurs propres simultan\'es de $H_{int}$
et de l'op\'erateur associ\'e au nombre quantique de saveur. Elle est
appel\'ee {\it base de saveur}.
La base appropri\'ee \`a la description de la propagation des particules
est celle dans laquelle $H_{propag}$ est diagonal. Elle est appel\'ee
{\it base de propagation} ou {\it base de masse}.
Elle ne co\"{\i}ncide pas avec la base de saveur puisque l'op\'erateur de
saveur ne commute pas avec $H_{propag}$.

Le cas le plus simple est celui des neutrinos pris comme particules stables,
o\`u l'hamiltonien $H_{propag}$ contient la matrice de masse tandis que
l'hamiltonien $H_{int}$ contient les interactions faibles des neutrinos avec les
bosons {\it W} et {\it Z}. La saveur est dans ce cas le nombre leptonique qui est
\'electronique, muonique ou tauique et \'eventuellement st\'erile.
Il est conserv\'e par l'interaction faible, mais est viol\'e par la matrice de
masse qui m\'elange les neutrinos de saveur d\'efinie.

Un cas plus compliqu\'e est celui des particules
instables, o\`u une partie des interactions est incorpor\'ee dans $H_{propag}$
pour d\'ecrire la d\'esint\'egration de la particule.
Consid\'erons par exemple les kaons neutres, qui peuvent \^etre produits par
les interactions fortes mais ne peuvent se d\'esint\'egrer que par interaction
faible. Dans une premi\`ere \'etape, l'hamiltonien total\footnote{Il s'agit en
fait d'un hamiltonien effectif qui peut \^etre construit dans le cadre de la
th\'eorie chirale perturbative. On d\'efinira plus pr\'ecis\'ement ce que
l'on entend par interactions fortes et faibles dans ce contexte lorsque l'on
abordera les m\'elanges en th\'eorie des champs au chapitre
\ref{melangesprop}.} contenant toutes les interactions des kaons
est exprim\'e comme la somme d'un hamiltonien $H_F$ contenant les interactions
fortes et d'un hamiltonien $H_f$ contenant les interactions faibles.
L'hamiltonien $H_f$ peut \^etre vu comme une perturbation de l'hamiltonien
$H_F$. La base de saveur est d\'efinie comme l'ensemble des vecteurs propres de
$H_F$. La saveur est ici le nombre quantique d'\'etranget\'e, pr\'eserv\'e par
les interactions fortes mais viol\'e par les interactions faibles. Il vaut
$S=+1$ pour le $K^0$ et $S=-1$ pour le $\overline{K^0}$. A l'ordre z\'ero de
la perturbation de $H_F$ par $H_f$, les kaons neutres ont la m\^eme masse,
sont stables et n'interagissent pas entre eux.
Dans une seconde \'etape, avec le but de d\'ecrire la propagation et la
d\'esint\'egration de ces particules, la m\'ethode de
Wigner-Weisskopf \cite{wigner} est utilis\'ee pour construire un hamiltonien effectif non
hermitien $H_{propag}$ restreint aux interactions entre les kaons neutres.
L'hamiltonien $H_{propag}$ contient d'une part la matrice de masse de ces
particules dans la limite o\`u $H_f$ est n\'eglig\'e par rapport \`a $H_F$ et
d'autre part les interactions faibles de $H_f$ qui permettent les transitions
entre kaons neutres dans la base de saveur.
La non hermiticit\'e de $H_{propag}$ exprime le fait que le syst\`eme des kaons
n'\'evolue pas en isolation en raison des d\'esint\'egrations possibles
des kaons vers les autres \'etats propres de $H_F$.
L'hamiltonien $H_{int}$ contient les interactions fortes des kaons avec les
autres \'etats propres de $H_F$.
La construction de $H_{propag}$ et de $H_{int}$ peut se d\'ecrire
sch\'ematiquement comme
\begin{eqnarray*}
  H_{total} &=& H_F + H_f
  \\
  &=& H_{int} + H_{masse} + H_f
  \quad \stackrel{WW}{\longrightarrow} \quad
  H_{total,\, effectif} \; = \; H_{int} + H_{propag} \, .
\end{eqnarray*}

La relation entre la base de propagation et la base de saveur peut s'\'ecrire
\begin{equation}
  |\nu_\alpha (0) \!> = \sum_{j} \, V_{\alpha j} \, |\nu_j (0) \!> \, .
  \label{chgtbase}
\end{equation}
o\`u $V$ est la matrice diagonalisant $H_{propag}$ si cet hamiltonien a \'et\'e
\'ecrit au d\'epart dans la base de saveur.
Les $|\nu_j (0) \!>$ sont les \'etats propres de propagation au temps $t=0$, de
masse et largeur d\'efinies, appartenant \`a la base de propagation
tandis que les $|\nu_\alpha (0) \!>$ sont les
\'etats propres appartenant \`a la base de saveur.

La forme de la matrice $V$ d\'epend du type d'hamiltonien $H_{propag}$
d\'ecrivant la propagation de la particule.
Dans le premier cas, o\`u les particules
oscillantes sont stables et leur lagrangien est connu, la matrice $V$ provient
de la diagonalisation de la matrice de masse du lagrangien et est unitaire
(une discussion plus \'elabor\'ee de cette diagonalisation pour les fermions
figure \`a la section \ref{oscifermion}). Dans le deuxi\`eme cas, o\`u les
particules oscillantes sont instables, nous avons vu ci-dessus que
l'hamiltonien $H_{propag}$ est n\'ecessairement non hermitien pour
engendrer une \'evolution non unitaire, c'est-\`a-dire pour permettre la
d\'esint\'egration des particules.
La matrice $V$ diagonalisant cet hamiltonien n'est en
g\'en\'eral pas unitaire, bien qu'elle puisse l'\^etre si certaines sym\'etries
imposent suffisamment de contraintes sur l'hamiltonien, de sorte que la matrice
correspondante soit normale (c'est-\`a-dire qu'elle commute avec son hermitienne
conjugu\'ee).

Il existe encore une autre origine possible de la matrice $V$,
qui consiste \`a tenir compte dans l'hamiltonien des interactions de la
particule avec le milieu dans lequel elle se propage (effet MSW \cite{msw}).
On obtient dans ce cas un hamiltonien effectif satisfaisant \`a une \'equation
de type Schr\"odinger. Nous ne discuterons pas de ce m\'ecanisme qui est tout
\`a fait diff\'erent, puisque la particule oscillante ne se propage pas
librement.

Pour calculer une amplitude, nous devons pouvoir calculer le produit scalaire
de deux vecteurs de base.
Qu'en est-il des relations d'orthogonalit\'e dans ces deux bases?
Dans la base de saveur, les relations d'orthogonalit\'e suivantes sont
satisfaites:
\begin{equation}
  <\! \nu_\beta(0) \, | \, \nu_\alpha(0) \!> = \delta_{\alpha\beta} \, ,
  \label{orthogsaveur}
\end{equation}
puisque $H_{int}$ est hermitien et qu'il pr\'eserve la saveur.
Dans la base de propagation, $H_{propag}$ n'est hermitien que si les particules
sont stables. Le produit scalaire entre les \'el\'ements de cette base doit
donc \^etre d\'efini en g\'en\'eral \`a partir de la base de saveur:
\begin{equation}
  <\! \nu_k(0) \, | \, \nu_j(0) \!>
  = \sum_\gamma \,  V^{-1}_{j\gamma} \, V^{-1 \, \dagger}_{\gamma k} \, .
   \label{orthogmasse}
\end{equation}
Ces produits scalaires deviennent orthogonaux si la matrice $V$ est unitaire, ce
qui est le cas pour des particules stables. Comme on l'a dit plus haut, la
matrice $V$ peut aussi \^etre unitaire pour des particules instables si une
sym\'etrie impose suffisamment de contraintes sur $H_{propag}$.
Dans le syst\`eme $K^0\overline{K^0}$, la sym\'etrie CP suffit pour que $V$
soit unitaire. Comme la violation de cette sym\'etrie est de l'ordre du
milli\`eme, le membre de droite de l'\'equation (\ref{orthogmasse}) est aussi
de l'ordre du milli\`eme pour les kaons neutres.

Notons cependant que les relations d'orthogonalit\'e dans la base de saveur
ne sont valables que si tous les \'etats propres de masse sont
\'energ\'etiquement permis \cite{bilenkygiunti}.
Par exemple, supposons qu'il existe quatre neutrinos stables tels que $m_i=0$
pour $i=1,2,3$ et $m_4\!>\! 1$ GeV. Pour des neutrinos d'\'energie inf\'erieure
\`a 1 GeV, on aura
$$
  <\! \nu_\beta(0) \, | \, \nu_\alpha(0) \!> =
  \delta_{\alpha\beta} - V_{\alpha4} \, V^*_{\beta4} \, .
$$ 
Ce probl\`eme est une premi\`ere indication des difficult\'es li\'ees \`a la
d\'efinition de la base de saveur. Par exemple, est-il possible d'associer
une particule observable \`a chaque \'etat de saveur?
Remettons \`a plus tard le cas des particules instables: elles ne
corres\-pondent en toute rigueur \`a aucun \'etat de l'espace de Hilbert des
\'etats physiques mais apparaissent en th\'eorie des champs comme des p\^oles
de la matrice $S$.
Par contre, \`a chaque particule stable est associ\'ee en m\'ecanique quantique
relativiste une repr\'esentation du groupe de Poincar\'e caract\'eris\'ee par
une masse et un spin d\'efinis.
Dans ce cadre, les \'etats de saveur ne correspondent pas \`a des particules,
\`a  moins d'\'etre d\'eg\'en\'er\'es.
Si l'on examine le probl\`eme en th\'eorie des champs, on remarque que
les op\'erateurs de cr\'eation et d'annihilation d'\'etats propres de masse sont
bien d\'efinis. Par contre, le changement de base (\ref{chgtbase}) appliqu\'e
\`a ces op\'erateurs ne fournit pas des op\'erateurs de cr\'eation et
d'annihilation d'\'etats de saveur satisfaisant aux relations de commutation
canoniques \cite{giunti1}. On pourrait se demander quelle r\'ealit\'e attribuer
aux \'etats de saveur apparaissant dans l'\'equation (\ref{chgtbase}).
Il se fait que dans un processus o\`u des oscillations
sont pr\'esentes dans la propagation, les particules oscillantes sont
uniquement identifi\'ees \`a la production et \`a la d\'etection dans une
superposition d'\'etats propres de masse formellement \'equivalente \`a un
\'etat de saveur. Dans ce sens-l\`a, les \'etats de saveur sont observ\'es
dans ce type d'exp\'erience tandis que les \'etats propres de masse ne le sont
pas.

\subsection{Oscillations de la probabilit\'e de d\'etection}

Partant d'un \'etat propre d'interaction de saveur $\alpha$, l'amplitude de
d\'etecter au temps $t$ un \'etat de saveur $\beta$ est donn\'ee selon les
r\`egles de la m\'ecanique quantique par
\begin{eqnarray*}
  {\cal A}(\alpha \to \beta,t)
  &=& <\! \nu_\beta(0) \, | \exp(-i \, H_{propag} \, t) | \, \nu_\alpha(0) \!>
  \\
  &=& <\! \nu_\beta(0) \, |
      \sum_{j} \, V_{\alpha j} \,
      \exp \left( -i \, E_j t - \Gamma_j t /2 \right)
      | \, \nu_j(0) \!>
  \\
  &=& <\! \nu_\beta(0) \, |
      \sum_{j,\gamma} \, V_{\alpha j} \,    
      \exp \left( -i \, E_j t - \Gamma_j t /2 \right)
      V_{j \gamma}^{-1}
      | \, \nu_\gamma(0) \!> \, .
\end{eqnarray*}
Les symboles $E_j$, $m_j$ et $\Gamma_j$ d\'esignent les
\'energie, masse et largeur de l'\'etat propre de masse $| \, \nu_j(t) \!>$.
L'amplitude peut \^etre r\'e\'ecrite comme
\begin{equation}
  {\cal A}(\alpha \to \beta,t) =
  \sum_j \, V_{\alpha j} \,    
  \exp \left( -i \, E_j t - \Gamma_j t /2\right)
  V_{j \beta}^{-1} \, .
  \label{oscMQ}
\end{equation}
On a utilis\'e la relation d'orthogonalit\'e (\ref{orthogsaveur}) dans la base
de saveur, puisque la base de propagation (\ref{orthogmasse}) n'est en
g\'en\'eral pas orthogonale.
Cette non orthogonalit\'e implique que les \'etats \mbox{$| \, \nu_i(0) \!>$}
peuvent \^etre normalis\'es de diff\'erentes fa\c{c}ons.
Cela n'a pas d'importance tant qu'ils n'apparaissent pas dans le r\'esultat
final du calcul de l'amplitude, comme on le voit clairement ci-dessus
o\`u la formule d'oscillation est calcul\'ee sans m\^eme poser la question de
cette normalisation.
Cependant, cette question r\'eappara\^{\i}t lorsque l'on tente
d'\'ecrire des amplitudes impliquant un \'etat propre de masse comme \'etat
initial ou final. Par exemple, l'amplitude
\mbox{${\cal M}(K_L\!\to\!\pi\pi)\equiv$}$<\!\pi\pi\, |H_{total}| \, K_L\!>$
d\'epend de la normalisation choisie pour l'\'etat $| \, K_L\!>$. On verra
au chapitre \ref{applications} la mani\`ere correcte de d\'efinir de telles
amplitudes en th\'eorie des champs.

La formule (\ref{oscMQ}) de l'amplitude d'oscillation est souvent \'ecrite
\cite{nachtmann} en termes des \mbox{$<\! \nu_k(0) \, | \, \nu_j(0) \!>$}:
\begin{equation}
  {\cal A}(\alpha \to \beta,t) =
  \sum_{j,k} \, V_{\alpha j} \,    
  \exp \left( -i \, E_j t - \Gamma_j t /2\right)
  V_{k \beta}^\dagger \,
  <\! \nu_k(0) \, | \, \nu_j(0) \!> \, ,
  \label{amplitrad}
\end{equation}
mais cette expression est \'equivalente \`a la formule (\ref{oscMQ}).
En effet, le remplacement dans l'\'equation ci-dessus des
\mbox{$<\! \nu_k(0) \, | \, \nu_j(0) \!>$} par leurs expressions
(\ref{orthogmasse}) donne
\begin{eqnarray*}
  {\cal A}(\alpha \to \beta,t)
  &=& \sum_{j,k,\gamma} \, V_{\alpha j} \,    
  \exp \left( -i \, E_j t - \Gamma_j t /2\right)
  V_{k \beta}^\dagger \,
   V^{-1}_{j\gamma} \, V^{-1 \, \dagger}_{\gamma k}
  \\
  &=& \sum_j \, V_{\alpha j} \,    
  \exp \left( -i \, E_j t - \Gamma_j t /2\right)
  V_{j \beta}^{-1} \, ,
\end{eqnarray*}
et l'on retrouve la formule (\ref{oscMQ}). Ce n'est pas surprenant, puisqu'\`a
travers l'\'equation (\ref{orthogmasse}), ce sont les relations
(\ref{orthogsaveur}) d'orthogonalit\'e dans la base de saveur qui sont
utilis\'ees.

Le r\'esultat est \'egalement identique \`a celui obtenu par l'utilisation de
deux bases de saveur
\mbox{$| \, \nu_\alpha(0) \!>_{in}$} et
\mbox{${}_{out}\!<\! \nu_\alpha(0) \, |$},
reli\'ees par l'op\'erateur de renversement dans le temps T
\cite{sachs,enz,alvarez}.
Les {\sl ket} appartiennent \`a la base {\it in} tandis que les {\it bra}
appartiennent \`a la base {\it out}. On ne se sert pas des vecteurs hermitiens
conjugu\'es.
De nouvelles bases
\mbox{$| \, \nu_j(0) \!>_{in}$} et
\mbox{${}_{out}\!<\! \nu_j(0) \, |$}
d'\'etats propres de masse sont d\'efinies telles
qu'elles diagonalisent l'hamiltonien $H_{propag}$, c'est-\`a-dire que les nombres
\mbox{${}_{out}\!<\! \nu_j(0) \, |H_{propag}| \, \nu_j(0) \!>_{in}$}
forment une matrice diagonale:
\begin{eqnarray}
  | \, \nu_\alpha(0) \!>_{in}
  &=& \sum_j \, V_{\alpha j} \, | \, \nu_j(0) \!>_{in} \, ,
  \nonumber \\
  {}_{out}\!<\! \nu_\alpha(0) \, |
  &=& \sum_j \, {}_{out}\!<\! \nu_j(0) \, | \, V_{j\alpha}^{-1} \, ,
  \label{doublebase}
\end{eqnarray}
mais les bases d'\'etats propres de masse ne sont pas reli\'ees par
l'op\'erateur T.
Les \'etats propres de masse {\it in} et {\it out} sont orthogonaux:
$$
  {}_{out}\!<\! \nu_j(0) \, | \, \nu_i(0) \!>_{in} = \delta_{i \, j} \, .
$$
La formule d'oscillation obtenue par cette m\'ethode est identique \`a
l'\'equation (\ref{oscMQ}). De nouveau, ce n'est pas \'etonnant puisque les
nouvelles bases {\it in} et {\it out} sont d\'efinies \`a partir de la base
de saveur.
Ce proc\'ed\'e peut sembler artificiel: quelle est la signification de cette
nouvelle base?
Peut-on prendre le produit scalaire entre ces deux bases qui ne sont m\^eme
pas duales sous l'op\'erateur T?
Pourquoi introduire une nouvelle base alors que les \'etats propres de
propagation ne sont jamais directement observ\'es?

Notons comme cas particulier que si $H_{propag}$ est hermitien,
alors $V$ est unitaire et $\Gamma_j=0$.
L'amplitude s'\'ecrit dans ce cas comme
$$
  {\cal A}(\alpha \to \beta,t) = \sum_{j}
   V_{\alpha \, j} \,    
   \exp \left( -i \, E_j t \right)
   V_{\beta \, j}^{*} \, .
$$
Cette expression est couramment utilis\'ee pour les m\'elanges de neutrinos stables. 

Dans le cas g\'en\'eral d'un hamiltonien $H_{propag}$ non hermitien,
la probabilit\'e de
d\'etection est donn\'ee par la norme au carr\'e de l'amplitude d'oscillation
(\ref{oscMQ}) et s'\'ecrit
\begin{eqnarray}
  {\cal P}(\alpha \to \beta \, , \, t )
  &=& \left| \sum_{j} \, V_{\alpha \, j} \,    
        \exp \left( -i \, E_j t - \Gamma_j t /2 \right)
        V_{j \, \beta}^{-1} \right|^2
  \nonumber
  \\
  &=& \sum_{j,k} \, V_{\alpha \, j}   \,  V_{j \, \beta}^{-1}    \,
                      V_{\alpha \, k}^* \, (V^{-1})_{k \, \beta}^* \,
        e^{ -i \Delta E_{jk} \, t} \, e^{ - \Gamma_{jk} t}
  \label{probaMQexpo}
\end{eqnarray}
o\`u $\Gamma_{j \, k} \equiv (\Gamma_j + \Gamma_k)/2$ et
$\Delta E_{jk} \equiv E_j - E_k$.
On peut aussi la r\'e\'ecrire comme
\parbox{\textwidth}{
\begin{eqnarray}
  & & {\cal P}(\alpha \to \beta \, , \, t )
  = \sum_{j} \, e^{- \, \Gamma_j t} \, |V_{\alpha \, j}|^2 \,
                                       |V_{j \, \beta}^{-1}|^2
  \nonumber
  \\
  & & {} + 2 \, \sum_{j < k} \, e^{- \, \Gamma_{jk} t} \,
          {\cal R}\!e \, \left( V_{\alpha \, j}   \,  V_{j \, \beta}^{-1}    \,
                      V_{\alpha \, k}^* \, (V^{-1})_{k \, \beta}^*
               \right) \,
         \cos \left( \Delta E_{jk} \, t \right)
  \nonumber
  \\
  & & {} + 2 \, \sum_{j < k} \, e^{- \, \Gamma_{jk} t}
          {\cal I}\!m \, \left( V_{\alpha \, j}   \,  V_{j \, \beta}^{-1}    \,
                      V_{\alpha \, k}^* \, (V^{-1})_{k \, \beta}^*
               \right) \,
         \sin \left( \Delta E_{jk} \, t \right) \, ,
  \label{probaMQcos}
\end{eqnarray}
}
Cette probabilit\'e de d\'etection est une somme de termes qui d\'ecroissent
tous comme des exponentielles, soit en $e^{-\Gamma_j t}$, soit en
$e^{-\Gamma_{jk} t}$. Certains de ces termes oscillent en
\mbox{$\cos ( \Delta E_{jk} \, t )$} et d'autres en
\mbox{$\sin ( \Delta E_{jk} \, t )$}. C'est \`a ces comportements
oscillatoires que l'on pense lorsque l'on parle d'oscillation de la
probabilit\'e de d\'etection.  

Quant \`a l'oscillation des antiparticules, le th\'eor\`eme CPT appliqu\'e
\`a l'amplitude (\ref{oscMQ}) donne la relation
$$
  {\cal A}(\bar\alpha \to \bar\beta,t) = {\cal A}(\beta \to \alpha,t)  \, ,
$$
avec la cons\'equence que
\mbox{${\cal A}(\bar\alpha \to \bar\alpha,t) = {\cal A}(\alpha \to \alpha,t)$}.
La violation CP dans les oscillations ne peut donc \^etre mesur\'ee que
s'il y a changement de saveur ($\alpha \neq \beta$), puisqu'elle se manifeste
par une diff\'erence entre \mbox{$|{\cal A}(\bar\alpha \to \bar\beta,t)|^2$}
et \mbox{$|{\cal A}(\alpha \to \beta,t)|^2$}. 
Si la matrice de diagonalisation $V$ est unitaire, seul le terme
${\cal I}\!m(V_{\alpha \, j} \, V_{j \, \beta}^{-1} \, V_{\alpha \, k}^* \,
   (V^{-1})_{k \, \beta}^*)$ (invariant de Jarlskog \cite{jarlskog})
viole la sym\'etrie CP dans l'expression de la probabilit\'e (\ref{probaMQcos}).
En effet, le passage de
${\cal P}(\alpha \to \beta,t)$ \`a
${\cal P}(\bar\alpha \to \bar\beta,t)$, qui revient \`a \'echanger $\alpha$
et $\beta$, est \'equivalent \`a remplacer $V$ par $V^*$ si $V$ est unitaire .
Cette substitution change le signe de
${\cal I}\!m(V_{\alpha \, j} \, V_{j \, \beta}^{-1} \, V_{\alpha \, k}^* \,
   (V^{-1})_{k \, \beta}^*)$
mais laisse les autres termes de la probabilit\'e invariants.

Les deux cas dans lesquels la formule (\ref{probaMQexpo}) est utilis\'ee,
- m\'esons {\it K} et {\it B}, neutrinos~-, n\'ecessitent un traitement relativiste,
que n'offre pas cette formule,
puisque le facteur $\Delta E_{jk} \, t$ n'est pas un invariant sous les
transformations de Lorentz. Cette non invariance r\'esulte du fait que l'on a
n\'eglig\'e la d\'ependance spatiale des \'etats dans la d\'erivation de la
probabilit\'e de d\'etection. Pourtant, la diff\'erence
entre le lieu de production et de d\'etection des particules introduit une
phase suppl\'ementaire dont il faut en principe tenir compte lorsque l'on
\'etudie des ph\'enom\`enes d'interf\'erence.
Voici les solutions couramment apport\'ees \`a ce probl\`eme.
Le calcul de la probabilit\'e (\ref{probaMQexpo}) est en tout cas valable si
la particule oscillante est consid\'er\'ee dans son rep\`ere au repos.
On remplace alors la phase \mbox{$-i \, E_j t$} par la phase
correspondante dans le rep\`ere au repos, c'est-\`a-dire
\mbox{$-i \, m_j \tau$}, o\`u $\tau$ est le temps propre de la particule.
Remarquons que si la particule a une tri-impulsion ${\bf p}_j$ bien
d\'efinie, cette phase peut aussi s'\'ecrire
\begin{equation}
  -i \, m_j \tau = -i \, E_j t + i \, {\bf p}_j \cdot {\bf x} \, .
  \label{phase}
\end{equation}
Le membre de droite peut aussi \^etre calcul\'e en r\'esolvant l'\'equation
de Schr\"odinger pour une onde plane, c'est-\`a-dire en tenant compte de la
d\'ependance spatiale d\`es le d\'epart dans le calcul de l'amplitude
(\ref{oscMQ}).
Traditionnellement, on utilise le membre de gauche de l'\'equation
(\ref{phase}) pour les kaons et le membre de droite pour les neutrinos.

Dans le cas des m\'esons {\it K} et {\it B}, on passe dans le rep\`ere du laboratoire par la
relation \mbox{$t=\gamma \tau = E\tau/m$}, o\`u $m$ est choisi arbitrairement
comme \mbox{$m=(m_j+m_k)/2$} et $E$ est l'\'energie de la particule.
Si l'on note la phase du terme d'interf\'erence par $\Phi$, la probabilit\'e de
d\'etection (\ref{probaMQexpo}) contient des termes oscillant en
\begin{eqnarray*}
  \exp \left( -i \Phi \right)
  &=& \exp \left( -i (m_j-m_k) \, \tau \right)
  \\
  &=& \exp \left( -i (m_j-m_k) \, \frac{m}{E} \, t \right)
  \\
  &=& \exp \left( -i \frac{m_j^2 - m_k^2}{2E} \, t \right) \, .
\end{eqnarray*}
et ces termes d\'ecroissent  en
$$
  \exp \left( -\Gamma_{jk} \tau \right)
  = \exp \left( -\Gamma_{jk} \, \frac{m}{E} \, t \right) \, .
$$
Comme les exp\'eriences ne mesurent que la d\'ependance spatiale de la
probabilit\'e de d\'etection, il faut passer \`a une expression ne d\'ependant
que de la distance.
On substitue donc dans ces expressions la formule de la trajectoire classique,
$t=EL/P$, o\`u $P$ est l'impulsion de la particule.
D\'efinissant la {\it longueur d'oscillation} pour le m\'elange $jk$ par
\begin{equation}
  l_{jk}^{osc} \equiv \frac{4\pi \, P}{m_j^2 - m_k^2} \, ,
  \label{longoscMQ}
\end{equation}
la probabilit\'e de d\'etection (\ref{probaMQexpo}) contient des termes
oscillant en
$$
  \exp \left( -i \Phi \right)
  =  \exp \left( -2\pi i \frac{L}{ l_{jk}^{osc} } \right) \, ,
$$
et ces termes d\'ecroissent en
$$
   \exp \left( -\Gamma_{jk} \, \frac{m}{P} \, L \right) \, .
$$
Les oscillations sont observables si $L/l_{jk}^{osc} \approx {\cal O}(1)$ et si
\mbox{${\cal O}(l_{jk}^{osc}) \le P/(m\Gamma_{jk})$}, ou encore si
\mbox{${\cal O}(m_j-m_k) \ge {\cal O}(\Gamma_{jk})$}. Cette derni\`ere condition
signifie banalement qu'on ne verra pas d'oscillations si la particule se
d\'esint\`egre avant d'osciller notablement.

On peut d\'ej\`a se rendre compte des probl\`emes li\'es \`a ce type d'extension
relativiste. Si l'on avait d\'efini un temps propre pour chaque \'etat propre de
masse, la phase $\Phi$ serait devenue
$$
  \Phi = m_j \tau_j - m_k \tau_k = (m_j^2-m_k^2) \, t/E \, ,
$$
o\`u l'on a pris la m\^eme \'energie pour les \'etats $j$ et $k$,
faute de mieux.
La longueur d'oscillation correspondante est alors {\it deux fois} plus courte
\cite{srivastava}! Le probl\`eme r\'eside dans la question de l'existence
d'un rep\`ere au repos d'une particule qui n'a pas de masse bien d\'efinie.
Il existe d'autres possibilit\'es de se tromper en prenant des \'energies
diff\'erentes, ou des temps diff\'erents, ou des vitesses diff\'erentes, ou
des impulsions diff\'erentes pour les \'etats $j$ et $k$. Les arguments utilis\'es
pour justifier le bon choix \cite{lipkin2,lowe,kayser1} ne sont pas vraiment
convaincants. Pour des processus du type
\mbox{$\phi(1020) \!\to\! K^0\overline{K^0}$}, o\`u deux particules oscillent
simultan\'ement, les ambigu\"{\i}t\'es se multiplient!

La situation ne s'am\'eliore pas si l'on travaille avec le membre de droite
de l'\'equation (\ref{phase}), ce qui est la proc\'edure couramment suivie dans
le cas des neutrinos.
En effet, il faut se donner des prescriptions pour manipuler l'\'energie-impulsion
\mbox{$(E_j,{\bf p}_j)$}, le temps $t$ et la distance ${\bf x}$.
La solution couramment adopt\'ee \cite{pal}, tout \`a fait {\it ad hoc},
est de supposer l'\'egalit\'e des tri-impulsions,
\mbox{${\bf p}_j={\bf p}_k\equiv {\bf p}$},
de sorte que
$$
  E_j = \sqrt{ {\bf p}^2 + m_j^2} \cong E + \frac{ m_j^2 - m^2}{2E} \, ,
$$
o\`u $m\equiv (m_j+m_k)/2$, $E\equiv\sqrt{ {\bf p}^2 + m^2}$ et l'on a suppos\'e
que $|m_j-m_k| \ll E$.
La probabilit\'e (\ref{probaMQexpo}) oscille alors en
$$
  \exp \left( -i\Delta E_{jk}\,t \right) \cong
  \exp \left( -i \, \frac{ m_j^2 - m_k^2}{2E} \,t \right) \, .
$$
Le raisonnement se poursuit alors comme pour la particule instable, avec la
m\^eme d\'efinition de la longueur d'oscillation.
Une deuxi\`eme \'ecole \cite{winter,goldman,dolgov} utilise des \'energies et
tri-impulsions diff\'erentes pour les diff\'erents \'etats propres de masse,
et les calcule par conservation de l'\'energie-impulsion \`a la production.
Le calcul de la phase $\Phi$ du terme d'interf\'erence donne la m\^eme
longueur d'oscillation que l'\'equation (\ref{longoscMQ}) si le temps $t$
est remplac\'e par $t=EL/P$ quel que soit l'\'etat propre de masse.
Les ambigu\"{\i}t\'es lors du passage d'une d\'ependance temporelle \`a une
d\'ependance uniquement spatiale persistent dans ces deux traitements.
On pourrait par exemple argumenter que les \'etats propres de masse ont
des vitesses diff\'erentes et arrivent donc au d\'etecteur en des temps $t_j$
et $t_k$ l\'eg\`erement diff\'erents. Si l'on pose l'\'egalit\'e des
tri-impulsions, la longueur d'oscillation est alors deux fois plus courte que
l'expression (\ref{longoscMQ}).
Enfin, une troisi\`eme possibilit\'e \cite{lipkin3} consiste \`a \'egaler les
\'energies et \`a calculer les tri-impulsions par la formule
\mbox{${\bf p}_j^2=E^2-m^2_j$}. Dans ce cas, on obtient directement une
d\'ependance spatiale dans le terme d'oscillation, mais la prescription
d'\'egalit\'e des \'energies para\^{\i}t aussi {\it ad hoc} que la
prescription d'\'egalit\'e des tri-impulsions.

Il est pratique d'\'ecrire la longueur d'oscillation pour les neutrinos sous la forme
$$
  l_{jk}^{osc} \cong 2.48 \, \mbox{m} \,
  \frac{ P \, (\mbox{MeV}) }{ \Delta m^2 \, (\mbox{eV}^2) } \, ,
$$
qui permet d'\'evaluer facilement la diff\'erence de masse n\'ecessaire pour
une longueur d'oscillation donn\'ee.

\section[Critique de la d\'erivation traditionnelle]
{Critique de la d\'erivation traditionnelle de la formule d'oscillation}

La d\'erivation expliqu\'ee \`a la section pr\'ec\'edente de la probabilit\'e
de d\'etection d'une particule en m\'elange  est incorrecte pour toute une
s\'erie de raisons. On peut \'emettre les objections suivantes:

\begin{enumerate}
\item
L'interf\'erence entre \'etats de masses diff\'erentes est interdite en
m\'ecanique quantique non relativiste. En effet, l'invariance de l'\'equation de
Schr\"odinger sous le groupe de
Galil\'ee fixe la loi de transformation d'un \'etat: il est
multipli\'e par un facteur de phase qui d\'epend de la masse de l'\'etat,
de la position et du temps. Des \'etats de masses diff\'erentes se
transforment donc diff\'eremment et la phase relative d'une superposition
d'\'etats n'est pas conserv\'ee sous le groupe de Galil\'ee.
Une superposition coh\'erente de ces \'etats est donc exclue (r\`egle de
superposition de Bargmann \cite{bargmann}). Pour la m\^eme raison, les
particules instables ne peuvent \^etre d\'ecrites en m\'ecanique quantique
non relativiste car il n'existe pas de transition entre \'etats de masses
diff\'erentes.

\item
Il faut de toute fa\c{c}on tenir compte des corrections
relativistes pour d\'ecrire les exp\'eriences.
Habituellement, on part de la formule non relativiste dans le
rep\`ere au repos de la particule et on effectue un boost pour passer dans
le rep\`ere du laboratoire comme expliqu\'e ci-dessus \cite{gottfried}.
Cette proc\'edure est ambigu\"e lorsque l'on examine l'interf\'erence entre
\'etats de masses et/ou de largeurs diff\'erentes: existe-t-il un rep\`ere au
repos de l'\'etat oscillant, ou en faut-il deux, un pour chaque \'etat propre de
masse? Cette ambigu\"{\i}t\'e a suscit\'e des controverses sur la question des
oscillations \cite{srivastava,lowe}.
Il vaut mieux utiliser un formalisme relativiste d\`es le d\'epart,
c'est-\`a-dire recourir \`a la th\'eorie des champs, puis effectuer des approximations
non relativistes si n\'ecessaire.

\item
Les \'etats propres d'interaction (ou de saveur) sont mal d\'efinis. Il
n'existe pas d'espace de Fock pour des \'etats qui n'ont pas de masse bien
d\'efinie, c'est-\`a-dire qu'il n'existe pas d'op\'erateurs de cr\'eation et d'annihilation
d'\'etats de saveur satisfaisant aux relations de commutation
canoniques \cite{giunti1}. Une tentative a \'et\'e faite en th\'eorie des champs
non perturbative, \`a l'aide de transformations de Bogoliubov mais elle m\`ene
\`a de s\'erieux probl\`emes d'interpr\'etation et brise la covariance sous
les transformations de Lorentz \cite{blasone}.

\item
Pour les particules instables, un probl\`eme suppl\'ementaire  appara\^{\i}t
puisque l'ha\-mil\-to\-nien effectif n'est pas hermitien.
Comme on l'a vu ci-dessus, les \'etats propres de masse ne sont plus orthogonaux
avec leurs hermitiens conjugu\'es et des probl\`emes de normalisation se posent.
Une solution possible (voir \'equation (\ref{doublebase})) consiste \`a utiliser
deux bases {\it in} et {\it out} \`a la place d'une
seule base (et son hermitienne conjugu\'ee) \cite{sachs,enz,alvarez}.
N\'eanmoins, cette m\'ethode peut para\^{\i}tre artificielle.
La question de la normalisation a de l'importance car
dans le syst\`eme $K^0\overline{K^0}$, l'incertitude sur la normalisation des
\'etats est du m\^eme ordre que la violation CP directe \cite{beuthe}.

\item
Les particules instables n'apparaissent pas comme \'etats asymptotiques: ceux-ci
ne sont donc pas contenus dans l'espace de Hilbert des \'etats physiques. En fait,
elles ne peuvent \^etre trait\'ees de fa\c{c}on coh\'erente que
comme \'etats interm\'ediaires dans le cadre de la th\'eorie quantique des
champs \cite{veltman}.

\item
Les \'etats oscillants \'etudi\'es jusqu'\`a pr\'esent (m\'esons {\it K} et {\it B},
neutrinos) ne sont pas observ\'es directement car ils sont \'electriquement
neutres. Seuls les particules avec lesquelles ils interagissent (ou dans
lesquelles ils se d\'esint\`egrent) par interaction faible sont observ\'es.
Dans une approche r\'ealiste, on ne devrait calculer que des amplitudes de
transition entre \'etats observables. 

\item
Les exp\'eriences \'etudient la propagation spatiale, et non temporelle.
L'utilisation de la formule classique $L=v \, t$ dans un calcul \'eminemment
quantique pour remplacer le temps par la distance laisse songeur. D'ailleurs, ne
faudrait-il pas utiliser des vitesses $v$ diff\'erentes pour les diff\'erents
\'etats propres de masse? Ce probl\`eme est li\'e \`a la question  mentionn\'ee
ci-dessus du passage \`a un formalisme relativiste.
Cette ambigu\"{\i}t\'e m\`ene \`a des diff\'erences
{\it du simple au double} dans le calcul de la fr\'equence
d'oscillation \cite{lipkin2,srivastava,kayser1}.

\item
La d\'erivation traditionnelle implique une \'energie-impulsion parfaitement
d\'e\-fi\-nie, c'est-\`a-dire que les \'etats sont des ondes planes.
Par le principe
d'incertitude, leur localisation spatio-temporelle est alors inconnue. Comment
parler dans ce cas de propagation sur une distance d\'efinie?

\item
Il n'y a pas de raison de penser que les \'energies ou les impulsions des
\'etats oscillants soient \'egales. Cette question est d'ailleurs ambigu\"e:
peut-on parler de l'\'energie ou de l'impulsion d'un \'etat propre de masse
lorsqu'il est superpos\'e lin\'eairement \`a un autre \'etat propre de masse?
L'hypoth\`ese qui consiste \`a \'egaler les tri-impulsions des \'etats propres
de masse des neutrinos est donc injustifi\'ee, comme d'ailleurs aussi
l'hypoth\`ese oppos\'ee qui consiste \`a \'egaler leurs \'energies.
D'ailleurs, la conservation de l'\'energie-impulsion \`a la production
n'impose-t-elle pas l'\'egalit\'e de l'\'energie-impulsion des diff\'erents
\'etats propres de masse?
Ou alors, faut-il consid\'erer que l'\'energie-impulsion des autres particules
\'emises \`a la production de la particule oscillante varie selon l'\'etat
propre de masse \'emis?
Cette question n'est pas purement acad\'emique. Tout d'abord elle remet en
question la formule d'oscillation.
Ensuite, l'utilisation d'une conservation stricte de l'\'energie-impulsion
\`a la production m\`ene \`a consid\'erer les particules associ\'ees \`a la
particule oscillante (par exemple le muon dans \mbox{$\pi\!\to\!\mu\nu$})
comme une superposition d'\'etats d'\'energies-impulsions diff\'erentes, avec
le r\'esultat qu'ils oscillent eux aussi \cite{widom,sassaroli}.
Nous verrons au chapitre \ref{applications} que ce n'est pas le cas.

\item
Les crit\`eres d'observabilit\'e des oscillations (\`a part
$L \approx l_{jk}^{osc}$) n'apparaissent pas explicitement dans la formule
d'oscillation (\ref{probaMQexpo}).
Par exemple, on s'attend \`a ce que les oscillations soient inobservables\\
- si la longueur d'oscillation est de l'ordre
de grandeur ou plus petite que l'incertitude sur la position de la source
et/ou du d\'etecteur.\\
- si l'\'energie et l'impulsion des \'etats oscillants peuvent \^etre mesur\'ees
avec une pr\'ecision suffisante pour d\'eterminer quel \'etat propre de masse se
propage.\\
Pour que la superposition quantique soit possible, il faut donc simultan\'ement
une localisation suffisamment pr\'ecise de la source et du d\'etecteur et une
incertitude suffisante sur l'\'energie et/ou l'impulsion \`a l'\'emission et \`a
la d\'etection \cite{kayser2}.

\end{enumerate}

Toutes ces objections m\`enent \`a abandonner un traitement direct des
\'etats oscillants. Le processus d'oscillation va \^etre consid\'er\'e
dans son ensemble: les \'etats oscillants deviennent des \'etats
interm\'ediaires, non observ\'es directement, se propageant entre une source et
un d\'etecteur. Ils sont repr\'esent\'es en th\'eorie des champs par leur
propagateur. La forme de celui-ci, consistante avec la relativit\'e restreinte
et la causalit\'e, d\'etermine l'\'evolution spatio-temporelle ainsi que la
probabilit\'e de d\'e\-sin\-t\'e\-gra\-tion si la particule est instable.
Etant donn\'e que la covariance sous les transformations de Lorentz est
implicite dans cette approche,
le recours \`a des rep\`eres particuliers et des boosts est rendu inutile.
Cette m\'ethode a \'et\'e propos\'ee pour la premi\`ere fois par Jacob et
Sachs \cite{jacob} pour les particules scalaires instables et
Giunti, Kim et Lee \cite{giunti2} pour les neutrinos.

Dans un premier temps, nous allons pr\'esenter un mod\`ele simplifi\'e
suffisant \`a d\'ecrire la propagation d'une seule particule stable. Dans cet
exemple, l'\'etude de la d\'ependance spatiale de l'amplitude globale de
propagation se r\'eduit \`a l'analyse de la transform\'ee de Fourier du
propagateur par rapport \`a sa tri-impulsion. L'esprit de cette m\'ethode
est proche de Schwinger \cite{schwinger} qui \'etudie la d\'ependance
temporelle de l'amplitude.
On appliquera aussi ce mod\`ele \`a la description de particules instables,
bien qu'en toute rigueur il ne convienne pas \`a cette situation.
On estimera les contributions non exponentielles \`a l'amplitude de propagation,
et l'on calculera aussi une probabilit\'e d'oscillation d\'ependant uniquement
de la distance.
Ses insuffisances vis-\`a-vis des particules instables mises \`a part,
le mod\`ele simplifi\'e ne permet pas non plus de
r\'epondre \`a toutes les questions pos\'ees ci-dessus \`a propos de la
d\'erivation traditionnelle des oscillations en m\'ecanique quantique. Je
pense d'une part au probl\`eme de l'\'egalit\'e ou non des
\'energies-impulsions des \'etats oscillants et d'autre part aux conditions
d'observabilit\'e de ce ph\'enom\`ene.

Dans un second temps, nous proposerons un mod\`ele plus sophistiqu\'e.
Pour tenir compte des restrictions exp\'erimentales sur les
oscillations, les \'etats entrants et sortants \`a la source et au d\'etecteur
seront repr\'esent\'es par des paquets d'ondes. On pourra ainsi d\'eriver les
conditions d'observabilit\'e des oscillations.
La propagation de particules instables pourra aussi \^etre d\'ecrite de
fa\c{c}on coh\'erente. Enfin, toutes les questions soulev\'ees
ci-dessus concernant les oscillations trouveront une r\'eponse.
Les formules th\'eoriques seront appliqu\'ees dans une s\'erie de cas concrets
concernant des exp\'eriences actuelles. 
Une param\'etrisation coh\'erente de la violation CP sera aussi \'etablie.
Tout au long de ce travail, la sym\'etrie CPT sera suppos\'ee exacte.
Param\'etriser une violation \'eventuelle de cette sym\'etrie dans un mod\`ele
comme le n\^otre bas\'e sur la th\'eorie des champs est incoh\'erent puisque
personne n'a r\'eussi jusqu'\`a pr\'esent \`a construire une th\'eorie des
champs relativiste violant CPT. Autant param\'etriser cette violation dans le
cadre de la m\'ecanique quantique, comme cela a d\'ej\`a \'et\'e fait de
nombreuses fois (toutefois, la premi\`ere param\'etrisation,
\`a ma connaissance, de la violation de CPT dans le syst\`eme
$K^0\overline{K^0}$ a \'et\'e effectu\'ee en th\'eorie des champs!
\cite{sachs}).

\chapter{Le propagateur complet}
\label{propagateur}

Il appara\^\i t dans ce chapitre que le propagateur complet constitue
l'\'el\'ement cl\'e de toute description d'un ph\'enom\`ene de propagation
macroscopique en th\'eorie des champs. Remarquons tout d'abord que
l'amplitude de propagation d'une particule d\'epend en g\'en\'eral des
processus de production et de d\'etection. Sous certaines conditions
analys\'ees dans le chapitre \ref{production}, il est possible de factoriser
l'amplitude globale en trois parties: production, propagation et d\'etection.
La propagation d'une particule peut alors \^etre \'etudi\'ee \`a partir du
seul propagateur, sans prendre en compte les \'etats entrants et sortants.
Dans ce cas, l'utilisation de la th\'eorie quantique des champs n'est pas plus
compliqu\'ee que celle de la m\'ecanique quantique et fournit directement
une formule d'oscillation spatiale, ce qui est d\'ej\`a un plus par rapport
\`a la formule (\ref{probaMQexpo}) d\'eriv\'ee en m\'ecanique quantique.
Notons cependant que nos calculs ne s'appliqueront pas \`a la propagation de
bosons de jauge charg\'es: le propagateur n'\'etant pas invariant de jauge,
il ne peut \^etre consid\'er\'e ind\'ependamment. Un tel processus ne peut
\^etre factoris\'e qu'en sous-processus invariants de jauge \cite{stuart}.

Apr\`es le passage en revue des hypoth\`eses du mod\`ele simplifi\'e,
nous \'etudierons la repr\'esentation spectrale du propagateur complet
et les cons\'equences de celle-ci sur l'amplitude de propagation spatiale,
tant pour une particule stable que pour une particule instable.
En particulier, les corrections non exponentielles dues aux seuils de
production de plusieurs particules r\'eelles seront examin\'ees. Le chapitre
se termine par un exemple de calcul de propagateur complet.

\section{Propagation: le mod\`ele simplifi\'e}
\label{modelesimplifie}

Dans cette section, nous pr\'esentons un mod\`ele simplifi\'e tel que
l'amplitude de propagation se factorise et l'\'etude de la propagation se
r\'esume \`a celle du propagateur.
Le processus de propagation d'une particule scalaire est symbolis\'e par

\parbox{\textwidth}{
\begin{eqnarray*}
  P_I(q)
 \stackrel{ (t_{\SS P},{\bf x}_{\SS P} ) }{\longrightarrow}
  P_F(k) + \nu(p) & &
  \\
  &\searrow&
  \\
  & & \nu(p) + D_I(q') 
  \stackrel{ (t_{\SS D},{\bf x}_{\SS D} ) }{\longrightarrow} D_F(k')
\end{eqnarray*}
}

$P_I$ repr\'esente l'ensemble des particules arrivant dans la r\'egion
de production centr\'ee autour du point $( t_{\SS P},{\bf x}_{\SS P} )$,
d'\'energie-impulsion totale $q$
tandis que $P_F$ repr\'esente l'ensemble des particules issues de la r\'egion
de production et d'\'energie-impulsion totale $k$, \`a l'exception de la particule
interm\'ediaire $\nu$ dont on \'etudie la propagation\footnote{Tout au long
de ce travail, les vecteurs en 3 dimensions
seront not\'es en gras tandis que les vecteurs en 4 dimensions seront not\'es
normalement. Les produits scalaires s'\'ecrivent respectivement
${\bf p} \cdot {\bf q}$ et $p \cdot q$. La m\'etrique utilis\'ee est
$g_{\mu\nu}=diag(1,-1,-1,-1)$.}.
$D_I$, $D_F$ et $( t_{\SS D},{\bf x}_{\SS D} )$ ont des d\'efinitions
similaires mais concernent le point de d\'etection.
Le point d'interaction \`a la production est not\'e par $x$ et le point
d'interaction \`a la d\'etection est not\'e $x'$.
On pose que l'\'etat interm\'ediaire se propageant de $x$ en $x'$ a les
nombres quantiques d'une particule, pas d'une antiparticule.
Toutes les particules du processus sont suppos\'ees scalaires.
Les conditions exp\'erimentales (\'energies-impulsions de $P_{I,F}$, $D_{I,F}$)
sont choisies telles que le processus ci-dessus, o\`u une particule quasi-r\'eelle
$\nu$ se propage macroscopiquement, est s\'electionn\'e parmi tous les processus
possibles. En d'autres termes, les \'energies-impulsions des \'etats initiaux et
finaux sont telles que l'\'el\'ement de matrice $S$ correspondant \`a ce
processus est \'evalu\'e pr\`es de la singularit\'e du propagateur de la
particule interm\'ediaire.

Ce processus pourrait aussi \^etre symbolis\'e par  le diagramme suivant, sur
lequel les r\'egions de production et de d\'etection sont indiqu\'ees par des
cercles en pointill\'e et les fl\`eches indiquent la direction des impulsions:
\begin{center}
\includegraphics[width=14cm]{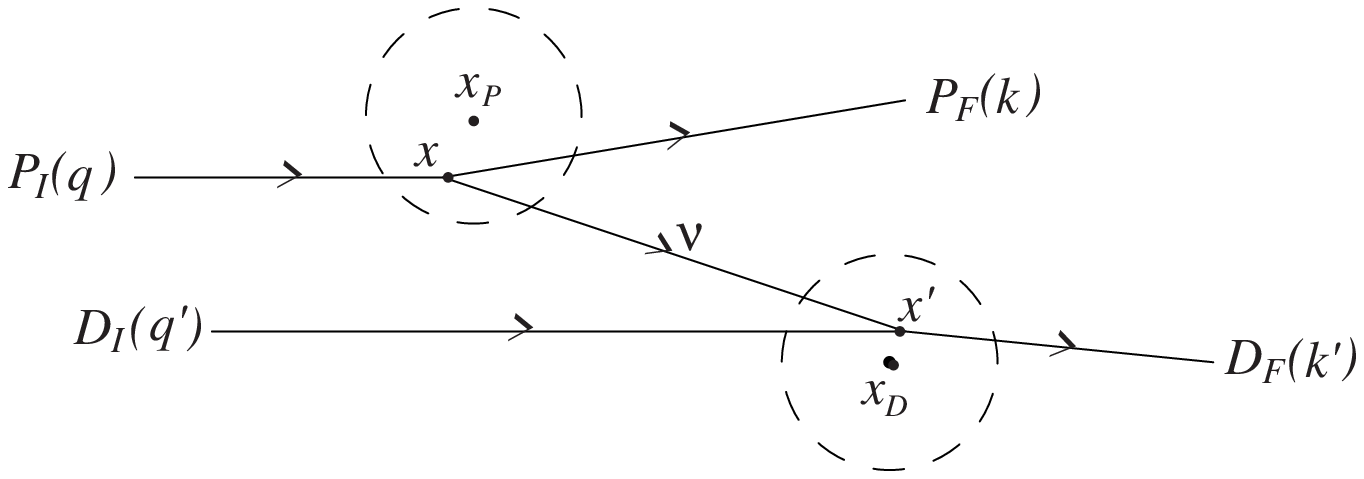}
\end{center}

Les hypoth\`eses simplificatrices de ce mod\`ele sont les suivantes:
\begin{itemize}
\item
tous les \'etats entrants et sortants sont stationnaires, c'est-\`a-dire
que leurs \'energies ont des valeurs d\'etermin\'ees.
\item
un \'etat au point de production et un \'etat au point de d\'etection
sont localis\'es spatialement \`a une pr\'ecision infinie.
\item
les autres \'etats ont des tri-impulsions d\'etermin\'ees \`a une
pr\'ecision infinie, c'est-\`a-dire qu'ils sont des ondes planes dans
l'espace de configuration.
\end{itemize}

Pour fixer les id\'ees, supposons que $P_I$ et $D_I$ soient chacun
constitu\'es d'un seul \'etat stationnaire d\'esign\'e respectivement
dans l'espace de configuration par
$$
  \widetilde{ \phi_{P_I} } \left( {\bf x} \!-\! {\bf x}_{\SS P} \right) \,
  e^{- i E_{P_I} t }
  \quad \mbox{et} \quad
  \widetilde{ \phi_{D_I} } \left( {\bf x'} \!-\! {\bf x}_{\SS D} \right) \,
  e^{ - i E_{D_I} t' } \, .
$$
On ne pourra parler de propagation macroscopique que si les interactions \`a
la production et \`a la d\'etection sont suffisamment localis\'ees par
rapport \`a la distance de propagation. C'est pourquoi on a pos\'e l'hypoth\`ese
extr\^eme que les fonctions d'onde des \'etats $P_I$ et $D_I$
sont des distributions delta de sorte que les interactions de
production et d\'etection soient localis\'ees spatialement \`a une pr\'ecision infinie:
$$
  \widetilde{ \phi_{P_I} } \left( {\bf x} \!-\! {\bf x}_{\SS P} \right) \,
  \sim \delta^{(3)}\left( {\bf x} \!-\! {\bf x}_{\SS P} \right)
  \quad \mbox{et} \quad
  \widetilde{ \phi_{D_I} } \left( {\bf x'} \!-\! {\bf x}_{\SS D} \right)
  \sim \delta^{(3)}\left( {\bf x'} \!-\! {\bf x}_{\SS D} \right) \, .
$$
Bien s\^{u}r, le prix \`a payer est une incertitude compl\`ete sur la
tri-impulsion de la particule interm\'ediaire.
Quant aux \'etats $P_F$ et $D_F$, on suppose qu'ils ne contiennent chacun
qu'une particule, qui est repr\'esent\'ee par une onde plane:
$$
  \widetilde{ \phi_{P_F} }({\bf x},t)
  \sim e^{- i E_{P_F} t  + i \, {\bf k  \cdot x} }
  \quad \mbox{et} \quad
  \widetilde{ \phi_{D_F} }({\bf x},t')
  \sim e^{-i E_{D_F} t' + i \, {\bf k' \cdot x'} } \, .
$$

L'amplitude du processus s'\'ecrit dans l'espace de configuration comme
\begin{eqnarray}
  {\cal A} &\sim& M_P M_D \,
  \int d^4x \int d^4x' \,
  \widetilde{ \phi_{P_I} }\left( {\bf x} \!-\! {\bf x}_{\SS P} \right) \,
  e^{ - i E_{P_I} t } \,
  \widetilde{ \phi_{D_I} } \left( {\bf x'} \!-\! {\bf x}_{\SS D} \right) \,
  e^{ - i  E_{D_I} t' }
  \nonumber
  \\ & & \times \;
  e^{i E_{P_F} t  - i \, {\bf k  \cdot x} } \,
  e^{i E_{D_F} t' - i \, {\bf k' \cdot x'} } \,
  \int \frac{d^4p}{(2\pi)^4} \,
  e^{- \, i \, p \cdot (x'-x)} \, G(p^2)
  \label{amplipro}
\end{eqnarray}
o\`u $M_P$, $M_D$ sont les couplages, suppos\'es constants et $\bf k$,
$\bf k'$ sont les impulsions totales des \'etats $P_F$ et $D_F$. Le
propagateur est symbolis\'e par $G(p^2)$ ou par $G(p^0,{\bf p})$ selon
les besoins.
Cette formule est justifi\'ee
rigoureusement au chapitre \ref{production}, dans le cadre du mod\`ele
sophistiqu\'e.
L'int\'egration sur $t$ et $t'$ donne
\begin{eqnarray*}
  {\cal A} &\sim& M_P M_D \,
  \int d^3x \int d^3x' \int \frac{d^4p}{(2\pi)^2} \,
  \delta \left( E_{P_I} \! - E_{P_F} \! - p^0 \right) \,
  \delta \left( E_{D_I} \! - E_{D_F} \! + p^0 \right)
  \\ & & \qquad \times \,
  e^{- i \, {\bf k  \cdot \bf x} - i \, {\bf k' \cdot \bf x'} } \,
  \widetilde{ \phi_{P_I} } \left( {\bf x} \!-\! {\bf x}_{\SS P} \right) \,
  \widetilde{ \phi_{D_I} } \left( {\bf x'} \!-\! {\bf x}_{\SS D} \right) \,
  e^{ i \, {\bf p} \cdot ({\bf x'}- {\bf x}) } \, G(p^2)
  \\
  &\sim& M_P M_D \, 2\pi \,
  \delta \left( E_{D_I} \! - E_{D_F} \! + E_{P_I} \! - E_{P_F} \right) \,
  \int d^3x \int d^3x' \int \frac{d^3p}{(2\pi)^3} \,
  \\ & & \qquad \times \,
  e^{ - i \, {\bf k  \cdot \bf x} - i \, {\bf k' \cdot \bf x'} } \,
  \widetilde{ \phi_{P_I} } \left( {\bf x} \!-\! {\bf x}_{\SS P} \right) \,
  \widetilde{ \phi_{D_I} } \left( {\bf x'} \!-\! {\bf x}_{\SS D} \right) \,
  e^{ i \, {\bf p} \cdot ({\bf x'} - {\bf x}) } \, G(E, {\bf p}) \, ,
\end{eqnarray*}
o\`u $ E \equiv E_{P_I} \! - E_{P_F} \!>\! 0 $.
L'int\'egration sur ${\bf x}$ et ${\bf x'}$ donne, apr\`es insertion des
fonctions d'onde delta
\begin{equation}
  \fbox{$ \displaystyle
  {\cal A} \sim M_P M_D \, 2\pi \,
  \delta \left( E_{D_I} \! -\! E_{D_F} \! +\! E_{P_I} \!  -\! E_{P_F} \right) \,
  \int \frac{d^3p}{(2\pi)^3} \,
   e^{ i \, {\bf p} \cdot {\bf L} } \, G(E,{\bf p})
  $}
  \label{amplisimpli}
\end{equation}
o\`u ${\bf L} \equiv {\bf x}_{\SS D} \!-\! {\bf x}_{\SS P}$.

On voit que dans ce mod\`ele simplifi\'e, l'amplitude de propagation est
directement proportionnelle \`a la {\it transform\'ee de Fourier du propagateur
sur sa tri-impulsion} et aura donc une d\'ependance spatiale en ${\bf L}$.
Notons que l'on \'etudie habituellement
la transform\'ee de Fourier du propagateur sur l'\'energie \cite{schwinger,brown}
ce qui donne une amplitude d\'ependant du temps.
Outre le fait qu'aucun
mod\`ele n'est donn\'e pour justifier ce choix, il est plus logique
d'\'etudier la d\'ependance spatiale du propagateur puisqu'aucune
exp\'erience n'observe directement la d\'ependance temporelle.

Ce mod\`ele-ci pr\'esente cependant plusieurs d\'efauts:
\begin{itemize}
\item
il ne s'applique pas \`a la d\'esint\'egration des particules instables,
pour lesquelles il ne peut \^etre question d'\'etat stationnaire lors de leur
d\'esint\'egration.
\item
l'approximation, soit d'\'etats stationnaires, soit d'ondes planes, est trop
forte car elle ne permet pas d'\'etudier l'influence des conditions de
production et de d\'etection. De plus, comme on le verra \`a la section
\ref{oscisimpli}, elle est trop impr\'ecise pour que l'on
puisse r\'epondre par la suite \`a la question d'\'egalit\'e ou non des
\'energies-impulsions des \'etats oscillants.
\item
l'amplitude est ind\'ependante de la direction de ${\bf L}$.
\end{itemize}

Ces d\'efauts ne deviennent pr\'eoccupants que pour l'\'etude de
ph\'enom\`enes tels que les oscillations de particules, qui r\'esultent de la
compensation quasi totale de diff\'erentes phases.
Nous y reviendrons dans le prochain chapitre et
d\'eve\-lop\-perons un mod\`ele plus complet par la suite.
Une autre imperfection de ce mod\`ele rel\`eve de la question de la
possibilit\'e d'une propagation libre. Dans une chambre \`a bulles, la
trajectoire d'une particule charg\'ee est signal\'ee par des bulles
r\'esultant des interactions \'electromagn\'etiques de la particule avec le
milieu de propagation \cite{fonda}. Il faudrait en principe tenir compte de
ces mesures successives dans le formalisme quantique. Ce probl\`eme ne sera pas
examin\'e ici puisque l'on vise \`a d\'ecrire la propagation de particules
neutres (kaons, neutrinos). La question de principe subsiste cependant, car
ces particules interagissent encore faiblement avec le milieu de propagation.
Le mod\`ele MSW \cite{msw} est un exemple o\`u ces interactions jouent un
r\^ole d\'eterminant.

Pour revenir au mod\`ele simplifi\'e, il convient maintenant d'\'etudier la
forme g\'en\'erale du propagateur et sa transform\'ee de Fourier.

\section{Deux formes de repr\'esentations spectrales}
\label{deuxformes}

Un champ relativiste d\'ecrit une excitation locale qui produit un spectre
d'\'ener\-gies ou de masses poss\'edant une limite inf\'erieure
caract\'eristique du type de champ. Par exemple, la limite th\'eorique
pour n'importe quel boson ou lepton \'electriquement charg\'e est la
masse de l'\'electron, tandis que pour les champs neutres de ces deux
types, le spectre commence en principe en z\'ero. Pour les baryons, cette
limite est la masse du proton. Ces limites sont fix\'ees par les masses des
particules stables restantes apr\`es que toutes les d\'esint\'egrations
possibles ont eu lieu.

La propagation de ces excitations est d\'ecrite par la fonction de Green
\` a deux points ordonn\'ee dans le temps, qui pour un
champ bosonique scalaire est d\'efinie par
$$
  G(x-y) = < \! 0 \, |T \left( \phi(x) \,\, \phi^*(y) \right) | \, 0 \!> \; .
$$
Elle est aussi appel\'ee {\it propagateur complet}.
En vertu des principes fondamentaux de la th\'eorie relativiste des champs,
cette fonction prend la forme suivante, appel\'ee {\it repr\'esentation
spectrale} ou de K\"{a}llen-Lehmann \cite{kallen}:
$$
  G(x-y) = \int^{\infty}_{M^2} \, ds \, \rho(s) \, \Delta_F (x-y,s)
$$
o\`u
$$
  \Delta_F (x-y,s) \equiv \int \, \frac{d^4p}{(2\pi)^4} \, e^{-i p (x-y)} \,
  \frac{i}{p^2 -s +i \epsilon}
$$
est le propagateur de Feynman du champ scalaire libre.
La {\it fonction spectrale} $\rho(s)$ est d\'efinie par
\begin{equation}
  \rho(s=p^2) \equiv (2\pi)^3 \, \sum_n \, \delta^{(4)} \left( p - p_n \right) \,
  |\!<\!0\,|\phi(0)|\,n\!>\!|^2 \, ,
  \label{fctspec}
\end{equation}
o\`u la sommation sur les \'etats $|n\!>$ \`a $n$ particules comprend aussi
une int\'egration sur leurs tri-impulsions.
Cette fonction peut \^etre interpr\'et\'ee comme \'etant
la densit\'e d'\'etats de masse $s$. $M$ est la masse invariante de l'\'etat
\`a plusieurs particules le plus l\'eger en interaction avec le champ. On a
$$
  \rho(s) = 0 \qquad \mbox{pour} \qquad s \le M^2 \, .
$$
La repr\'esentation spectrale du propagateur complet dans l'espace des
impulsions s'obtient par transform\'ee de Fourier et s'\'ecrit
\begin{equation}
  \fbox{$\displaystyle
  G(p^2) =  \int^{\infty}_{M^2} \, ds
  \, \rho(s) \,\frac{i}{p^2 - s + i \epsilon}
  $}
  \label{spectral1}
\end{equation}
o\`u l'on reconna\^\i t une sommation pond\'er\'ee selon la masse sur le
propagateur de la particule libre.

Le propagateur $G(p^2)$ peut \^etre vu comme la valeur limite de la fonction
$G(z)$ d\'efinie dans le plan complexe:
$$
  G(p^2) = \lim_{z\to p^2 +i\epsilon} G(z)
       = \lim_{z\to p^2 +i\epsilon} \,
        \int^{\infty}_{M^2} \, ds \, \rho(s) \, \frac{i}{z-s} \; .
$$
La fonction $G(z)$ est analytique partout sauf sur une coupure le long de
l'axe r\'eel positif \`a partir de $z=M^2$.

Pour que les int\'egrales d\'efinies ci-dessus existent, il faut que la
fonction $\rho(s)$ d\'ecroisse lorsque $s$ tend vers l'infini. Par
cons\'equent, le propagateur a un comportement asymptotique en $G(p^2) \sim
i \, p^{-2}$ lorsque $p^2 \rightarrow \infty$. Cette propri\'et\'e
ca\-rac\-t\'e\-ristique d'un champ local peut \^etre utilis\'ee pour \'etablir une
repr\'esentation spectrale \'equivalente du propagateur \cite{schwinger},
qui est par\-ti\-cu\-li\`e\-rement pratique pour \'etudier les particules instables
et faire le lien avec l'approche perturbative des diagrammes de Feynman.
Observons d'abord que $G(z)$ n'a pas de z\'eros complexes puisque
$$
  G(x+iy) = i \, \int^{\infty}_{M^2} \, ds \,
                 \frac{(x-s) \, \rho(s)}{(x-s)^2 + y^2} \, 
           + y \, \int^{\infty}_{M^2} \, ds \,
                 \frac{\rho(s)}{(x-s)^2 + y^2}
          = 0
$$
implique $y=0$.
La fonction $P(z) \equiv i \, G^{-1}(z) - z$ est donc r\'eguli\`ere dans le
plan complexe, \`a l'exception de la coupure sur l'axe r\'eel positif
d\'ej\`a mentionn\'ee. La fonction $z^{-1} \, P(z)$ tend vers z\'ero \`a
l'infini dans le plan complexe et pr\'esente un p\^{o}le \`a l'origine. Elle
peut s'\'ecrire
$$
  z^{-1} \, P(z) = - \, \frac{M_0^2}{z}
                   -  \int^{\infty}_{M^2} \, ds \, \frac{\sigma(s)}{z-s} \, ,
$$
o\`u $\sigma(s)$ est une fonction dont on va d\'eterminer les propri\'et\'es.
On peut encore \'ecrire
$$
  i \,  G^{-1}(z) = z - M_0^2
             - z \, \int^{\infty}_{M^2} \, ds \, \frac{\sigma(s)}{z-s} \, .
$$
En comparant avec la formule pr\'ec\'edente pour $G(z)$
(\'equation (\ref{spectral1})), on a
$$
  i \, G(0) = M_0^{-2}
            = \int^{\infty}_{M^2} \, ds \, \frac{\rho(s)}{s}
            > 0 \, .
$$
De plus, on v\'erifie que
\begin{eqnarray}
  {\cal R}\!e \, G(x + i \, \epsilon)
  &=& - \, {\cal R}\!e \, G(x - i \, \epsilon)
  = \pi \, \rho(x) \ge 0 \, ,
  \label{coupure}
  \\
  {\cal I}\!m \, G(x + i \, \epsilon)
  &=& {\cal I}\!m \, G(x - i \, \epsilon) \, , \nonumber
\end{eqnarray}
donc 
\begin{eqnarray*}
  G^{-1}(x + i \epsilon) - G^{-1}(x - i \epsilon) &\ge& 0
  \quad \mbox{pour} \quad x > M^2 \, ,
  \\
  G^{-1}(x + i \epsilon) - G^{-1}(x - i \epsilon)  &=&  0
  \quad \mbox{pour} \quad x \le M^2 \, ,
\end{eqnarray*}
On v\'erifie aussi que
$$
  G^{-1}(x + i \epsilon) - G^{-1}(x - i \epsilon)
  = 2\pi \, x \, \sigma(x) \, ,
$$
donc $\sigma(x) \ge 0$ avec $\sigma(x) = 0$ pour $x \le M^2$.
Ces deux conditions, $M_0^2 > 0$ et $\sigma(x) \ge 0 \,$, garantissent que
$G(z)$ n'a pas de p\^{o}le pour $z$ r\'eel n\'egatif.
En conclusion, la deuxi\`eme repr\'esentation spectrale du propagateur
s'\'ecrit
\begin{equation}
  \fbox{$\displaystyle
  G(z) = \frac{i}{z - M_0^2 - \Pi(z)}
  $}
  \label{spectral2}
\end{equation}
o\`u
\begin{equation}
  \Pi(z) \equiv z \, \int^{\infty}_{M^2} \, ds \, \frac{\sigma(s)}{z-s} \, .
  \label{spectralenergie}
\end{equation}
Remarquons que
\begin{equation}
  {\cal I}\!m \, \Pi(p^2) = \lim_{z\to p^2 +i\epsilon} \,  {\cal I}\!m \, \Pi(z)
                 = - \,\pi \, p^2 \, \sigma(p^2)
                 \le 0 \, .
  \label{imagenergie}
\end{equation}

Cette forme du propagateur se retrouve aussi \`a partir des diagrammes de
Feynman. Soit la fonction $-i \, \Pi(p^2)$, appel\'ee {\it \'energie propre},
d\'esignant la somme des diagrammes irr\'eductibles \`a une particule (1PI),
\begin{center}
\includegraphics[width=14cm]{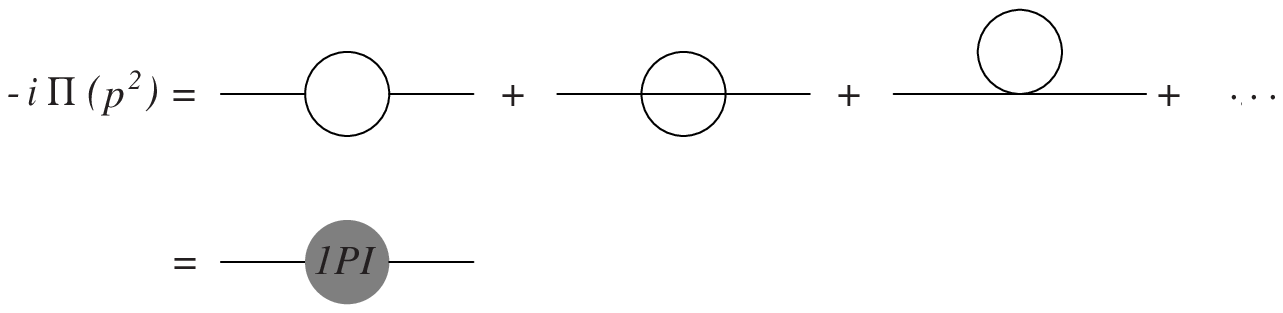}
\end{center}
Le propagateur complet est la somme de toutes les insertions possibles de
l'\'energie propre dans le propagateur:
\begin{center}
\includegraphics[width=14cm]{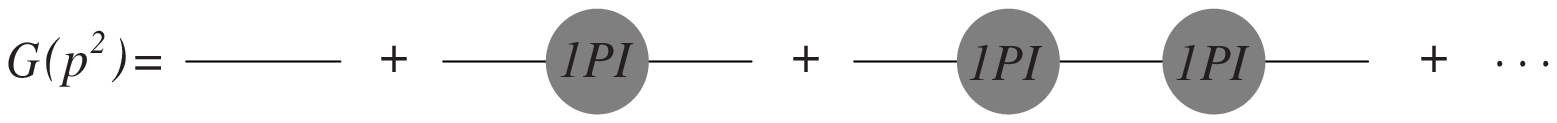}
\end{center}
et peut s'exprimer comme une s\'erie g\'eom\'etrique en $-i \, \Pi(p^2)$:
\begin{eqnarray}
  G(p^2)
  &=& \frac{i}{p^2 - M^2_0} +
  \frac{i}{p^2 - M^2_0} \left( -i \, \Pi(p^2) \right) \frac{i}{p^2 - M^2_0}
  + \quad...
  \nonumber \\
  &=& \frac{i}{ p^2 - M^2_0 - \Pi(p^2) } \, .
  \label{seriegeom}
\end{eqnarray}
On retrouve la deuxi\`eme forme de la repr\'esentation spectrale.
La fonction $\Pi(p^2)$ \'etant en g\'en\'eral divergente, cette s\'erie ne
converge pas et il faut renormaliser la masse $M_0$ par une constante
infinie qui absorbe la divergence de $\Pi(p^2)$. Cette proc\'edure faite,
la s\'erie converge pour un couplage faible et loin des p\^{o}les \'eventuels
de $G(p^2)$.
Soit $M$ la masse renormalis\'ee et $f(p^2)$ la partie restante de $\Pi(p^2)$
apr\`es renormalisation de la masse (son expression exacte d\'epend du point de
renormalisation). Le propagateur complet devient
\begin{equation}
  G(p^2) = \frac{i}{ p^2 - M^2 - f(p^2) }
  \label{propren}
\end{equation}
La fonction $f(p^2)$ acquiert une partie imaginaire pour $p^2 > M^2_{seuil}$,
qui est l'\'energie-impulsion minimale pour cr\'eer des particules
interm\'ediaires r\'eelles (c'est-\`a-dire sur leur couche de masse).
En prolongeant
$G(p^2)$ dans le plan complexe en $G(z)$, on retrouve bien la coupure sur l'axe
des r\'eels, pour $p^2 > M^2_{seuil}$, pr\'esente dans la repr\'esentation
spectrale.

Remarquons enfin que le p\^{o}le du propagateur complet en
$p^2 \! = \! m^2$ appara\^ \i t
dans la deuxi\`eme repr\'esentation spectrale comme un z\'ero du d\'enominateur,
c'est-\`a-dire comme solution de
$$
  p^2 - M^2 -f(p^2) = 0 \, .
$$
On va voir imm\'ediatement que ce p\^ole peut \^etre interpr\'et\'e comme la 
{\it masse physique} de l'\'etat \`a une particule cr\'e\'e par le champ.
Cette masse physique $m$ n'est pas n\'ecessairement identique \`a la masse
renormalis\'ee $M$.

\section{Particule stable: repr\'esentation spectrale}

Supposons que le champ cr\'ee un \'etat \`a une particule de masse $m$,
ainsi que des \'etats \`a plusieurs particules.
La fonction spectrale (\ref{fctspec}) s'\'ecrit dans ce cas
\begin{eqnarray*}
  \rho(p^2) &=& \int \frac{d^3q}{(2\pi)^3 \, 2 \sqrt{ {\bf q}^2 + m^2} } \,
   (2\pi)^3 \, \delta^{(4)} (p-q) \, | \! <\!0\,|\phi(0)|\,q\!> \! |^2
  \\ & &
  + \sum_{n \ge 2} (2\pi)^3 \,
  \delta^{(4)} \left(p-p_n \right) \, | \! <\!0\,|\phi(0)|\,p_n\!> \! |^2 \, ,
\end{eqnarray*}
ou encore
\begin{equation}
  \rho(s) = Z \, \delta \left( s - m^2 \right) + \rho_{multi} (s) \, ,
  \label{rho1part}
\end{equation}
o\`u $Z \equiv | \! <\!0\,|\phi(0)|\,q \!> \! |^2$ est une constante d\'ependant
uniquement de $q^2=m^2$ par invariance de Lorentz. Elle peut s'interpr\'eter
comme la probabilit\'e que le champ $\phi$ cr\'ee un \'etat \`a une particule
\`a partir du vide. 
La fonction $\rho_{multi}(s)$ est une fonction continue et repr\'esente la
contribution des \'etats \`a plusieurs particules. Ces \'etats ne contribuent
que si l'\'energie disponible est sup\'erieure au seuil de production de ces
\'etats sur leur couche de masse. Le seuil le plus bas est not\'e $M^2_{seuil}$.
$$
  \rho_{multi} (s)=0 \quad \mbox{ pour } \quad s \le M^2_{seuil} \, .
$$
$M_{seuil}$ doit \^etre sup\'erieur \`a $m$ sinon la particule de masse
$m$ se d\'esint\`egrerait dans les particules associ\'ees \`a $M_{seuil}$.
Dans ce cas, cette particule n'appara\^ \i trait pas dans les \'etats
asymptotiques et la repr\'esentation spectrale ne serait plus valide. On
verra plus tard comment modifier cette repr\'esentation pour y inclure la
description des particules instables.

En y ins\'erant l'expression de la fonction spectrale (\ref{rho1part}),
le propagateur complet (\ref{spectral1}) devient
\begin{eqnarray}
  G(p^2) &=& \frac{i \, Z}{p^2 - m^2 + i\epsilon} \, 
  + \, \int^{ \infty}_{M^2_{seuil} } ds \,
  \frac{i \, \rho_{multi} (s)}{p^2 -s + i\epsilon}
  \nonumber
  \\
  &\equiv& G_{1part}(p^2) \, + \, G_{multi}(p^2) \, .
  \label{propstab}
\end{eqnarray}
Le p\^ole du propagateur complet s'interpr\`ete donc comme la masse physique
de l'\'etat \`a une particule cr\'e\'e par le champ. La constante $Z$ est le
r\'esidu au p\^ole du propagateur complet.

\section{Particule stable: propagation spatiale}

Rappelons que le but est d'\'etudier l'amplitude de propagation
(\ref{amplisimpli}), ce qui revient \`a analyser la d\'ependance spatiale du
propagateur en calculant sa transform\'ee de Fourier par rapport
\`a la tri-impulsion. Celle-ci sera not\'ee $G(p^0,\bf L)$:
$$
  G(p^0,{\bf L}) \equiv
   \int \frac{d^3p}{(2\pi)^3} \, e^{i \, {\bf p \cdot \bf L} } \, G(p^2)
  = \int \frac{d^3p}{(2\pi)^3} \, e^{i \, {\bf p \cdot \bf L} } \,
    \left( G_{1part}(p^2) \, + \, G_{multi}(p^2) \right) \, .
$$
A partir du propagateur (\ref{propstab}) d'un champ cr\'eant une particule
stable, on obtient
\begin{eqnarray}
  G_{1part}(p^0,\bf L)
  &=& \int \frac{d^3p}{(2\pi)^3} \, e^{i \, \bf p \cdot \bf L} \,
  \frac{i \, Z}{p^2 - m^2 + i\epsilon}
  \nonumber
  \\ &=&
  \int_0^{2\pi} d\varphi \int_0^\pi d\theta \sin \theta
  \int_0^\infty \frac{ {\bf p}^2 \, d|{\bf p}| }{(2\pi)^3} \,
  e^{i \, |{\bf p}| \, L \, \cos \theta } \,
  \frac{i \, Z}{ {p^0}^2 - {\bf p}^2 - m^2 + i\epsilon}
  \nonumber
  \\ &=&
  \frac{Z}{4\pi^2 \, L} \, \int^\infty_0 |{\bf p}| \, d|{\bf p}| \,
  \frac{ e^{i \, |{\bf p}| L} - e^{- i \, |{\bf p}| \, L} }
       { {p^0}^2 - {\bf p}^2 - m^2 + i\epsilon }
  \nonumber
  \\ &=& 
  \frac{Z}{4\pi^2 \, L} \, \int^\infty_{-\infty} q dq \,
  \frac{ e^{i q L} }{ {p^0}^2 - q^2 - m^2 + i\epsilon}
  \nonumber
  \\ &=&
  \frac{-i Z}{4\pi L} \, e^{i p_m L} \, ,
  \label{propstabpole}
\end{eqnarray}
o\`u $L \equiv |{\bf L}|$ et $p_m \equiv \sqrt{ {p^0}^2 \!- m^2 + i \epsilon}$.
La derni\`ere \'etape a \'et\'e calcul\'ee par int\'egrale de contour en
fermant le contour par le haut.

Si ${p^0}^2 \! - m^2 < 0$, $p_m = i \, \sqrt{m^2 - {p^0}^2 } $ et l'amplitude
d\'ecro\^\i t exponentiellement en fonction de la distance.

Si ${p^0}^2 \! - m^2 > 0$, la d\'ecroissance est en $L^{-1}$ qui est le facteur
g\'eom\'etrique attendu, de sorte que la probabilit\'e de
d\'etection sur un \'el\'ement de sph\`ere de rayon $L$ d\'ecroisse en
$L^{-2}$.

De m\^eme,
\begin{equation}
  G_{multi}(p^0,{\bf L}) = \frac{-i}{4\pi \, L} \, \int^\infty_{ M^2_{seuil} }
  ds \, \rho_{multi}(s) \, e^{i p_s L} \, ,
  \label{propstabmulti}
\end{equation}
o\`u $p_s \equiv \sqrt{ {p^0}^2 \! - s + i \epsilon}$.

En rassemblant les deux termes (\ref{propstabpole}) et (\ref{propstabmulti}),
on obtient
$$
  G(p^0,{\bf L}) =  \frac{-i}{4\pi \, L}
  \left( Z \, e^{i p_m L} \,
        + \int^\infty_{ M^2_{seuil} } ds \, \rho_{multi}(s) \,
          e^{i p_s L} \right) \, .
$$
Par le lemme de Riemann, le deuxi\`eme terme tend vers z\'ero lorsque la
distance $L$ tend vers l'infini. En effet, on peut effectuer la d\'ecomposition
suivante:
\begin{eqnarray*}
  G_{multi}(p^0,{\bf L})
  &=& -\frac{i}{4\pi \, L} \, \int^{{p^0}^2}_{ M^2_{seuil} }
      ds \, \rho_{multi}(s) \, e^{i p_s L}
      -\frac{i}{4\pi \, L} \, \int^\infty_{{p^0}^2}
      ds \, \rho_{multi}(s) \, e^{-\sqrt{s-{p^0}^2} L}
  \\
  &\equiv& G_{multi,I}(p^0,{\bf L}) + G_{multi,II}(p^0,{\bf L}) \, .
\end{eqnarray*}
Notons que $G_{multi,I}=0$ si ${p^0}^2 \le M^2_{seuil}$ puisque
$\rho_{multi}(s)=0$ pour $s \le M^2_{seuil}$.
Sous le changement de variable $s \rightarrow \omega\equiv p_s$,
$G_{multi,I}$ devient
$$
  G_{multi,I}(p^0,{\bf L}) =
   -\frac{i}{4\pi \, L} \, \int^{p_{seuil}}_0
      d\omega \, f(\omega) \, e^{i \omega L} \, ,
$$
o\`u $f(\omega) \equiv 2\omega \rho_{multi}({p^0}^2-\omega^2)$ et
$p_{seuil} \equiv \sqrt{ {p^0}^2 - M^2_{seuil} }$.
Une int\'egrale par parties donne
$$
  G_{multi,I}(p^0,{\bf L}) = \left( \frac{-i}{4\pi \, L} \right) \,
  \left( \frac{i}{L} \right) \,
  \int_0^{ p_{seuil} } d\omega \, \frac{df(\omega)}{d\omega}
  \, e^{i \omega L} \, ,
$$
o\`u l'on a utilis\'e le fait que
$ f(p_{seuil}) = 2 p_{seuil} \rho_{multi}(M^2_{seuil}) =0 $.
De m\^eme, sous le changement de variable
$s \rightarrow \alpha \equiv \sqrt{ s - {p^0}^2 }$, $G_{multi,II}$ devient
$$
  G_{multi,II}(p^0,{\bf L}) =
   -\frac{i}{4\pi \, L} \, \int^\infty_0
      d\alpha \, g(\alpha) \, e^{-\alpha L} \, ,
$$
o\`u $g(\alpha) \equiv 2\alpha \rho_{multi}( \alpha^2 + {p^0}^2 )$.
Une int\'egrale par parties donne
$$
  G_{multi,II}(p^0,{\bf L}) = \left( \frac{-i}{4\pi \, L} \right) \,
  \left( \frac{1}{L} \right) \,
  \int_0^\infty d\alpha \, \frac{dg(\alpha)}{d\alpha}
  \, e^{-\alpha L} \, ,
$$
o\`u l'on a utilis\'e le fait que $ g(0) = 0$ et que $\rho_{multi}(s)$ ne
diverge pas \`a l'infini.
En r\'ep\'etant le processus $J$ fois, on obtient
$$
  G_{multi}(p^0,{\bf L}) = \frac{-i}{ 4\pi \, L^{J+1} }
  \left(
           i^J \, \int_0^{ p_{seuil} } d\omega \, \frac{d^J f(\omega)}{d\omega^J}
            \, e^{i \omega L}
          + \int_0^\infty d\alpha \, \frac{d^J g(\alpha)}{d\alpha^J}
           \, e^{-\alpha L}
  \right) \, .
$$ 
Ce proc\'ed\'e est
r\'ep\'etable autant de fois que les fonctions $f(\omega)$ et $g(\alpha)$
(c'est-\`a-dire la fonction $\rho_{multi}$) sont d\'erivables.
La d\'ecroissance en $L$ est donc plus rapide que n'importe quelle puissance
inverse de $L$ si ces fonctions sont infiniment d\'erivables.
Mais $\rho_{multi}$ n'est pas infiniment d\'erivable aux seuils
$s_{\SS N}$, qui
sont les points o\`u les \'etats \`a $N$ particules commencent \`a contribuer.
La valeur de $J$ correspondant au seuil $s_{\SS N}$ se calcule en observant
que pr\`es de ce seuil \cite{brown},
$$
  \rho_{multi}(s) \sim \left( \sqrt{s - s_{\SS N} } \right)^{3N-5} \, .
$$
En d\'eveloppant $p_s$ autour de ce seuil, on calcule que la contribution \`a
$G_{multi}(p^0,\bf L)$ due au seuil $s_{\SS N}$ a un comportement asymptotique
pour $L$ grand en
\begin{equation}
  G_{Npartic}(p^0, {\bf L} ) \sim 
  \frac{const.}{L} \,
  (\mu L)^{ -\frac{3}{2} \, (N-1) } \exp ( i p_{ s_{\SS N} } L ) \, ,
  \label{stabcorr}
\end{equation}
o\`u
$$
  p_{ s_{\SS N} } \equiv \sqrt{{p^0}^2 \! - s_{\SS N} + i\epsilon} \, .
$$
Si $s_{\SS N} \!>\! {p^0}^2$, la racine choisie a une partie imaginaire
positive.
La masse $\mu$ carac\-t\'erisant l'\'echelle d'\'energie a \'et\'e
introduite pour raison dimensionnelle.
La contribution des seuils n'est importante qu'en de\c{c}\`a de
$\mu L \approx 1$, ce qui pour
$\mu \approx 500 $ MeV correspond \`a
$L \approx 10^{-16}$ m. 
Ce terme est donc n\'egligeable par rapport \`a $G_{1part}(p^0,\bf L)$
lorsque l'on \'etudie la propagation sur des distances macroscopiques.

Retournant \`a la formule (\ref{amplisimpli}) de l'amplitude de propagation et
y ins\'erant l'expression (\ref{propstabpole}), on
conclut que la propagation d'une particule stable sur une distance
macroscopique $L$ est donn\'ee par
\begin{equation}
  \fbox{$\displaystyle
  {\cal A}(L) \sim M_P M_D \, 2\pi \,
  \delta \left( E_{D_I} \! -\! E_{D_F} \! +\! E_{P_I} \!  -\! E_{P_F} \right) \,
  \frac{-i Z}{4\pi L} \, e^{i p_m L} \, ,
  $}
\end{equation}
o\`u $p_m \equiv \sqrt{E^2 - m^2}$ et
\mbox{$E \equiv E_{P_I} \!  -\! E_{P_F} \ge m$}.

\section{Particule instable: repr\'esentation spectrale}

On a vu plus haut que les p\^{o}les du propagateur $G(p^2)$ correspondent \`a
des \'etats stables \`a une particule. Il est indispensable pour la
stabilit\'e de la particule qu'ils soient inf\'erieurs au seuil de
contribution des \'etats \`a plusieurs particules. Pourrait-on d\'ecrire les
particules instables en violant cette condition? Notons tout de suite que ce
p\^{o}le ne peut \^etre r\'eel sinon la propagation spatiale serait du m\^eme
type que celle de la particule stable, alors que l'on s'attend \`a une
d\'ecroissance de l'amplitude de propagation, due \`a la d\'esint\'egration
possible de la particule. Supposons donc que le propagateur $G(z)$
ait un p\^{o}le complexe $z_0$ dont la partie r\'eelle est
sup\'erieure \`a $M_{seuil}$. Pour des valeurs r\'eelles de $z=p^2$, le
propagateur ne pr\'esentera pas de p\^{o}le mais un pic en $p^2={\cal R}\!e \, z_0$
d'autant plus \'elev\'e que la partie imaginaire du p\^{o}le est petite. Il est
tentant d'interpr\'eter la partie r\'eelle du p\^{o}le comme la masse de la
particule instable tandis que la partie imaginaire devrait \^etre reli\'ee au
temps de vie (puisque son annulation implique un temps de vie infini).

Il reste un probl\`eme: comme prouv\'e plus haut, la repr\'esentation spectrale
du propagateur n'admet pas de p\^{o}les complexes! N\'eanmoins, la coupure sur
l'axe des r\'eels permet un prolongement analytique
de $G(z)$ \`a travers la coupure sur d'autres feuilles de Riemann
\cite{peierls}. Il suffit que le p\^{o}le se trouve sur une de ses feuilles
de Riemann  suppl\'ementaires et la repr\'esentation spectrale n'est plus
viol\'ee\footnote{Toutefois, la correspondance univoque entre les particules
instables et les p\^oles complexes de la matrice $S$ (qui apparaissent ici dans
le propagateur) n'est pas d\'emontr\'ee dans l'approche axiomatique de la
matrice $S$. Il est possible \cite{fonda} de construire un mod\`ele
o\`u une r\'esonance de type Breit-Wigner ne correspond \`a aucun p\^ole et
vice versa. Par exemple, on pourrait appliquer ce mod\`ele \`a la r\'esonance
scalaire $f_0(400-1200)$, si l'on ne veut pas l'ins\'erer dans l'octet scalaire
du groupe $SU(3)_f$.}.

Partons de la deuxi\`eme repr\'esentation spectrale
(\'equation (\ref{spectral2}))
du propagateur qui s'\'ecrit
$$
  G(z) = \frac{i}{z - M_0^2 - \Pi(z)}
$$
et rappelons que (\'equation (\ref{imagenergie}))
$$
   {\cal I}\!m \, \Pi(p^2)
   = \lim_{z\to p^2 +i\epsilon} \,  {\cal I}\!m \, \Pi(z)
   = - \pi \, p^2 \, \sigma(p^2)
   \le 0 \, .
$$
Si la particule n'est pas extr\^emement instable, la partie imaginaire du
p\^{o}le sera petite, c'est-\`a-dire que la partie imaginaire de la solution de
$z - M_0^2 - \Pi(z)$ sera petite et donc n\'egative en vertu de la
propri\'et\'e ${\cal I}\!m \, \Pi(p^2) \le 0 $. On va donc effectuer le prolongement
de $G(z)$ \`a travers la coupure {\it vers le bas} ce qui facilite les
choses puisque le propagateur est connu juste au-dessus de la coupure.

Le prolongement analytique de $G(z)$ est d\'efini tel que la fonction
analytiquement continu\'ee $G_{II}(z)$ juste en dessous de l'axe r\'eel soit
\'egale \`a la fonction originale $G(z)$ juste au-dessus de l'axe r\'eel. On
prolonge donc de fa\c {c}on analogue $\Pi(z)$ en $\Pi_{II}(z)$ et
$G_{II}(z)$ s'\'ecrit
$$
  G_{II}(z) = \frac{i}{ z - M_0^2 - \Pi_{II}(z) } \, .
$$
Le p\^{o}le $z_0$ est solution de l'\'equation
$$
  z_0 - M_0^2 - \Pi_{II}(z_0) = 0 \, .
$$
Le propagateur complet renormalis\'e peut aussi \^etre prolong\'e analytiquement
et l'\'equation (\ref{propren}) devient
\begin{equation}
   G_{II}(z) = \frac{i}{ z - M^2 - f_{II}(z) } \, .
  \label{propinstab}
\end{equation}
o\`u $M$ est la masse renormalis\'ee et $f_{II}(z)$ le prolongement
analytique de la fonction $f(p^2)$ qui est la partie restante de l'\'energie
propre apr\`es renormalisation de la masse (voir \'equation (\ref{propren})).

On aimerait disposer d'une expression \'equivalente o\`u le p\^ole complexe
du propa\-gateur appara\^{\i}t explicitement au d\'enominateur.
Dans ce but, l'\'energie propre est d\'evelopp\'ee autour du
p\^{o}le $z_0$:
$$
  \Pi_{II}(z) = \Pi_{II}(z_0) + \left( z-z_0 \right) \, \Pi'_{II}(z_0)
                + h(z) \, \left( 1 - \Pi'_{II}(z_0) \right) \, ,
$$
avec $h(z_0)=0$ et $h'(z_0)=0$. Les primes indiquent les d\'eriv\'ees par rapport
\`a $z$.

Le d\'enominateur du propagateur se r\'e\'ecrit
\begin{eqnarray*}
  z - M_0^2 - \Pi_{II}(z)
  &=& \left( 1 - \Pi'_{II}(z_0) \right) \, \left( z - z_0 - h(z) \right)
  \\
  &=& Z^{-1} \left( z - m^2 + i m \Gamma - h(z) \right) \, ,
\end{eqnarray*}
o\`u
\begin{equation}
  Z^{-1} \equiv 1 - \Pi'_{II}(z_0)
 \qquad \mbox{et} \qquad
  z_0 \equiv m^2 - i m \Gamma \, .
  \label{zrenorm}
\end{equation}
Avec ces d\'efinitions, le propagateur complet peut se r\'e\'ecrire pr\`es du p\^{o}le comme
$$
  G_{II}(z) = \frac{i \, Z}{ z - m^2 + i m \Gamma - h(z) } \, .
$$
La constante $Z$, appel\'ee
{\it constante de renormalisation de la fonction d'onde}, est le r\'esidu du
propagateur complet au p\^ole. Bien que le r\'esidu du propagateur de la
particule stable ait pu \^etre interpr\'et\'e comme la probabilit\'e de cr\'eer
un \'etat \`a une particule, ce n'est pas le cas pour la particule instable
puisque la constante $Z$ est complexe.

\section{Particule instable: propagation spatiale}

Le mod\`ele simplifi\'e ne convient pas tout \`a fait \`a la description
de la propagation d'une particule instable en raison de l'approximation
d'\'etat stationnaire. Il sera rem\'edi\'e \`a cet inconv\'enient dans le prochain
chapitre. Cependant, cette inconsistance n'invalide pas les le\c{c}ons que
l'on peut tirer de l'analyse de l'amplitude (\ref{amplisimpli}) du mod\`ele
simplifi\'e qui,
rappelons-le, est proportionnelle \`a la transform\'ee de Fourier du propagateur
complet. Bien s\^ur, ce n'est que du point de vue du mod\`ele plus sophistiqu\'e
que l'on pourra en juger.

L'\'etude de la propagation spatiale ressemble \`a celle de la particule stable.
Avec l'objectif d'utiliser le th\'eor\`eme des r\'esidus, on va r\'e\'ecrire le
propagateur comme une somme d'une fonction singuli\`ere au p\^{o}le mais
analytique ailleurs et une fonction r\'eguli\`ere au p\^{o}le mais non
analytique:
$$
  G_{II}(z) = G_{1part}(z) + G_{multi}(z)
$$
o\`u
\begin{eqnarray*}
  G_{1part}(z) &\equiv& \frac{ i \, Z}{ z - m^2 + i m \Gamma }
  \\
  G_{multi}(z) &\equiv&   \frac{i \, Z}{ z - m^2 + i m \Gamma - h(z) }
                     - \frac{ i \, Z}{ z - m^2 + i m \Gamma } \, .
\end{eqnarray*}
$G_{multi}(z)$ est une fonction r\'eguli\`ere au p\^{o}le
$z_0=m^2-i \, m \Gamma$ mais non analytique \`a cause des seuils o\`u les
\'etats \`a plusieurs particules commencent \`a contribuer: pour un seuil
donn\'e, il existe un $J$ tel que la
$J${\footnotesize\`eme} d\'eriv\'ee de $G_{multi}(z)$ est
discontinue.

La transform\'ee de Fourier du propagateur par rapport \`a sa tri-impulsion est
$$
  G(p^0,{\bf L}) = \int \frac{d^3p}{(2\pi)^3} \, e^{i \, {\bf p \cdot \bf L} }
  \left( \frac{ i \, Z}{ p^2 - m^2 + i m \Gamma } \,
  + \, G_{multi}(p^2) \right) \, .
$$
En faisant les m\^emes calculs que pour la particule stable
(\'equations (\ref{propstabpole}) et (\ref{stabcorr})), on obtient
\begin{equation}
  G(p^0,{\bf L}) =
    \frac{-i Z}{4\pi L} \, e^{i \sqrt{ {p^0}^2 - m^2 +im\Gamma } L }
  + \sum_{N \ge 2} d_{\SS N} \, \frac{e^{i p_{ s_{\SS N} } L } }{L} \,
    (\mu_{\SS N} L)^{ -\frac{3}{2} \, (N-1) } \, ,
  \label{instabcorr}
\end{equation}
o\`u les d\'efinitions de $s_{\SS N}$, $p_{ s_{\SS N} }$ et $L$ sont
les m\^emes que pour la particule stable et les $d_{\SS N}$ sont des coefficients
qui ne peuvent \^etre calcul\'es que dans un mod\`ele particulier. $\mu_{\SS N}$
correspond \`a $\mu$ pour l'\'etat $N$.

On a utilis\'e l'hypoth\`ese que la particule n'est pas extr\^emement instable
pour n\'egliger les termes en $\Gamma/m$, sauf dans l'exponentielle o\`u ils
sont multipli\'es par $L$ et contribuent notablement \`a grande distance.
Sinon, il n'y a de
toute fa\c {c}on pas de propagation macroscopique \`a observer car la particule
se d\'esint\`egre trop vite. Par contre, si
la particule a un long temps de vie, le p\^{o}le se situe pr\`es de l'axe
r\'eel et $\Gamma/m \ll 1$.

Le premier terme peut \^etre identifi\'e comme l'amplitude de propagation de la
particule instable: on a bien une d\'ecroissance exponentielle avec la distance.
Sous quelles conditions ce terme est-il dominant par rapport \`a la contribution
des \'etats \`a plusieurs particules? Sans perte de g\'en\'eralit\'e, posons
$\mu_{\SS N} = |p^0|$.

\begin{enumerate}

\item
Pour $m^2 - {p^0}^2 \gg m \Gamma$, ce terme a une d\'ecroissance exponentielle
tr\`es rapide selon la distance en $\exp(- \sqrt{ m^2 -{p^0}^2 } L)$ et la
propagation spatiale macroscopique de la particule instable est inobservable:
$G_{multi}$ domine $G_{1 part}$.

\item
Pour
$|{p^0}^2 - m^2 | \approx {\cal O}(m \Gamma)$
ou plus petit, la d\'ecroissance de ce terme est exponentielle en
$\exp(- \alpha \sqrt{ m \Gamma } L)$,
o\`u $\alpha$ est un nombre d'ordre un. La d\'ecroissance est plus lente que
dans le premier cas.
Ce terme domine $G_{multi}$ sur des distances interm\'ediaires (voir cas suivant) 
mais l'interpr\'etation de $\Gamma$ comme l'inverse du temps de vie est ici
incorrecte. Ce domaine est cependant tr\`es petit dans la majorit\'e des cas
puisque
$\Gamma/m \ll 1$.

\item
Pour
${p^0}^2 \! - m^2 \gg m \Gamma$, on peut d\'evelopper l'argument de
l'exponentielle en
$m\Gamma/p_m$, o\`u
$p_m \equiv \sqrt{ {p^0}^2 - m^2 }$.
La d\'ecroissance selon la distance est une exponentielle plus mod\'er\'ee:
$$
  G_{1part}(p^0,{\bf L}) =
    \frac{-i Z}{4\pi L} \,
    \exp \left(i p_m L - \frac{m\Gamma }{2 p_m} L \right) \, .
$$
Comme dans le cas d'une particule stable, ce terme sera domin\'e par $G_{multi}$
pour des distances petites ($\mu L \le 1$), mais il sera aussi
domin\'e \`a grande distance en raison de sa d\'ecroissance exponentielle, alors
que $G_{multi}$ ne d\'ecro\^\i t qu'en puissance de $L$.
Pour des distances interm\'ediaires, $G_{1part}$ est dominant.
Si l'on utilise la relation $L = v \, T$  valable pour une trajectoire classique,
avec $v = p_m/|p^0|$,
on retrouve la forme habituelle de la d\'ecroissance de l'amplitude de
propagation en fonction du temps de propagation $T$, mais sous forme relativiste:
$$
  G_{1part}(p^0,{\bf L})
  \sim \exp \left( - \frac{m\Gamma}{2 p_m} L \right)
  \sim \exp \left( - \frac{m\Gamma}{2 |p^0|} T\right)
  \sim \exp \left(- \frac{\Gamma \tau}{2} \right)  \, ,
$$
o\`u $\tau$ est le temps propre de la particule.
La largeur $\Gamma$ s'interpr\`ete dans cas comme l'inverse du temps de
vie \cite{peierls}.
\end{enumerate}

La contribution de $G_{multi}$ est-elle observable \`a grande distance?
Consid\'erons la contribution du seuil de production de deux particules.
$G_{multi}$ commence \`a dominer $G_{1part}$ lorsque
$$
  C_1 \, \exp \left(-\frac{m\Gamma}{2 p_m} L \right)
  \le C_2 \, ( |p^0| \, L)^{ -\frac{3}{2} } \, ,
$$
o\`u $C_1$ et $C_2$ regroupent des constantes.

Prenant le logarithme de cette expression, les facteurs constants $C_1$ et $C_2$
auront une contribution n\'egligeable pour $L$ grand et l'on obtient la
condition
$$
  \frac{m\Gamma L}{p_m} \ge 3 \ln (|p^0| L) \, .
$$
Pour des particules se d\'esint\'egrant faiblement (ou {\it quasi-stables}),
on peut se restreindre au domaine spatial
$$
  \frac{\Gamma}{|p^0|} \ll \Gamma L \ll \frac{m}{\Gamma}
$$
car pour $\Gamma L \approx m/\Gamma \approx {\cal O}(10^{14})$ (kaons neutres),
$G_{1part}$ et $G_{multi}$ seront tous les deux inobservables.
La borne inf\'erieure du domaine vient de la condition $\mu L \gg 1$.
D\`es lors
$$
  3 \ln (|p^0| \,  L) = 3 \ln \left( \frac{|p^0|}{\Gamma} \Gamma L \right)
        \cong 3 \ln \left( \frac{|p^0|}{\Gamma} \right) \, .
$$
$G_{multi}$ domine $G_{1part}$ pour
$$
  \Gamma L \ge 3 \, \frac{p_m}{m} \ln (|p^0|/\Gamma) \, .
$$
Cette borne d\'epend de la valeur de $p_m$: moins la particule va vite, moins
elle va loin! Il est plus significatif de passer au temps de propagation par
$L=v \, T$. La condition devient
$$
  \Gamma \, T \ge 3 \, \frac{|p^0|}{m} \ln (|p^0|/\Gamma)
           \ge  3 \ln (M/\Gamma) \, .
$$
Si $M/\Gamma \approx  {\cal O}(10^{14})$, $\Gamma \, T > 97$, ce qui est
inobservable.

La contribution des \'etats \`a plusieurs particules est donc
inobservable pour les particules quasi-stables. Le r\'egime o\`u la
propagation de la particule est bien d\'ecrite par le propagateur $G_{1part}$
est fix\'e par les conditions
\begin{eqnarray*}
  \mu L &\ge& 1 \, ,
  \\
  \Gamma \, T &\le&  3 \ln (M/\Gamma) \, .
\end{eqnarray*}

En conclusion, l'amplitude de propagation (\ref{amplisimpli}) d'une particule
quasi-stable sur une
distance macroscopique $L$ est donn\'ee en tr\`es bonne approximation par
\begin{equation}
  \fbox{$\displaystyle
  {\cal A} \sim M_P M_D \, 2\pi \,
  \delta \left( E_{D_I} \! -\! E_{D_F} \! +\! E_{P_I} \!  -\! E_{P_F} \right) \,
  \frac{-i Z}{4\pi L} \,
  \exp \left(i p_m L - \frac{m\Gamma }{2 p_m} L \right) \, ,
  $}
  \label{instabpole}
\end{equation}
o\`u $p_m \equiv \sqrt{E^2-m^2}$ et
\mbox{$E \equiv E_{P_I} \!  -\! E_{P_F} \ge m$}.

\section{Exemple: propagateur complet du kaon}
\label{exemplepropagkaon}

Il sera utile par la suite de disposer de formules explicites dans un
cas concret.
Consid\'erons le kaon neutre, de masse $M_0$, en interaction avec des pions
charg\'es de masse $m$ ($M_0 > 2m$). Le lagrangien d'interaction s'\'ecrit
$$
  {\cal L} = - g K\pi\pi^*  \, .
$$
Le calcul \`a une boucle de l'\'energie propre du kaon donne, pour $s$
quelconque
$$
  -i \, \Pi(s) = -i \,\delta M^2 - \frac{i g^2}{4 \pi^2}
  \left( 1 - \frac{4 m^2_\pi}{s} \right)^\frac{1}{2}
  \ln \frac{ ( 1-4 m^2_\pi/s )^\frac{1}{2} + 1 }
           { ( 1-4 m^2_\pi/s )^\frac{1}{2} - 1 } \, ,
$$
o\`u $\delta M^2$ est une constante divergente qui vaut en r\'egularisation
dimensionnelle
$$
  \delta M^2 \equiv - \frac{2}{\epsilon} - \ln 4\pi + \gamma
  + \ln \frac{m^2_\pi}{\mu^2_R} - 2 \, ,
$$
Dans cette expression, $\epsilon = 2-d/2$, $\mu_R$ est le point de
renormalisation et $\gamma$ la constante d'Euler.
$\delta M^2$ est incorpor\'ee dans la masse renormalis\'ee:
$$
  M^2 \equiv M^2_0 + \delta M^2 \, .
$$ 
Le propagateur renormalis\'e (\ref{propren}) s'\'ecrit
$$
  G(s) = \lim_{z \to s + i \epsilon} \frac{i}{z - M^2 - f(z)}
$$
avec
$$
  f(z) =  \frac{g^2}{4 \pi^2}
  \left( 1 - \frac{4 m^2_\pi}{z} \right)^\frac{1}{2}
  \ln \frac{ (1-4 m^2_\pi/z)^\frac{1}{2} +1 }
           { (1-4 m^2_\pi/z)^\frac{1}{2}-1 } \, .
$$
Pour $z$ r\'eel, la fonction $f(z)$ a une partie imaginaire non nulle si
$z$ est sup\'erieur \`a $4 m^2$ et on a une coupure sur l'axe r\'eel pour
$s \ge 4 m^2_\pi$:
$$
  f(s \pm i \epsilon) = u(s) \mp i \, v(s) \, ,
$$
avec
\begin{eqnarray*}
  u(s) &=& \frac{g^2}{4 \pi^2}
  \left( 1 - \frac{4 m^2_\pi}{s} \right)^\frac{1}{2}
  \ln \frac{ ( 1-4 m^2_\pi/s )^\frac{1}{2} + 1 }
           { ( 1-4 m^2_\pi/s )^\frac{1}{2} - 1 } \, ,
  \\
  v(s) &=& \frac{g^2}{4 \pi}
           \left( 1 - \frac{4 m^2_\pi}{s} \right)^\frac{1}{2} \, .
\end{eqnarray*}
La condition (\ref{imagenergie}) est satisfaite: 
\begin{equation}
   {\cal I}\!m \, \Pi(s)  = - \frac{g^2}{4 \pi} \,
                  \left( 1 - \frac{4 m^2_\pi}{s} \right)^\frac{1}{2}
                  \theta(s - 4 m^2_\pi)
                \le 0 \, .
  \label{defimpi}
\end{equation}
La fonction spectrale peut se calculer explicitement \`a partir de l'\'equation
(\ref{coupure}):
\begin{eqnarray}
  \rho(s+i\epsilon)
  &=& \frac{1}{\pi} \,  {\cal I}\!m \left( i \, G(s+i\epsilon) \right) \nonumber
  \\
  &=& - \frac{1}{ \pi} \,  {\cal I}\!m \frac{1}{s - M^2 - f(s+i\epsilon) } \nonumber
  \\
  &=& \frac{1}{ \pi} \, \frac{v(s)}{ \left( s-M^2 -u(s) \right)^2 + v^2(s) } \, .
  \label{rhokaon}
\end{eqnarray}
Selon l'\'equation (\ref{propinstab}), le prolongement analytique sur la
deuxi\`eme feuille de Riemann est donn\'e par
$$
  G_{II}(z) = \frac{i}{ z-M^2 - u(z) + i \, v(z) } \, .
$$
Le p\^{o}le $z_0$ est solution de
$$
  z_0 - M^2 - u(z_0) + i \, v(z_0) = 0 \, .
$$
Pour un couplage $g$ petit, $u(z_0)$ et $v(z_0)$ sont petits par rapport \`a
$M^2$ et $z_0 \cong M^2$.

Dans ce cas, l'expression du p\^{o}le au premier ordre en $g^2$ est donn\'ee par
$$
  z_0 \cong  M^2 + u(M^2) - i \, v(M^2) + {\cal O}(g^2)
      \equiv m^2_K - i \, m_K \, \Gamma \, .
$$
On a donc
\begin{eqnarray*}
  m^2_K &=& M^2 + u(M^2) + {\cal O}(g^2) \, ,
  \\
  m_K \, \Gamma &=& v(M^2) + {\cal O}(g^2) = v(m^2_K) + {\cal O}(g^2) \, ,
\end{eqnarray*}
c'est-\`a-dire
\begin{equation}
  \Gamma = \frac{g^2}{4 \pi \, m^2_K}
  \left( m^2_K - 4 m^2_\pi \right)^\frac{1}{2} \, .
  \label{largeurK} 
\end{equation}
On pourrait aussi calculer la constante $Z$ \`a l'aide de l'\'equation
(\ref{zrenorm}).
 
En ne tenant en compte que du seuil \`a deux pions, on peut \'ecrire
$$
   {\cal I}\!m \, \Pi(s+i \, \epsilon)
  = - v(s)
  = - \frac{ m^2_K \, \Gamma }{ \sqrt{s} }
   \left( \frac{s-4m^2_\pi}{m^2_K-4m^2_\pi} \right)^\frac{1}{2} \,
   \theta(s-4m^2_\pi) \, .
$$
Remarquons que si l'on veut seulement conna\^{\i}tre ${\cal I}\!m \, \Pi(s)$, il est plus
facile de le calculer \`a l'aide des r\`egles de
Veltman-Cutkosky \cite{veltman,cutkosky}.
Je trouve cependant int\'eressant de montrer comment tous les param\`etres de la repr\'esentation
spectrale peuvent \^etre \'evalu\'es dans un exemple concret.

\chapter{M\'elanges de propagateurs}
\label{melangesprop}

Au chapitre \ref{oscillationsMQ}, nous avons d\'efini en m\'ecanique quantique
la notion de m\'elange de particules dans le cas des oscillations par
l'impossibilit\'e  de faire co\"{\i}ncider la base d'\'etats en interaction
(\'etats propres de saveur) et la base d'\'etats propres de masse.
Les d\'efinitions de m\'elange et d'oscillation en th\'eorie des champs sont
analogues, except\'e le fait que ce ne sont plus les \'etats physiques qui sont
m\'elang\'es mais les champs. Cette distinction permettra d'\'eviter tous les
ennuis li\'es \`a la d\'efinition de bases d'\'etats propres de saveur et de
propagation.

Le lagrangien total est subdivis\'e en un lagrangien $L_{propag}$, d\'ecrivant la
propagation des particules, et un lagrangien $L_{int}$, d\'ecrivant les
interactions produisant ces particules. Ces deux parties du lagrangien total
peuvent \^etre distingu\'ees s'il existe une transformation dite
{\it de saveur}, laissant $L_{int}$ invariant mais modifiant $L_{propag}$,
\`a laquelle on associe le nombre quantique de saveur.
Il y a {\it m\'elange de particules}, si le propagateur construit \`a partir de 
$L_{propag}$ et repr\'esentant la cr\'eation d'une particule de saveur
$\alpha$ au point $x$ et l'annihilation d'une particule de saveur $\beta$
au point $x'$ est non diagonal, c'est-\`a-dire non nul pour $\alpha\neq\beta$.
Le lagrangien $L_{propag}$ contient toujours le terme cin\'etique et le terme de
masse. Dans le cas de particules instables, on y incorpore aussi l'interaction
provoquant la d\'esint\'egration.

Le type de lagrangien utilis\'e d\'epend du syst\`eme de particules \'etudi\'e.
La cons\-truction du propagateur la plus simple correspond toujours \`a des
particules stables dont on conna\^{\i}t le lagrangien fondamental.
Ce cas est illustr\'e par le syst\`eme
compos\'e des diff\'erents types de neutrinos. Le lagrangien $L_{propag}$
contient la matrice de masse (g\'en\'er\'ee par les interactions de Yukawa)
tandis que le lagrangien $L_{int}$ contient les interactions faibles (groupe
$SU(2)_L \times U(1)_Y$). Si l'on se place dans la base de saveur des champs,
le propagateur du syst\`eme des particules m\'elang\'ees sera non diagonal.
Comme on va le voir dans ce chapitre, il peut \^etre reli\'e au propagateur
diagonal de la base de masse par une transformation unitaire.

Une construction plus compliqu\'ee du propagateur est n\'ecessaire si les
particules auxquelles on s'int\'eresse sont instables.
Il se peut que l'on connaisse le lagrangien fondamental, comme cela pourrait
se produire si les neutrinos s'av\`erent instables et que l'on conna\^{\i}t les
interactions \`a l'origine de la d\'esint\'egration. Dans ce cas, il faut
incorporer dans $L_{propag}$ les interactions \`a l'origine de l'instabilit\'e.
Le propagateur libre est alors remplac\'e par le propagateur complet obtenu par
sommation sur l'\'energie propre comme nous l'avons vu au chapitre
pr\'ec\'edent. Ce propagateur complet est non diagonal et doit \^etre
diagonalis\'e par une transformation qui cette fois n'est pas n\'ecessairement
unitaire, puisque la matrice de l'\'energie propre du syst\`eme n'est pas
hermitienne quand les particules sont instables.

Une derni\`ere complication surgit si le lagrangien du syst\`eme n'est pas
compl\`e\-te\-ment connu, comme c'est le cas pour le syst\`eme des kaons neutres
ou des m\'esons {\it B}. Nous sommes en effet incapables de d\'eriver le lagrangien
de ces particules \`a partir du lagrangien fondamental de la QCD et des
interactions faibles. N\'eanmoins, il est possible de construire un lagrangien
effectif d\'ecrivant ces particules en employant comme contraintes les
diff\'erentes
sym\'etries du lagrangien fondamental de la QCD et des interactions faibles
ainsi que la fa\c{c}on dont elles sont \'eventuellement bris\'ees. Le terme
{\it lagrangien effectif} signifie que ce lagrangien ne contient que les
particules observables \`a basse \'energie que sont les m\'esons pseudoscalaires
et \'eventuellement les m\'esons vectoriels.
Le lagrangien effectif contient un nombre infini de termes qui
sont class\'es en un d\'eveloppement perturbatif en l'\'energie du processus.
On parle de {\it th\'eorie chirale perturbative}.
Chaque terme de ce lagrangien repr\'esente une sommation sur un
nombre infini de diagrammes de Feynman de la th\'eorie fondamentale sous-jacente,
chaque diagramme respectant la structure du terme en question.
Les diff\'erentes interactions fondamentales sont donc m\'elang\'ees dans chaque
terme du lagrangien effectif. On d\'efinit la {\it partie forte} du lagrangien
effectif comme l'ensemble des termes fournissant les contributions dominantes
aux processus respectant les sym\'etries de la QCD. Le {\it terme de masse} est
le terme non d\'erivatif quadratique en les champs. Il est g\'en\'er\'e au
niveau fondamental \`a la fois par les interactions \'electrofaibles et les
interactions de la QCD. Dans la m\^eme logique, la {\it partie faible} du
lagrangien effectif est d\'efinie comme l'ensemble des termes g\'en\'erant les
processus violant les sym\'etries de la QCD et de l'\'electromagn\'etisme mais
respectant les sym\'etries des interactions faibles au niveau fondamental. La
partie \'electromagn\'etique du lagrangien effectif est d\'efinie de fa\c{c}on
analogue.
Les param\`etres de ce lagrangien effectif sont fix\'es par des r\'esultats
exp\'erimentaux car leur pr\'ediction th\'eorique n'est pas tr\`es satisfaisante
en raison de la difficult\'e \`a \'evaluer les \'el\'ements de
matrice hadroniques ainsi que les corrections \`a longue distance
\cite{buchalla}.
Le traitement du syst\`eme $B^0\overline{B^0}$ se fait \`a une autre \'echelle
d'\'energie o\`u la th\'eorie chirale perturbative n'est plus valable.
N\'eanmoins, la partie \`a courte distance de la machinerie utilis\'ee pour
calculer les param\`etres du syst\`eme $K^0\overline{K^0}$ (op\'erateurs
effectifs \`a quatre quarks et coefficients de Wilson) est imm\'ediatement
transposable aux $B^0\overline{B^0}$.

Dans le but de d\'ecrire les oscillations des kaons, le lagrangien effectif peut
\'egalement \^etre 
s\'epar\'e en un lagrangien $L_{int}$ contenant la partie forte du lagrangien
effectif et un lagrangien $L_{propag}$ contenant le terme cin\'etique, le terme
de masse (contenant des masses d\'eg\'en\'er\'ees puisque le lagrangien d\'ecrit
une particule et son antiparticule) ainsi que la partie faible du lagrangien
effectif \`a l'origine de la d\'esint\'egration des kaons.
De nouveau, le propagateur complet est construit par sommation sur l'\'energie
propre. Une diff\'erence de masse appara\^{\i}t alors entre
les kaons se propageant et se manifeste par la non d\'eg\'en\'erescence des
parties r\'eelles des p\^oles du propagateur complet du syst\`eme des kaons.
Cette diff\'erence de masse peut \^etre attribu\'ee aux interactions faibles
(m\'ecanisme GIM inclus!) avec le bon ordre de grandeur \cite{okun}.

Tout d'abord, nous expliquerons comment diagonaliser le propagateur de
particules stables si le lagrangien est connu.
Ensuite nous montrerons comment calculer le propagateur complet pour des
particules instables en m\'elange et comment le dia\-go\-na\-liser.
Le syst\`eme $K^0\overline{K^0}$ sera examin\'e comme exemple.
Enfin, la probabilit\'e d'oscillation sera calcul\'ee dans le
cadre du mod\`ele simplifi\'e pr\'esent\'e au chapitre pr\'ec\'edent et les
d\'eficiences du mod\`ele seront identifi\'ees.

\section{Propagateur de particules stables en m\'elange}

Dans cette section, nous expliquons comment diagonaliser le propagateur
de particules scalaires stables si leur lagrangien est connu (le cas des
fermions est discut\'e \`a la section \ref{oscifermion}).
Le lagrangien $L_{propag}$ contient uniquement le terme cin\'etique et le terme de masse.
La matrice de masse est hermitienne et peut \^etre diagonalis\'ee par une
transformation unitaire $V$ sur les champs\footnote{Remarquons que si l'on
d\'efinissait des \'etats comme au chapitre \ref{oscillationsMQ}
(\'equation (\ref{chgtbase})), la relation ci-dessus donnerait 
$
  |\nu_\alpha (0) \!> = \sum_{j} \, V^t_{\alpha j} \, |\nu_j (0) \!> \, ,
$
c'est-\`a-dire que la matrice $V$ diagonalisant le terme de masse est la
transpos\'ee de celle du \mbox{chapitre \ref{oscillationsMQ}}.}:
\begin{equation}
  \nu_\alpha = \sum_j \, V^\dagger_{\alpha j} \, \nu_j \, ,
  \label{relchamp}
\end{equation}
o\`u les indices grecs d\'esignent toujours la base de saveur et les indices
latins la base de masse.

Le propagateur est d\'efini par la fonction \`a deux points ordonn\'ee dans le
temps:
$$
  G_{\beta\alpha}(x'-x) \equiv
  <\! 0 \,|T \left( \nu_\beta(x') \: \nu^*_\alpha(x) \right) |\, 0 \!> \; .
$$
Comme la contraction de champs (th\'eor\`eme de Wick) ne s'applique que sur
des champs repr\'esentant des particules de masses d\'etermin\'ees, il faut
substituer dans la fonction \`a deux points la relation (\ref{relchamp}):
$$
  G_{\beta\alpha}(x'-x) =
  \sum_{j,k} \, V^\dagger_{\beta k} \, V_{j\alpha} \,
  <\! 0 \,|T \left( \nu_k(x') \: \nu^*_j(x) \right) |\, 0 \!> \; .
$$
La contraction des champs donne
\begin{equation}
  G_{\beta\alpha}(x'-x) =
  \sum_j \, V^\dagger_{\beta j} \,
  G_{D,jj}(x'-x)
   \, V_{j\alpha} \, ,
  \label{stablenondiag}
\end{equation}
o\`u $G_{D,jj}(x'-x)$ est le propagateur ($D$ pour diagonal) d'une particule
scalaire libre de masse $m_j$:
\begin{equation}
  G_{D,jj}(x'-x) \equiv
  \int \, \frac{d^4p}{(2\pi)^4} \, e^{-i p (x'-x)} \,
  \frac{i}{p^2 -m_j^2 +i \epsilon} \, .
  \label{stablediag}
\end{equation}

\section{Propagateur de particules instables en m\'elange}

Le propagateur complet de particules m\'elang\'ees est une matrice non
diagonale. Il s'obtient \`a partir du propagateur en l'absence de m\'elange par
sommation sur l'\'energie propre comme dans le cas des particules non
m\'elang\'ees (\'equation (\ref{seriegeom})). Cette sommation
s'effectue en passant par l'\'equation de Dyson \cite{baulieu},
qui s'exprime sous forme diagrammatique par
\begin{center}
\includegraphics[width=14cm]{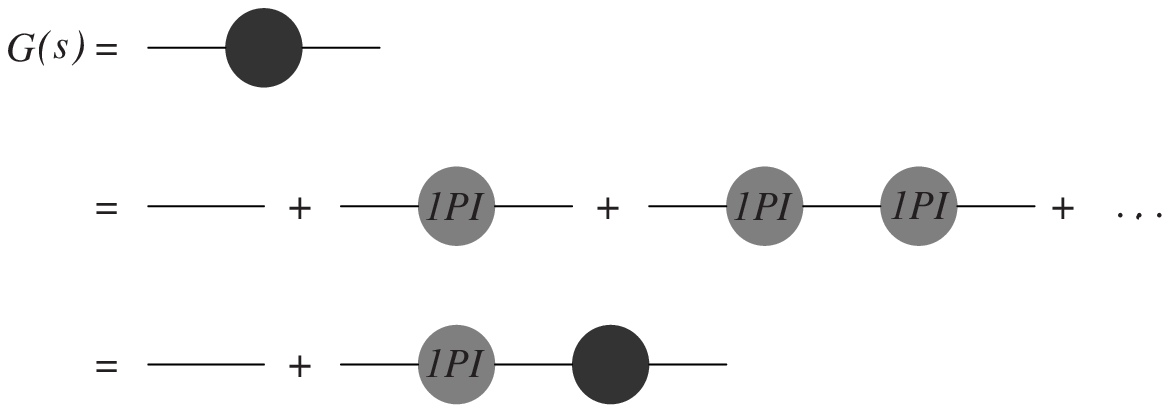}
\end{center}
ou sous forme matricielle
$$
  G^{ij} = G^{ij}_F + G^{ik}_F \left( -i \, \Pi^{kl} \right) G^{lj} \, ,
$$
o\`u $G^{ij}_F$ est le propagateur de Feynman des particules libres (donc
non m\'elang\'ees) et $-i \, \Pi^{kl}$ est l'\'energie propre du syst\`eme des
particules m\'elang\'ees dans la base  de saveur.
En sous-entendant la notation matricielle, l'\'equation pr\'ec\'edente s'\'ecrit
$$
  G = G_F + G_F \, (-i \, \Pi) \, G
$$
ce qui donne pour un propagateur inversible
$$
  i \, G^{-1} =i \, G^{-1}_F - \Pi \, .
$$

Posons que le propagateur des particules libres est donn\'e par la matrice
$$
  i \, G^{-1}_F(s) = s \, I - M^2_0 \, ,
$$
o\`u $I$ est la matrice unit\'e et $M^2_0$ la matrice de masse non
renormalis\'ee.
Le propagateur complet devient
$$
  i \, G^{-1}(s) = s \, I - M^2_0 - \Pi(s) \, .
$$

Consid\'erons le m\'elange de deux particules scalaires, par exemple le
syst\`eme $K^0\bar{K}^0$. Apr\`es avoir effectu\'e le prolongement analytique
n\'ecessaire dans la deuxi\`eme feuille de Riemann, les \'etats propres de
propagation peuvent \^etre 
caract\'eris\'es par les p\^{o}les complexes du propagateur
complet, ou encore par les z\'eros de l'inverse du propagateur.
Nous allons diagonaliser
ce propagateur pour \'etudier l'\'evolution spatio-temporelle des particules se
propageant sur des distances et intervalles de temps macroscopiques.
S\'eparons l'analyse en trois \'etapes.

\begin{enumerate}

\item
La diagonalisation est toujours possible si les valeurs propres du propagateur
complet sont distinctes deux \`a deux:
\begin{eqnarray}
  -i \, G(z)
  &=& \left( z \, I - M^2_0 - \Pi(z) \right)^{-1}
  \label{diag1} \\
  &=& \left( z \, I - M^2 - f(z) \right)^{-1}
  \nonumber \\
  &=& W(z)
  \left(
  \begin{array}{cc}
  ( z - m_1^2 - f_1(z) )^{-1}  &  0                          \\
  0                            &  ( z - m_2^2 - f_2(z) )^{-1}
  \end{array}
  \right)
  W^{-1}(z) \, .
  \nonumber
\end{eqnarray}
Pour $z$ fix\'e, $( z - m_j^2 - f_j(z) )^{-1}$ sont les valeurs propres
de $-i \, G(z)$.
Chaque p\^{o}le $z_j$ est la solution de l'\'equation $\det G^{-1}(z_j) = 0$.
La matrice de dia\-go\-na\-li\-sa\-tion $W(z)$ contient les vecteurs propres
correspondants.
Si le m\'elange entre les particules est faible, les parties r\'eelles
des p\^oles du propagateur seront approximativement \'egales aux masses
avant calcul de l'\'energie propre et seront positives.
Dans l'analyse de la propagation spatiale, la d\'ependance en $z$ des matrices
$W(z)$ engendrera des contributions \`a l'amplitude en puissances inverses
de la distance macroscopique de propagation $L$.
Elles sont dues aux seuils de production d'\'etats \`a plusieurs
particules et s'ajoutent aux contributions de m\^eme type provenant des
seuils des \'energies propres diagonalis\'ees $f_1(z)$ et $f_2(z)$.

\item
On aimerait que les matrices de diagonalisation soient ind\'ependantes de $z$,
pour pouvoir les sortir des int\'egrales lors du calcul d'une amplitude.
On pose
$$
  -i \, G(z) \equiv U
  \left(
  \begin{array}{cc}
  ( z-z_1 )^{-1}  &  0                 \\
  0                  &  ( z-z_2 )^{-1}
  \end{array}
  \right)
  V 
  + Q(z) \, .
$$
Les r\'esidus $Z_{1,2}$ aux p\^oles sont incorpor\'es dans les matrices de
diagonalisation. $U$ et $V$ sont des matrices constantes \`a sp\'ecifier,
$z_{1,2}$ sont les p\^{o}les du propagateur complet $G(z)$ tandis que $Q(z)$
est une matrice r\'eguli\`ere aux p\^{o}les.

Cette expression doit satisfaire \`a deux conditions:
$G^{-1}G \!=\! I$ et $GG^{-1} \!=\! I$, ou encore
\begin{eqnarray*}
  i \, G^{-1}(z)
  \left( U
    \left(
    \begin{array}{cc}
    ( z-z_1 )^{-1}  &  0                 \\
    0                  &  ( z-z_2 )^{-1}
    \end{array}
    \right)
    V + Q(z)
  \right) &=& I \, ,
  \\
  \left( U
    \left(
    \begin{array}{cc}
    ( z-z_1 )^{-1}  &  0                 \\
    0                  &  ( z-z_2 )^{-1}
    \end{array}
    \right)
    V + Q(z)
  \right) i \, G^{-1}(z) &=& I \, ,
\end{eqnarray*}
Pour que ces \'equations soient satisfaites au voisinage des p\^{o}les,
il faut que les matrices $U$ et $V$ soient compos\'ees des vecteurs propres de
$M^2_0 + \Pi(z)$ \'evalu\'es en $z_1$ et $z_2$.

Plus pr\'ecis\'ement, si
$$
  U \equiv \left( {\bf u}_1 \quad {\bf u}_2 \right)
  \qquad \mbox{et} \qquad
  V \equiv
  \left(
    \begin{array}{c}
    {\bf v}_1^t \\
    {\bf v}_2^t
    \end{array}
  \right)
$$
alors
\begin{eqnarray*}
  \left( M^2 + f(z_{1,2}) \right) {\bf u}_{1,2} = z_{1,2} \, {\bf u}_{1,2}
  \\
  {\bf v}_{1,2}^t \left( M^2 + f(z_{1,2}) \right) = z_{1,2} \, {\bf v}_{1,2}^t
\end{eqnarray*}
de sorte que l'on ait, pr\`es du p\^{o}le $z_j$,
$$
  G^{-1}(z) \, U \sim z - z_j
  \qquad \mbox{et} \qquad
  V \, G^{-1}(z) \sim z - z_j \, ,
$$
sinon le membre de gauche des \'equations \`a satisfaire serait divergent et ne
pourrait \^etre \'egal \`a la matrice unit\'e du membre de droite.
Cette diagonalisation exacte est \`a l'origine des m\'ethodes utilisant deux
angles de m\'elange pour un m\'elange \`a deux particules \cite{harte}.

\item
Un cas particuli\`erement int\'eressant est celui de p\^{o}les quasiment
d\'eg\'en\'er\'es. Des ph\'enom\`enes d'oscillation de la probabilit\'e de
d\'etection d'\'etats propres d'interaction se r\'ev\`elent alors dans
l'\'evolution spatio-temporelle.
Par exemple, les p\^{o}les des particules $K^0\bar{K}^0$ sont d\'eg\'en\'er\'es
si l'interaction faible est n\'eglig\'ee.

A l'ordre $g^2$ en l'interaction faible, on peut approximer \cite{sachs}
\begin{equation}
  f(z) \approx f(m^2) \, ,
  \label{energieprcste}
\end{equation}
o\`u $m$ est la masse d\'eg\'en\'er\'ee (c'est-\`a-dire la
masse si les interactions faibles sont absentes). On a alors
$$
  U = V^{-1}
  \qquad \mbox{et} \qquad
  Q(z) = 0
$$
ou encore
\begin{eqnarray}
  -i \, G(z) &\cong& V^{-1}
  \left(
  \begin{array}{cc}
  ( z-z_1 )^{-1}  &  0                 \\
  0                  &  ( z-z_2 )^{-1}
  \end{array}
  \right)
  V
  \nonumber
  \\
  &\equiv& -i \, V^{-1} \, G_D(z) \, V \, .
  \label{propdiag}
\end{eqnarray}
Les r\'esidus $Z_{1,2}$ sont incorpor\'es \`a la matrice $V$.
Cette approximation est analogue \`a celle qui consiste \`a prendre un
hamiltonien effectif constant dans
l'approche non relativiste de Wigner-Weisskopf \cite{wigner}: l'\'energie propre
apparaissant dans le propagateur correspond \`a la matrice de masse effective
de Wigner-Weisskopf si l'on approxime cette \'energie propre par une constante.

\end{enumerate}

En conclusion, les approximations ci-dessus m\`enent \`a analyser l'\'evolution
d'un m\'elange de particules comme la superposition des \'evolutions des
particules non m\'elang\'ees.
Notons que les matrices de diagonalisation $U^{-1}=V$ ne sont pas unitaires si
le propagateur n'est pas hermitien, comme dans le cas du
syst\`eme $K^0\bar{K}^0$ lorsque la violation $CP$ est prise en compte.

Ce traitement peut \^etre imm\'ediatement \'etendu \`a des m\'elanges de
particules vectorielles, en particulier le syst\`eme $\rho^0\!-\!\omega\!-\!\phi$
qui a une grande importance dans la description des interactions en-dessous de
1 GeV (mod\`ele de dominance vectorielle). Comme il s'agit de particules
vectorielles, le propagateur contient un facteur de la forme
\mbox{$g_{\mu\nu} -k_\mu k_\nu / m^2$}, qui n'influence pas la diagonalisation
puisque la contribution du terme $k_\mu k_\nu / m^2$ est toujours nulle en raison
du couplage du $\rho^0$ \`a des courants conserv\'es. Le m\'elange des propagateurs
a \'et\'e \'etudi\'e dans \cite{harte}. Par contre, la composante longitudinale
du propagateur doit \^etre prise en compte dans le m\'elange photon-boson Z,
qui est d'ailleurs compliqu\'e par la question de l'invariance de jauge et de
la brisure spontan\'ee de la sym\'etrie $SU(2)_L \times U(1)_Y$ \cite{baulieu}.

\section{Exemple: le syst\`eme $K^0 \overline{K^0}$}
\label{exempleKK}

Quand il s'agit de m\'elange de particules, le cas d'\'ecole est le syst\`eme 
$K^0 \overline{K^0}$ car il pr\'esente \`a la fois toutes les complications
possibles et toutes les vertus qui les rendent observables:
diff\'erence de masse et oscillation, violation CP directe et indirecte,
quasi-stabilit\'e donc propagation macroscopique, m\^eme ordre de grandeur
de la longueur d'oscillation et de la longueur de d\'esint\'egration.
L'analyse qui suit s'applique sans modification aux syst\`emes
$D^0 \overline{D^0}$ et $B^0 \overline{B^0}$.

L'\'energie propre du syst\`eme $K^0 \overline{K^0}$ ne peut \^etre
calcul\'ee que dans une th\'eorie effective, dont un exemple est le
lagrangien ${\cal L}=-gK\pi\pi^*$ qui appara\^{\i}t dans le calcul de
$K\!\to\!\pi\pi$ vu \`a la section \ref{exemplepropagkaon}.
On se limite ici \` a param\'etriser le propagateur.
La sym\'etrie CPT impose l'\'egalit\'e des \'el\'ements
diagonaux dans la base de saveur \cite{sachs}.
Une sym\'etrie CP imposerait en plus l'\'egalit\'e des
\'el\'ements non diagonaux mais nous savons qu'elle est viol\'ee
dans ce syst\`eme. La petitesse de la violation ($\!\sim\! {\cal O}(10^{-3})$)
impose n\'eanmoins que les \'el\'ements non diagonaux soient presque \'egaux.
L'inverse du propagateur complet pour des kaons neutres d'impulsion $p$
est param\'etris\'e  \cite{beuthe} par
\begin{equation}
  i G^{-1}(p^2) =
  \left(
  \begin{array}{cc}
  \langle K^0 |\hat G^{-1}| K^0 \rangle &
  \langle K^0 |\hat G^{-1} |\overline{K^0} \rangle \\
  \langle \overline{K^0} |\hat G^{-1}| K^0 \rangle &
  \langle \overline{K^0} |\hat G^{-1}| \overline{K^0} \rangle
  \end{array}
  \right)
  \equiv
  \left(
  \begin{array}{cc}
  d   &  a+b  \\
  a-b &  d
  \end{array}
  \right)
  \label{propKK}
\end{equation}
avec les d\'efinitions
\begin{eqnarray*}
  d   &\equiv& p^2 - m^2 - f_{00}(p^2)
  \\
  a+b &\equiv& - f_{0\bar0}(p^2)
  \\
  a-b &\equiv& - f_{\bar0 0}(p^2)
\end{eqnarray*}
o\`u $m$ est la masse renormalis\'ee et les $-if_{\alpha\beta}(p^2)$
sont les \'energies propres complexes  renormalis\'ees du syst\`eme des
kaons neutres. $\hat G^{-1}$ est l'op\'erateur correspondant au
propagateur. Notons que les termes non diagonaux d\'ependent de la
convention de phase choisie pour les kaons. Nous y reviendrons au chapitre
\ref{applications}.

La base des vecteurs propres de l'op\'erateur CP est reli\'ee \`a la
base de saveur par
\begin{equation}
  \left(
  \begin{array}{c}
  |K_1 \rangle \\ |K_2 \rangle
  \end{array}
  \right)
  =\frac{1}{ \sqrt{2} }
  \left(
  \begin{array}{rr}
  1 & 1 \\
  1 & -1
  \end{array}
  \right)
  \left(
  \begin{array}{c}
  |K^0 \rangle \\ |\overline{K^0} \rangle
  \end{array}
  \right)
  \label{baseCP}
\end{equation}
o\`u CP$|K^0\rangle=|\overline{K^0}\rangle$.
Le $K_1$ est donc pair sous CP tandis que $K_2$ est impair sous CP.
Dans cette base, l'inverse du propagateur s'\'ecrit
$$
  i \, \tilde{G}^{-1} (p^2) \equiv
  \frac{1}{2}
  \left(
  \begin{array}{rr}
  1 & 1 \\
  1 & -1
  \end{array}
  \right)
  iG^{-1} (p^2)
  \left(
  \begin{array}{rr}
  1 & 1 \\
  1 & -1
  \end{array}
  \right)
  =\left(
        \begin{array}{cc}
        d+a  &  -b  \\
        b &  d-a
        \end{array}
  \right)
$$
On voit clairement dans cette base que la violation CP se traduit par le
param\`etre $b$.

Dans le but de diagonaliser ce propagateur, d\'efinissons un param\`etre
complexe $\hat \epsilon$ tel que
\begin{equation}
  \frac{\hat \epsilon}{1+\hat \epsilon^2} \equiv \frac{b}{2a} \, .
  \label{defepsilon}
\end{equation}
L'inverse du propagateur est diagonalis\'e en
$$
  i \, \tilde{G}^{-1} (p^2) =
  \frac{1}{1-\hat\epsilon^2}
  \left(
  \begin{array}{cc}
  1  &  \hat\epsilon  \\
  \hat\epsilon &  1
  \end{array}
  \right)
  \left(
  \begin{array}{cc}
  d + a \frac{1-\hat\epsilon^2}{1+\hat\epsilon^2}  & 0  \\
  0 &  d - a \frac{1-\hat\epsilon^2}{1+\hat\epsilon^2}
  \end{array}
  \right)
  \left(
  \begin{array}{rr}
  1  &  -\hat\epsilon  \\
  -\hat\epsilon &  1
  \end{array}
  \right)
$$
Par cons\'equent, la base physique des kaons neutres consiste en deux \'etats
$K_{L,S}$ de masses $m_{\SS L,S}$ et largeurs $\Gamma_{L,S}$ d\'efinies par
\begin{eqnarray*}
  d_{\SS S} &\equiv& p^2 - m_{\SS S}^2 + i m_{\SS S} \Gamma_S
  = d + a \frac{1-\hat\epsilon^2}{1+\hat\epsilon^2}
  \\
  d_{\SS L} &\equiv& p^2 - m_{\SS L}^2 + i m_{\SS L} \Gamma_S
  = d - a \frac{1-\hat\epsilon^2}{1+\hat\epsilon^2}
\end{eqnarray*}
o\`u l'on a fait l'approximation expliqu\'ee dans la section pr\'ec\'edente
d'\'evaluer l'\'energie propre \`a la masse d\'eg\'en\'er\'ee ($f(z) \approx
f(m^2)$). Dans cette notation, le propagateur s'\'ecrit
$$
  -i \, \tilde{G} (p^2) =
  \frac{1}{1-\hat\epsilon^2}
  \left(
  \begin{array}{rr}
  1  &  \hat\epsilon  \\
  \hat\epsilon &  1
  \end{array}
  \right)
  \left(
  \begin{array}{cc}
  d_{\SS S}^{-1}  & 0  \\
  0 &  d_{\SS L}^{-1}
  \end{array}
  \right)
  \left(
  \begin{array}{rr}
  1  &  -\hat\epsilon  \\
  -\hat\epsilon &  1
  \end{array}
  \right)
$$
Les matrices de diagonalisation $V$ et $V^{-1}$ de la base de saveur
$K^0 \overline{K^0}$ sont donn\'ees par\footnote{La normalisation de $V$ est
en fait arbitraire car les normalisations de $V$ et $V^{-1}$ se
compensent, puisqu'elles apparaissent toujours par paire dans notre approche
o\`u seuls les \'etats $K^0$ et $\overline{K^0}$ sont observables.
Nous choisissons ici des normalisations \'egales.}
\begin{eqnarray}
  V &=& \frac{1}{ \sqrt{ 2(1- \hat \epsilon^2)} }
  \left(
  \begin{array}{rr}
  1 & -\hat \epsilon \\
  -\hat \epsilon & 1
  \end{array}
  \right)
  \left(
  \begin{array}{rr}
  1 & 1 \\
  1 & -1
  \end{array}
  \right)
  \nonumber \\
  V^{-1} &=& \frac{1}{ \sqrt{ 2(1- \hat \epsilon^2)} }
  \left(
  \begin{array}{rr}
  1 & 1 \\
  1 & -1
  \end{array}
  \right)
  \left(
  \begin{array}{rr}
  1 & \hat \epsilon \\
  \hat \epsilon & 1
  \end{array}
  \right)
  \label{matriceV}
\end{eqnarray}
Notons que si $\hat \epsilon$ est purement imaginaire, la matrice $V$ est
unitaire.
Dans ce cas, il est possible de choisir une convention de phase \'etrange
(voir section \ref{determparam}) dans laquelle $\hat \epsilon=0$. Les \'etats
$K_{1,2}$ sont alors \'etats propres de propagation. On verra aussi que dans
ce cas il n'y a pas de violation CP dans les oscillations.

Le propagateur dans la base de saveur est donc diagonalis\'e en tr\`es
bonne approximation par
$$
  -i \, G(p^2)  \cong
  V^{-1}
  \left(
  \begin{array}{cc}
  d_{\SS S}^{-1}  & 0  \\
  0 &  d_{\SS L}^{-1}
  \end{array}
  \right)
  V \, .
$$
Comme pr\'evu, ces matrices sont non unitaires puisque le propagateur
dans la base de saveur est non hermitien.
Le lien avec le formalisme couramment utilis\'e dans la litt\'erature se voit
en d\'efinissant une {\sl double}
base physique orthogonale et normalis\'ee (voir \'equation (\ref{doublebase})),
c'est-\`a-dire des \'etats entrants
$|ket \rangle$ et sortants $\langle bra |$ ind\'ependants qui sont vecteurs propres
\` a droite et \`a gauche du propagateur \cite{sachs,enz,alvarez}:
$$
  \left(
  \begin{array}{c}
  |K_S \rangle \\ |K_L \rangle
  \end{array}
  \right)
  \equiv\frac{1}{ \sqrt{1-\hat\epsilon^2} }
  \left(
  \begin{array}{rr}
  1 & \hat\epsilon \\
  \hat\epsilon & 1
  \end{array}
  \right)
  \left(
  \begin{array}{c}
  |K_1 \rangle \\ |K_2 \rangle
  \end{array}
  \right)
$$
ainsi que
$$
\left(
  \begin{array}{c}
  \langle K_S | \\ \langle K_L |
  \end{array}
  \right)
  \equiv \frac{1}{ \sqrt{1-\hat\epsilon^2} }
  \left(
  \begin{array}{rr}
  1 & -\hat\epsilon \\
  -\hat\epsilon & 1
  \end{array}
  \right)
  \left(
  \begin{array}{c}
  \langle K_1 | \\ \langle K_2 |
  \end{array}
  \right)
$$
A moins que $\hat\epsilon$ ne soit purement imaginaire
(c'est-\`a-dire qu'il n'y a pas de violation CP), ces deux bases ne sont pas
hermitiennes conjugu\'ees l'une de l'autre.
Si l'on tient \`a travailler uniquement avec la base de vecteurs propres \`a
droite $|K_{S,L} \rangle$, la question de la normalisation
est incontournable puisque les vecteurs de cette base ne sont pas orthogonaux
\`a leurs hermitiens conjugu\'es. On choisit couramment \cite{nachtmann} de
normaliser par $1/\sqrt{1+|\hat\epsilon|^2}$ au lieu de
$1/\sqrt{1-\hat\epsilon^2}$ comme ici, de sorte que le produit scalaire de
$|K_S \rangle$ avec son hermitien conjugu\'e soit normalis\'e \`a 1
(on proc\`ede de m\^eme fa\c{c}on pour $|K_L \rangle$).
N\'eanmoins, $|K_S \rangle$ n'est toujours pas orthogonal \`a
l'hermitien conjugu\'e de $|K_L \rangle$.
Comme il a d\'ej\`a \'et\'e mentionn\'e au chapitre \ref{oscillationsMQ},
le choix de la normalisation influence la forme des expressions du genre
\mbox{${\cal M}(K_L\!\to\!\pi\pi)\equiv<\!\pi\pi\, |T| \, K_L\!>$}, o\`u $T$
est la matrice de transition.
On verra au chapitre \ref{applications} la mani\`ere correcte de d\'efinir de
telles amplitudes en th\'eorie des champs.
L'ambigu\"{\i}t\'e de normalisation en $\hat\epsilon^2$ est beaucoup plus
petite que la violation CP indirecte ($\sim \hat\epsilon$)
et n'a donc pas pos\'e de probl\`eme pour l'interpr\'etation des exp\'eriences,
en tout cas jusqu'aux premi\`eres mesures de la violation CP directe
($\sim \epsilon'\sim\hat\epsilon^2$).

Les quantit\'es $m_{\SS S,L}$ et $\Gamma_{S,L}$ sont mesur\'ees
exp\'erimentalement tandis que les param\`etres $a$, $b$, $m$ et $f(m^2)$
peuvent en principe \^etre calcul\'es th\'eoriquement \cite{buchalla}.
Les relations entre les deux ensembles de param\`etres sont
\begin{eqnarray}
  & & a = \frac{1}{2} \,
  \frac{1 +\hat\epsilon^2}{1 -\hat\epsilon^2} \,
  \left(
        m_{\SS L}^2 - m_{\SS S}^2 - i (m_{\SS L} \Gamma_L - m_{\SS S} \Gamma_S)
  \right) \; ,
  \label{relationa}
  \\
  & & m^2 - f(m^2) = \frac{1}{2} \,
  \left(
        m_{\SS L}^2 + m_{\SS S}^2 - i (m_{\SS L} \Gamma_L + m_{\SS S} \Gamma_S)
  \right) \; ,
  \nonumber
  \\
  & & b=\frac{\hat\epsilon}{1 -\hat\epsilon^2}
  \left(
        m_{\SS L}^2 - m_{\SS S}^2 - i (m_{\SS L} \Gamma_L - m_{\SS S} \Gamma_S)
  \right) \; .
  \nonumber
\end{eqnarray}

\section{Oscillations dans le mod\`ele simplifi\'e}
\label{oscisimpli}

La propagation de particules en m\'elange va \^etre d\'ecrite par le
processus de la section \ref{modelesimplifie} l\'eg\`erement modifi\'e pour
tenir compte de la saveur.
Le processus de propagation d'une particule $\nu_\alpha$ de saveur
$\alpha$ produite \`a la source en une particule $\nu_\beta$ de saveur $\beta$
identifi\'ee dans le d\'etecteur est symbolis\'e par:

\parbox{\textwidth}{
\begin{eqnarray*}
  P_I(q)
 \stackrel{ (t_{\SS P},{\bf x}_{\SS P} ) }{\longrightarrow}
  P_F(k) + \nu_\alpha(p) & &
  \\
  &\searrow&
  \\
  & & \nu_\beta(p) + D_I(q') 
  \stackrel{ (t_{\SS D},{\bf x}_{\SS D} ) }{\longrightarrow} D_F(k')
\end{eqnarray*}
}
On suppose que l'on peut identifier la saveur $\alpha$ de la particule
interm\'ediaire $\nu$ produite dans la r\'egion de
$(t_{\SS P},{\bf x}_{\SS P} )$ au moyen de l'\'etat sortant $P_F(k)$
et la saveur $\beta$ du $\nu$ d\'etect\'e dans la r\'egion de
$(t_{\SS D},{\bf x}_{\SS D} )$ au moyen de l'\'etat sortant $D_F(k')$.
S'il est impossible d'identifier la saveur sortante
(ex: $K^0, \overline{K^0} \to \pi^+\pi^-$), il suffit de sommer sur les
diff\'erentes saveurs.
On garde les hypoth\`eses simplificatrices de repr\'esenter $P_I$ et $D_I$
par des \'etats stationnaires infiniment bien localis\'es dans l'espace et
$P_F$ et $D_F$ par des ondes planes.

Soit $G_{\beta\alpha}(x'-x)$ le propagateur complet symbolisant la propagation
de particules de saveur $\alpha$ produites en $x$ en particules de saveur
$\beta$ d\'etect\'ees en $x'$.
Nous venons de voir (\'equations (\ref{stablenondiag}) et (\ref{propdiag}))
que le propagateur peut \^etre diagonalis\'e
en tr\`es bonne approximation par des matrices $V$ constantes\footnote{Rappelons
une fois de plus que les
matrices $V$ utilis\'ees ici sont les transpos\'ees des matrices
$V$ diagonalisant les kets dans le traitement de m\'ecanique quantique
(voir \'equation (\ref{chgtbase})).}:
$$
  G_{\beta\alpha}(p^2)= (V^{-1} \, G_D(p^2) \, V)_{\beta\alpha}
                       = \sum_j V_{\beta j}^{-1} \, G_{D,jj} \, V_{j\alpha} \, .
$$
L'amplitude du processus global est donn\'ee par l'\'equation (\ref{amplipro})
de l'amplitude de propagation d'une particule isol\'ee, o\`u le propagateur
$G(p^2)$ est remplac\'e par le propa\-gateur $G_{\beta\alpha}(p^2)$.
Elle peut donc s'\'ecrire comme une superposition
lin\'eaire d'amplitudes de propagation d'\'etats propres de masse:
$$
  {\cal A}(\alpha \!\to\! \beta,T,{\bf L})
  = \sum_j V_{\beta j}^{-1} \, {\cal A}_j \, V_{j\alpha} \, ,
$$
o\`u l'amplitude partielle ${\cal A}_j$ s'\'ecrit, apr\`es int\'egration sur $x$ et $x'$
(voir \'equation (\ref{amplisimpli})),
$$
  {\cal A}_j \sim M_P M_D \, 2\pi \,
  \delta \left( E_{D_I} \!-\! E_{D_F} \!+\! E_{P_I} \!-\! E_{P_F} \right) \,
  \int \frac{d^3p}{(2\pi)^3} \,
   e^{ i \, {\bf p} \cdot {\bf L} } \, G_{D,jj}(E,{\bf p})
$$
o\`u ${\bf L} \equiv {\bf x}_{\SS D} \!-\! {\bf x}_{\SS P}$ et
\mbox{$ E \equiv E_{P_I} \! - E_{P_F} \!>\! 0 $}.
Si l'on int\`egre sur la tri-impulsion, l'amplitude partielle devient
(voir \'equation (\ref{instabpole}))
$$
  {\cal A}_j \sim M_P M_D \, 2\pi \,
  \delta \left( E_{D_I} \!-\! E_{D_F} \!+\! E_{P_I} \!-\! E_{P_F} \right) \,
  \frac{-i Z}{4\pi L} \,
  \exp \left(i p_{m_j} L - \frac{m_j\Gamma_j }{ 2 p_{m_j} } L \right) \, ,
$$
o\`u $p_{m_j} \equiv \sqrt{ E^2 - m_j^2 }$.

Les oscillations entre les \'etats propres de masse $i$ et $j$ surgiront des
termes d'interf\'erence ${\cal A}_i {\cal A}_j^*$ dans la probabilit\'e
donn\'ee par la norme au carr\'e de l'amplitude:
$$
  {\cal A}_i {\cal A}_j^* \sim
  \exp \left(
              i (p_{m_i} -p_{m_j}) \, L
             - \frac{m_i\Gamma_i}{2 p_{m_i}}  - \frac{m_j\Gamma_j}{2 p_{m_j}}
       \right) \, .
$$
Le facteur d'oscillation pour le m\'elange $ij$ est donc un cosinus
d'argument
$$
  (p_{m_i} -p_{m_j}) \, L \cong \frac{\Delta m_{ij}^2}{2 p_m} \, L
$$
o\`u $\Delta m_{ij}^2 \equiv m_i^2-m_j^2$ et
$p_m \equiv \sqrt{E^2-m^2}$ avec $m$ une masse de r\'ef\'erence choisie
par exemple comme $m\equiv (m_i+m_j)/2$.
On a fait l'approximation \mbox{$|m_i -m| \!\ll\! p_m$} puisque dans les cas
o\`u une oscillation macroscopique est observable, \mbox{$|m_i-m_j| \ll p_m$}.

Si l'on d\'efinit la {\it longueur d'oscillation} $L_{ij}^{osc}$ pour le
m\'elange $ij$ par
$$
  L_{ij}^{osc} \equiv \frac{4\pi p_m}{\Delta m_{ij}^2} \, ,
$$
on retrouve la formule d'oscillation standard (\ref{longoscMQ}).

Ce mod\`ele r\'epond \`a plusieurs des objections soulev\'ees au
chapitre \ref{oscillationsMQ} (section (\ref{oscillationtempMQ})),
en consid\'erant dans un cadre relativiste (objections 1 et 2)
la particule comme un \'etat interm\'ediaire non observ\'e directement
(objections 3,4,5 et 6) et en engendrant une
d\'ependance spatiale (objections 7 et 8).

Tout n'est pas rose pour autant.
Par exemple, il est tentant d'interpr\'eter $E$ et  $p_{m_i}$ comme
l'\'energie et l'impulsion de l'\'etat propre de masse $i$.
Cette interpr\'etation implique curieusement que les
diff\'erents \'etats propres de masse de la particule oscillante ont la m\^eme
\'energie mais des tri-impulsions diff\'erentes alors que dans la d\'erivation
en m\'ecanique quantique on consid\`ere le plus souvent le cas inverse
(\'energies diff\'erentes mais tri-impulsions \'egales), sauf dans la
prescription de Lipkin \cite{lipkin2,lipkin3}.
L'origine de notre r\'esultat est \'evidemment le choix de proc\'eder \`a
l'analyse spatiale du propagateur. L'analyse temporelle du propagateur
\cite{schwinger} implique de prendre des \'energies diff\'erentes et des tri-impulsions
\'egales mais n'est pas justifi\'ee par un mod\`ele de th\'eorie
des champs.

Il ne faudrait pas en conclure pour autant que la question de l'\'egalit\'e ou
non de l'\'energie-impulsion des \'etats oscillants est r\'egl\'ee par notre
mod\`ele simplifi\'e.  L'approximation d'\'etats stationnaires impose qu'il n'y
ait qu'une seule valeur de l'\'energie possible pour les \'etats oscillants,
tandis que l'hypoth\`ese de localisation \`a une pr\'ecision infinie de la
source et du d\'etecteur n'impose, par contre, aucune contrainte sur les valeurs des
tri-impulsions des \'etats oscillants, qui sont alors fix\'ees par les p\^oles
des propagateurs.

La solution de ce probl\`eme n\'ecessite donc le recours \`a un mod\`ele plus
r\'ealiste, qui abandonne les hypoth\`eses de stationnarit\'e et de localisation
infiniment pr\'ecises. Il sera du m\^eme coup possible d'\'etudier les conditions
d'observabilit\'e des oscillations, alors que le mod\`ele simplifi\'e n'en
fournit aucune. Par exemple, les oscillations sont interdites si tous les
\'etats entrants et sortants sont des ondes planes, car il n'y a alors qu'une seule
valeur possible pour l'\'energie-impulsion des \'etats oscillants. L'analyse des
situations interm\'ediaires devrait donc permettre d'\'etudier l'influence des
conditions exp\'erimentales.

Finalement, on voudrait d\'ecrire de fa\c{c}on coh\'erente la
propagation d'une parti\-cule instable. De plus, comme mentionn\'e ci-dessus,
les corrections n\'e\-gli\-g\'ees en $\Gamma/m$ sont du m\^eme ordre de grandeur
que le rapport $\Delta m/m$ dans le syst\`eme des kaons neutres. Il faudrait
s'assurer qu'elles n'aient pas d'influence sur le facteur d'oscillation.
Dans le prochain chapitre, un mod\`ele plus sophistiqu\'e de propagation
d'une seule particule est d\'evelopp\'e dans le but
de r\'epondre \`a ces questions.

\chapter{Propagation: le mod\`ele sophistiqu\'e}
\label{production}

Ce chapitre est consacr\'e \`a l'\'etude d'un mod\`ele de propagation
sophistiqu\'e, te\-nant compte dans la mesure du possible des conditions de
production et de d\'etection. Abandonnant la simplification extr\^eme
de consid\'erer les particules entrantes et sortantes soit comme
des ondes planes soit comme des \'etats stationnaires localis\'es \`a une
pr\'ecision infinie, on mod\'elisera ces \'etats
de mani\`ere plus r\'ealiste par des paquets d'ondes \cite{jacob,giunti2}.
La particule interm\'ediaire est repr\'esent\'ee par son
propagateur relativiste.

Ce chapitre est plut\^ot technique, d\'ebutant par quelques
consid\'erations sur les paquets d'ondes, suivies de la formulation de
l'amplitude. Il continue par l'analyse temporelle de celle-ci, avec une
\'etude d\'etaill\'ee des contributions non exponentielles \`a la propagation.
On verra par exemple que les contributions des \'etats
\`a plusieurs particules vues au chapitre pr\'ec\'edent sont souvent
absentes pour les processus de propagation macroscopiques, car les seuils de
production de ces \'etats se trouvent souvent en dehors de la r\'egion
d'int\'egration. Ces contributions sont remplac\'ees par celles des
seuils dus \`a la limitation dans l'espace des impulsions de la grandeur
des paquets d'ondes entrants et sortants. Le chapitre se termine par
l'analyse spatiale de l'amplitude et le calcul de la probabilit\'e du
processus, en ne conservant que la d\'ependance spatiale.

Les bases du calcul ont \'et\'e jet\'ees par Jacob et Sachs \cite{jacob}.
La notation a \'et\'e d\'efinie au chapitre \ref{propagateur}.
Le processus global
de propagation \'etait symbolis\'e par
\begin{eqnarray*}
  P_I(q)
 \stackrel{ (t_{\SS P},{\bf x}_{\SS P} ) }{\longrightarrow}
  P_F(k) + \nu(p) & &
  \\
  &\searrow&
  \\
  & & \nu(p) + D_I(q') 
  \stackrel{ (t_{\SS D},{\bf x}_{\SS D} ) }{\longrightarrow} D_F(k')
\end{eqnarray*}
Rappelons la signification des notations.
$P_I$ repr\'esente l'ensemble des particules arrivant dans la r\'egion
de production centr\'ee autour du point $( t_{\SS P},{\bf x}_{\SS P} )$,
d'impulsion totale $q$
tandis que $P_F$ repr\'esente l'ensemble des particules issues de la r\'egion
de production et d'impulsion totale $k$, \`a l'exception de la particule
interm\'ediaire $\nu$ dont on \'etudie la propagation. $D_I$, $D_F$ et 
$( t_{\SS D},{\bf x}_{\SS D} )$ ont des d\'efinitions
similaires mais concernent le point de d\'etection.
Le point d'interaction \`a la production est not\'e par $x$ et le point
d'interaction \`a la d\'etection est not\'e $x'$.
On pose que l'\'etat interm\'ediaire se propageant de $x$ en $x'$ a les
nombres quantiques d'une particule, pas d'une antiparticule.
On ne tiendra pas compte des instabilit\'es \'eventuelles des particules
ext\'erieures. Seule l'instabilit\'e de la particule interm\'ediaire sera
prise en compte\footnote{Certaines tentatives ont \'et\'e entreprises pour
tenir compte de l'instabilit\'e de la source dans les conditions de
production, d'une part dans le cadre de la m\'ecanique quantique non relativiste
\cite{rich,mohanty}, et d'autre part en th\'eorie des champs
\cite{campagne} }. Si elle se d\'esint\`egre dans le processus \'etudi\'e,
alors $D_I(q')$ est en r\'ealit\'e
un \'etat sortant. L'\'ecriture d'une amplitude tenant compte de
l'instabilit\'e des particules entrantes et sortantes ne pose en elle-m\^eme
aucun probl\`eme. Il suffit de consid\'erer un processus plus global o\`u
toutes les particules instables sont vues comme interm\'ediaires. Les
\'etats initiaux et finaux sont stables. Les r\`egles de Feynman permettent
d'\'ecrire facilement l'amplitude correspondant \`a un tel processus en
cascade. Par contre, l'\'evaluation des int\'egrales est la plupart du temps
un obstacle insurmontable. C'est pourquoi on se limite \`a consid\'erer
comme \'etat interm\'ediaire la particule dont on veut \'etudier les
propri\'et\'es.

Toutes les particules sont suppos\'ees sans spin.
Pour localiser l'interaction de production en $(t_{\SS P},{\bf x}_{\SS P} )$,
les particules entrantes et sortantes au point $x$
sont mod\'elis\'ees par des paquets d'ondes qui ne se recouvrent (dans
l'espace de configuration) que dans la r\'egion de production centr\'ee
autour de $(t_{\SS P},{\bf x}_{\SS P} )$.
En vue d'\'etudier l'influence des facteurs de production, les paquets
ont une certaine largeur dans l'espace de configuration. De plus, comme les
exp\'eriences mesurent g\'en\'eralement aussi l'impulsion, on peut supposer
qu'ils sont bien localis\'es aussi dans l'espace des impulsions autour de leurs
impulsions moyennes.
L'interaction de d\'etection est localis\'ee de
m\^eme mani\`ere.
Notons qu'il n'y aurait aucune difficult\'e \`a remplacer certains paquets
d'ondes par des \'etats li\'es, si la mod\'elisation d'un processus particulier
l'exige.

A ce processus correspond le diagramme de Feynman suivant
\begin{center}
\includegraphics[width=8cm]{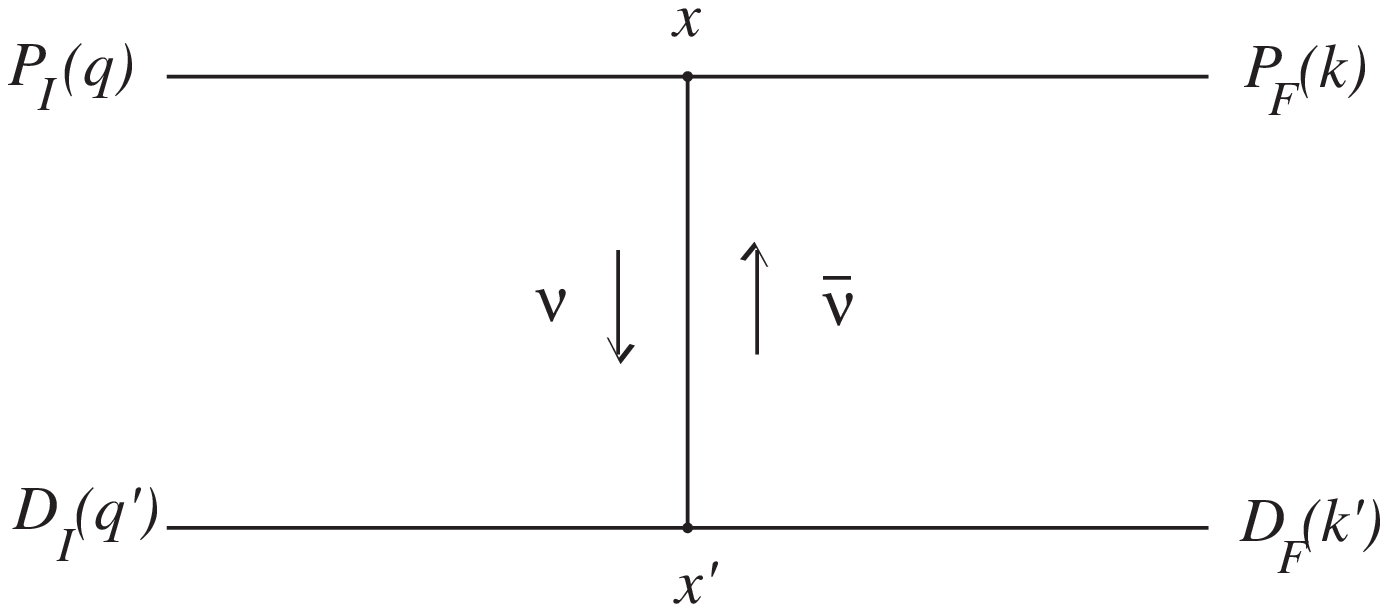}
\end{center}
Ce diagramme repr\'esente toute une s\'erie de processus. Le n\^otre se
distingue par des conditions exp\'erimentales telles que l'interaction en
$x$ pr\'ec\`ede toujours l'interaction en $x'$ et telles qu'il y ait un transfert
d'\'energie (positive) de $x$ en $x'$. Ces situations seront s\'electionn\'ees
automatiquement par la forme choisie des paquets d'ondes, avec pour
cons\'equence que seule la particule $\nu$ contribuera \`a la propagation, et
non l'antiparticule $\bar \nu$.

On ne sp\'ecifiera pas ici la forme des interactions de production et de d\'etection
mais il n'y aurait aucun probl\`eme \`a le faire. Par exemple, si la
particule interm\'ediaire est un neutrino, les m\'ethodes de d\'etection sont
de trois types {\cite{bilenky}:
\begin{eqnarray*}
  \nu_l + X &\to& l + Y \, ,
  \\
  \nu_l + X &\to& \nu_l + Y \, ,
  \\
  \nu_l + l &\to& \nu_l + l \, ,
\end{eqnarray*}
o\`u $l$ symbolise un lepton et $X$ et $Y$ ne contiennent pas de leptons.
Pour calculer l'amplitude du troisi\`eme processus \cite{grimus}, il faut
sommer sur les
deux diagrammes repr\'esentant l'interaction par courant neutre et par
courant charg\'e. Dans les cas o\`u des neutrinos sont produits dans le
d\'etecteur (deuxi\`eme et troisi\`eme processus), il faut sommer sur leurs
diff\'erents \'etats propres de masse possibles, puisque ce sont les \'etats
physiques asymptotiques. Le neutrino final doit alors \^etre r\'eexprim\'e
dans la base de masse:
\mbox{$\nu_l = \sum_j V^\dagger_{lj} \nu_j$}, ce qui ajoute une matrice de
m\'elange suppl\'ementaire dans l'amplitude. Cette matrice n'a cependant
rien \`a voir avec les oscillations de la particule interm\'ediaire.

\section{Paquets d'onde}

Un paquet d'ondes repr\'esentant un \'etat donn\'e $|\,\phi \!\!>$ de
masse $m$ s'\'ecrit dans l'espace des impulsions \cite{peskin}
$$
  |\,\phi \!> = \int [d{\bf k}] \, \phi({\bf k}) \, |\,{\bf k} \!> \, ,
$$
o\`u 
$\phi({\bf k})$ est la fonction d'onde dans l'espace des impulsions et
$|\,{\bf k} \!>$ est un \'etat \`a une particule d'impulsion ${\bf k}$ dans la
th\'eorie en interaction. On utilise la notation
$$  
  [d{\bf k}] \equiv \frac{d {\bf k}}{(2\pi)^3 } \,
  \frac{1}{ \sqrt{2 \, E({\bf k}) } } \, ,
$$
o\`u
$E({\bf k}) \equiv \sqrt{ {\bf k}^2 + m^2}$.
La normalisation des \'etats libres est donn\'ee par
$$
  <\! {\bf k} \, | \, {\bf p} \!>
  = 2 E({\bf k}) \, (2\pi)^3 \, \delta^{(3)}( {\bf p} \! - \! {\bf k} ) \, .
$$
On a
$$
  <\! \phi \,|\,\phi \!> = 1
  \quad \mbox {si} \quad
  \int \frac{d {\bf k}}{(2\pi)^3 } \,
  |\phi({\bf k})|^2 =1 \, .
$$
Dans l'espace de configuration, la fonction d'onde s'\'ecrit
$$
  \widetilde{\phi}({\bf x},t) = \int \frac{d {\bf k}}{(2\pi)^3 } \,
  \phi({\bf k}) \, e^{ - i E({\bf k}) t  + i \, {\bf k} \cdot {\bf x} } \, ,
$$
de sorte qu'elle satisfasse \`a l'\'equation de Klein-Gordon.
La relation r\'eciproque est donn\'ee par
$$
  \phi({\bf k}) = \int d{\bf x} \, \widetilde{\phi}({\bf x},t) \,
   e^{ i E({\bf k}) t - i \, {\bf k} \cdot {\bf x}  }
$$
Si le paquet d'ondes repr\'esente une particule dot\'ee d'une impulsion
approximativement connue, il a un seul maximum global, disons en
${\bf k} \!=\! {\bf K}$. Le maximum de
$\widetilde{\phi}( {\bf x},t \!=\! 0 )$ se situera en ${\bf x} \!=\! {\bf 0}$
pour autant que
$\phi({\bf k})$ ait une sym\'etrie centrale autour de ${\bf K}\,$:
$$
  \phi({\bf K} \!+\! {\bf k'}) = \phi({\bf K} \!-\! {\bf k'})
$$
Nous ferons l'hypoth\`ese que les paquets d'ondes ont une impulsion bien
d\'efinie: leur maximum est bien localis\'e et $|\phi({\bf k})|$ a
approximativement une sym\'etrie centrale. On va faire l'hypoth\`ese
suppl\'ementaire que la phase de $\phi({\bf k})$ a aussi une sym\'etrie
centrale, de sorte que le paquet soit centr\'e en ${\bf x} \!=\! {\bf 0}$.
Sous ces conditions, le paquet d'ondes est not\'e $\phi({\bf k},{\bf K})$.

Les paquets centr\'es en ${\bf x} \!\neq\! {\bf 0}$ sont construits par
translation. A l'aide de l'op\'erateur de translation
$\exp (i \hat{P} \!\cdot\! x)$, o\`u
$x \!=\! (T,{\bf X})$, on construit des paquets d'ondes centr\'es en
${\bf X}$ au temps $T$. Si le paquet d'onde dans l'espace des impulsions est
donn\'e par
\begin{equation}
  \Phi({\bf k},{\bf K},{\bf X},T)
  \equiv \phi({\bf k},{\bf K}) \,
  \exp \left( -i \, {\bf k} \!\cdot\! {\bf X} + i E({\bf k}) T \right) \, ,
  \label{paquetonde}
\end{equation}
alors le paquet d'ondes dans l'espace de configuration
$$
  \widetilde{\Phi}({\bf x},t,{\bf K},{\bf X},T)
  = \int \frac{d {\bf k}}{(2\pi)^3 } \; \phi({\bf k},{\bf K}) \,
  \exp \left(  i\, {\bf k} \!\cdot\! ( {\bf x} \!-\! {\bf X} )
             - i E({\bf k})(t\!-\!T)
       \right) \, ,
$$ sera centr\'e en ${\bf X}$ au temps $T$. 

Sans perte de g\'en\'eralit\'e, on va travailler avec une seule particule dans 
$P_I(q)$, dans $P_F(k)$, dans $D_I(q')$ et dans $D_F(k')$. L'extension \`a
un nombre plus grand de particules est imm\'ediate et ne fait que compliquer la
notation.
Les paquets d'ondes correspondant aux diff\'erentes particules sont
construits de sorte que ceux concernant la production du $\nu$ sont
centr\'es en ${\bf x}_{\SS P}$ au temps $t_{\SS P}$ tandis que ceux
concernant la d\'etection du $\nu$ sont centr\'es en ${\bf x}_{\SS D}$ au
temps $t_{\SS D}$. Ils sont not\'es
\begin{eqnarray*}
  |\, P_I \!> &=& \int [d{\bf q}] \,
  \Phi_{ P_{\SS I} } \left( {\bf q},{\bf Q},{\bf x}_{\SS P},t_{\SS P} \right)
  |\, P_{\SS I} ({\bf q}) \!>
  \\
  |\, P_F \!> &=& \int [d{\bf k}] \,
  \Phi_{ P_{\SS F} } \left( {\bf k},{\bf K},{\bf x}_{\SS P},t_{\SS P} \right)
  |\, P_{\SS F} ({\bf k}) \!>
  \\
  |\, D_I \!> &=& \int [d{\bf q'}] \,
  \Phi_{ D_{\SS I} } \left( {\bf q'},{\bf Q'},{\bf x}_{\SS D},t_{\SS D} \right)
  |\, D_{\SS I} ({\bf q'}) \!>
  \\
  |\, D_F \!> &=& \int [d{\bf k'}] \,
  \Phi_{ D_{\SS F} } \left( {\bf k'},{\bf K'},{\bf x}_{\SS D},t_{\SS D} \right)
  |\, D_{\SS F} ({\bf k'}) \!> \, .
\end{eqnarray*}

On retombe sur le mod\`ele simplifi\'e des chapitres pr\'ec\'edents en
imposant
\begin{enumerate}
  
  \item
  la condition de stationnarit\'e sur $P_I$ et $D_I$, c'est-\`a-dire que les
  \'energies $E_{ P_{\SS I} }$ et $E_{ D_{\SS I} }$ sont constantes.

  \item
  la localisation infiniment pr\'ecise de $P_I$ et $D_I$, c'est-\`a-dire que
  $\phi_{ P_{\SS I} }({\bf q},{\bf Q})$ et
  $\phi_{ D_{\SS I} }({\bf q'},{\bf Q'})$ sont
  des fonctions constantes.

  \item
  que les \'etats $P_F$ et $D_F$ soient des ondes planes, c'est-\`a-dire que les
  $\phi_{ P_{\SS F} }({\bf k},{\bf K})$ et
  $\phi_{ D_{\SS F} }({\bf k'},{\bf K'})$ sont des
  fonctions delta:
  $$
    \phi_{ P_{\SS F} }({\bf k},{\bf K}) \sim \delta^{(3)}({\bf k} - {\bf K})
    \quad \mbox{et} \quad
    \phi_{ D_{\SS F} }({\bf k'},{\bf K'}) \sim \delta^{(3)}({\bf k'} - {\bf K'}) \, .
  $$
  Les \'energies de ces \'etats sont fix\'ees par la relation
  \mbox{$E^2={\bf p}^2 + m^2$} si l'on impose qu'ils satisfassent \`a
  l'\'equation de Klein-Gordon.
  
\end{enumerate}

\section{Amplitude}
\label{amplitude}

La formule g\'en\'erale de l'amplitude est donn\'ee par
$$
  {\cal A} =
  <\! P_F,D_F \,| \,
  T \left( \exp \left( - i \int d^4x \, {\cal H}_I \right)  \right) - {\bf 1}
  |\, P_I,D_I \!> \, ,
$$ 
o\`u ${\cal H}_I$ est le lagrangien d'interaction de la particule $\nu$ qui se
propage et $T$ l'op\'erateur qui ordonne dans le temps. Soit $g$ la constante de
couplage du champ $\nu$ avec les autres champs. D\'eveloppant cette amplitude \`a
l'ordre $g^2$, et y ins\'erant les expressions des paquets d'ondes, on obtient
$$
  {\cal A} =
  \int [d{\bf q}] \, \Phi_{ P_{\SS I} }
  \int [d{\bf k}] \, \Phi_{ P_{\SS F} }^*
  \int [d{\bf q'}] \, \Phi_{ D_{\SS I} }
  \int [d{\bf k'}] \, \Phi_{ D_{\SS F} }^* \;
  {\cal A}_{ondes \; planes}(q,k,k',q')
$$
avec
\begin{eqnarray*}
  {\cal A}_{ondes \; planes}(q,k,q',k') &\equiv&
  \int d^4x \, M_P(q,k) \, e^{ -i (q-k) \cdot x }
  \int d^4x' \, M_D(q',k') \, e^{ -i (q'-k') \cdot x' }
  \\ & & \times \; G(x'-x)
\end{eqnarray*}
o\`u $M_P(q,k)$ et $M_D(q',k')$ sont les amplitudes des processus de
production et de d\'etection; la contraction des champs a donn\'e le
propagateur
\begin{equation}
  G(x'-x) =
  \int \frac{d^4p}{(2\pi)^4} \, e^{ -ip \cdot (x'-x) } \, G(p^2) \, .
  \label{proppartic}
\end{equation}
o\`u $G(p^2)= i (p^2 - M_0^2 +i\epsilon)^{-1}$ est le propagateur de la
particule libre dans l'espace des impulsions.
Les particules ext\'erieures sont sur leur couche de masse:
$$
  q^0 = E_{ P_{\SS I}}({\bf q}) = \sqrt{ {\bf q}^2 + m^2_{P_{\SS I} } } \, ,
$$
et ainsi de suite.

On a suppos\'e que la particule ($p^0 \!>\! 0$) se propage de $x$ en $x'$ et
l'antiparticule ($p^0 \!<\! 0$) de $x'$ en $x$. Si les interactions \`a la
source et au d\'etecteur sont telles que la particule se propage de $x'$ en $x$
et l'antiparticule de $x$ en $x'$,
la contraction des champs $\nu$ aurait donn\'e le propagateur
\begin{equation}
  \overline{G}(x'-x) =
  \int \frac{d^4p}{(2\pi)^4} \, e^{ ip \cdot (x'-x) } \, G(p^2) \, .
  \label{propantipartic}
\end{equation} 
Le signe diff\'erent de l'exponentielle s\'electionnerait le p\^ole de
l'antiparticule lors de l'int\'egrale de contour ult\'erieure.

Si l'objectif est de
d\'ecrire des particules instables, il est n\'ecessaire que ce soit le
propagateur {\it complet} qui apparaisse dans l'amplitude et non le propagateur
de la particule libre. Il suffit pour cela de d\'evelopper la formule
g\'en\'erale de l'amplitude \`a tous les ordres et de garder tous les diagrammes
repr\'esentant une insertion d'\'energie dans le propagateur de la particule
interm\'ediaire. La sommation de tous ces diagrammes est une s\'erie
g\'eom\'etrique en l'\'energie propre (voir \'equation (\ref{seriegeom})).
Elle fournit la m\^eme formule que ci-dessus except\'e le fait que $G(p^2)$
symbolise maintenant le propagateur complet (\'equation (\ref{spectral2})):
$$
  G(p^2) = \frac{i}{p^2 - m^2_0 - \Pi(p^2) + i\epsilon} \, .
$$
On proc\`ede ensuite \`a un changement de variable:
$$
  x \to x + x_P \qquad \mbox{et} \qquad x' \to x' + x_D \, ,
$$
o\`u $x_P = (t_P,{\bf x}_P)$ et  $x_D = (t_D,{\bf x}_D)$, ainsi qu'\`a des
int\'egrations sur les variables $x$ et $x'$ qui donnent des fonctions delta.
On arrive \`a la formule suivante de l'amplitude
\begin{equation}
  \fbox{$ \displaystyle
  {\cal A} = \int d^4p \, \varphi(p) \, G(p^2) \, e^{ -ip \cdot (x_D-x_P) }
  $}
 \label{ampli}
\end{equation}
o\`u la fonction-poids $\varphi(p)$ est une int\'egrale de recouvrement des
paquets d'ondes entrants et sortants. Elle est d\'efinie par
\begin{eqnarray}
  \varphi(p) &\equiv&
  \int [d{\bf q}]  \, \phi_{ P_{\SS I} }   ({\bf q},{\bf Q})
  \int [d{\bf k}]  \, \phi_{ P_{\SS F} }^* ({\bf k},{\bf K})
  \int [d{\bf q'}] \, \phi_{ D_{\SS I} }   ({\bf q'},{\bf Q'})
  \int [d{\bf k'}] \, \phi_{ D_{\SS F} }^* ({\bf k'},{\bf K'})
  \nonumber \\ & & \times \;
  (2\pi)^4 \, \delta^{(4)}(k+k'-q-q') \, \delta^{(4)}(p-q+k) \,
  M_P(q,k) \, M_D(q',k') \, .
  \label{recouvrement}
\end{eqnarray}
Insistons  sur le fait que ce sont bien les paquets d'ondes $\phi$
ind\'ependants de $x_P$ et $x_D$ qui figurent dans la fonction $\varphi(p)$.
Les fonctions delta y figurant imposent la conservation de
l'\'energie et de l'impulsion \`a la production et \`a la d\'etection de la
particule interm\'ediaire $\nu$. Notons aussi que cette fonction d\'epend de
$p$ (et donc de la direction de l'impulsion totale des \'etats entrants et
sortants) et non de $p^2$. La conservation de l'\'energie-impulsion introduit un
lien entre $p$ et les valeurs moyennes ${\bf Q}$, ${\bf K}$, ${\bf Q'}$, ${\bf
K'}$ des impulsions des \'etats entrants et sortants de sorte que $\varphi(p)$
d\'epende \`a la fois de la grandeur et de la direction de $p$.
L'exception est le mod\`ele simplifi\'e du chapitre \ref{propagateur}
puisque la tri-impulsion des \'etats stationnaires est tout \`a fait
ind\'etermin\'ee.
Dans ce cas, la fonction-poids  vaut
\mbox{$\varphi(p) \sim \delta(p^0\!-\! E)$} et est ind\'ependante de
${\bf p}$.

La plupart des calculs en th\'eorie des champs concernent des processus
microscopiques o\`u les \'etats entrants et sortants peuvent \^etre approxim\'es
par des ondes planes. Dans ce cas, les fonctions $\phi_{ P_{\SS I} }$,
$\phi_{ P_{\SS F} }$, $\phi_{ D_{\SS I} }$ et $\phi_{ D_{\SS F} }$ sont
toutes des fonctions delta et la fonction-poids $\varphi(p)$ est \'egale \`a
$$
  \varphi(p) = (2\pi)^4 \, \delta^{(4)}(K+K'-Q-Q') \, \delta^{(4)}(p-Q+K) \,
  M_P(Q,K) \, M_D(Q',K') \, ,
$$
o\`u $Q \equiv \left( \sqrt{ {\bf Q}^2 + m_{P_{\SS I}}^2 },{\bf Q} \right)$ et
ainsi de suite. L'int\'egration dans l'amplitude (\ref{ampli}) est alors
imm\'ediate et l'on obtient 
$$
  {\cal A} = (2\pi)^4 \, \delta^{(4)}(K+K'-Q-Q') \, G\left( (Q-K)^2 \right) \,
  e^{-i (Q-K) \cdot ( x_D - x_P ) } \, .
$$
Ce type d'expression ne peut mener \`a des oscillations dans une superposition
de plusieurs amplitudes puisque la phase de l'exponentielle ne d\'epend pas de
la masse de l'\'etat interm\'ediaire. Par contre, dans le cas d'une amplitude
mod\'elisant des processus macroscopiques, la fonction-poids $\varphi(p)$ n'est
pas une fonction delta et l'int\'egration dans l'amplitude n'est plus aussi simple.
L'int\'egration dans l'amplitude (\ref{ampli}) ne peut alors se faire sans recourir
\`a des m\'ethodes d'approximation. D'ailleurs, la forme exacte des paquets
d'ondes est inconnue.
Remarquons enfin que l'amplitude  (\ref{ampli}) d\'epend \`a la fois de la
distance et du temps macroscopiques de propagation.
Il faudra int\'egrer sur le temps pour se d\'ebarrasser de la d\'ependance
temporelle et obtenir ainsi une expression applicable aux exp\'eriences .

\section{Analyse temporelle de l'amplitude}
\label{analysetemp}

On va d'abord \'etudier la d\'ependance temporelle de l'amplitude en
int\'egrant sur l'\'energie $p^0$.
Soit l'int\'egrale
$$
  I(T) \equiv \int dp^0 \, \varphi(p) \, G(p^2) \, e^{-ip^{\SS 0} T} \, .
$$
Pour \'etudier la propagation macroscopique de la particule, il suffit
d'examiner le comportement asymptotique $T \!\to\! \infty$ de cette int\'egrale.
Cela ne peut cependant se faire sans une connaissance minimale de la
fonction-poids $\varphi(p)$. On va supposer que le spectre d'\'energie des
particules entrantes et sortantes couvre un domaine limit\'e. Par cons\'equent,
les fonctions delta figurant dans l'expression de la fonction-poids limitent le
domaine de $p$ pour lequel la fonction-poids est non nulle:
$$
  \varphi(p) \neq 0
  \quad \mbox{pour} \quad
  0 \!<\! E_1 \!<\! p^0 \!<\! E_2
  \quad \mbox{et} \quad
  {\bf p} \in {\cal D} \, ,
$$
o\`u ${\cal D}$ est un domaine \`a support compact.
Ceci implique qu'il existe des bornes $M_1$ et $M_2$ pour lesquelles
$$
  \varphi(p) \neq 0
  \quad \mbox{pour} \quad
  M_1^2 \!<\! p^2 \!<\! M_2^2 \, .
$$
On a pos\'e des conditions exp\'erimentales telles que $\varphi(p) \neq 0$
pour $p^0$ positif puisque la propagation macroscopique ($T\!>\!0$) de $x$ en $x'$ signifie
qu'il y a un transfert d'\'energie positive de $x$ en $x'$.
Comme l'amplitude a \'et\'e d\'eriv\'ee telle que la particule transf\`ere
une \'energie $p^0$ de $x$ en $x'$ (et l'antiparticule une \'energie de
$-p^0$ de $x'$ en $x$), il s'agit bien de poser des conditions exp\'erimentales
telles que $p^0 \!>\!0$. On supposera que le domaine $(E_1,E_2)$ n'inclut pas
z\'ero.

Si $\varphi(p)$ \'etait non nul pour $E_1 \!<\! p^0 \!<\! E_2 \!<\! 0$,
l'int\'egrale
de contour ult\'erieure donnerait z\'ero car la condition $T\!>\!0$ de
propagation macroscopique de $x$ en $x'$ impose de fermer le contour par le
bas. Pour \'etudier la propagation macroscopique d'une antiparticule,
il suffit de remplacer $G(x'-x)$ (\'equation \ref{proppartic})
par $\overline{G}(x'-x)$ (\'equation \ref{propantipartic}) et de poser des
conditions exp\'erimentales telles que
$\varphi(p) \neq 0$ pour \mbox{$E_1 \!<\! p^0 \!<\! E_2 \!<\! 0$}. 

Comme pour l'\'etude de la
contribution des seuils au propagateur complet, c'est le comportement de
l'int\'egrand aux bornes du domaine qui d\'eterminera le comportement
asymptotique de l'int\'egrale.
Il sera utile par la suite de faire la distinction entre les particules se
d\'esint\'egrant faiblement, appel\'ees {\it quasi-stables} et celles se
d\'esint\'egrant fortement, appel\'ees {\it r\'e\-so\-nances}.
Soit $z_0$ le p\^ole du propagateur complet.
En utilisant la m\'ethode de Jacob et Sachs \cite{jacob}, on montre que,
pour des particules stables et quasi-stables, $I(T)$ est donn\'e en tr\`es bonne
approximation pour $T$ grand par
\begin{equation}
  \fbox{$ \displaystyle
  I(T) \approx Z \, \pi \, \left( z_0 + {\bf p}^2 \right)^{ - \frac{1}{2} } \,
        \varphi(z_0,{\bf p}) \,
        \exp \left( -i \sqrt{ z_0 \!+\! {\bf p}^2 } \, T \right)
   $}
  \label{amplitemp}
\end{equation}
pour autant que les conditions exp\'erimentales soient telles que
$$
  M_1^2 \!<\! {\cal R}\!e \, z_0 \!<\! M_2^2
$$
sinon la propagation macroscopique serait inobservable. En effet, les conditions
exp\'e\-ri\-men\-tales ne seraient pas r\'eunies pour qu'une particule interm\'ediaire
se propage sur sa couche de masse, comme une particule r\'eelle.
La d\'erivation d\'etaill\'ee de cette formule figure dans l'appendice du
chapitre. Notons qu'une int\'egrale de contour sur un contour de type
demi-cercle ne converge pas pour la plupart des types de paquets d'ondes. Par
exemple, une fonction-poids gaussienne diverge sur ce contour. Le contour
propos\'e par Jacob et Sachs permet d'int\'egrer une classe beaucoup plus large
de fonctions-poids; entre autres, les gaussiennes y sont incluses. Ce point est
n\'eglig\'e dans l'\'etude des oscillations de kaons par Sudarsky {\it et al}
\cite{sudarsky}.

Si ${\cal I}\!m \, z_0 \ll {\cal R}\!e \, z_0$, le coefficient de variation de la phase de
l'amplitude en fonction du temps est
$\sqrt{ {\cal R}\!e \, z_0 \!+\! {\bf p}^2 }$.
Interpr\'etant cette expression comme l'\'energie $E$ de la particule, on peut \'ecrire
\begin{equation}
  E^2 - {\bf p}^2 = {\cal R}\!e \, z_0 \, .
  \label{onshell}
\end{equation}
Comme la masse de la particule au carr\'e est \'egale \`a la partie r\'eelle
du p\^ole du propagateur complet, la particule est, dans ce sens-l\`a, sur sa
couche de masse.
Quant \`a la partie imaginaire du p\^ole (dans le cas o\`u elle est non nulle),
elle fixe la rapidit\'e de la d\'ecroissance exponentielle de l'amplitude et est
proportionnelle \`a la largeur de la particule instable.

Les corrections \`a cette formule sont en puissances inverses de $T$. Elles
sont de deux types:

\begin{enumerate}

  \item
  Des corrections dues au spectre fini de l'\'energie des particules
  entrantes et sortantes. Elles sont donc aussi li\'ees \`a l'incertitude sur
  la localisation temporelle des interactions de production et d\'etection.
  La grandeur de ces corrections est proportionnelle \`a
  $(\Delta M T)^{-n-1}$, o\`u \mbox{$\Delta M \equiv M_j - m$} et $n$ est un nombre
  positif. Ces corrections sont importantes pour des temps tr\`es petits
  et sont en pratique inobservables dans la propagation des les particules
  stables et quasi-stables.
  Dans le cas des particules quasi-stables, elles dominent aussi l'exponentielle
  d\'ecroissante pour des temps tr\`es grands mais \`a ce moment les deux
  termes sont trop faibles pour \^etre observables dans la propagation.

  \item
  Des corrections dues \`a un seuil de production $z=b^2$ de plusieurs
  particules r\'eelles, si celui-ci se trouve dans la r\'egion o\`u le
  spectre de
  l'\'energie de la particule interm\'ediaire est non nul. Ce sont les
  corrections que nous avons examin\'e qualitativement au chapitre
  \ref{propagateur} (\'equations (\ref{stabcorr}) et (\ref{instabcorr})). 

  \begin{enumerate}
    
      \item
      Si la particule est stable, la grandeur de ces corrections est
      proportionnelle \`a $(mT)^{-3/2} g^2/Q^2$, o\`u $Q \equiv m - b$ et $g$
      est la constante de couplage avec les particules produites au seuil.
      Ces corrections ne sont importantes que pour des temps tr\`es petits
      et sont en pratique inobservables dans la propagation.
        
      \item
      Si la particule est instable, la grandeur de ces corrections est
      proportionnelle \`a $(QT)^{-3/2} \Gamma/Q$ (o\`u $Q \equiv m - b$).
      Pour des
      temps tr\`es petits, ces corrections modifient la d\'ecroissance
      exponentielle de l'amplitude. La formule habituelle de passage de la
      largeur au temps de vie est donc modifi\'ee \`a petit temps si les deux
      conditions
      suivantes ne sont pas satisfaites:
      $$
         (m - b) T \gg 1 \quad \mbox{et} \quad \Gamma/(m - b) \ll 1 \, .
      $$
      Ces conditions sont respect\'ees pour les particules quasi-stables mais
      vio\-l\'ees pour les r\'esonances.
      A grand temps, ces corrections sont aussi dominantes mais si les
      particules sont quasi-stables, les deux
      termes sont alors trop faibles pour \^etre observ\'es dans la propagation.
  
  \end{enumerate}

\end{enumerate}
  
On constate que ces deux types de correction sont importants \`a tout temps
dans le cas des r\'esonances. Ce probl\`eme ne nous concerne cependant pas ici 
puisque les r\'esonances ne se propagent pas macroscopiquement.
Dans l'appendice, figure la d\'erivation du comportement non exponentiel \`a
grand temps, qui est en puissance inverse de $T$. Par contre, bien que l'on
ait d\'elimit\'e ici l'extension du r\'egime exponentiel, le comportement non
exponentiel \`a petit temps n'a pas \'et\'e calcul\'e. Certains calculs effectu\'es en
m\'ecanique quantique montrent que, dans ce domaine temporel, la probabilit\'e
d\'ependrait quadratiquement du temps \cite{peres}, mais cette pr\'ediction n'a
pas encore pu \^etre test\'ee.

\section{Analyse spatiale de l'amplitude}
\label{analysespatiale}

Rappelons que l'amplitude de propagation (\ref{ampli}) est donn\'ee par
$$
  {\cal A} = \int d^3p \, I(T) \, e^{ i \, {\bf p} \cdot {\bf L} } \, ,
$$
o\`u ${\bf L} \equiv {\bf x}_{\SS D} - {\bf x}_{\SS P}$,
$T \equiv t_{\SS D} - t_{\SS P}$,
et $I(T)$ est donn\'e asymptotiquement
par (\ref{amplitemp}). Si l'on y substitue l'expression
asymptotique de $I(T)$, l'amplitude devient
$$
  {\cal A}(T,{\bf L}) \approx
  \int d^3p \,\psi(z_0,{\bf p}) \,
  \exp \left( - i \, \sqrt{ z_0 + {\bf p}^2 } \, T 
              + i \, {\bf p} \cdot {\bf L}
       \right) \, ,
$$
o\`u
$\psi(z_0,{\bf p}) \equiv Z \,\pi ( z_0 + {\bf p}^2 )^{-1/2} \, \varphi(z_0,{\bf p})$
est l'int\'egrale de recouvrement des paquets d'ondes entrants et sortants.
Comme ceux-ci ont des \'energies et impulsions bien d\'efinies, cette fonction
aura un pic prononc\'e.
Si l'on se souvient de la d\'efinition du p\^ole,
$z_0 = m^2 - i m \Gamma$, la condition $\Gamma/m \ll 1$
permet de r\'e\'ecrire l'amplitude sous la forme
\begin{equation}
  {\cal A}(T,{\bf L}) \approx
  \int d^3p \,\psi(m^2,{\bf p}) \,
  \exp \left(
              -\frac{m \Gamma}
                    { 2 \sqrt{ m^2 + {\bf p}^2 } } \, T
       \right) \,
  \exp \left( - i \, \sqrt{ m^2 + {\bf p}^2 } \, T 
              + i \, {\bf p} \cdot {\bf L}
       \right)
  \label{amplispatemp}
\end{equation}
Cette int\'egrale ne peut pas \^etre calcul\'ee exactement et va donc \^etre
\'evalu\'ee par approximation autour du point-selle, puisque la fonction
$\psi(z_0,{\bf p})$ a par hypoth\`ese un maximum prononc\'e.

Il est n\'ecessaire de v\'erifier pr\'ealablement que si
$\Gamma \!\neq\! 0$, l'exponentielle d\'e\-crois\-sante a une influence
n\'egligeable
sur la position du maximum dans le domaine temporel observ\'e en pratique.
Dans le cas contraire, la position du maximum d\'ependrait du temps $T$!
Restreignons-nous \`a une dimension et prenons pour mod\`ele
$$
  \psi \sim \exp \left( - \frac{ (p-P)^2 }{4 \sigma^2} \right) \, .
$$
L'int\'egrand est maximal est $p=P_{max}$, qui est la solution de
$$
  P_{max} - P = a \, P_{max} \, (P_{max}^2 + m^2)^{-3/2}
  \qquad \mbox{o\`u} \qquad
  a \equiv \sigma^2 m \Gamma T \, .
$$
Si $T=0$, $P_{max} = P$.\\
Si $T \!>\! 0$, $P_{max} = (1+\varepsilon) \,P$ avec
$\epsilon \cong a \, (P_{max}^2 + m^2)^{-3/2}$, qui est beaucoup plus petit que
1 tant que $\Gamma T \ll m^2/\sigma^2$, c'est-\`a-dire pour un grand nombre de temps
de vie. On exige que la position du maximum ne
soit pas modifi\'ee par l'exponentielle d\'ecroissante \`a une pr\'ecision de
$\bar \varepsilon$. On aura $\varepsilon \ge \bar \varepsilon$ lorsque
$$
  \Gamma T \ge
  \frac{\bar \varepsilon (P_{max}^2 + m^2)^{3/2}}{\sigma^2 m}\, .
$$
Pour un kaon $K_S$, on a par exemple $P \cong m$ dans l'exp\'erience CPLEAR
\cite{fry} et
$\sigma \cong 3 \times 10^{-2}$ MeV (incertitude sur la masse).
D\`es lors, le niveau de pr\'ecision voulu sera viol\'e lorsque
$$
  \Gamma T \ge 7 \times 10^8 \, \bar \varepsilon \, ,
$$ 
Par exemple, si l'on ne d\'esire pas une pr\'ecision sup\'erieure \`a
$\bar \varepsilon = 10^{-7}$ sur la loca\-lisation du maximum, sa position ne
changera pas sur 70 vies moyennes. Ce chiffre est nettement plus grand que les
20 vies moyennes observ\'ees dans CPLEAR \cite{fry}.

Notons par ${\bf P}$ la position du maximum de $|\psi(m^2,{\bf p})|$.
L'hypoth\`ese de sym\'etrie centrale de la phase de $\psi$ autour du maximum
implique les d\'eriv\'ees premi\`eres de $\psi$ sont nulles au maximum
${\bf P}$.
Dans le domaine o\`u elle est non nulle, la fonction $\psi$ est approxim\'ee
par une gaussienne:
\begin{equation}
  \psi(m^2,{\bf p}) \approx \psi(m^2,{\bf P}) \,
  \exp \left(
              - ({\bf p}-{\bf P}) \, W \, ({\bf p}-{\bf P})
       \right) \, .
  \label{defW}
\end{equation}
$W$ est une matrice complexe sym\'etrique contenant les d\'eriv\'ees secondes
de $\ln \psi$ \'evalu\'ees en ${\bf P}$. Le produit matriciel est d\'esormais
implicite dans les expressions du type ${\bf p}\,W\,{\bf p}$.
On d\'eveloppe aussi les autres termes de l'int\'egrand autour de ${\bf P}$:
\begin{equation}
  \sqrt{ m^2 + {\bf p}^2 } \cong
  E + {\bf v} \cdot ({\bf p}-{\bf P})
  + ({\bf p}-{\bf P}) \, R \, ({\bf p}-{\bf P}) \, ,
  \label{defR}
\end{equation}
o\`u $E \equiv \sqrt{ m^2 + {\bf P}^2 }$ et $ {\bf v} \equiv {\bf P}/E$.
La matrice $R$ contenant les d\'eriv\'ees secondes est r\'eelle et sym\'etrique.
On n\'eglige la d\'ependance en ${\bf p}$ du terme en $\Gamma$ puisque sa
contribution \`a la localisation du maximum a pu \^etre n\'eglig\'ee. Il est
donc \`a peu pr\`es constant dans le domaine o\`u $\psi$ est maximal.

Apr\`es l'insertion de ces d\'eveloppements, l'amplitude de propagation
(\ref{amplispatemp}) se r\'e\'ecrit
\begin{eqnarray*}
  {\cal A}(T,{\bf L})
  &\cong& \psi(m^2,{\bf P}) \; \exp ( -m \Gamma T /2E ) \; \exp (-iET)
  \\
  &\times& \int d^3p \;
  \exp \left[ - i {\bf v} \! \cdot \! ({\bf p}-{\bf P}) \, T
              + i \,{\bf p} \cdot {\bf L}
       \right] \;
  \exp \left( - ({\bf p}-{\bf P}) \, ( W + i R T) \, ({\bf p}-{\bf P})
       \right)
  \\
  &\cong& \psi(m^2,{\bf P}) \; \exp ( -m \Gamma T /2E ) \;
  \exp (-iET + i \,{\bf P} \cdot {\bf L} )
  \\
  &\times& \int d^3p \;
  \exp \left[ \, i {\bf p} \cdot ({\bf L} - {\bf v} T )  
              - {\bf p} \, ( W + i R T) \, {\bf p}
       \right] \, ,
\end{eqnarray*}
o\`u l'on a proc\'ed\'e \`a un changement de variable sur ${\bf p}$.
L'int\'egration sur ${\bf p}$ se fait par la formule habituelle de
l'int\'egrale gaussienne:
\begin{eqnarray}
  {\cal A}(T,{\bf L})
  &\cong& \pi^{3/2} \, \psi(m^2,{\bf P}) \; \exp ( -m \Gamma T /2E ) \;
  \left( \det (W+iRT) \right)^{-1/2}
  \label{ampliintegree}
  \\
  &\times& \exp (-iET + i \,{\bf P} \cdot {\bf L} ) \;
  \exp \left( -\frac{1}{4} \,
             ({\bf L} - {\bf v} T) \, ( W + i R T)^{-1} \, ({\bf L} - {\bf v} T)
       \right) \, .
  \nonumber
\end{eqnarray}
Si l'on mesure \`a la fois la distance ${\bf L}$ et le temps $T$ de
propagation, cette expression peut \^etre analys\'ee comme suit:

\begin{enumerate}
  \item
  La premi\`ere exponentielle contient la d\'ecroissance temporelle de
  l'amplitude due \`a la d\'esint\'egration possible de la particule.
  \item
  La deuxi\`eme exponentielle contient les termes d'oscillation
   spatio-temporelle.
  \item
  La troisi\`eme exponentielle impose la relation approximative
  ${\bf L} \!-\!{\bf v} T \le \sigma_x$ sinon l'amplitude est quasiment
  nulle; $\sigma_x$ est la largeur de la fonction-poids
  $\widetilde \psi(m^2,{\bf x})$
  dans l'espace de configuration, c'est-\`a-dire qu'elle d\'epend de la pr\'ecision
  avec laquelle on peut localiser la source et le d\'etecteur.

\end{enumerate}

On va s'int\'eresser \`a la valeur de l'amplitude dans la direction de
${\bf v}$ puisque c'est la seule direction dans laquelle elle est non n\'egligeable.
Le temps $T$ n'\'etant pas mesur\'e dans les exp\'eriences, on calcule
la probabilit\'e int\'egr\'ee sur un grand intervalle de temps dont
l'\'etendue peut \^etre prise de $-\infty$ \`a $+\infty$ sans grande erreur
gr\^ace \`a la gaussienne en $T$.
La probabilit\'e int\'egr\'ee sur le temps, dans la direction $z$
(${\bf L} = L \,\hat{\bf e}_z$) fix\'ee par
${\bf v} = v \, \hat{\bf e}_z = (P/E) \, \hat{\bf e}_z$,
se calcule \`a partir de la formule
$$
  {\cal P}(L \hat {\bf e}_z)
  \sim
  \int^{+\infty}_{-\infty} \, dT \; |{\cal A}(T,L \hat {\bf e}_z)|^2 \, .
$$
En y ins\'erant l'expression de l'amplitude, on obtient
\begin{eqnarray}
  {\cal P}(L \hat {\bf e}_z)
  &\sim& \pi^3 \, |\psi(m^2,{\bf P})|^2 \; |\det (W+iRL/v)|^{-1}
  \\
  &\times&
  \int^{+\infty}_{-\infty} dT \;
  \exp \left( - \frac{m \Gamma}{E} \, T
              - \frac{ W^{zz} + W^{zz\,*} }
                     { 4 \, | W^{zz} + i R^{zz} L/v |^2 } \,
              (L-vT)^2
       \right) \, .
  \nonumber
  \label{proba1part}
\end{eqnarray}
Les indices $zz$ sont des indices matriciels.
La dispersion a \'et\'e n\'eglig\'ee dans les termes en $RT$ en y substituant
$T=L/v$. On n\'eglige aussi les termes en $\Gamma^2$. Les facteurs de
normalisation sont omis car ils ne jouent pas de r\^ole dans notre analyse.
L'int\'egration sur $T$ donne
\begin{equation}
  \fbox{$ \displaystyle
   {\cal P}(L \hat {\bf e}_z) \sim \exp \left( -\frac{m \Gamma}{P} \, L\right)
  $}
  \label{desint}
\end{equation}
On reconna\^{\i}t la formule de la probabilit\'e de
d\'esint\'egration d'une particule en fonction de la distance, dans sa forme
relativiste.

Remarquons que gr\^ace aux fonctions delta figurant dans la fonction-poids
$\varphi(p)$, l'\'energie $E$ et l'impulsion ${\bf P}$ sont \'egales \`a
l'\'energie totale et l'impulsion totale entrantes (et sortantes), \`a une
incertitude $\sigma_p$ pr\`es d\'ependant de la taille des paquets d'ondes
dans l'espace de configuration. On a donc
$$
  {\bf v} \equiv \frac{{\bf P}}{E}
          \cong  \frac{{\bf P}_{in}}{E_{in}}
          \cong  \frac{{\bf P}_{out}}{E_{out}} \, ,
$$
o\`u
\begin{eqnarray*}
  {\bf P}_{in} &\equiv& {\bf Q} - {\bf K} \;
                \cong \; {\bf P}_{out} \;
                \equiv \; {\bf K'} - {\bf Q'}
  \\
   E_{in} &\equiv& E_{P_{\SS I}}({\bf Q}) -  E_{P_{\SS F}}({\bf K}) \;
           \cong \; E_{out} \;
           \equiv \; E_{D_{\SS F}}({\bf K'}) -  E_{D_{\SS I}}({\bf Q'})
\end{eqnarray*}

La prochaine \'etape consiste \`a appliquer cette formule de l'amplitude
et de la probabilit\'e de propagation \`a un m\'elange de particules. Ce
travail fait l'objet du prochain chapitre.

\section{Appendice: int\'egration sur $p^0$ dans l'amplitude}
\label{appendice}

L'objectif est de calculer la valeur asymptotique $T \!\to\! \infty$ de
l'int\'egrale (\ref{amplitemp}):
$$
  I(T) \equiv \int dp^0 \, \varphi(p) \, G(p^2) \, e^{-ip^{\SS 0} T} \, .
$$
Le propagateur est donn\'e par l'\'equation (\ref{spectral2}) ou
(\ref{propren}).
La fonction-poids $\varphi(p)$ n'est diff\'erente de z\'ero que sur un
intervalle fini. Nous prendrons comme mod\`ele
\begin{eqnarray*}
  \varphi(p) &=& (p^2 - M_1^2)^n \, (p^2 - M_2^2)^n \,
  \Omega(p^2,{\bf p})
  \quad \mbox{pour} \quad
  0 < M_1^2 < p^2 < M_2^2 \, ,
  \\
  \varphi(p) &=& 0 \quad \mbox{sinon} \, .
\end{eqnarray*}
$p^0$ est positif ainsi que le nombre $n$. La fonction $\Omega$ est
analytique en $p^2$. Le comportement sym\'etrique de $\varphi(p)$ \`a ses deux
bornes a \'et\'e choisi pour simplifier le calcul mais n'est pas indispensable.
L'exigence d'analyticit\'e de $\Omega$ n'est pas trop contraignante car on peut
toujours approximer une fonction r\'eguli\`ere sur un intervalle fini par un
polyn\^ome (qui est analytique).

Sous le changement de variable $z=p^2={p^0}^2 \!-\! {\bf p}^2$,
l'int\'egrale $I(T)$ devient
$$
  I(T) = \frac{1}{2} \int_{M_1^2}^{M_2^2} dz \,
  (z \!+\! {\bf p}^2 )^{ -\frac{1}{2} } \, \varphi(z,{\bf p}) \, G(z) \,
  e^{ -i \sqrt{ z + {\bf p}^2 } T } \, .
$$
Le propagateur $G(z)$ a des points de branchement aux seuils de contribution
des \'etats \`a plusieurs particules. Soit $z\!=\!b^2$ le premier point de
branchement qui correspond au seuil de production de plusieurs
particules r\'eelles. La coupure sur l'axe r\'eel commence donc en
$z\!=\!b^2$. On fait l'hypoth\`ese que les autres points de branchement sont
sup\'erieurs \`a $M_2^2$.
La continuation analytique de $G(z)$ sur la deuxi\`eme
feuille de Riemann est not\'ee comme au chapitre pr\'ec\'edent $G_{II}(z)$ et a
un p\^ole en
$z \!=\! z_0 \!=\! m^2 - i m \Gamma$.
On va diff\'erencier les cas selon les positions du p\^ole et du seuil $b$, et
suivant que la particule est ou non instable.

\begin{enumerate}

\item
Particule instable: $b < M_1 < m < M_2$ et $\Gamma \neq 0$ \\
On choisit le contour d'int\'egration d\'efini par
\begin{center}
\includegraphics[width=10cm]{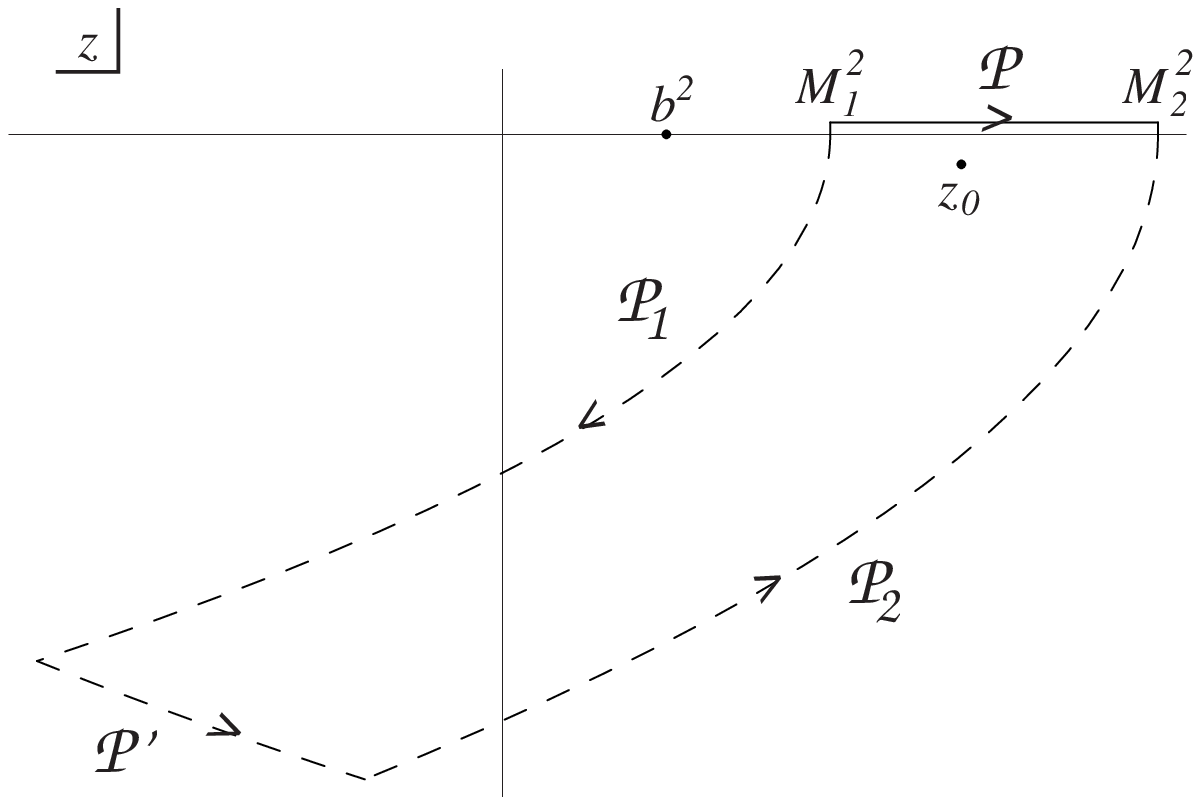}
\end{center}
L'expression analytique des contours ${\cal P}_1$ et ${\cal P}_2$ est
$$
  {\cal P}_j : \, z =
  \left(
        -i \omega + \sqrt{ M_j^2 + {\bf p}^2 }
  \right)^2
  - {\bf p}^2
  \quad (j=1,2) \, ,
$$
o\`u $\omega$ varie de z\'ero \`a $\omega_\infty$ sur ${\cal P}_1$ et
de $\omega_\infty$ \`a z\'ero sur ${\cal P}_2$.

L'expression analytique du contour ${\cal P'}$ est
$$
  {\cal P'} : \, z =
  \left(
        -i \omega_\infty + \sqrt{ M^2 + {\bf p}^2 }
  \right)^2
  - {\bf p}^2 \, ,
$$
o\`u $M$ varie de $M_1$ \`a $M_2$.
On prend la limite $\omega_\infty \to \infty$.

L'int\'egrale recherch\'ee est \'egale \`a
$$
  I(T) = J + J_1 + J_2 + J' \, .
$$
$J$ est la contribution du p\^ole $z_0$ tandis que $J_1$, $J_2$ et $J'$ sont
les contributions respectives de ${\cal P}_1$, ${\cal P}_2$, ${\cal P'}$. Leurs
expressions analytiques s'\'ecrivent

  \begin{enumerate}
  
  \item
  Contribution du p\^ole:
  \begin{eqnarray*}
    J &=& Z \, \pi \, \left( z_0 + {\bf p}^2 \right)^{ - \frac{1}{2} } \,
        \varphi(z_0,{\bf p}) \,
        \exp \left(
                -i \sqrt{ z_0 \!+\! {\bf p}^2 } \, T
             \right)
    \\
      &=& Z \, \pi \, \left( z_0 + {\bf p}^2 \right)^{ - \frac{1}{2} } \,
        \varphi(z_0,{\bf p}) \,
        \exp \left(
                -i \sqrt{ m^2 \!+\! {\bf p}^2 } \, T
                - \frac{m \Gamma}{ 2 \, \sqrt{ m^2 \!+\! {\bf p}^2 } } \, T
             \right) \, .
  \end{eqnarray*}
  On a d\'efini $z_0 = m^2 - i m \Gamma$ et fait l'hypoth\`ese que
  $\Gamma/m \ll 1$. Comme on l'a d\'ej\`a dit, cette relation est tr\`es bien
  v\'erifi\'ee dans la majorit\'e des cas. Par exemple,
  $\Gamma/m \approx {\cal O}(10^{-14})$ pour le $K^0_S$.
  Posant $\Delta M \cong |m-M_{1,2}|$, l'ordre de grandeur de $J$ est donn\'e par
  \mbox{${\cal O}(J) \sim m^{2n-1} \, (\Delta M)^{2n} \, e^{-\Gamma T/2}$}.
  
  \item
  Contribution des contours en faucille:
  $$
    J_j = i (-1)^j \,
          \exp \left(
                     -i \sqrt{ M_j^2 \!+\! {\bf p}^2 } \, T
               \right)
          \int_0^\infty d\omega \, \varphi \left( z(\omega),{\bf p} \right) \,
          G_{II}\left( z(\omega) \right) \, e^{-\omega T} \, .
  $$
  Pour $T$ grand , la contribution dominante \`a l'int\'egrale vient des valeurs
  de $\omega$ proches de z\'ero \`a cause de l'exponentielle d\'ecroissante.
  Comme l'int\'egrand tend vers z\'ero lorsque $\omega$ tend vers z\'ero
  ($\varphi(M_j^2)=0)$), le comportement asymptotique de l'int\'egrale
  d\'epend du type de convergence vers z\'ero de l'int\'egrand.
  On y substitue $y=\omega T$ et l'on effectue un d\'eveloppement en $1/T$:
  \begin{eqnarray*}
    J_j &=& -(1)^{n(2-j)+j} \, 2^n \, n! \,
            \left( M_2^2 \!-\! M_1^2 \right)^n \,
            \left( M_j^2 \!+\! {\bf p}^2 \right)^{n/2} \,
            \Omega \left( M_j^2,{\bf p} \right) \,
            G_{II}\left( M_j^2 \right)
    \\
    & & \times \; (iT)^{-(n+1)} \,
        \exp \left(
                    -i \sqrt{ M_j^2 \!+\! {\bf p}^2 } \, T
             \right) \, .
  \end{eqnarray*}
  Les corrections \`a cette formule sont en $(\Delta M T)^{-(n+2)}$, o\`u
  $\Delta M \approx M_j \!-\! m$,
  c'est-\`a-dire de l'ordre de grandeur de
  l'incertitude sur la masse de la particule.
  On fait implicitement l'hypoth\`ese que la fonction $\Omega (z,{\bf p})$
  diverge moins vite que $\exp (-\omega T)$ sur les contours ${\cal P}_j$ quand
  $\omega$ tend vers l'infini, ce qui recouvre une large classe de fonctions.
  Par exemple, les gaussiennes conviennent, alors que leur int\'egrale diverge
  sur des contours de type demi-cercle \`a de rayon infini.
  L'ordre de grandeur de $J_j$ est donn\'e par
  \mbox{${\cal O}(J_j) \sim m^{2n-1} \, (\Delta M)^{n-1}\, T^{-(n+1)}$}. 
  
  \item
  Contribution du contour \`a l'infini:
  $$
    J' = e^{-\omega_\infty T} \int_{M_1^2}^{M_2^2} dM
        \frac{M}{ \sqrt{M^2 \!+\! {\bf p}^2 } } \,
        \varphi \left( z(M) \right) \,
        G_{II} \left( z(M) \right) \,
        \exp \left( -i \sqrt{ M^2 \!+\! {\bf p}^2 } \, T \right) \, .
  $$
  Si $\Omega$ satisfait aux m\^emes conditions \`a l'infini que ci-dessus,
  $J' \!\sim\! \exp(-\omega_\infty T)$ et tend vers z\'ero lorsque
  $\omega_\infty \to \infty$. 
  
  \end{enumerate}

En conclusion, la contribution du p\^ole est une exponentielle d\'ecroissante
en $T$ tandis que les contributions dues aux bornes du spectre d'\'energie sont
en puissances inverses de $T$.

Pour des temps petits, on a ${\cal O}(J) \sim m^{2n-1} \, (\Delta M)^{2n}$ donc
\mbox{${\cal O}(J_j/J) \sim (\Delta M \, T)^{-(n+1)}$}.
Comme les interf\'erences
entre $J_j$ et $J$ disparaissent par moyenne sur de petits intervalles de temps,
la contribution de $J_j$ par rapport $J$ \`a la probabilit\'e sera plut\^ot
donn\'ee par $(J_j/J)^2 \sim (\Delta M \, T)^{-2(n+1)}$, qui est notable pour
$\Delta M \, T \le 1$. En de\c{c}\`a de cette borne, le calcul asymptotique de
$J_j$ n'est plus valable car on a n\'eglig\'e lors du calcul des termes en
$(\Delta M T)^{-(n+2)}$. 

Dans le cas d'une particule quasi-stable, par exemple le $K_S^0$, la masse est
mesur\'ee \`a une pr\'ecision de $10^{-2}$ MeV.
On a donc $\Delta M \approx 10^{-2}$ MeV et les termes
non exponentiels contribueront notablement pour $T \le 10^{-19}$ s, ce qui est
inobservable dans la propagation puisque le temps de vie du $K_S^0$ est de
$0.89 \, \times \, 10^{-10}$ s.
Pour le $B^0$, $\Delta M \approx 2$ MeV et les termes
non exponentiels contribueront notablement pour $T \le 10^{-22}$ s, ce qui est
inobservable dans la propagation puisque le temps de vie du $B^0$ est de
$1.29 \, \times \, 10^{-13}$ s.

Dans le cas d'une r\'esonance, par exemple le $\Delta(1232)$,
$\Delta M \approx 2$ MeV et les termes non exponentiels contribueront
notablement pour $T \le 10^{-21}$ s, qui est grand par rapport \`a l'inverse
de la largeur \'egale \`a $5 \, \times \, 10^{-24}$ s. Quel que soit le
domaine temporel, la propagation des r\'esonances ne peut jamais  mod\'elis\'ee 
par la seule contribution du p\^ole: il faut tenir compte des conditions de
production et de d\'etection. Cependant, les r\'esonances ne se
propagent pas macroscopiquement donc il n'est pas crucial de pouvoir
calculer par exemple leur temps de vie en fonction de leur largeur.   

Dans le cas de la particule quasi-stable, la contribution du p\^ole ne domine
pas non plus \`a grand temps. Les termes non exponentiels
prennent le dessus lorsque
$$
  \Gamma T \, > \, 2(n+1) \, \ln(\Delta M T) \,
  = \, 2(n+1) \left( \ln (\Gamma T) + \ln(\Delta M /\Gamma) \right) \, .
$$
Pour le $K_S^0$, soit $n=1/2$ et
$\Delta M / \Gamma \approx {\cal O}(10^{10})$, de sorte que $\Gamma T > 82$, ce
qui est inobservable.
Pour le $B^0$, soit $n=1/2$ et
$\Delta M / \Gamma \approx {\cal O}(10^{9})$, de sorte que $\Gamma T > 75$, ce
qui est inobservable.

En fin de compte, $I(T)$ est donn\'ee en tr\`es bonne approximation pour les
particules quasi-stables par
\begin{equation}
  \fbox{$\displaystyle
  I(T) \approx Z \, \pi \, \left( z_0 + {\bf p}^2 \right)^{ - \frac{1}{2} } \,
        \varphi(z_0,{\bf p}) \,
        \exp \left( -i \sqrt{ z_0 \!+\! {\bf p}^2 } \, T \right)
  $}
  \label{integIT}
\end{equation}

\item
Particule instable: $M_1 < b < m < M_2$ et $\Gamma \neq 0$\\
Dans ce cas-ci, le seuil de production de plusieurs particules r\'eelles se
trouve dans la domaine d'\'energies permises. Il va g\'en\'erer des
contributions en puissances inverses du temps, qui sont les
corrections que nous avons examin\'ees qualitativement au chapitre
\ref{propagateur} (\'equations (\ref{stabcorr}) et (\ref{instabcorr})).

On choisit le contour d\'efini par
\begin{center}
\includegraphics[width=10cm]{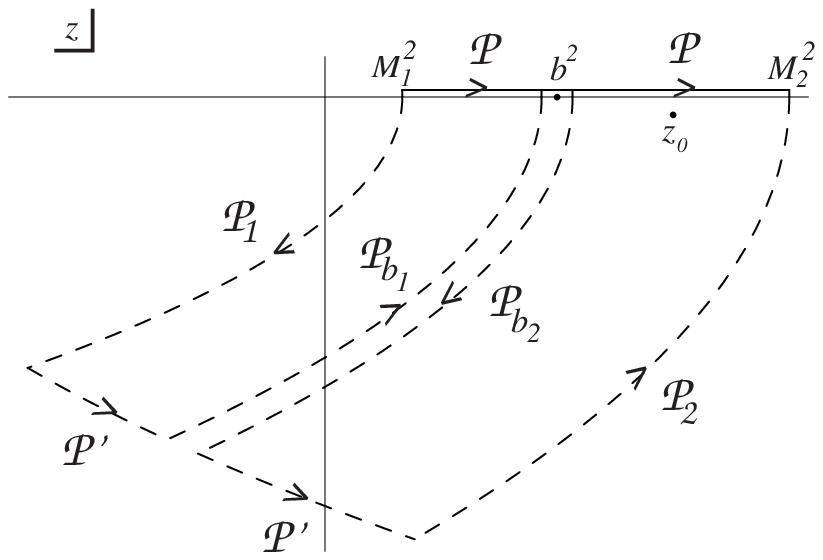}
\end{center}
Les contributions des contours ${\cal P}_j$ sont les m\^emes que dans le cas
pr\'ec\'edent, sauf que l'on substitue $G(z)$ \`a $G_{II}(z)$ sur ${\cal P}_1$
puisque ce trajet se trouve sur la premi\`ere feuille de Riemann. La valeur
asymptotique $T \!\to\! \infty$ de $J_j$ ne changera pas.

La somme des contributions des trajets ${\cal P}_{b_j}$ serait nulle si ces
trajets se trouvaient sur la m\^eme feuille de Riemann mais ce n'est pas le
cas. Leur contribution totale $J_b$ s'\'ecrit
\begin{eqnarray}
  J_b &=& \frac{1}{2} \int_{ {\cal P}_b } dz \,
  (z \!+\! {\bf p}^2 )^{ -\frac{1}{2} } \, \varphi(z,{\bf p}) \,
  \left( G_{II}(z) - G(z) \right) \,
  e^{ -i \sqrt{ z + {\bf p}^2 } T }
  \nonumber \\
  &=& -i \, e^{ -i \sqrt{ b^2 + {\bf p}^2 } T }
  \int_0^\infty d\omega \, \varphi(z(\omega),{\bf p}) \,
  \left( G_{II}(z) - G(z) \right) \,
  e^{-\omega T} \, .
  \label{exprJb}
\end{eqnarray}
L'expression analytique du contour ${\cal P}_b$ est donn\'ee par
\begin{equation}
  {\cal P}_b : \, z =
  \left(
        -i \omega + \sqrt{ b^2 + {\bf p}^2 }
  \right)^2
  - {\bf p}^2 \, ,
  \label{contourb}
\end{equation}
o\`u $\omega$ varie de z\'ero \`a $\omega_\infty$ et l'on prend la limite
$\omega_\infty \to \infty$.
L'int\'egrale $I(T)$ est la somme de toutes les contributions des contours:
$$
  I(T) = J+J_1+J_2+J_b+J' \, .
$$

Le comportement asymptotique de $J_b$ est \'etudi\'e de
m\^eme mani\`ere que celui des $J_j$. Pour $T$ grand, la
contribution dominante \`a l'int\'egrale vient des valeurs de
$\omega$ proches de z\'ero \`a cause de l'exponentielle d\'ecroissante. Comme
l'int\'egrand tend vers z\'ero lorsque $\omega$ tend vers z\'ero
($G_{II}(z) \!\to\! G(z)$), le comportement asymptotique de l'int\'egrale
d\'epend du type de convergence vers z\'ero de l'int\'e\-grand. On substitue
$y=\omega T$ et on d\'eveloppe l'int\'egrand en $1/T$.
Si l'on se souvient que $G_{II}(z)$ est d\'efini par prolongement analytique de
$G(x+i\epsilon)$ en dessous de la coupure (\'equation (\ref{propinstab}))
et que sa discontinuit\'e \`a travers la coupure est donn\'ee par l'\'equation
(\ref{coupure}),
$$
  G(x+i\epsilon) - G(x-i\epsilon)  = 2\pi \, \rho(x) \, ,
$$
o\`u $\rho(x)$ est la fonction spectrale, on a que
$$
  G_{II}(z) - G(z) = 2\pi \, \rho(z) \, .
$$
Pour fixer les id\'ees, prenons le syst\`eme $K\pi\pi$ dont nous avons
explicitement calcul\'e la fonction spectrale $\rho(z)$
(\'equation (\ref{rhokaon})).
Le seuil \`a deux particules est en $z=b^2$, avec $b=2 m_\pi$. 
$$
  \rho(z) =\frac{1}{\pi}
  \frac{v(z)}{ \left( z - m^2 - u(z) \right)^2 + v^2(z) } \, ,
$$
o\`u $m$ est la masse du kaon et
$$
  v(z) = \frac{g^2}{4\pi}
  \sqrt{ 1- b^2/z } \, .
$$
L'\'evaluation de l'int\'egrale $J_b$ donne
$$
  J_b = -i \, (-2i\pi)^\frac{1}{2} \, T^{-3/2} \, \frac{g^2}{4\pi b} \,
  \frac{ \left( b^2 + {\bf p}^2 \right)^{1/4} }
       { \left( b^2 - m^2 - u(b^2) \right)^2 } \, 
  \varphi(b^2,{\bf p}) \,
  e^{ -i \sqrt{ b^2 + {\bf p}^2 } T } \, .
$$
Les corrections \`a cette formule sont de l'ordre de $(QT)^{-5/2}$ o\`u 
$Q \equiv m-b$ est l'\'energie disponible lors de la d\'esint\'egration de la
particule.
Puisque nous avons calcul\'e la largeur de $K\!\to\!2\pi$
(\'equation (\ref{largeurK})), on peut remplacer $g^2$ par son expression en
fonction de $\Gamma$, $m$ et $Q$. L'ordre de grandeur de $J_b$ est donn\'e par
$$
  {\cal O}(J_b) \sim
  (QT)^{-3/2} \, \frac{\Gamma}{Q} \, \frac{1}{m} \, \varphi(b^2,{\bf p}) \, .
$$ 
Pour des temps petits, l'ordre de grandeur de $J$ est donn\'e par
\mbox{${\cal O}(J) \sim \varphi(m^2,{\bf p})$} et l'on obtient
$$
  {\cal O}\left(\frac{J_b}{J}\right) \sim (Q T)^{-3/2} \, \frac{\Gamma}{Q}
$$
qui est beaucoup plus petit que 1 si $QT \gg 1$ et $\Gamma/Q \ll 1$.
C'est le cas pour les particules quasi-stables. Par exemple,
$\Gamma/Q \approx 10^{-14}$ pour le $K_S^0$.
Par contre, ce n'est pas vrai pour les r\'esonances. Par exemple,
$\Gamma/Q \approx 0.8$ pour le $\Delta(1232)$.
En de\c{c}\`a de la borne $QT \approx 1$, le calcul asymptotique de $J_b$ n'est
plus valable, puisque l'on a n\'eglig\'e des corrections en
$(QT)^{-5/2}$.

Pour des grands temps, $J_b$ prend le dessus pour
$$
  \Gamma T - 3 \ln(\Gamma T) > 5 \ln(Q/\Gamma) \, .
$$
qui se traduit par $\Gamma_S T \!>\! 165$ pour le $K_S^0$,
$\Gamma_L T \!>\! 202$ pour le $K_L^0$ et
$\Gamma_{L,H} T \!>\! 157$ pour le $B^0_{L,H}$ (pour ce dernier, on a pris
$Q \cong 1$ MeV). Apr\`es un aussi grand nombre de temps de vie, la probabilit\'e
de d\'etection d'une particule est inobservable et ces corrections non
exponentielles sont donc ind\'etectables.

Pour les r\'esonances, la contribution de $J_b$ est comparable ou sup\'erieure
\`a celle du p\^ole pour tout domaine temporel.

Dans le cas des particules quasi-stables, l'int\'egrale $I(T)$ est donc donn\'ee
en tr\`es bonne approximation par la m\^eme expression qu'au premier cas
(\'equation (\ref{integIT})):
$$
  I(T) \approx Z \, \pi \, \left( M_j^2 + {\bf p}^2 \right)^{ - \frac{1}{2} } \,
        \varphi(z_0,{\bf p}) \,
        \exp \left( -i \sqrt{ z_0 \!+\! {\bf p}^2 } \, T \right)
$$  
Quant aux r\'esonances, elles ne se propagent pas macroscopiquement, leur
largeur \'etant du m\^eme ordre de grandeur que les \'energies typiques des
processus.

\item
Particule stable: $M_1 < m < M_2 < b $ et $\Gamma=0$\\
Ce cas se traite similairement au premier cas sauf que le p\^ole est r\'eel. On
peut reprendre le r\'esultat (\ref{integIT}). Il n'y a pas de
d\'ecroissance exponentielle de la contribution du p\^ole puisque celui-ci est r\'eel.
Les corrections \`a grand temps ne dominent jamais la contribution du p\^ole.

\item
Particule stable: $M_1 < m < b < M_2$ et $\Gamma=0$\\
Ce cas se traite similairement au deuxi\`eme cas sauf que le p\^ole est r\'eel.
On peut reprendre le r\'esultat en posant $\Gamma=0$.
Le rapport $J_b/J$ vaut
$J_b/J \sim (m T)^{-3/2} g^2/Q^2$, o\`u g est la constante de couplage avec les
particules produites au seuil. La contribution de $J_b$ est inobservable en
pratique \`a petit temps et sous-dominante \`a grand temps. 

\item
$m < M_1$ ou $m > M_2$\\
Le p\^ole se trouve en dehors du contour d'int\'egration et ne contribuera pas.
La propagation macroscopique de la particule sera inobservable. Les
\'energies-impulsions des \'etats initiaux ou finaux ne permettent pas que la
particule interm\'ediaire soit sur sa couche de masse et cela n'a pas de sens
d'\'etudier sa propagation comme celle d'une particule quasi-r\'eelle.

\end{enumerate}

\chapter{Oscillations en th\'eorie des champs}

\section{Propagation d'un m\'elange de particules}

Maintenant que la propagation d'une seule particule a \'et\'e \'etudi\'ee en
d\'etail, nous sommes pr\^ets \`a attaquer le probl\`eme de la propagation
d'un m\'elange de particules dans le cadre du mod\`ele sophistiqu\'e o\`u
les \'etats entrants et sortants sont mod\'elis\'es par des paquets d'ondes.
Remarquons que l'\'etat interm\'ediaire n'est pas un paquet d'ondes.
Cette m\'ethode n'appartient donc pas \`a la cat\'egorie dite du
{\it traitement des oscillations par paquet d'ondes}, qui consiste \`a
consid\'erer la particule oscillante comme un paquet d'ondes dans le
cadre de la m\'ecanique quantique \cite{kayser2,giunti3}.
Notre m\'ethode rel\`eve bien de la
th\'eorie des champs, o\`u les ondes planes utilis\'ees dans le calcul
d'amplitudes de Feynman ne sont en fait que des
approximations rempla\c{c}ant les paquets d'ondes \cite{peskin}.

Apr\`es quelques rappels de notation, nous appliquerons la formule de
propagation calcul\'ee au chapitre pr\'ec\'edent \`a une superposition de
propagations de diff\'erents \'etats propres de masse. Les \'etapes de l'analyse
seront identiques. Suite \`a l'\'etablissement de la formule de
l'amplitude, nous en effectuons une analyse temporelle, lors de laquelle nous
v\'erifierons la non observabilit\'e de nouvelles corrections non exponentielles
dues au m\'elange. L'\'etude de la propagation du m\'elange continue par une
analyse spatiale de l'amplitude et le calcul de la probabilit\'e de d\'etection
int\'egr\'ee sur le temps.
La formule de probabilit\'e obtenue sera analys\'ee terme par terme.
Nous retrouverons les termes de d\'esint\'egration et d'oscillation,
quoique leur forme sera diff\'erente de celle obtenue dans le mod\`ele
simplifi\'e.
Par contre, les termes de {\it d\'ecoh\'erence} et de {\it localisation des
interactions} n'apparaissaient pas dans le mod\`ele pr\'ec\'edent. Ils
imposeront une s\'erie de conditions d'observabilit\'e des oscillations.
Pour terminer, nous passerons en revue les \'eclaircissement que ce
mod\`ele apporte aux diff\'erentes questions pos\'ees au chapitre
\ref{oscillationsMQ} ainsi que les points communs et diff\'erences de notre
approche avec les autres traitements existant dans la litt\'erature. 

Il est peut-\^etre utile de rappeler les
notations utilis\'ees dans la section \ref{oscisimpli} pour d\'ecrire
un processus de propagation avec m\'elange.

\section{Amplitude pour un m\'elange}

Le processus de propagation garde la m\^eme forme que celui du mod\`ele
simplifi\'e de la section \ref{oscisimpli}:
\begin{eqnarray*}
  P_I(q)
 \stackrel{ (t_{\SS P},{\bf x}_{\SS P} ) }{\longrightarrow}
  P_F(k) + \nu_\alpha(p) & &
  \\
  &\searrow&
  \\
  & & \nu_\beta(p) + D_I(q') 
  \stackrel{ (t_{\SS D},{\bf x}_{\SS D} ) }{\longrightarrow} D_F(k')
\end{eqnarray*}
La saveur $\alpha$ de la particule interm\'ediaire $\nu_\alpha$ produite dans
la r\'egion de $(t_{\SS P},{\bf x}_{\SS P} )$ est identifi\'ee par l'\'etat
sortant $P_F(k)$ tandis que la saveur $\beta$ de la particule $\nu_\beta$
d\'etect\'ee dans la r\'egion de $(t_{\SS D},{\bf x}_{\SS D} )$ est
identifi\'ee par l'\'etat sortant $D_F(k')$.
Si ce n'est pas possible (ex:
$K^0, \overline{K^0} \to \pi^+\pi^-$), on proc\`ede \`a une sommation sur
les diff\'erentes saveurs.

Les notations ne changent pas: le propagateur complet $G_{\beta\alpha}(x'-x)$ 
symbolise la propagation de particules de saveur $\alpha$ produites en $x$ en
particules de saveur $\beta$ d\'etect\'ees en $x'$.
Ce propagateur est diagonalis\'e
(\'equations (\ref{stablediag}) et (\ref{propdiag})) en tr\`es
bonne approximation par des matrices $V$ constantes\footnote{Rappelons que la
matrice $V$ telle qu'elle est d\'efinie ici est la transpos\'ee de la matrice
$V$ diagonalisant les kets dans la d\'erivation en m\'ecanique quantique
pr\'esent\'ee au chapitre \ref{oscillationsMQ}.}:
$$
  G_{\beta\alpha}(p^2)
  = (V^{-1} \, G_D(p^2) \, V)_{\beta\alpha}
  = \sum_j V_{\beta j}^{-1} \, G_{D,jj}(p^2) \, V_{j\alpha} \, ,
$$
o\`u $G_{D,jj}(p^2) \equiv i \, (p^2-z_j)^{-1}$. Le nombre complexe $z_j$ est le
p\^ole du propagateur complet associ\'e \`a l'\'etat $j$, de masse $m_j$ et de
largeur $\Gamma_j$, auxquelles il est reli\'e par
$$
  z_j \equiv m_j^2 - i m_j \Gamma_j \, .
$$
La propagation macroscopique implique qu'il s'agit de particules quasi-stables,
c'est-\`a-dire que $m_j \!\ll\! \Gamma_j$.

L'amplitude du processus global peut s'exprimer comme une superposition
lin\'eaire d'amplitudes ${\cal A}_j$ de propagation d'\'etats propres de masse:
\begin{equation}
  {\cal A}(\alpha \!\to\! \beta,T,{\bf L})
  = \sum_j V_{\beta j}^{-1} \, {\cal A}_j \, V_{j\alpha} \, ,
  \label{amplitot}
\end{equation}
o\`u l'amplitude partielle ${\cal A}_j$ s'\'ecrit comme l'amplitude de
propagation d'une particule isol\'ee (\'equation (\ref{ampli})):
\begin{equation}
  {\cal A}_j = \int d^4p \; \varphi(p) \; G_{D,jj}(p^2) \;
  e^{-i p \cdot ( x_{\SS D} - x_{\SS P} ) } \, .
  \label{defAj}
\end{equation}
La fonction-poids $\varphi(p)$ symbolise comme auparavant
(\'equation(\ref{recouvrement})) l'int\'egrale de recouvrement des paquets
d'ondes:
\begin{eqnarray}
  \varphi(p) &\equiv&
  \int [d{\bf q}]  \, \phi_{ P_{\SS I} }   ({\bf q},{\bf Q})
  \int [d{\bf k}]  \, \phi_{ P_{\SS F} }^* ({\bf k},{\bf K})
  \int [d{\bf q'}] \, \phi_{ D_{\SS I} }   ({\bf q'},{\bf Q'})
  \int [d{\bf k'}] \, \phi_{ D_{\SS F} }^* ({\bf k'},{\bf K'})
  \nonumber \\ & & \times \;
  (2\pi)^4 \, \delta^{(4)}(k+k'-q-q') \, \delta^{(4)}(p-q+k) \,
  M_P(q,k) \, M_D(q',k') \, .
  \label{fonctionpoids}
\end{eqnarray}
En ce qui concerne la signification des indices dans ce chapitre,
les indices inf\'erieurs $i$ et $j$ seront des indices de saveur tandis que les
indices sup\'erieurs seront des indices vectoriels ou matriciels (dans
l'espace des impulsions ou dans l'espace de configuration). Notons aussi que
le produit matriciel sera implicite dans les expressions du genre
${\bf L} W {\bf L}$.

\section{Analyse temporelle de l'amplitude de m\'elange}

Suivant la proc\'edure \'etablie pour \'etudier la propagation d'une
particule isol\'ee, nous effectuerons l'analyse temporelle et spatiale de
l'amplitude avant de calculer la probabilit\'e correspondante.

Il est plus facile pour commencer de ne consid\'erer que l'amplitude partielle
${\cal A}_j$ (\'equation (\ref{defAj})).
L'analyse temporelle consiste \`a int\'egrer sur $p^0$. On recourt
\`a la m\^eme m\'ethode d'int\'egration de contour qu'au chapitre pr\'ec\'edent
(voir l'\'equation (\ref{amplispatemp})) et l'on obtient
\begin{equation}
  {\cal A}_j = \int d^3p \; \psi(m^2_j,{\bf p}) \;
  \exp \left(
             - \frac{m_j \Gamma_j}{ 2 \sqrt{ m^2_j + {\bf p}^2 } } \, T
       \right) \;
  \exp \left(
             - i \sqrt{ m^2_j + {\bf p}^2 } \, T
             + i {\bf p} \!\cdot\! {\bf L}
       \right) \, ,
  \label{amplitempAj}
\end{equation}
o\`u
$\psi(m^2_j,{\bf p}) \equiv Z \,\pi ( z_j + {\bf p}^2 )^{-1/2} \, \varphi(z_j,{\bf p})$.

A ce stade, on pourrait se demander ce que sont devenues les corrections non
exponentielles \`a l'amplitude de propagation, qui ont \'et\'e \'etudi\'ees en d\'etail
au chapitre pr\'ec\'edent dans le cas de la propagation d'une particule
isol\'ee.
En premier lieu, nous avons \'et\'e confront\'es \`a des corrections dues
aux conditions de production et de d\'etection, d\'ependant de la forme des paquets
d'ondes initiaux et finaux. Dans le cas du m\'elange de particules \'etudi\'e dans ce
chapitre, ces corrections peuvent se calculer pour chaque amplitude partielle selon la
m\'ethode expliqu\'ee au chapitre pr\'ec\'edent et restent n\'egligeables.
En second lieu, nous avons rencontr\'e des corrections dues aux seuils de production
des \'etats \`a plusieurs particules.
Rappelons qu'elles se mani\-festent par des
discontinuit\'es (seuils) de la d\'eriv\'ee de l'\'energie propre en fonction
de $p^2$. Or la m\'ethode de diagonalisation par des matrices constantes
approxime l'\'energie propre par une constante (\'equation
(\ref{energieprcste})). Par cons\'equent ces seuils et ces corrections
disparaissent dans cette approximation.

Il serait n\'eanmoins bon de tester l'influence des corrections dues aux seuils
de production sur la propagation de particules m\'elang\'ees, pour v\'erifier
qu'elles ne perturbent pas les oscillations. Revenons donc un pas en
arri\`ere, \`a la diagonalisation exacte par des matrices d\'ependant de $z=p^2$.
En particulier, consid\'erons la diagonalisation exacte du propagateur complet
du $K^0\overline{K^0}$ (\'equation (\ref{diag1})).
$$
  -i \, G(z) =
  W(z)
  \left(
  \begin{array}{cc}
  ( z - m_1^2 - f_1(z) )^{-1}  &  0                          \\
  0                            &  ( z - m_2^2 - f_2(z) )^{-1}
  \end{array}
  \right)
  W^{-1}(z) \, .
$$
Nous allons n\'egliger la violation CP et approximer les \'etats de propagation
$K_S$ et $K_L$ par
les \'etats propres sous CP, $K_1$ et $K_2$. 
L'incidence sur le m\'elange des corrections dues \`a un seuil en $z=b^2$ peut
\^etre \'etudi\'ee en effectuant, par exemple, l'analyse temporelle de l'\'el\'ement
non diagonal $G_{0\bar0}(p^2)$. En effet, si nous d\'efinissons un
$I_{\beta\alpha}(T)$ similaire au $I(T)$ apparaissant \`a la
section \ref{analysetemp} par
$$
  {\cal A}(\alpha \!\to\! \beta,T,{\bf L}) \equiv
  \int d^3p \, I_{\beta\alpha}(T) \, e^{ i {\bf p} \cdot {\bf L} } \, ,
$$
on peut \'ecrire
$$
  I_{0\bar0}(T) \equiv \int dp^0 \, \varphi(p) \, G_{0\bar0}(p^2) \, e^{-ip^0T} \, ,
$$ 
o\`u $\varphi(p)$ est la fonction-poids d\'efinie en (\ref{fonctionpoids}) et l'on a extrait de
l'expression matricielle ci-dessus l'\'el\'ement $0\bar0$ du propagateur:
\begin{equation}
  G_{0\bar0}(p^2) =
    \frac{i \, W_{01}(p^2) W_{1\bar0}^{-1}(p^2)}{p^2 - m_1^2 -f_1(p^2)}
  + \frac{i \, W_{02}(p^2) W_{2\bar0}^{-1}(p^2)}{p^2 - m_2^2 -f_2(p^2)} \, .
  \label{G00}
\end{equation}
On reprend la m\'ethode du chapitre pr\'ec\'edent, section \ref{appendice}.
L'int\'egration s'effectue sur le m\^eme contour dans le plan
complexe avec la diff\'erence que le contour contient maintenant deux p\^oles
quasiment d\'eg\'en\'er\'es. L'int\'egrale $I_{0\bar0}(T)$ se d\'ecompose en la
contribution $J$ des p\^oles $z_{1,2}$, la contribution $J_{1,2}$ des seuils de
$\varphi(p)$, et la contribution $J_b$ du seuil de production de plusieurs
particules (on suppose qu'il n'y en a qu'un dans le domaine d'\'energie
permis par les conditions exp\'erimentales):
$$
  I_{0\bar0}(T) = J+J_1+J_2+J_b \, .
$$
$J$ est donn\'e cette fois par les r\'esidus de deux p\^oles:
\begin{eqnarray*}
  J &=& W_{01}(z_1) W_{1\bar0}^{-1}(z_1) \, \pi \,
        (z_1 + {\bf p}^2 )^{ -\frac{1}{2} } \, \varphi(z_1,{\bf p}) \,
        \exp \left( -i \sqrt{ z_1 + {\bf p}^2} \, T \right)
  \\
    & & {} + W_{02}(z_2) W_{2\bar0}^{-1}(z_2) \, \pi \,
         (z_2 + {\bf p}^2 )^{ -\frac{1}{2} } \, \varphi(z_2,{\bf p}) \,
      \exp \left( -i \sqrt{ z_2 + {\bf p}^2} \, T \right) \, .
\end{eqnarray*}
Comme la violation CP est n\'eglig\'ee, les matrices de diagonalisation
{\it \'evalu\'ees aux p\^oles} sont donn\'ees par la formule (\ref{baseCP})
de passage \`a la base propre de CP:
$$
  W(z_1) \cong W(z_2) \cong \frac{1}{ \sqrt{2} }
  \left(
  \begin{array}{rr}
  1 & 1 \\
  1 & -1
  \end{array}
  \right) \, .
$$
Ces expressions ne sont pas valables loin des p\^oles. On en tiendra compte dans
l'\'evaluation de $J_b$.
Pour \'evaluer l'ordre de grandeur de $J$, on d\'eveloppe l'argument des
exponentielles en utilisant la d\'efinition des p\^oles,
\mbox{$z_j \equiv m_j^2 - i m_j \Gamma_i$}:
$$
   -i \sqrt{ z_j + {\bf p}^2 } \, T  \cong
  - i E T - i \frac{m_j^2 - m^2}{2E} - \frac{m\Gamma_j T}{2E} \, ,
$$
o\`u $m$ est la masse d\'eg\'en\'ee
et $E \equiv \sqrt{ m^2+{\bf p}^2 }$.
L'ordre de grandeur de $J$ est alors donn\'e par
\begin{equation}
  {\cal O}(J) \sim \frac{1}{m} \, \varphi(m^2,{\bf p}) \, e^{-\Gamma_2 T/2}
  \left| \,
         e^{ - i \Delta m T - \Delta \Gamma T/2 }  - 1 \,
  \right| \, ,
  \label{grandeurJ}
\end{equation}
o\`u $\Delta m \equiv m_1-m_2$, $\Delta \Gamma \equiv \Gamma_1-\Gamma_2$ et
l'on a approxim\'e $E\cong m$.

Les contributions $J_1$ et $J_2$ des seuils de $\varphi(p)$ peuvent se calculer s\'epar\'ement
sur les deux termes de la somme (\ref{G00}) et ce calcul n'apporte rien de nouveau par
rapport au chapitre pr\'ec\'edent: ces corrections ne dominent l'exponentielle
que pour des temps tr\`es grands, quand l'amplitude totale est devenue
ind\'etectable.

Examinons la contribution de $J_b$. On a vu que $J_b$ d\'epend de la
diff\'erence entre le propagateur complet et son prolongement analytique
(\'equation (\ref{exprJb})):
$$
   J_b = -i \, e^{ -i \sqrt{ b^2 + {\bf p}^2 } T }
  \int_0^\infty d\omega \, \varphi(z(\omega),{\bf p}) \,
  \left( G_{0\bar0 \,,\, II}(z) - G_{0\bar0}(z) \right) \,
  e^{-\omega T} \, .
$$
L'expression $G_{II}(z) - G(z)$ peut \^etre calcul\'ee dans la deuxi\`eme
repr\'esentation spectrale (\'equation (\ref{spectral2})):
$$
  G_{0\bar0 \,,\, II}(z) - G_{0\bar0}(z) =
  -i \left( \Pi_{0\bar0 \,,\, II}(z) - \Pi_{0\bar0}(z) \right)
   G_{0\bar0}(z) \, G_{0\bar0 \,,\, II}(z) \, ,
$$
o\`u $\Pi_{0\bar0}(z)$ est l'\'el\'ement $0\bar0$ de la matrice d'\'energie
propre.
Lors de l'\'evaluation asymptotique de $J_b$ ($T$ grand), seul le domaine proche
de $\omega=0$ (c'est-\`a-dire $z=b^2$) va
contribuer notablement, en raison de l'exponentielle d\'ecroissante.
Il suffit donc de conna\^{\i}tre l'expression de l'int\'egrand pr\`es du seuil.
Juste en dessous de l'axe r\'eel ($z=x-i\epsilon$), la repr\'esentation
spectrale de l'\'energie propre (\'equation (\ref{spectralenergie}))
implique que
\begin{eqnarray*}
  \Pi_{0\bar0 \,,\, II}(z) - \Pi_{0\bar0}(z)
  &=& \Pi_{0\bar0}(x+i\epsilon) - \Pi_{0\bar0}(x-i\epsilon)
  \\
  &=& 2i \, {\cal I}\!m \, \Pi_{0\bar0}(x+i\epsilon) \, .
\end{eqnarray*}
Il reste \`a calculer ${\cal I}\!m \, \Pi_{0\bar0}(x+i\epsilon)$.
Comme la renormalisation de la masse ne concerne que la partie r\'eelle de
l'\'energie propre, on a l'\'egalit\'e
$$
  {\cal I}\!m \, \Pi_{0\bar0}(x+i\epsilon) =
  {\cal I}\!m \, f_{0\bar0}(x+i\epsilon) \, .
$$
Lors de la diagonalisation du propagateur du $K^0\overline{K^0}$
\`a la section \ref{exempleKK}, la relation entre l'\'el\'ement
non diagonal de l'\'energie propre \'evalu\'e au p\^ole et les quantit\'es
mesur\'ees exp\'erimentalement a \'et\'e donn\'ee dans l'\'equation
(\ref{relationa}):
\begin{eqnarray*}
  {\cal I}\!m \, f_{0\bar0}(x+i\epsilon)
  &\equiv& - {\cal I}\!m \, a
  \\
  &=& \frac{1}{2} \, \left( m_{\SS L} \Gamma_L - m_{\SS S} \Gamma_S \right)
  \\
  &\cong& \frac{1}{2} \, m (\Gamma_L-\Gamma_S)
  \\
  &\cong& - \frac{1}{2} \,m \, \Delta \Gamma \, ,
\end{eqnarray*}
puisque suivant les notations de l'\'equation (\ref{relationa}), $b=0$,
$\hat\epsilon=0$, $K_1 = K_S$ et $K_2=K_L$ en l'absence de violation CP
(ce $b$-ci n'est pas la valeur du seuil mais un des param\`etres du
propagateur).
Comme les canaux de d\'esint\'egration principaux des $K^0$ et $\overline{K^0}$
sont identiques (d\'esint\'egrations en deux pions), on mod\'elise
${\cal I}\!m \, f_{0\bar0}(x+i\epsilon)$ par
la m\^eme d\'ependance fonctionnelle que celle trouv\'ee pour
${\cal I}\!m \, f_{00}(x+i\epsilon)$ (\'equation (\ref{defimpi})), de sorte que
la valeur au p\^ole soit celle indiqu\'ee ci-dessus et que la valeur au seuil
soit nulle:
$$
  {\cal I}\!m \, f_{0\bar0}(x+i\epsilon)
  \equiv - \frac{1}{2} \,m \, \Delta \Gamma \,
  \sqrt{ \frac{x-b^2}{m^2-b^2} } \, .
$$
D\`es lors, on peut \'ecrire pr\`es du seuil
$$
  G_{0\bar0 \,,\, II}(z) - G_{0\bar0}(z)
  \cong - m \, \Delta \Gamma \, \sqrt{ \frac{z-b^2}{m^2-b^2} } \,
  G_{0\bar0}(z) \, G_{0\bar0 \,,\, II}(z) \, .
$$
Substituons $y=\omega T$ dans l'\'equation de $J_b$ et
effectuons le d\'eveloppement en $1/T$ de l'int\'egrand, en utilisant
la param\'etrisation de $z$ en fonction de $\omega$ donn\'ee par l'\'equation (\ref{contourb}):
\begin{eqnarray*}
  z &=& - \omega^2 + b^2 - 2 i \omega \, \sqrt{ b^2 + {\bf p}^2 }
  \\
  &=& b^2 - 2 i \frac{y}{T} \, \sqrt{ b^2 + {\bf p}^2 }
  + {\cal O}\left( \frac{1}{T^2} \right) \, .
\end{eqnarray*}
On obtient
\begin{eqnarray*}
  J_b
  &\cong& i \, T^{-3/2} \, e^{-i \sqrt{ b^2 + {\bf p}^2 } T } \,
  \varphi(b^2,{\bf p}) \,  m \, \Delta \Gamma \,
  \frac{(-2i)^{1/2} \, ( b^2 + {\bf p}^2 )^{1/4} }{ \sqrt{ m^2 - b^2 } }
  \\
  & & \times \,
  G_{0\bar0}(b^2) \, G_{0\bar0 \,,\, II}(b^2) \,
  \int_0^\infty dy \, \sqrt{y} \, e^{-y}
  \\
  &\cong&
  -i (-i\pi/2)^{1/2} \; T^{-3/2} \;
  \frac{ m^{3/2} \, \Delta \Gamma }{ (m^2 - b^2)^{5/2} } \;
  \varphi(b^2,{\bf p}) \;
  e^{-i \sqrt{ b^2 + {\bf p}^2 } T } \, ,
\end{eqnarray*}
o\`u l'on a approxim\'e dans la deuxi\`eme ligne $b^2 + {\bf p}^2 \cong m^2$
et l'on a n\'eglig\'e l'\'energie propre dans les propagateurs \'evalu\'es
en $b^2$ car elle est n\'egligeable loin du p\^ole.
On en conclut que l'ordre de grandeur de $J_b$ est donn\'e par
\begin{equation}
  {\cal O}(J_b) \sim
  (QT)^{-3/2} \, \frac{ \Delta \Gamma }{Q}\, \frac{1}{m} \,
  \varphi(b^2,{\bf p}) \, ,
  \label{grandeurJb}
\end{equation}
o\`u $b$ est l'\'energie du seuil et $Q\equiv m-b$ est l'\'energie cin\'etique
des produits de d\'esint\'egration.

On peut maintenant comparer les ordres de grandeur de $J$
(\'equation (\ref{grandeurJ})) et de $J_b$ (\'equation (\ref{grandeurJb})).

{\bf Pour $T$ petit}

${\cal O}(J_b) \sim {\cal O}(J)$ si
$$
  (QT)^{-3/2} \, \frac{ \Delta \Gamma }{Q}
  \sim | \sin \frac{\Delta m \, T}{2} |
  \sim \frac{|\Delta m| \, T}{2} \, ,
$$
ou encore si $\, T \sim  Q^{-1}$, puisque l'exp\'erience dans le cas du
$K^0\overline{K^0}$ \cite{pdg} et la th\'eorie dans le cas du
$B^0\overline{B^0}$ \cite{fleischer} donnent
\mbox{${\cal O}(\Delta \Gamma) \le {\cal O}(\Delta m)$}.
Comme $Q$ varie de 220 MeV pour les kaons \`a 1 GeV pour les m\'esons $B$,
les corrections non exponentielles ne dominent \`a petit temps que
pour $T \le 10^{-24}$ s. La propagation spatio-temporelle est inobservable \`a
des temps aussi petits.
Le calcul asymptotique de $J_b$ n'est plus valable non plus en de\c{c}\`a de
cette borne puisque l'int\'egrale $J_b$ a \'et\'e d\'evelopp\'ee en $1/QT$.
Les corrections non exponentielles dues aux seuils de production sont donc
inobservables \`a petit temps dans la propagation spatio-temporelle des
particules quasi-stables.

{\bf Pour $T$ grand}

Il faut traiter s\'epar\'ement les cas des kaons et des m\'esons $B$.

\begin{enumerate}

  \item
  Kaons: $\Gamma_S \gg \Gamma_L$\\
  ${\cal O}(J) \sim {\cal O}(J_b)$ si
  \begin{eqnarray*}
    e^{-\Gamma_L T/2}
    &\sim& (QT)^{-3/2} \, \frac{ \Gamma_S }{Q}
    \\
    &\sim& (\Gamma_L \, T)^{-3/2} \,
    \frac{ \Gamma_S \, \Gamma_L^{3/2} }{ Q^{5/2} } \, ,
  \end{eqnarray*}
  ou encore si
  $$
    \Gamma_L \, T - 3 \, \ln (\Gamma_L \, T)
    \sim 2 \, \ln \left( \frac{ Q^{5/2} }{ \Gamma_S \, \Gamma_L^{3/2} }  \right)
    \sim 174 \, ,
  $$
  c'est-\`a-dire si $\Gamma_L \, T \sim 190$, ce qui est inobservable.

  \item
  M\'esons $B$: $\Gamma_L \approx \Gamma_H$\\
   ${\cal O}(J) \sim {\cal O}(J_b)$ si
  \begin{eqnarray*}
    e^{-\Gamma_H T/2}
    &\sim& (QT)^{-3/2} \, \frac{ \Delta \Gamma }{Q}
    \\
    &\sim& (\Gamma_H \, T)^{-3/2} \,
    \frac{ \Delta \Gamma\, \Gamma_H^{3/2} }{ Q^{5/2} } \, ,
  \end{eqnarray*}
  o\`u l'on a approxim\'e le sinus de l'oscillation par sa valeur moyenne.
  Cette condition se r\'e\'ecrit
   \begin{eqnarray*}
    \Gamma_H \, T - 3 \, \ln (\Gamma_H \, T)
    &\sim& 2 \, \ln \left(
                         \frac{ Q^{5/2} }{ \Delta \Gamma \, \Gamma_H^{3/2} }
                  \right)
    \\
    &\sim&   5 \, \ln \left( \frac{Q}{\Gamma_H} \right)
          - 2 \, \ln \alpha_q \, ,
  \end{eqnarray*}
  o\`u $\alpha_q \equiv \Delta \Gamma/\Gamma_H$ est estim\'e \`a
  $\alpha_d \cong 4 \times 10^{-3}$ pour le $B^0_d$ et \`a
  $\alpha_s \cong 10^{-1}$ pour le $B^0_s$ \cite{fleischer}.
  On calcule que
  \begin{eqnarray*}
  B^0_d: \quad T - 3 \, \ln (\Gamma_H \, T) \sim 153
  \quad \Rightarrow \quad \Gamma_H \, T \cong 168 \, ,
  \\
  B^0_s: \quad T - 3 \, \ln (\Gamma_H \, T) \sim 147
  \quad \Rightarrow \quad \Gamma_H \, T \cong 162 \, . 
  \end{eqnarray*}
  Ces temps sont beaucoup trop grands pour que l'on puisse esp\'erer une
  correction non exponentielle mesurable: la probabilit\'e de d\'etection
  aura d\'ecru \`a une valeur ind\'etectable.
\end{enumerate}
Nos estimations des corrections non exponentielles en th\'eorie des champs
sont identiques aux formules
th\'eoriques obtenues par Chiu et Sudarshan \cite{chiu} (formule (4.50) de cet
article; leurs \'evaluations num\'eriques sont par contre assez fantaisistes)
et par Wang et Sanda \cite{wang} (formule (59) de cet article, mais leur formule
(61) d'ordre de grandeur des corrections est incompr\'ehensible et incorrecte
du point de vue des unit\'es).
Ces auteurs calculent ces corrections dans le cadre de la
m\'ecanique quantique et proposent des extensions du formalisme de
Wigner-Weisskopf.

En conclusion, les corrections non exponentielles ne sont pas plus visibles
dans la propagation macroscopique d'un m\'elange de particules que dans la propagation d'une
seule particule. Les seules contributions importantes pour la propagation
macroscopique viennent des p\^oles des propagateurs.
Par cons\'equent, la formule de l'amplitude partielle
${\cal A}_j$ (\ref{amplitempAj}) peut servir de base rigoureuse pour l'analyse
spatiale de l'amplitude.

\section{Analyse spatiale de l'amplitude de m\'elange}

L'int\'egration sur ${\bf p}$ de l'amplitude partielle (\ref{amplitempAj})
demande une certaine prudence. En effet, nous sommes int\'eress\'es par les
termes d'interf\'erence
${\cal A}_i \, {\cal A}_j^*$ ($i \neq j$) qui apparaissent dans la probabilit\'e
car ces termes
contiennent les oscillations d\'ependant de l'espace-temps et de la
diff\'erence de masse $\Delta m_{ij} = |m_i - m_j|$. Il s'agit
d'\'evaluer ces termes d'oscillation \`a une pr\'ecision sup\'erieure \`a
$\Delta m_{ij}/P$. Rappelons que $P$ est la position du maximum de
$\psi(m^2_j,{\bf p})$.
Cependant, les oscillations ne sont observables sur des distances macroscopiques
que si $\Delta m_{ij}/P$ est extr\^emement petit.
Par exemple,
$\Delta m_{ij}/P \approx {\cal O}(10^{-14})$ pour les kaons neutres,
$\Delta m_{ij}/P \approx {\cal O}(10^{-8})$ pour les neutrinos dans
l'exp\'erience LSND et des valeurs encore plus petites apparaissent dans les
oscillations des neutrinos atmosph\'eriques et solaires.

L'\'evaluation directe de ${\cal A}_j$ par une approximation autour du
point-selle ne fournira en principe pas la pr\'ecision voulue dans les termes
d'oscillation pour deux raisons:
\begin{enumerate}
 \item
  Si les positions des maxima de $\psi(m^2_i,{\bf p})$ et $\psi(m^2_j,{\bf p})$ sont not\'ees
  respectivement ${\bf P}_i$ et ${\bf P}_j$, la diff\'erence
  $|{\bf P}_i -{\bf P}_j|$ sera de l'ordre de grandeur de $\Delta m_{ij}$. Or
  la largeur du pic de la fonction $\psi$ est beaucoup plus grande que
  $\Delta m_{ij}$. Peu importe que l'on utilise ${\bf P}_i$ ou ${\bf P}_j$
  comme point-selle: la m\'ethode d'int\'egration n'atteint pas une pr\'ecision
  sup\'erieure \`a $\Delta m_{ij}/P$.
  \item
  Pour les particules instables, le terme en $\Gamma$ modifie la position du
  maximum de $\Delta P \cong \varepsilon \, P$, o\`u $\varepsilon$ ne peut \^etre
  inf\'erieur \`a $10^{-8}$ (voir section \ref{analysespatiale}),
  ce qui est beaucoup plus grand que
  $\Delta m_{ij}/P \approx {\cal O}(10^{-14})$ pour les kaons neutres. 
\end{enumerate}
Malgr\'e ces obstacles, ce n'est pas la fin des haricots.
Nous allons voir que les
termes d'oscillation sont ind\'ependants au premier ordre en $\Delta m_{ij}$
de la position des points-selle ${\bf P}_i$ et ${\bf P}_j$, de sorte que
l'approximation autour du point-selle peut quand m\^eme \^etre utilis\'ee.

En suivant les m\^emes \'etapes que pour la particule isol\'ee (voir \'equation
(\ref{ampliintegree})), on int\`egre par approximation autour du point-selle et
l'on obtient
\begin{eqnarray}
  {\cal A}_j
  &\cong& \pi^{3/2} \, \psi(m^2_j,{\bf P}_j) \; \exp ( -m_j \Gamma_j T /2E_j ) \;
  \left( \det ( W_j + i R_j T ) \right)^{-1/2}
  \label{expressionAj} \\
  & & \times \, \exp (-iE_jT + i \,{\bf P}_j \cdot {\bf L} ) \;
  \exp \left(
              -\frac{1}{4} \, ({\bf L} - {\bf v}_j T) \,
                              ( W_j + i R_j T)^{-1} \, ({\bf L} - {\bf v}_j T)
       \right) \, ,
  \nonumber 
\end{eqnarray}
o\`u
$$
  E_j = \sqrt{ m^2_j + {\bf P}_j^2 }
  \qquad \mbox{et} \qquad
 {\bf v}_j = \frac{ {\bf P}_j }{E_j} \, .
$$

La probabilit\'e int\'egr\'ee sur le temps se calcule \`a partir de la norme au
carr\'e de l'amplitude totale (\ref{amplitot}) et s'\'ecrit
$$
  {\cal P}( \alpha \to \beta,{\bf L} ) \sim
  \sum_{i,j} V_{i\alpha} \, V_{\beta i}^{-1} \,
         V_{j\alpha}^* \, V_{\beta j}^{-1 \, *} \,
  \int dT \, {\cal A}_i(T,{\bf L}) \; {\cal A}_j^*(T,{\bf L}) \, .
$$
En raison des gaussiennes d'argument $({\bf L}-{\bf v}_{i,j} T)^2$, la
probabilit\'e est tr\`es faible si les deux conditions suivantes ne sont pas
satisfaites:
$$
  {\bf L} - {\bf v}_i T \cong 0
  \qquad \mbox{et} \qquad
  {\bf L} - {\bf v}_j T \cong 0 \, .
$$
Il est possible de satisfaire ces deux conditions \`a l'ordre
$\epsilon=\Delta m_{ij}/|{\bf P}_j|$ puisque les positions des maxima
${\bf P}_i$ et ${\bf P}_j$ sont identiques \`a $\epsilon$ pr\`es.
On a aussi
$$
  {\bf v}_i \cong {\bf v}_j + {\cal O}(\epsilon)
   \qquad \mbox{et} \qquad
   E_i \cong E_j + {\cal O}(\epsilon) \, .
$$
Soit ${\bf P}$ la position du maximum de $\psi(m^2,{\bf p})$, pour une
masse de r\'ef\'erence $m$ qui est \'egale \`a $m_j$ \`a $\epsilon$ pr\`es.
Appelons $z$ la direction de ${\bf P}$. 
Pour les kaons, on peut, par exemple, prendre $m=(m_i+m_j)/2$ et $m=0$ pour les
neutrinos.
La valeur exacte de $m$ n'a pas d'importance.
Nous allons \'etudier la probabilit\'e dans cette direction $z$, c'est-\`a-dire que
${\bf L}$ sera approximativement parall\`ele aux ${\bf v}_{i,j}$.

On param\'etrise la d\'eviation des ${\bf P}_{i,j}$ par rapport \`a ${\bf P}$
par
$$
  {\bf P}_j \equiv P
   \left(
   (1+\epsilon_j^z) \, \hat{\bf e}_z + \epsilon_j^x \, \hat{\bf e}_x
   + \epsilon_j^y \, \hat{\bf e}_y
   \right)
$$
o\`u $\epsilon_j^{x,y,z} \cong {\cal O}(\epsilon)$.
Les diverses quantit\'es apparaissant dans l'amplitude peuvent \^etre
\'evalu\'ees avec cette param\'etrisation et s'\'ecrivent
\begin{eqnarray*}
  {\bf P}_j \cdot {\bf L} &=& (1+\epsilon_j^z) \,PL + {\cal O}(\epsilon^2)
  \\
  E_j &=& \sqrt{ m^2_j + {\bf P}_j^2 }
       = E + \frac{\Delta m_j^2}{2E} + \epsilon_j^z \, v \, P + {\cal O}(\epsilon^2)
  \\
  {\bf v}_j &=& v \left(
   ( 1 + \delta_j^z ) \, \hat{\bf e}_z + \epsilon_j^x \, \hat{\bf e}_x
   + \epsilon_j^y \, \hat{\bf e}_y
                  \right) + {\cal O}(\epsilon^2)
\end{eqnarray*}
o\`u
\begin{eqnarray*}
  \Delta m_j^2 &\equiv& m_j^2 - m^2
  \\
  E &\equiv& \sqrt{ m^2 + {\bf P}^2 }
  \\
  \delta_j^z &\equiv& -\frac{\Delta m_j^2}{2E^2} + (1-v^2) \, \epsilon_j^z
\end{eqnarray*}
On va maintenant \'evaluer l'int\'egrale sur $T$ dans la probabilit\'e.
La d\'ependance en $T$ de la dispersion est n\'eglig\'ee en rempla\c cant $T$
par $L/v$ dans les termes en $RT$.

Pour \^etre tout \`a fait explicite, nous donnons d'abord l'expression de la
probabilit\'e int\'egr\'ee sur $T$ pour une direction quelconque ${\bf L}$,
sans utiliser la param\'etrisation d\'efinie ci-dessus:
\begin{eqnarray}
  \lefteqn{
  {\cal P}( \alpha \!\to\! \beta,{\bf L} ) \sim
  \sum_{i,j} V_{i\alpha} \, V_{\beta i}^{-1} \,
         V_{j\alpha}^* \, V_{\beta j}^{-1 \, *}
          }
  \nonumber
  \\ & & \times \;
  \pi^3 \, \psi(m^2_i,{\bf P}_i) \; \psi^*(m^2_j,{\bf P}_j) \;
  \left( \det X_i + \det X_j^* \right)^{-1/2} \;
  2 \sqrt{ \frac{\pi}{Y} }
  \nonumber
  \\ & & \times \;
   \exp \left(
              - \frac{\Gamma_{ij}}{Y} \,
              \left(
                    {\bf v}_i \, X_i \, {\bf L} + {\bf v}_j \, X_j^* \, {\bf L}
              \right)
        \right)
  \nonumber
  \\ & & \times \;
   \exp \left(
                i \, ( {\bf P}_i - {\bf P}_j ) \cdot {\bf L}
              - i \, \frac{E_i - E_j}{Y} \,
              \left(
                    {\bf v}_i \, X_i \, {\bf L} + {\bf v}_j \, X_j^* \, {\bf L}
              \right) 
       \right)
  \nonumber
  \\ & & \times \;
  \exp \left(
             - \frac{1}{4} \, {\bf L} \left( X_i +  X_j^* \right) {\bf L}
             + \frac{1}{4Y} \,
               \left(
                     {\bf v}_i \, X_i \, {\bf L} + {\bf v}_j \, X_j^* \, {\bf L}
               \right)^2 
       \right)
  \nonumber
  \\ & & \times \;
  \exp \left(
              -\frac{1}{Y} \, (E_i - E_j)^2
              + 2 i (E_i - E_j) \, \frac{\Gamma}{Y}
       \right)
  \label{probaqcq}
\end{eqnarray}
o\`u
\begin{eqnarray}
  X_j &\equiv& \left(
                          W_j + i R_j |{\bf L}|/|{\bf v}_j|
                   \right)^{-1}
  \label{defx}
  \\
  Y  &\equiv& 
  {\bf v}_i \, X_i \, {\bf v}_i + {\bf v}_j \, X_j^* \, {\bf v}_j
  \\
  \Gamma_{ij}  &\equiv& \frac{m_i \Gamma_i}{2 E_i} + \frac{m_j \Gamma_j}{2 E_j}
  \label{defgamma}
\end{eqnarray}
Les termes en $\Gamma^2$ sont bien entendu n\'eglig\'es.

Nous donnons ensuite la probabilit\'e dans la direction
${\bf L}=L \, \hat{\bf e}_z$ o\`u la probabilit\'e est maximale, en utilisant
la param\'etrisation d\'efinie plus haut des d\'eviations par rapport \`a
cette direction:
\begin{eqnarray}
  \lefteqn{ {\cal P}( \alpha \!\to\! \beta, L \, \hat{\bf e}_z ) \sim
  \sum_{i,j} V_{i\alpha} \, V_{\beta i}^{-1} \,
         V_{j\alpha}^* \, V_{\beta j}^{-1 \, *} }
  \label{probaz}
  \\
  & & \times \;
  \pi^3 \, \psi(m^2_i,{\bf P}_i) \; \psi^*(m^2_j,{\bf P}_j) \;
  \left( \det X_i + \det X_j^* \right)^{-1/2} \;
  2 \frac{ \sqrt{\pi} }{v} \;
  \left( X_i^{zz} + X_j^{zz \, *} \right)^{-1/2} \;
  \nonumber
  \\
  & & \times \;
  \exp \left( -\frac{\Gamma_{ij} L}{v} \right) \;
  \exp \left(
             i \, L
             \left(
                   p_i^z - p_j^z - \frac{E_i - E_j}{v}
             \right)
       \right)
  \nonumber
  \\
  & & \times \;
  \exp \left(
             - \frac{L^2}{4} \,
               \frac{ X_i^{zz} X_j^{zz \, *} ( v_i^z - v_j^z)^2 }
                    { X_i^{zz} (v_i^z)^2 + X_j^{zz \, *} (v_j^z)^2 }
       \right) \;
  \exp \left(
              \frac{ - (E_i - E_j)^2 + 2 i \Gamma \, (E_i - E_j) }
                   { \left( X_i^{zz} + X_j^{zz \, *} \right) v^2 }
       \right)
  \nonumber
\end{eqnarray}
On a effectu\'e un d\'eveloppement en l'ordre  ${\cal O}(\epsilon^2)$ dans les
facteurs d\'ependant de la distance, en ne gardant que les termes en $\epsilon
L$ ou $\epsilon^2 L^2$ qui peuvent \^etre non n\'egligeables \`a grande
distance.

\section{Analyse de la probabilit\'e}

Dans cette section les diff\'erents facteurs apparaissant dans
la probabilit\'e d'interf\'erence (\ref{probaz}) sont analys\'es
successivement.
Posons
\begin{eqnarray}
  \lefteqn{
  {\cal P}( \alpha \!\to\! \beta, L \, \hat{\bf e}_z ) \sim
  \sum_{i,j} V_{i\alpha} \, V_{\beta i}^{-1} \,
         V_{j\alpha}^* \, V_{\beta j}^{-1 \, *}
         }
  \nonumber
  \\ & & \times \;
  \pi^3 \, \psi(m^2_i,{\bf P}_i) \; \psi^*(m^2_j,{\bf P}_j) \;
  \left( \det X_i + \det X_j^* \right)^{-1/2} \;
  2 \frac{ \sqrt{\pi} }{v} \;
  \left( X_i^{zz} + X_j^{zz \, *} \right)^{-1/2} \;
  \nonumber
  \\ & & \times \;
  \exp A \; \exp B \; \exp C \; \exp D
  \label{probabcd}
\end{eqnarray}
Les exponentielles $e^A$, $e^B$, $e^C$ et $e^D$ correspondent dans l'ordre
aux exponentielles de l'\'equation (\ref{probaz}).

\subsection{D\'esint\'egration}

Si $\Gamma_i \neq 0$ et/ou $\Gamma_j \neq 0$, alors $\Gamma_{ij} \neq 0$
(\'equation (\ref{defgamma})).
Dans ce cas, la premi\`ere exponentielle $e^A$
exprime la d\'ecroissance de la probabilit\'e de d\'etection
de la particule en raison de sa d\'esint\'egration possible.
Notre r\'esultat est relativiste et s'\'ecrit en fonction de $\Gamma_{i,j}$
comme
\begin{equation}
  \fbox{$\displaystyle
  \exp A =
  \exp \left(
           -\left( \frac{m_i\Gamma_i}{2E_i} + \frac{m_j\Gamma_j}{2E_j} \right)
            \frac{L}{v}
       \right)
  $}
  \label{facteurdesint}
\end{equation}
Comme il s'exprime directement en fonction de la distance, il n'y a pas
d'ambigu\"\i t\'e de passage du temps \`a la distance. On peut le comparer au
traitement non relativiste en m\'ecanique quantique qui donnait
$\exp \left( -(\Gamma_i + \Gamma_j)T/2 \right)$ dans le rep\`ere au repos de
la particule. L'extension relativiste de
cette derni\`ere formule ainsi que sa transformation en une expression d\'ependant de
la distance peut se faire suivant plusieurs prescriptions aboutissant \`a
des r\'esultats diff\'erents.

\subsection{Oscillation}

La deuxi\`eme exponentielle $e^B$ de l'\'equation (\ref{probabcd}) fait osciller
la probabilit\'e de d\'etection de la particule en fonction de la distance.
Au premier ordre en $\epsilon$, l'argument de l'exponentielle vaut
\begin{eqnarray}
  B =
  i L \left( p_i^z - p_j^z - \frac{E_i - E_j}{v} \right)
  &\cong&
  i L \left(
            P \, (\epsilon_i^z - \epsilon_j^z)
            - \left(  \frac{ \Delta m_{ij}^2 }{2Ev}
                    +  P \, (\epsilon_i^z - \epsilon_j^z) 
              \right)
      \right)
  \nonumber
  \\
  &\cong&
  -i \, \frac{ \Delta m_{ij}^2 \, L}{2 P}
  \nonumber
  \\
  &\equiv& -2 i \pi \, \frac{L}{ L_{ij}^{osc} }
  \label{facteurB}
\end{eqnarray}
o\`u $\Delta m_{ij}^2 \equiv m_i^2 - m_j^2$ et l'on a d\'efini la {\it
longueur d'oscillation} $L_{ij}^{osc}$ pour le m\'elange $ij$
par\footnote{Si $\Delta m_{ij}^2$ est n\'egatif, on change le signe de la
d\'efinition de $L_{ij}^{osc}$.}
\begin{equation}
  \fbox{$ \displaystyle
   L_{ij}^{osc}   \equiv \frac{4\pi \, P}{ \Delta m_{ij}^2 }
  $}
  \label{longosc}
\end{equation}
On voit que le facteur d'oscillation ne d\'epend pas au premier ordre en
$\epsilon$ du choix de ${\bf P}_i$ ou ${\bf P}_j$ comme point-selle
pour \'evaluer l'int\'egrale! La raison en est
que les \'etats propres de masse sont sur leur couche de masse (comme on l'a
vu \`a la section \ref{analysetemp} lors de l'analyse temporelle de
l'amplitude, \'equation (\ref{onshell}) )
puisqu'\`a grande distance seul le p\^ole contribue \`a la propagation.
L'exponentielle $e^B$ se r\'e\'ecrit donc comme
\begin{equation}
  \fbox{$\displaystyle
  \exp B =
  \exp \left(
              -2 i \pi \, \frac{L}{ L_{ij}^{osc} }
       \right)
  $}
  \label{facteurosc}
\end{equation}

\subsection{D\'ecoh\'erence}

La troisi\`eme exponentielle $e^C$ de l'\'equation (\ref{probabcd}) montre que
l'interf\'erence dispara\^\i t \`a grande distance. Ce ph\'enom\`ene est
appel\'e {\it d\'ecoh\'erence} \cite{nussinov,giunti2}.

\subsubsection{Calcul de la longueur de coh\'erence}

Montrons d'abord comment appara\^{\i}t cette troisi\`eme exponentielle.
Partons de la formule de l'amplitude dans une direction quelconque
(\'equation (\ref{probaqcq})) et consid\'erons sa troisi\`eme exponentielle.
Si l'on y introduit la param\'etrisation autour de la direction d'amplitude
maximale, l'argument de cette exponentielle devient

\parbox{\textwidth}{
\begin{eqnarray}
  C &=& - \frac{1}{4} \, {\bf L} \left( X_i +  X_j^* \right) {\bf L}
  + \frac{1}{4Y} \,
    \left(
          {\bf v}_i \, X_i \, {\bf L} + {\bf v}_j \, X_j^* \, {\bf L}
    \right)^2
  \label{expodecoh}
  \\ &=&
  -\frac{L^2}{4} \,
  \frac{ X_i^{zz} X_j^{zz \, *} ( v_i^z - v_j^z)^2}
       {X_i^{zz} (v_i^z)^2 + X_j^{zz \, *} (v_j^z)^2 }
  \nonumber
  \\ & &
  - \frac{L^2}{4} \,
    \frac{\left(
                (v_i^x)^2 X_i^{xx} + (v_i^y)^2 X_i^{yy} +
                (v_j^x)^2 X_j^{xx \, *} +(v_j^y)^2 X_j^{yy \, *}
         \right)
         \left( X_i^{zz} + X_j^{zz \, *} \right) }
         {X_i^{zz} (v_i^z)^2 + X_j^{zz \, *} (v_j^z)^2 }
  \nonumber
  \\ & &
  - \frac{L^2}{2} \,
    \frac{( v_i^z - v_j^z)
          \left(
                  ( v_i^x X_i^{xz} + v_i^y X_i^{yz} ) X_j^{zz \, *}
                - ( v_j^x X_j^{xz \, *} + v_j^y X_j^{yz \, *} ) X_i^{zz}
          \right)
         }
         {X_i^{zz} (v_i^z)^2 + X_j^{zz \, *} (v_j^z)^2 } \, ,
  \nonumber       
\end{eqnarray}
}
o\`u l'on a conserv\'e les termes en $\epsilon$ d'ordre ${\cal O}(\epsilon
L)$ et ${\cal O}(\epsilon^2 L^2)$
car ils peuvent \^etre non n\'egligeables \`a grande distance.

Comme cas de figure, posons
$$
  \psi(m_j^2,{\bf P}_j) \sim
  \exp \left( - \frac{({\bf p} - {\bf P}_j )^2}{4\sigma^2_p} \right) \, .
$$
Si l'on va rechercher les d\'efinitions de $W_j$,
\'equation (\ref{defW}), et de $R_j$, \'equation (\ref{defR}), on voit que
$W_j$ est li\'e \`a l'incertitude sur la localisation des interactions tandis
que $R_j$ repr\'esente la dispersion en \'energie:
\begin{eqnarray*}
  W_j^{ab} &=& \frac{ \delta^{ab} }{4 \sigma^2_p}
              \equiv \delta^{ab} \, \sigma^2_x
  \\
  R_j^{ab} &=& \frac{ \delta^{ab} }{2E_j}
                  - \frac{P_j^a P_j^b}{2E_j^3} \, \delta^{az} \, \delta^{bz}
                  + {\cal O}(\epsilon^2)
\end{eqnarray*}
ce qui donne en utilisant la d\'efinition (\ref{defx}) de $X_j$:
\begin{eqnarray*}
  X_j^{xx} &\cong& X_j^{yy}
  \cong \left( \sigma^2_x + \frac{i L}{2 E_j v} \right)^{-1}
   + {\cal O}(\epsilon^2)
  \\
  X_j^{zz} &\cong&
  \left( \sigma^2_x + \frac{i L m_j^2}{2 E_j^3 v} \right)^{-1}
   + {\cal O}(\epsilon^2)
\end{eqnarray*}

En examinant les diff\'erents termes de l'argument de l'exponentielle
(\'equation (\ref{expodecoh})), on remarque que le premier terme
(c'est-\`a-dire la premi\`ere ligne) de la somme domine les deux autres,
puisqu'on a toujours la relation
$L/2Ev \gg \sigma_x^2$ si la distance $L$ est macroscopique.
L'argument (\ref{expodecoh}) de l'exponentielle $e^C$ se r\'e\'ecrit donc
$$
  C \cong -\frac{L^2}{4} \,
  \frac{ (v_i^z - v_j^z)^2 }
       {X_j^{zz \, * \, -1} (v_i^z)^2 + X_i^{zz \, -1} (v_j^z)^2 }
$$
Examinons s\'epar\'ement le num\'erateur et le d\'enominateur de cette fraction.\\
D'une part,
$$
  v_i^z - v_j^z
  = v \left(
            -\frac{ \Delta m_{ij}^2 }{2E^2}
            + (1-v^2) \, ( \epsilon_i^z - \epsilon_j^z )
      \right)
  \equiv v \kappa \frac{\Delta m_{ij}^2}{2E^2}
$$
o\`u $\kappa$ est un nombre d'ordre ${\cal O}(1)$.\\
D'autre part,
$$
  X_j^{zz \, * \, -1} (v_i^z)^2 + X_i^{zz \, -1} (v_j^z)^2
  =
  2 v^2 \sigma_x^2 
  \left(
        1 + \frac{ i \tilde \eta L \Delta m_{ij}^2 }{2 v \sigma_x^2 E^3}
  \right)
  + {\cal O}(\epsilon^2)
$$
o\`u $\tilde \eta$ est un nombre d'ordre ${\cal O}(1)$.
L'argument de l'exponentielle $e^C$ vaut alors
$$
   C \cong -\frac{L^2}{4} \,
  \frac{ (v_i^z - v_j^z)^2 }
       {X_j^{zz \, * \, -1} (v_i^z)^2 + X_i^{zz \, -1} (v_j^z)^2 }
  \cong
  - \left( \frac{L}{ L_{ij}^{coh} } \right)^2
    \left( 1 + i \eta \frac{L}{ L_{ij}^{coh} } \right)^{-1}
$$
o\`u la {\it longueur de coh\'erence} $L_{ij}^{coh}$ pour le m\'elange $ij$
est d\'efinie par\footnote{Comme pour la longueur d'oscillation, on change le
signe de la d\'efinition de $ L_{ij}^{coh}$ si $\Delta m_{ij}^2$ est n\'egatif.}
\begin{equation}
  \fbox{$\displaystyle
  L_{ij}^{coh} \equiv
  \frac{ 4\sqrt{2} }{\kappa } \,
  \frac{ E^2 }{\Delta m_{ij}^2} \, \sigma_x
  $}
  \label{longcoh}
\end{equation}
et o\`u
\begin{equation}
  \eta \equiv \frac{ 2\sqrt{2} \tilde \eta }{ \kappa E v \sigma_x } \, .
  \label{defeta}
\end{equation}
L'exponentielle $e^C$ se r\'e\'ecrit
\begin{equation}
  \fbox{$\displaystyle
  \exp C =
  \exp \left(
              - \left( \frac{L}{ L_{ij}^{coh} } \right)^2
                \left( 1 + i \eta \frac{L}{ L_{ij}^{coh} } \right)^{-1}
       \right)
  $}
  \label{facteurdecoh}
\end{equation}

\subsubsection{Interpr\'etation de la longueur de d\'ecoh\'erence}

Comment interpr\'eter cette longueur de coh\'erence? Montrons d'abord qu'elle
est beaucoup plus grande que la longueur d'oscillation:
$$
  \frac{ L_{ij}^{coh} }{ L_{ij}^{osc} } =
  \frac{1}{ \sqrt{2} \pi \kappa v } \, \frac{E}{\sigma_p} \, .
$$
M\^eme dans le pire des cas, qui est celui d'une particule relativiste pour
laquelle $\sqrt{2} \pi \kappa v  \approx {\cal O}(1)$, la longueur d'oscillation
reste beaucoup plus petite que la longueur de coh\'erence:
$$
  \frac{ L_{ij}^{coh} }{ L_{ij}^{osc} } \approx \frac{E}{\sigma_p} \gg 1 \, .
$$

L'interf\'erence commence \`a dispara\^\i tre apr\`es un nombre d'oscillations
$(L_{ij}^{coh}/ L_{ij}^{osc})$ \'egal \`a $E/\sigma_p$. On peut donner une
justification intuitive de ce ph\'enom\`ene par l'ima\-ge suivante \cite{kiers}.
L'\'etat interm\'ediaire est repr\'esent\'e par la superposition de deux
paquets d'ondes correspondant \`a deux \'etats propres de masse. Si leur vitesse
est relativiste, la dispersion dans la direction du mouvement est n\'egligeable
\`a cause de la contraction de Lorentz. En raison de leurs masses
diff\'erentes, leurs vitesses de groupe diff\`erent de
$$
  \Delta v \equiv v^z_i - v^z_j = v \kappa \frac{ \Delta m^2_{ij} }{2 E^2} \, .
$$
Apr\`es un temps $T$ et une distance $L \approx vT$, les paquets d'ondes se
sont  d\'eplac\'es l'un par rapport \`a l'autre de
$$
  \Delta L = \Delta v \, T \approx \kappa \frac{ \Delta m^2_{ij} }{2 E^2} L \, .
$$
Dans cette image, les oscillations sont les battements des deux paquets d'ondes
lors de leur d\'eplacement relatif de $\Delta L = \lambda$, o\`u
$\lambda = 2\pi/E$ est la longueur d'onde associ\'ee \`a l'\'etat oscillant.
La longueur d'oscillation s'obtient en imposant que le d\'eplacement relatif des
paquets d'ondes soit d'une longueur d'onde:
$$
  \Delta L = \lambda
  \;\; \Rightarrow \;\; L^{osc}_{ij}
  = \frac{4\pi E}{\Delta m_{ij}^2}
  \qquad (\kappa=1)  \, ,
$$
Pour des \'etats relativistes, $E \cong P$ et l'on retrouve la longueur
d'oscillation d\'eriv\'ee ci-dessus.

Si la taille du paquet est donn\'ee par $\sigma_x$, le nombre maximal
d'oscillations est le nombre de longueurs d'onde dans le paquet,
$N_{max}= \sigma_x/\lambda = E/4\pi\sigma_p$. Si le nombre d'oscillations est
sup\'erieur \`a $N_{max}$,
les paquets ne se recouvrent plus et les oscillations disparaissent. La
longueur de coh\'erence est donc donn\'ee par
$$
  L^{coh}_{ij} = N_{max} \, L^{osc}_{ij}
               = \frac{E}{4\pi\sigma_p} \, L^{osc}_{ij} \, . 
$$
On retrouve la formule de la longueur de coh\'erence d\'eriv\'ee plus haut \`a
une constante pr\`es.

Par contre, pour des vitesses non relativistes, la dispersion dans la direction
du mouvement est importante et la d\'ecoh\'erence est beaucoup plus lente.

Revenons \`a notre facteur de d\'ecoh\'erence d\'eriv\'e en th\'eorie des
champs et exa\-minons les diff\'erentes situations possibles:
\begin{enumerate}

  \item
  Pour les particules instables, la d\'ecoh\'erence n'a souvent pas de
  signification  car les particules se d\'esint\`egrent sur une distance beaucoup
  plus courte. Par exemple, pour le syst\`eme $K^0_S \!-\! K^0_L$,
  $$
    \frac{ L^{coh} }{ L^{des}} = \frac{ \sqrt{2} }{\kappa v} \,
    \frac{\Gamma}{m} \,\frac{E}{\sigma_p} \,\frac{E}{m_{\SS L} - m_{\SS S} }
    \cong 10^4 \, ,
  $$
  o\`u l'on a d\'efini $L^{des} = v/\Gamma$.
  \item
  Pour les particules stables relativistes, $v \gg \sigma_p/E$ donc
  $\eta \ll 1$ et la d\'ecoh\'erence est en
  $\exp \left( - (L/ L^{coh})^2 \right)$.
  \item
  Pour les particules non relativistes telles que $v \ll \sigma_p/E$,
  le coefficient $\eta \sim 1/v$ est tr\`es grand et il n'y a pas de
  d\'ecoh\'erence. Le facteur de d\'ecoh\'erence
  devient une oscillation spatiale de p\'eriode beaucoup plus grande que la
  longueur d'oscillation. Nous avons expliqu\'e dans l'image des
  paquets d'ondes que cette absence de d\'ecoh\'erence est due \`a la
  dispersion des paquets d'ondes.
\end{enumerate}
L'approche de la th\'eorie des champs ne permet, cependant, que de fixer une borne
inf\'erieure pour la longueur de coh\'erence. En effet, elle d\'epend de la
position des maxima ${\bf P}_i$ et ${\bf P}_j$. Par exemple, on peut avoir
$\kappa=0$ donc $L^{coh} = \infty$ si
$|{\bf P}_i| m_j = |{\bf P}_j| m_i$!
D'autres longueurs de coh\'erence existent mais elles sont sup\'erieures \`a
celle examin\'ee ici \cite{rich,kiers,mohanty}. 

Notons que la d\'ecoh\'erence fait dispara\^{\i}tre l'oscillation \`a une
certaine distance en supprimant le terme d'interf\'erence, mais qu'elle ne
supprime pas les transitions d'un \'etat de saveur vers un autre. La probabilit\'e de
d\'etection est modifi\'ee par une constante au del\`a de la longueur de
coh\'erence et cette modification est observable. Par exemple, pour un
m\'elange \`a deux saveurs, d'angle de m\'elange $\theta$
(la matrice de m\'elange $V$ est ici tout simplement la matrice de rotation
d'angle $\theta$), la probabilit\'e de survie de l'\'etat $\alpha$ devient
$$
  {\cal P}(\alpha \!\to\! \alpha,L) = 1 - \frac{1}{2} \, \sin^2 \, 2 \theta \, .
$$

\subsubsection{Exemples de d\'ecoh\'erence}

Donnons quatre exemples o\`u la d\'ecoh\'erence pourrait en principe jouer
un r\^ole. Le facteur crucial est le rapport entre la longueur de
d\'ecoh\'erence et la longueur d'oscillation,
$F \equiv L^{coh}/L^{osc}$,
approximativement \'egal \`a $E/\sigma_p$ pour des particules
relativistes.
\begin{enumerate}
  \item
  Dans l'exp\'erience LSND \cite{lsnd}, l'\'energie des
  neutrinos tourne autour de 30 MeV, tandis que l'incertitude
  $\sigma_p$ pourrait \^etre estim\'ee \`a 5 MeV \cite{mohanty}. Le
  facteur $F$ vaut dans ce cas seulement quelques longueurs d'oscillation
  et pourrait donc jouer un r\^ole dans les exp\'eriences futures.
  \item
  Pour la d\'etection des neutrinos solaires, l'\'energie des neutrinos
  tourne autour de 1 MeV et l'incertitude $\sigma_p$ peut \^etre
  estim\'ee \`a 0.002 MeV \cite{kiers}. Le facteur $F$ vaut ici $500$.
  On ne peut bien s\^ur faire varier \`a sa guise la distance
  Terre-Soleil donc l'oscillation n'est observable que sur la variation
  saisonni\`ere de $3\%$ de la distance \cite{pontecorvo}.
  Le mod\`ele d'oscillations dans le
  vide  fournit une longueur d'oscillation de $3 \, \times \, 10^7$ km donc
  la d\'ecoh\'erence n'affecte pas les oscillations observables dans ce
  mod\`ele. Par contre, dans le mod\`ele MSW \cite{msw} la diff\'erence de
  masse entre les neutrinos est telle que la longueur d'oscillation dans le vide
  tourne autour de 250 km. Le nombre d'oscillations sur la distance Terre-Soleil
  est de 60 000, bien sup\'erieur au facteur F. Cette d\'ecoh\'erence n'est
  n\'eanmoins pas v\'erifiable, car l'ignorance du lieu de production exact
  et la moyenne sur l'\'energie du neutrino impliquent une moyenne de la
  probabilit\'e sur un grand nombre de longueur d'oscillations. L'interf\'erence
  est \'elimin\'ee non seulement \`a cause de la d\'ecoh\'erence mais aussi
  en raison de cette moyenne. Seule la d\'ecroissance de la probabilit\'e est
  visible. Bien que cela ne constitue pas un test de la pr\'esence de la
  d\'ecoh\'erence, l'observation ou non d'une variation de la probabilit\'e de
  d\'etection correspondant \`a la variation annuelle de la distance Terre-Soleil
  \'eliminerait soit le mod\`ele MSW soit le mod\`ele des oscillations dans le
  vide.
  Notons cependant que la moyenne sur les \'energies des neutrinos peut aussi
  rendre invisible l'influence de la variation de la distance Terre-Soleil dans
  le cas du mod\`ele d'oscillations dans le vide
  \cite{frautschi,glashow,wilczek,kiers}.
  \item
  Si l'on n\'eglige les contraintes des mod\`eles cosmologiques sur la masse des
  neutrinos, la masse du neutrino tau n'est born\'ee que par 18.2 MeV \cite{pdg}.
  Si la masse du troisi\`eme \'etat propre de masse est de l'ordre de plusieurs
  MeV, tandis que celles des deux autres \'etats sont de l'ordre de l'eV, les
  longueurs de coh\'erence $L_{3j}^{coh}$ pour les m\'elanges du troisi\`eme
  \'etat peuvent \^etre inf\'erieures \`a la longueur d'oscillation
  $L_{12}^{osc}$ des deux autres \'etats dans les exp\'eriences o\`u l'\'energie
  des neutrinos est de quelques MeV. Un tel m\'elange de neutrinos relativistes
  et non relativistes a \'et\'e \'etudi\'e par Ahluwalia et Goldman
  \cite{ahluwalia}, qui identifient le troisi\`eme neutrino avec la particule de
  33.9 MeV propos\'ee pour expliquer une anomalie dans l'exp\'erience KARMEN. 
  \item
  Pour terminer, l'\'emission de neutrinos par des supernovae offre une
  possibilit\'e d'\'etudier soit des diff\'erences de masse encore plus faibles
  ($\Delta m^2 \sim 10^{-20} \, \mbox{eV}^2$), soit d'observer l'effet de la
  d\'ecoh\'erence par l'observation de vagues de neutrinos correspondant aux
  diff\'erents \'etats propres de masse \cite{reinartz}.
\end{enumerate}

\subsection{Localisation des interactions}

La quatri\`eme exponentielle $e^D$ de la formule de la probabilit\'e
(\ref{probabcd}) exprime une condition sur $\sigma_x$ ou une
longueur de coh\'erence suivant les cas. En effet,
$$
  E_i - E_j \cong
  \frac{ \Delta m_{ij}^2 }{2E} + v P (\epsilon_i^z - \epsilon_j^z)
  \equiv \tilde \kappa \frac{ \Delta m_{ij}^2 }{2E} \, ,
$$
o\`u $\tilde \kappa$ est un nombre d'ordre ${\cal O}(1)$.
Supposons que nous ayons affaire \`a des particules telles
que le coefficient $\eta$ d\'efini dans le calcul de la longueur
de coh\'erence (\'equation (\ref{defeta})) est beaucoup plus petit que 1,
c'est-\`a-dire que $v \!\gg\! \sigma_p/E$.
On peut montrer dans ce cas que l'argument de la quatri\`eme exponentielle
devient\footnote{Le terme en $\Gamma$ est omis car il ne
donne pas d'information int\'eressante.}
$$
  D = - \frac{ (E_i - E_j)^2  }
       { \left( X_i^{zz} + X_j^{zz \, *} \right) v^2 }
  \cong - 2 \pi^2 \tilde \kappa^2 \left( \frac{1}{ L^{osc}_{ij} } \right)^2
         \left(
               \sigma_x^2 + \left( \frac{L m^2}{2 E^3 v \sigma_x } \right)^2
         \right) \, .
$$
Il est malheureusement difficile de dire quel terme de la somme domine sans
conna\^{\i}tre l'ordre de grandeur des masses.

Si le terme en $L$ est n\'egligeable, on obtient
\begin{equation}
  D \cong - 2 \pi^2 \tilde \kappa^2
  \left( \frac{\sigma_x}{ L^{osc}_{ij} } \right)^2 \, .
  \label{local}
\end{equation}
Pour que l'amplitude soit non nulle, il faut donc que
$\sigma_x \ll L^{osc}_{ij}$, c'est-\`a-dire que l'incertitude sur la localisation
de la source et du d\'etecteur doit \^etre inf\'erieure \`a la longueur
d'oscillation pour que cette derni\`ere soit observable.

Si le terme en $L$ domine, on obtient
\begin{equation}
  D \cong - \left( \frac{\tilde \kappa}{\kappa} \, \frac{m^2}{P^2} \right)^2
  \left( \frac{L}{ L^{coh}_{ij} } \right)^2 \, .
  \label{decohloc}
\end{equation}
Il s'agit d'un nouveau terme de d\'ecoh\'erence qui peut acc\'el\'erer
la d\'ecoh\'erence vue \`a la section pr\'ec\'edente si les particules sont
non relativistes (tout en respectant la condition $\eta \!\ll\! 1$).

Pour des particules non relativistes telles que $\eta \!\gg\!1$, une analyse
similaire peut \^etre men\'ee mais n'est pas concluante en raison du
grand nombre de param\`etres inconnus.

Le facteur $D$ d\'epend aussi du choix des points-selle ${\bf P}_i$ et ${\bf P}_j$. Par
exemple, si $E_i \!=\! E_j$, l'exponentielle vaut 1. En tout \'etat de cause,
la contrainte $\sigma_x \!\ll\! L^{osc}_{ij}$ n'est jamais perdue car la
condition $\Delta m_{ij}^2/E \!\ll\! \sigma_p$ est n\'ecessaire lors de
l'int\'egration de l'amplitude sur $p^0$, sinon le p\^ole correspondant \`a
une des particules $i,j$ se trouverait hors du contour d'int\'egration.
On peut donc toujours effectuer la substitution
\begin{equation}
  \fbox{$\displaystyle
  \exp D \rightarrow
  \exp \left( -2 \pi^2 \left( \frac{\sigma_x}{L_{ij}^{osc}} \right)^2 \right)
  $}
  \label{facteurlocal}
\end{equation}
tout en gardant en m\'emoire les contributions possibles de $e^D$ \`a la
d\'ecoh\'erence.

Cette condition r\'esout aussi la question de savoir si une mesure
suffisamment pr\'ecise de l'\'energie et de
l'impulsion des \'etats oscillants pourrait d\'eterminer quel \'etat propre
de masse se propage. Une mesure \`a la pr\'ecision de $\Delta m_{ij}^2$ viole la
condition ci-dessus et l'oscillation ne sera plus observable.

Elle montre aussi pourquoi on ne pourrait observer l'oscillation de leptons
charg\'es, mis \`a part le fait que leurs interactions \'electromagn\'etiques
les rendent plus facilement identifiables et pourraient supprimer l'interf\'erence.
La diff\'erence entre leurs masses est trop importante et viole la condition
ci-dessus. De plus, la longueur d'oscillation est si petite qu'elle serait de
toute fa\c{c}on inobservable. Par exemple, la longueur d'oscillation du
syst\`eme muon-\'electron est d'environ $10^{-14}$ m pour une impulsion
typique de 100 MeV.

Finalement, notons que si le neutrino mu ou tau est trop lourd,
il ne pourra pas osciller avec les
neutrinos l\'egers \cite{rich}. Soit une source de neutrinos constitu\'ee par
un noyau radioactif dans un r\'eseau. L'incertitude sur la position du
noyau est d'environ $10^{-10}$ m. L'oscillation de neutrinos
sera donc inobservable si $\Delta m_{ij}^2$ est sup\'erieur \`a
$4 \, \times \, 10^9 \, \mbox{eV}^2 \, (E_\nu/1\mbox{MeV})$.
Si le neutrino le plus l\'eger a une
masse n\'egligeable, la masse du neutrino le plus lourd ne peut d\'epasser
$60 \, \mbox{keV} \, \sqrt{E_\nu/1\mbox{MeV}}$.

\subsection{Oscillations sophistiqu\'ees et oscillations simplifi\'ees}

Sous quelles conditions retrouve-t-on la formule d'oscillation
(\ref{probaMQexpo}) habituellement utilis\'ee dans la litt\'erature ?
Par hypoth\`ese, $\epsilon=\Delta m_{ij}/P$ est un param\`etre tr\`es petit.
On peut donc d\'evelopper tous les facteurs pr\'ec\'edant les exponentielles
en $\epsilon$ et ne garder que les termes d'ordre z\'ero.
Ces termes sont ind\'ependants des masses $m_i$, $m_j$ et peuvent donc
sortir de la somme sur $i,j$.
Si l'on ins\`ere dans la formule (\ref{probabcd}) de la probabilit\'e les
expressions des exponentielles $e^A$, $e^B$, $e^C$ et $e^D$
(\'equations (\ref{facteurdesint}), (\ref{facteurosc}), (\ref{facteurdecoh})
et (\ref{facteurlocal})), on obtient
\begin{eqnarray}
  & & {\cal P}( \alpha \!\to\! \beta, L \, \hat{\bf e}_z ) \sim
  \sum_{i,j} \, V_{i\alpha} \, V_{\beta i}^{-1} \,
                V_{j\alpha}^* \, V_{\beta j}^{-1 \, *} \;
  \exp \left(
             - \Gamma_{ij} \frac{L}{v}
             -2i\pi \frac{L}{L^{osc}_{ij} }
       \right)
  \nonumber
  \\  & & \times
  \exp \left(
            - \left( \frac{L}{ L_{ij}^{coh} } \right)^2
              \left( 1 + i \eta \frac{L}{ L_{ij}^{coh} } \right)^{-1}
        \right) \;
  \exp \left(
             -2 \pi^2 \left( \frac{\sigma_x}{L_{ij}^{osc}} \right)^2
       \right)
\end{eqnarray}
Les d\'efinitions pour le m\'elange $ij$ de la largeur $\Gamma_{ij}$,
de la longueur d'oscillation $L^{osc}_{ij}$, et de la longueur de
coh\'erence $L^{coh}_{ij}$ sont donn\'ees respectivement par les \'equations
(\ref{defgamma}), (\ref{longosc}) et (\ref{longcoh}).

Il s'ensuit que si les conditions d'observabilit\'e des oscillations sont
satisfaites\footnote{Si les particules sont instables, il faut \'evidemment
aussi que $L^{osc}_{ij} \! \le \! v/\Gamma_{ij}$, c'est-\`a-dire que les particules
aient le temps d'osciller avant de se d\'esint\'egrer. Cette condition est incluse
dans l'\'equation (\ref{formuleoscsimple})},
c'est-\`a-dire $L \!\ll\! L^{coh}$ et $\Delta m_{ij} \!\ll\! \sigma_p$,
seuls les facteurs de d\'esint\'egration et d'oscillation ont une contribution
fortement d\'ependante en $i,j$. 
La probabilit\'e d'oscillation s'\'ecrit alors
\begin{equation}
  \fbox{$\displaystyle
  {\cal P}( \alpha \!\to\! \beta, L \, \hat{\bf e}_z ) =
  \sum_{i,j} \, V_{i\alpha} \, V_{\beta i}^{-1} \,
                V_{j\alpha}^* \, V_{\beta j}^{-1 \, *} \,
  \exp \left(
             - \Gamma_{ij} \frac{L}{v}
             - 2i\pi \frac{L}{L^{osc}_{ij} }
       \right)
  $}
  \label{formuleoscsimple}
\end{equation}
Le signe d'\'egalit\'e provient de la condition de conservation de la probabilit\'e
dans le cas d'une \'evolution unitaire, o\`u $\Gamma_{ij}=0$ et $V^{-1}=V^\dagger$:
$$
  \sum_\beta {\cal P}( \alpha \!\to\! \beta, L \, \hat{\bf e}_z ) = 1 \, ,
$$
ce qui est bien le r\'esultat attendu: la somme des probabilit\'es de
transition d'une particule stable dans les autres particules du m\'elange
doit \^etre \'egale \`a l'unit\'e.

La longueur d'oscillation que nous avons d\'eriv\'ee
(\'equation (\ref{longosc}) correspond donc \`a celle de
la formule d\'eriv\'ee dans le traitement de m\'ecanique quantique
(\'equa\-tion (\ref{longoscMQ})), mais il n'y a plus de
tour de passe-passe! Les r\'eponses fournies par notre formule aux objections
soulev\'ees contre la d\'erivation traditionnelle sont pass\'ees en revue
dans la conclusion.

\section{Oscillations de fermions}
\label{oscifermion}

Quelles sont les modifications \`a apporter pour les oscillations de fermions?
Les paquets d'ondes repr\'esentant les fermions contiendront des spineurs.
Le propagateur de la particule interm\'ediaire est le
propagateur fermionique. Le propagateur libre d'un \'etat propre de masse
de type Dirac ou Majorana (si la transition ne viole pas le nombre leptonique
total) s'\'ecrit
$$
  G_F(x'-x) = i \int \frac{d^4p}{(2\pi)^4}
             \frac{ {\p} + m}{p^2 - m^2 + i \epsilon} \,
             e^{-i p \cdot (x'-x)}
$$
et d\'ecrit la propagation d'un fermion de masse $m$ de $x$ en $x'$, ainsi que
d'un anti-fermion de m\^eme masse de $x'$ en $x$. Comme pour les particules
scalaires, soit la particule, soit l'antiparticule a une contribution
n\'egligeable \`a l'amplitude de propa\-gation lorsque la source et le
d\'etecteur sont s\'epar\'es par une distance macroscopique.

Si l'on s'int\'eresse \`a des fermions stables dont le lagrangien est connu,
par exem\-ple des neutrinos, la matrice de masse peut \^etre diagonalis\'ee
\cite{pal,bilenky} par une transformation unitaire sur les champs. Dans le cas
de termes de masse de Dirac, qui m\'elangent les champs de chiralit\'e gauche
et droite, la transformation unitaire est diff\'erente pour les champs gauches
et droits \mbox{($M_{diag}=U_L^\dagger M U_R$)} mais seule la matrice de
m\'elange $U_L$ des champs gauches appara\^{\i}t dans les amplitudes de
transition.
En effet, dans le mod\`ele standard, les courants droits n'interagissent qu'en
tant que courants neutres et le terme les contenant est invariant sous une transformation
unitaire, contrairement au terme contenant les courants charg\'es. 
Si le lagrangien inclut des termes de masse de Majorana, on peut consid\'erer les
termes de masse de Dirac comme une superposition de termes de masse de Majorana
et les regrouper avec les termes de masse de Majorana d\'ej\`a pr\'esents.
La matrice de masse correspondante est toujours sym\'etrique et peut donc \^etre
diagonalis\'ee par une transformation unitaire $U$ en une matrice \`a entr\'ees
r\'eelles positives \mbox{($M_{diag}=U^t M U$)}.
La matrice de m\'elange $U$ relie aussi bien les champs de saveur (appartenant 
\`a des doublets sous $SU(2)_L$) que les champs st\'eriles (singulets sous
$SU(2)_L$) aux champs de masse. Comme les champs st\'eriles ne subissent pas
d'interaction faible, seule la partie de la matrice $U$ reliant les champs de
saveur aux champs de masse interviendra dans les termes d'interaction. Cette
sous-matrice de la matrice $U$ appara\^{\i}tra aussi dans les courants neutres
car elle n'est pas n\'ecessairement unitaire \cite{schechter}.

Les amplitudes de propagation pour des neutrinos de Majorana ou de Dirac ne
diff\`erent que par des termes en $m_\nu/E$ \cite{perrier} qui ne sont pas
d\'etectables dans les oscillations, pas plus
d'ailleurs que les phases suppl\'ementaires violant CP dans la matrice de
m\'elange des neutrinos de Majorana, en raison d'une invariance bien connue
de la probabilit\'e d'oscillation sous une reparam\'etrisation de la matrice de
m\'elange \cite{doi}.

Si l'on s'int\'eresse \`a des fermions instables,
il faut d'abord calculer l'\'energie propre
puis obtenir le propagateur complet par sommation sur les diagrammes d'\'energie
propre, en passant par l'\'equation de Dyson (comme pour le propagateur
matriciel d'un m\'elange de particules):
$$
  G(p^2) = \frac{iZ}{ {\p} - m - f({\p}) + i\epsilon}
$$
Le propagateur complet pour un m\'elange de particules se diagonalise
similairement \`a celui des particules scalaires \`a l'aide de matrices $V$ et
$V^{-1}$. Si le m\'elange entre les particules est faible, les parties r\'eelles
des p\^oles du propagateur seront approximativement \'egales aux masses
avant le calcul de l'\'energie propre et seront positives.

L'\'evaluation de l'amplitude de propagation ne diff\`ere pas notablement de
celle d'un m\'elange de particules scalaires. Les termes contenant des
matrices $\gamma^\mu$, sandwich\'es entre les spineurs des \'etats entrants et
sortants, sont incorpor\'es dans la fonction de recouvrement des paquets
d'ondes.

Contrairement \`a ce que l'on dit souvent \cite{giunti2,cardall}, il
n'est pas n\'ecessaire de prendre la limite relativiste pour retrouver la
formule d'oscillation classique. Il suffit que les diff\'erences entre les
masses des \'etats interm\'ediaires soient beaucoup plus petites que leurs
impulsions. De plus, il n'est pas n\'ecessaire non plus que l'interaction
soit chirale, sauf si l'on tient \`a factoriser explicitement le
processus \cite{cardall} en une amplitude de production fois une amplitude de propagation
fois une amplitude de d\'etection (dans ce cas, il faut aussi prendre la limite
relativiste).

\section{Conclusion}

L'\'etude des oscillations de particules en th\'eorie quantique des
champs a d\'ej\`a \'et\'e entreprise dans plusieurs articles, dans les cas
sp\'ecifiques des kaons \cite{sachs,sudarsky,beuthe} et des
neutrinos \cite{giunti2,grimus,campagne,shtanov,weiss,cardall,ioannisian}. Ces
articles diff\`erent entre eux par les mod\`eles id\'ealis\'es qu'ils proposent
et par les approximations utilis\'ees pour le calcul. Aucun mod\`ele
parfaitement fid\`ele et aucun calcul parfaitement exact ne sont possibles,
malgr\'e les affirmations que l'on lit parfois ici et l\`a.
Certains auteurs prennent trop au s\'erieux les mod\`eles simplifi\'es et les
m\'ethodes de calcul qu'ils ont choisis de sorte que leurs conclusions sont \`a
prendre avec un grain de sel.

Notre analyse est la premi\`ere \`a traiter de fa\c{c}on unifi\'ee les cas
stable/instable et relativiste/non relativiste. Cette approche conduit \`a
une s\'erie de conclusions, parfois nouvelles, que nous \'enum\'erons ci-dessous.

\begin{enumerate}

\item
Les corrections non exponentielles \`a l'amplitude de propagation ne sont pas
plus apparentes pour des particules m\'elang\'ees que pour des particules
isol\'ees. La d\'emonstration n'en avait pas encore \'et\'e faite, \`a ma
connaissance, dans le cadre de la th\'eorie des champs, bien que des \'etudes
\`a ce sujet existent en m\'ecanique quantique \cite{chiu,wang}.

\item
Une deuxi\`eme conclusion, d\'ej\`a connue par ailleurs \cite{giunti2,rich}, est
que le traitement du processus de propagation dans sa totalit\'e permet de
calculer directement une probabilit\'e d\'ependant de la distance  et non du
temps de propagation. L'application de la formule aux exp\'eriences est ainsi
possible sans passage ambigu du temps \`a l'espace par une formule de physique
classique ext\'erieure au formalisme. Cette ambigu\"{\i}t\'e a conduit \`a
certaines conclusions fausses dans la litt\'erature
\cite{srivastava,widom,sassaroli}
concernant les oscillations de particules.
Nous y reviendrons dans le prochain chapitre, lors de l'\'etude du processus
$\phi(1020) \to K^0\overline{K^0}$.

\item
Un troisi\`eme r\'esultat qui a d'abord \'et\'e \'etabli par des
arguments intuitifs en m\'ecanique quantique \cite{nussinov,kayser2,rich,kiers},
puis en th\'eorie des champs \cite{giunti2} concerne les conditions
d'observabilit\'e des oscillations. D'une part, la longueur d'oscillation doit
\^etre sup\'erieure \`a l'incertitude sur les positions de la source et du
d\'etecteur. D'autre part, les oscillations des particules stables ne
peuvent \^etre d\'etect\'ees au del\`a d'une certaine distance appel\'ee
longueur de coh\'erence. Une nouveaut\'e de notre analyse est
la prise en compte de la dispersion, qui supprime la d\'ecoh\'erence
pour des particules stables non relativistes. Malheureusement, les
oscillations \'etudi\'ees pour l'instant (kaons, B , neutrinos) ne
rel\`event pas de cette cat\'egorie.

\item
Notre \'etude a \'eclairci la question de l'\'egalit\'e ou non des
\'energies et/ou impulsions des \'etats oscillants. Cette question est en
fait mal pos\'ee, car il y a plusieurs fa\c{c}ons d'identifier a posteriori
les \'energies-impulsions de ces \'etats.
Il serait tentant dans notre approche de les faire correspondre \`a
$E_j = \sqrt{m_j^2 + {\bf P}_j^2 }$ et ${\bf P}_j$, o\`u ${\bf P}_j$ est la
position du maximum de la fonction-poids lors de l'\'evaluation de l'int\'egrale
(\ref{amplitempAj}). Ces 
\'energies-impulsions sont \'egales ou diff\'erentes selon les valeurs de
${\bf P}_i$ et ${\bf P}_j$.
Notons que
$E_i-E_j \sim P_i - P_j \sim {\cal O}(\Delta m_{ij})$. Cette diff\'erence
\'etant tr\`es petite, il est d\'elicat d'imposer une valeur bien pr\'ecise
\`a $E_i$, $E_j$, ${\bf P}_i$ et ${\bf P}_j$ par des \'evaluations
approximatives d'int\'egrales.

D'ailleurs, l'identification que nous venons de proposer n'est pas
univoque. Modifions un peu la proc\'edure de calcul en prenant la norme au
carr\'e de l'amplitude avant d'int\'egrer sur la tri-impulsion.
On peut alors choisir d'ef\-fec\-tuer d'abord la
moyenne sur le temps $T$ avant d'int\'egrer sur les tri-impulsions ${\bf p}$ et
${\bf p'}$. On obtiendra en tr\`es bonne approximation l'\'egalit\'e
$$
  \sqrt{m_i^2 + {\bf p}^2 } \cong \sqrt{m_j^2 + {\bf p'}^2 } \, .
$$
La pr\'ecision \`a laquelle est satisfaite l'\'egalit\'e d\'epend de
l'intervalle d'int\'egration de $T$ mais cette pr\'ecision est meilleure
que $\Delta m_{ij}$ pour des temps macroscopiques.
Cette \'egalit\'e imposera que l'int\'egration sur
${\bf p}$ et ${\bf p'}$ devra se faire avec la contrainte $E_i=E_j$.
De cette fa\c{c}on, on sera conduit \`a la conclusion que les \'energies des
\'etats oscillants sont \'egales. Ce raisonnement s'applique \`a d'autres
calculs que le n\^otre, par exemple \cite{giunti2}.

Il se fait qu'il n'est pas n\'ecessaire de r\'epondre \`a cette question
d'\'egalit\'e ou non des \'energies-impulsions pour calculer sans
ambigu\"{\i}t\'e le facteur d'oscillation, puisque celui-ci est ind\'ependant
au premier ordre en $\epsilon=\Delta m_{ij}/P_j$ des valeurs de
$E_i$, $E_j$, ${\bf P}_i$ et ${\bf P}_j$:
$$
  i L \left( p_i^z - p_j^z - \frac{E_i - E_j}{v} \right) =
  - 2 i\pi \, \frac{L}{ L_{ij}^{osc} }
  + {\cal O}(\epsilon^2) \, .
$$
Cette compensation entre \'energie et impulsion se fait parce que
les \'etats propres de masse interm\'ediaires sont sur leur
couche de masse (dans le sens de l'\'equation (\ref{onshell})).
Ils ne peuvent en effet franchir une distance macroscopique
que s'ils sont quasi-r\'eels.

\item
Pour le cas de particules interm\'ediaires stables (concr\`etement des
neutrinos), des \'etudes assez compl\`etes ont d\'ej\`a \'et\'e
r\'ealis\'ees \cite{giunti2,grimus}.
Notre traitement montre de plus que\\
- la limite relativiste n'est pas n\'ecessaire, seule la condition
$\Delta m_{ij}/P \!\ll\! 1$ compte.\\
- il faut tenir compte de la dispersion pour \'etudier la longueur de
coh\'erence.\\
- l'\'egalit\'e ou non des \'energies et/ou impulsions associ\'ees aux
particules interm\'ediaires est un artefact de calcul (voir point pr\'ec\'edent).

\item
Notre approche a abouti \`a un seul type de comportement asymptotique en $L$ de
l'amplitude, contrairement \`a l'article \cite{ioannisian}, qui d\'erive un
r\'egime interm\'ediaire pour l'amplitude, sous la condition
$\sigma_x^2 P \gg L$ (qui est probablement impos\-sible \`a satisfaire pour
des distances macroscopiques). Dans ce r\'egime, toute la d\'ependance angulaire de
l'amplitude est contenue dans le facteur oscillant
\mbox{$\exp(i {\bf P}_j \cdot {\bf L})$} avec la cons\'equence illogique que
l'amplitude a le m\^eme ordre de grandeur dans toutes les directions, quelles
que soient les impulsions entrantes.
En comparaison, notre amplitude (\'equation (\ref{expressionAj})) contient,
en plus du facteur oscillant, une gaussienne avec une d\'ependance angulaire en
\mbox{$ \exp \left( - ( {\bf L} - {\bf v}_j T )^2 / 4\sigma_x^2 \right) $}.
Cette gaussienne rend notre amplitude n\'egligeable dans toute direction
diff\'erant notablement du vecteur ${\bf v}_j$ reli\'e directement aux
impulsions entrantes. L'absence de ce terme dans
l'amplitude figurant dans \cite{ioannisian} n'est pas \'etonnante car la
m\'ethode d'int\'egration utilis\'ee dans cet article recourt \`a une
m\'ethode incorrecte d'int\'egration sur des angles complexes.

\item
Pour le cas des particules instables, les analyses existantes sont
incompl\`etes, ne tenant pas compte soit de la possibilit\'e
$P_i \neq P_j$ \cite{sachs}, soit de $E_i \neq E_j$ \cite{sudarsky}.
(les auteurs de ce dernier article se restreignent \`a une dimension spatiale
et calculent d'ailleurs des int\'egrales de contour non convergentes).
Ces articles n'\'etudient pas non plus les conditions d'observabilit\'e des
oscillations.
Dans le cadre du traitement unifi\'e stable/instable, nous avons discut\'e de la
longueur de coh\'erence associ\'ee aux particules instables, tout en constatant
qu'elle est en g\'en\'eral bien sup\'erieure \`a la distance typique de
d\'esint\'egration.

\item
Quelques lacunes restent \`a combler. L'effet du temps de vie fini d'une
source de neutrinos au repos a \'et\'e consid\'er\'e dans \cite{rich,mohanty},
mais seulement en m\'ecanique quantique. Une \'etude des oscillations de
neutrinos r\'esultant de la d\'e\-sin\-t\'e\-gra\-tion d'un pion en mouvement
a \'et\'e tent\'ee en th\'eorie des champs \cite{campagne} mais n'est pas
convaincante. Bien qu'il soit ais\'e en th\'eorie des champs d'\'ecrire
l'amplitude d'un processus complexe (par exemple en cascade), l'\'evaluation des
int\'egrales constitue souvent une barri\`ere infranchissable. Par exemple,
la description d'une oscillation r\'esultant d'une source instable au repos ou
en mouvement implique une int\'egration sur deux propagateurs (celui de la source et
celui de la particule oscillante) dont il est difficile de
tirer des informations sans effectuer des simplifications abusives.

\end{enumerate}

Il serait dommage de ne pas appliquer la formule d'oscillation que nous avons
d\'eriv\'ee \`a quelques cas repr\'esentatifs. Cette formule devra \^etre
adapt\'ee \`a chaque cas. Pour ce faire, quelques prescriptions de calcul
seraient bienvenues pour \'eviter d'encombrantes manipulations de la formule de
la probabilit\'e. C'est le sujet du prochain chapitre.

\chapter{Applications}
\label{applications}

\section[Prescription de calcul]
{Prescription de calcul et\\matrice de masse effective}

La th\'eorie des champs nous a donn\'e les moyens de d\'eriver la probabilit\'e
de d\'etection d'une particule en m\'elange se propageant sur une distance
macroscopique.
Cette formule d'oscillation n'est cependant pas tr\`es commode \`a utiliser,
m\^eme si l'on suppose que les conditions d'observabilit\'e sont satisfaites.
Par exemple, l'application de cette formule aux oscillations simultan\'ees du
$K^0$ et du $\overline{K^0}$ produits par la d\'esint\'egration du
m\'eson $\phi(1020)$ n\'ecessite un retour \`a l'amplitude pour y inclure le
propagateur de la seconde particule oscillante. Sous certaines conditions \`a
d\'efinir, les calculs ult\'erieurs ne sont pas compliqu\'es mais la
formule r\'esultante pour la probabilit\'e est extr\^emement peu pratique \`a
manipuler.

Le ph\'enom\`ene d'oscillation est pourtant d\'ej\`a identifiable dans
l'amplitude. La somme dans l'amplitude d'exponentielles oscillant selon des
\'energies-impulsions l\'eg\`erement diff\'erentes est \`a l'origine
des interf\'erences oscillant selon les petites diff\'erences
d'\'energies-impulsions.
Pourquoi devrions-nous calculer la probabilit\'e?
Le probl\`eme g\^{\i}t en la d\'ependance de l'amplitude en le
temps de propagation, d\'ependance qui ne dispara\^\i t qu'apr\`es int\'egration de la
probabilit\'e sur le temps. Il serait pratique de travailler sur une amplitude
d\'ependant uniquement de la longueur $L$ (mesur\'ee dans la direction o\`u la
probabilit\'e est maximale), comme celle d\'eriv\'ee dans le mod\`ele
simplifi\'e, \`a la section \ref{oscisimpli}.
V\'erifions donc si les approximations appliqu\'ees \`a la formule de la probabilit\'e
ne sont pas transf\'erables \`a la formule de l'amplitude.

Tout d'abord, si l'on suppose que les conditions d'observabilit\'e de l'oscillation
(d\'ecoh\'erence et localisation des interactions) sont v\'erifi\'ees, seuls
les facteurs d'oscillation et de d\'esint\'egration restent \`a l'int\'erieur
de la somme sur les \'etats propres dans l'amplitude, comme on l'a \'ecrit
dans l'\'equation (\ref{formuleoscsimple}). 

Ensuite, la comparaison de la formule de l'amplitude partielle ${\cal A}_j$ de
propagation d'un \'etat $j$ (\'equation (\ref{expressionAj})),
avec la formule de la probabilit\'e int\'egr\'ee sur le temps,
${\cal P}(\alpha \!\to\! \beta, L \hat{\bf e}_z)$ (\'equation (\ref{probaz})),
montre qu'en ce qui concerne les facteurs de d\'esint\'egration et
d'oscillation, l'int\'egration sur le temps revient \`a remplacer $T$ par $L/v$
dans l'amplitude ${\cal A}_j$.
La substitution $T \!\to \!L/v$ change l'expression de l'amplitude partielle en
$$
  {\cal A}_j \sim \exp
  \left(
        -i \left(
                E_j - v P_j - i \frac{m_j \Gamma_j}{2 E_j}
           \right) \, \frac{L}{v}
  \right) \, ,
$$
o\`u $E_j = \sqrt{ m_j^2 + {\bf P}_j^2 }$ et ${\bf P}_j$ est la valeur pour laquelle la
fonction-poids $\varphi(m_j^2,{\bf p})$ est maximale.
En d\'e\-ve\-lop\-pant comme au chapitre pr\'ec\'edent $E_j$ et ${\bf P}_j$ en $\Delta
m_j^2=m_j^2-m^2$, o\`u $m$ est une masse de r\'ef\'erence, on obtient
$$
  E_j - v P_j
  \cong E - v P + \frac{\Delta m_j^2}{2E}
  \cong \frac{m^2}{E} + \frac{\Delta m_j^2}{2E}
  \cong \frac{ m^2 + m_j^2 }{2E} \, .
$$
On peut, de plus, approximer $m_j \Gamma_j/E_j$ par $m \Gamma_j/E$ car la relation
$\Delta m/m \!\ll\!\Delta \Gamma/\Gamma$ est toujours v\'erifi\'ee, puisque
$\Gamma/m \!\ll\!\Delta \Gamma/\Delta m$.

Par cons\'equent, on peut \'etablir la prescription suivante: {\sl si les
conditions d'observabilit\'e des oscillations (d\'ecoh\'erence et
localisation des interactions) sont v\'erifi\'ees, l'amplitude de propagation,
dans la direction $z$ o\`u elle est maximale, prend la forme}
\begin{equation}
\fbox{$ \displaystyle
  {\cal A}(\alpha \!\to\! \beta,L \hat{\bf e}_z)
  \sim
  \sum_j V_{\beta j}^{-1} \,
    \exp \left( -i 
         \left(
               \frac{m^2 + m_j^2}{2E} - i \frac{m \Gamma_j}{2 E}
         \right) \frac{L}{v}
         \right) \,
    V_{j\alpha}
  $}
  \label{presc1}
\end{equation}
On retrouve facilement la longueur d'oscillation dans la diff\'erence de deux
facteurs oscillants:
$$
  \left(
        \frac{m^2 + m_i^2}{2E} - \frac{m^2 + m_j^2}{2E}
  \right) \frac{L}{v}
  = \frac{\Delta m_{ij}^2 \, L}{2P}
  = 2\pi \frac{L}{ L^{osc}_{ij} }
$$
Si l'on d\'efinit $M$ comme une matrice diagonale dont les termes diagonaux
sont
$$
  M_{jj} = (m^2 + m_j^2)/2E - i m \Gamma_j/2 E \, ,
$$
on peut r\'e\'ecrire l'amplitude sous les formes matricielles \'equivalentes:
\begin{eqnarray}
  {\cal A}(\alpha \!\to\! \beta,L \hat{\bf e}_z) &\sim&
  \left(
        V^{-1} \, \exp \left\{ -i M \, L/v \right\} \, V
  \right)_{\beta\alpha}
  \nonumber
  \\
  &\sim&
  \left(
       \exp \left\{
                   -i V^{-1} M V \, L/v
            \right\}
  \right)_{\beta\alpha}
  \nonumber
  \\
  &\sim&
  \left(
       \exp \left\{
                   -i M_{saveur}  \, L/v
            \right\}
  \right)_{\beta\alpha}
  \label{presc2}
\end{eqnarray}
o\`u l'on a d\'efini la matrice $M_{saveur} \equiv V^{-1} \, M V$.\\
Comme $m^2 + m_j^2 \cong  2 m m_j$, les \'el\'ements de $M$
deviennent, pour des particules non relativistes,
$$
  M_{jj} \cong m_j - i \Gamma_j/2  \, .
$$
Ces \'el\'ements sont identiques \`a ceux trouv\'es lors de la diagonalisation
de la matrice de masse effective d\'eriv\'ee par la m\'ethode non relativiste de Wigner-Weisskopf
\cite{wigner}. 
Dans la limite non relativiste, la matrice $M_{saveur}$ est la matrice de masse
effective. Notre calcul prescrit que l'extension relativiste de
cette formule se fait en rempla\c{c}ant le temps $T$ par le temps propre commun
aux deux particules $\tau = mT/E$.
Dans les sections suivantes, nous appliquerons la prescription de calcul (\ref{presc1})
ou (\ref{presc2}) aux exp\'eriences CPLEAR et DA$\Phi$NE qui ont \'et\'e d\'ecrites pour
la premi\`ere fois en th\'eorie des champs, mais de fa\c{c}on diff\'erente, dans \cite{beuthe}.

\section{L'exp\'erience CPLEAR}

Dans l'exp\'erience CPLEAR
\cite{angelopoulos1,angelopoulos2,adler+-,adler00,fry}, une particule $K^0$ ou
$\overline{K^0}$ est produite au point $x_{\SS P}$ par l'annihilation par
interaction forte de $p\bar p$, et se d\'esint\`egre au point $x_{\SS D}$ par
interaction faible.
Les m\'ecanismes de production du $K^0$ ou $\overline{K^0}$ sont
$p\bar p \!\to\! K^0 K^- \pi^+ , \, \overline{K^0} K^+ \pi^-$.
Le rapport de branchement total est de $0.4\%$ et $10^6$ antiprotons par seconde
sont dirig\'es vers la cible d'hydrog\`ene.
Comme les interactions fortes conservent l'\'etranget\'e, les amplitudes
suivantes sont nulles:
$$
  {\cal M}(p\bar p \!\to\! K^0 K^+ \pi^-) =
  {\cal M}(p\bar p \!\to\!\overline{K^0} K^- \pi^+) = 0 \, .
$$
 L'\'etranget\'e des kaons neutres produits peut donc
\^etre connue en identifiant le kaon charg\'e associ\'e. Apr\`es leur
production, le $K^0$ ou $\overline{K^0}$ oscille entre ses deux composantes
$K_L$ et $K_S$ avant de se d\'esint\'egrer. Nous voudrions
d\'ecrire l'\'evolution spatiale de l'amplitude de d\'esint\'egration et ses
ph\'enom\`enes d'interf\'erence. Notons que malgr\'e que les kaons charg\'es et
les pions contenus dans l'\'etat final aient des temps de vie comparables au
$K_L$, ils seront trait\'es
comme des \'etats asymptotiques. L'invariance CPT donne aussi
$$
  {\cal M}(p\bar p \!\to\! K^0 K^- \pi^+) =
  {\cal M}^*(p\bar p \!\to\! \overline{K^0} K^+ \pi^-) \equiv {\cal C} \, .
$$
L'impulsion moyenne des kaons neutres est de $550$ MeV, de sorte que la
distance moyenne parcourue par un $K_S$ pendant une vie moyenne est de 3 cm.
Le d\'etecteur permet d'\'etudier les d\'esint\'egrations pendant 20 vies
moyennes du $K_S$. Environ $3\%$ des $K_L$ se d\'esint\`egrent dans cette
r\'egion.

\subsection{D\'esint\'egrations semi-leptoniques}

Consid\'erons d'abord des \'etats finaux contenant un pion et des leptons.
Il y a en principe quatre d\'esint\'egrations semi-leptoniques pour les kaons:
\begin{eqnarray*}
  K^0 \to e^+ \pi^- \nu  \, , \\
  \overline{K^0} \to e^- \pi^+ \bar\nu  \, , \\
  K^0 \to e^- \pi^+ \bar\nu  \, , \\
  \overline{K^0} \to e^+ \pi^- \nu \, .
\end{eqnarray*}
Les deux premi\`eres sont caract\'eris\'ees par $\Delta S = \Delta Q$ et sont
permises dans le mod\`ele standard tandis que les deux derni\`eres sont
caract\'eris\'ees par $\Delta S = -\Delta Q$ et sont n\'egligeables dans ce
mod\`ele car elles n'apparaissent qu'au second ordre dans l'interaction faible. 
Notons cependant que cette r\`egle n'est test\'ee exp\'erimentalement qu'\`a $10^{-3}$ pr\`es
\cite{angelopoulos1}. La d\'esint\'egration des kaons neutres n'\'etant pas
imm\'ediate, les oscillations introduisent une composante de
$\overline{K^0}$ dans le $K^0$ et vice versa et les d\'esint\'egrations
interdites se produisent.
En comptant les \'electrons produits par les $K^0$ et
les positrons produits par les $\overline{K^0}$ ({\it wrong-sign leptons}),
on peut calculer l'asym\'etrie $A_T(L)$ d\'efinie par
\begin{equation}
  A_T(L) \equiv
  \frac{
    |{\cal M} \left( \overline{K^0} (x=0) \to e^+ \pi^- \nu (x=L) \right)|^2 -
    |{\cal M} \left( K^0 (x=0) \to e^- \pi^+ \bar\nu (x=L)        \right)|^2 }
  { |{\cal M} \left( \overline{K^0} (x=0) \to e^+ \pi^- \nu (x=L) \right)|^2 +
    |{\cal M} \left( K^0 (x=0) \to e^- \pi^+ \bar\nu (x=L)        \right)|^2 } \, .
  \label{defasymAT}
\end{equation}
Les processus dont on calcule l'asym\'etrie sont reli\'es par la transformation
CP. Par cons\'equent, une asym\'etrie $A_T$ non nulle signifie une violation CP. Cette
asym\'etrie est souvent interpr\'et\'ee \cite{angelopoulos2,alvarez} comme
param\'etrisant la violation du renversement du temps T, ce qui revient au
m\^eme dans notre formalisme o\`u CPT est conserv\'e.
En effet, les violations de CP et de
T ne peuvent \^etre diff\'erenci\'ees que si la violation de CPT est aussi
possible. En m\'ecanique quantique, on se permet de param\'etriser 
simultan\'ement les violations de CP, T et CPT. Proc\'eder de m\^eme dans
notre mod\`ele mettrait en jeu sa coh\'erence, puisqu'il est bas\'e sur la
th\'eorie des champs dont le th\'eor\`eme CPT est un pilier.

Enfin, notons que l'invariance CPT impose que
$$
  {\cal M} \left( K^0 (x=L) \to e^+ \pi^- \nu (x=L) \right) =
  {\cal M}^* \left( \overline{K^0} (x=L) \to e^- \pi^+ \bar\nu (x=L)  \right)
  \equiv {\cal D} \, .
$$

Calculons d'abord l'amplitude ${\cal T}_{0\bar0}(L)$ du processus complet de la
production d'un $K^0$ se d\'esint\'egrant apr\`es propagation (et oscillation)
sur une distance $L$ en $e^- \pi^+ \bar\nu$.
La saveur $\alpha$ \`a la source est $K^0$ ($S=+1$).
La saveur $\beta$ du kaon juste avant sa d\'esint\'egration est identifi\'ee
par le signe du lepton charg\'e produit; dans ce cas-ci, on veut
$\beta=\overline{K^0}\, (S=-1)$. Notre prescription de calcul (\ref{presc2}) donne,
dans la direction o\`u l'amplitude est maximale,
$$
  {\cal T}_{0\bar0}(L) \sim
  \left( 0 \quad {\cal D}^* \right) \,
   V^{-1} \, \exp \left\{ -i M \frac{L}{v} \right\} \, V \,
  \left(
  \begin{array}{c}
  {\cal C} \\ 0
  \end{array}
  \right) \, .
$$
Le vecteur colonne \`a l'extr\^eme droite s\'electionne la saveur $\alpha=K^0$,
tandis que le vecteur ligne \`a l'extr\^eme gauche s\'electionne la saveur
$\beta=\overline{K^0}$.
La matrice de masse effective $M$ est donn\'ee par
\begin{equation}
  M =
  \left(
  \begin{array}{cc}
  m m_{\SS S}/E - i m \Gamma_S/2E & 0 \\
  0 & m m_{\SS L}/E - i m \Gamma_L/2E
  \end{array}
  \right)
  \label{matriceM}
\end{equation}
o\`u $m=(m_{\SS S}+m_{\SS L})/2$ et $E$ est l'\'energie du syst\`eme
$e^- \pi^+ \bar\nu$ r\'esultant de la d\'esint\'egration du $K^0$.
Les matrices $V$ et $V^{-1}$ ont \'et\'e calcul\'ees au chapitre
\ref{melangesprop} (\'equations (\ref{matriceV})) et
sont donn\'ees par
\begin{eqnarray}
  V &=& \frac{1}{ \sqrt{ 2(1- \hat \epsilon^2}) }
  \left(
  \begin{array}{rr}
  1 & -\hat \epsilon \\
  -\hat \epsilon & 1
  \end{array}
  \right)
  \left(
  \begin{array}{rr}
  1 & 1 \\
  1 & -1
  \end{array}
  \right)
  \nonumber
  \\
  V^{-1} &=& \frac{1}{ \sqrt{ 2(1- \hat \epsilon^2}) }
  \left(
  \begin{array}{rr}
  1 & 1 \\
  1 & -1
  \end{array}
  \right)
  \left(
  \begin{array}{rr}
  1 & \hat \epsilon \\
  \hat \epsilon & 1
  \end{array}
  \right)
  \label{matriceVV}
\end{eqnarray}
L'amplitude devient
\begin{eqnarray*}
  & & {\cal T}_{0\bar0}(L) \sim
  \frac{ {\cal C} \, {\cal D}^* }{2 (1-\hat \epsilon^2)}
  \\ & & \times
  \left( 0 \quad 1 \right) \,
  \left(
  \begin{array}{rr}
  1 & 1 \\
  1 & -1
  \end{array}
  \right)
  \left(
  \begin{array}{rr}
  1 & \hat \epsilon \\
  \hat \epsilon & 1
  \end{array}
  \right)
  \exp \left\{ -i M \frac{L}{v} \right\} \,
  \left(
  \begin{array}{rr}
  1 & -\hat \epsilon \\
  -\hat \epsilon & 1
  \end{array}
  \right)
  \left(
  \begin{array}{rr}
  1 & 1 \\
  1 & -1
  \end{array}
  \right)
  \left(
  \begin{array}{c}
  1 \\ 0
  \end{array}
  \right) \, .
\end{eqnarray*}
Apr\`es injection de la matrice $M$, l'amplitude devient
\begin{equation}
  {\cal T}_{0\bar0}(L) \sim
  \frac{ {\cal C}  \, {\cal D}^* }{2} \,
  \frac{1 - \hat \epsilon}{1 + \hat \epsilon} \,
  \left[
  \exp \left(
             \left(- i m_{\SS S} - \frac{ \Gamma_{\SS S}}{2} \right)
             \frac{m }{ P} \, L
       \right) - 
  \exp \left(
             \left(- i m_{\SS L} - \frac{ \Gamma_{\SS L}}{2} \right)
             \frac{m }{ P} \, L
       \right)
  \right] \, ,
  \label{ampli0bar0}
\end{equation}
o\`u $P$ est l'impulsion totale du syst\`eme
$e^- \pi^+ \bar\nu$ r\'esultant de la d\'esint\'egration du $K^0$.

Calculons maintenant l'amplitude ${\cal T}_{\bar0 0}(L)$ du processus complet de la
production d'un $\overline{K^0}$ se d\'esint\'egrant apr\`es propagation
(et oscillation) sur une distance $L$ en $e^+ \pi^- \nu$.
La saveur $\alpha$ \`a la source est $\overline{K^0}$ ($S=-1$).
La saveur $\beta$ du kaon juste avant sa d\'esint\'egration est identifi\'ee
par le signe du lepton charg\'e produit; dans ce cas-ci on veut
$\beta=K^0 \, (S=+1)$. Un calcul analogue donne
\begin{equation}
  {\cal T}_{\bar0 0}(L) \sim
  \frac{ {\cal C}^*  \, {\cal D} }{2} \,
  \frac{1 + \hat \epsilon}{1 - \hat \epsilon} \,
  \left[
  \exp \left(
             \left(- i m_{\SS S} - \frac{ \Gamma_{\SS S}}{2} \right)
             \frac{m }{ P} \, L
       \right) - 
  \exp \left(
             \left(- i m_{\SS L} - \frac{ \Gamma_{\SS L}}{2} \right)
             \frac{m }{ P} \, L
       \right)
  \right] \, .
  \label{amplibar00}
\end{equation}

On peut enfin calculer l'asym\'etrie $A_T(L)$.
L'insertion des expressions de ${\cal T}_{0 \bar 0}(L)$
(\'equation (\ref{ampli0bar0})) et ${\cal T}_{\bar0 0}(L)$
(\'equation (\ref{amplibar00})) dans la d\'efinition de $A_T$
(\'equation (\ref{defasymAT})), donne
une asym\'etrie ind\'ependante de la distance:
\begin{equation}
  A_T = \frac{ 1 - |\sigma|^4 }{ 1 + |\sigma|^4 } \, .
  \label{asymlepto}
\end{equation}
o\`u
\begin{equation}
  \sigma \equiv \frac{1 - \hat \epsilon}{1 + \hat \epsilon} \, ,
  \label{defsigma}
\end{equation}
Il y a violation CP si le param\`etre $\sigma$ n'est pas une phase pure:
$|\sigma| \neq 1$.
La condition correspondante pour $\hat \epsilon$ s'obtient \`a l'aide de la
relation
$$
   {\cal R}\!e \, \hat \epsilon =
  \frac{ 1 - |\sigma|^2 }{ 1 + 2 {\cal R}\!e \sigma + |\sigma|^2} \, .
$$
L'asym\'etrie $A_T$ est non nulle si le param\`etre $\hat \epsilon$ n'est pas purement
imaginaire: ${\cal R}\!e \, \hat \epsilon \neq 0$. On parle de violation CP {\it indirecte}
pour souligner le fait qu'elle est uniquement due au m\'elange et qu'elle n'appara\^it
pas dans les amplitudes de production et de d\'esint\'egration ${\cal C}$ et ${\cal D}$.
L'exp\'erience CPLEAR donne
$A_T = (6.6 \pm 1.6) \times 10^{-3}$ \cite{angelopoulos2}.\\

\subsection{D\'esint\'egrations pioniques}

Consid\'erons maintenant des \'etat finaux qui sont des \'etats propres sous CP,
en particulier des \'etats constitu\'es de deux pions charg\'es.
Ces \'etats sont pairs sous CP.
Calculons d'abord l'amplitude associ\'ee au processus complet de la production
d'un $K^0$ se d\'e\-sin\-t\'e\-grant ensuite en $\pi^+\pi^-$.
La saveur $\alpha$ \`a la source est $K^0$ ($S=+1$).
La saveur $\beta$ \`a la d\'etection n'est pas identifi\'ee.
On d\'etecte des \'etats finaux $\pi^+\pi^-$ qui peuvent \^etre produits aussi
bien par la d\'esint\'egration de la saveur $\beta=K^0$ ($S=+1$) que de la saveur
$\beta=\overline{K^0}$ ($S=-1$). On doit donc sommer sur la saveur finale
$\beta$. En utilisant notre prescription de calcul (\ref{presc2}), l'amplitude
totale d\'ependant de la distance, dans la direction o\`u elle est maximale,
s'\'ecrit
\begin{eqnarray*}
  {\cal T}_{\SS +-}(L) &=&
  \sum_\beta {\cal A}(\alpha \!\to\! \beta,L \hat{\bf e}_z)
  \\
  &\sim&
  \left( 
         {\cal M}(K^0 \!\to\! \pi^+\pi^- ) \quad
         {\cal M}(\overline{K^0} \!\to\! \pi^+\pi^- ) 
  \right) \,
   V^{-1} \, \exp \left\{ -i M \frac{L}{v} \right\} \, V \,
  \left(
  \begin{array}{c}
  {\cal C} \\ 0
  \end{array}
  \right)
\end{eqnarray*}
L'indice ${\SS +-}$ de ${\cal T}_{\SS +-}$ symbolise la charge des pions
r\'esultant de la d\'esint\'egration du $K^0$. Le vecteur colonne \`a
l'extr\^eme droite s\'electionne la saveur $\alpha=K^0$.
Les matrices $V$ et $V^{-1}$ figurent \`a la section pr\'ec\'edente
(\'equations \ref{matriceVV})),
ainsi que la matrice $M$ (\'equation (\ref{matriceM})),
o\`u $E$ repr\'esente maintenant l'\'energie des
deux pions de la d\'esint\'egration du $K^0$.
Le lien entre la base des vecteurs propres de l'op\'erateur CP et la base de
saveur a \'et\'e donn\'e chapitre \ref{melangesprop} (\'equation
(\ref{baseCP})) et s'\'ecrit
$$
  \left(
  \begin{array}{c}
  |K_1 \rangle \\ |K_2 \rangle
  \end{array}
  \right)
  =\frac{1}{ \sqrt{2} }
  \left(
  \begin{array}{rr}
  1 & 1 \\
  1 & -1
  \end{array}
  \right)
  \left(
  \begin{array}{c}
  |K^0 \rangle \\ |\overline{K^0} \rangle
  \end{array}
  \right)
$$

L'amplitude devient
\begin{eqnarray*}
  & & {\cal T}_{\SS +-}(L) \sim 
  \frac{ {\cal C} }{ \sqrt{2} (1-\hat\epsilon^2) }
  \\ & & \times
  \left( 
         {\cal M}(K_1 \!\to\! \pi^+\pi^- ) \quad
         {\cal M}(K_2 \!\to\! \pi^+\pi^- ) 
  \right) \,
  \left(
  \begin{array}{rr}
  1 & \hat \epsilon \\
  \hat \epsilon & 1
  \end{array}
  \right)
  \exp \left\{ -i M \frac{L}{v} \right\}
  \left(
  \begin{array}{rr}
  1 & -\hat \epsilon \\
  -\hat \epsilon & 1
  \end{array}
  \right)
  \left(
  \begin{array}{c}
  1 \\ 1
  \end{array}
  \right)
\end{eqnarray*}
Elle peut se mettre sous la forme
\begin{eqnarray}
  & & {\cal T}_{\SS +-}(L) \sim
  \frac{ {\cal C} }{ \sqrt{2} (1+\hat\epsilon) } \,
  {\cal M}(K_1 \!\to\! \pi^+\pi^- )
  \label{ampli+-}
  \\ & & \times
  \left[
  ( 1 + \chi_{\SS +-} \hat\epsilon ) \,
  \exp \left(
             \left(- i m_{\SS S} - \frac{ \Gamma_{\SS S}}{2} \right)
             \frac{m }{ P} \, L
       \right)
  + ( \hat\epsilon + \chi_{\SS +-} ) \,
  \exp \left(
             \left(- i m_{\SS L} - \frac{ \Gamma_{\SS L}}{2} \right)
             \frac{m }{ P} \, L
       \right)
  \right]
  \nonumber
\end{eqnarray}
o\`u $P$ est la norme de l'impulsion totale des pions finaux et le param\`etre
$$
  \chi_{\SS +-} \equiv
  \frac{ {\cal M}(K_2 \!\to\! \pi^+\pi^- ) }
       { {\cal M}(K_1 \!\to\! \pi^+\pi^- ) }
$$
d\'ecrit la violation CP dans les d\'esint\'egrations, puisque
$CP(K_1)=CP(2\pi)= +1$ et $CP(K_2)= -1$. On parle de violation CP {\it directe}.

Consid\'erons maintenant le processus analogue de production d'un
$\overline{K^0}$ se d\'e\-sin\-t\'e\-grant ensuite en $\pi^+\pi^-$. En suivant la
m\^eme m\'ethode que dans le cas de la production d'un $K^0$, on obtient
l'expression suivante pour l'\'evolution spatiale du $\overline{K^0}$:
\begin{eqnarray}
  & & \overline{ {\cal T}_{\SS +-} }(L) \sim
  \frac{ {\cal C}^* }{ \sqrt{2} (1-\hat\epsilon) } \,
  {\cal M}(K_1 \!\to\! \pi^+\pi^- )
  \label{amplibar+-}
  \\ &\times&
  \left[
  ( 1 + \chi_{\SS +-} \hat\epsilon ) \,
  \exp \left(
             \left(- i m_{\SS S} - \frac{ \Gamma_{\SS S}}{2} \right)
             \frac{m }{ P} \, L
       \right) -
  ( \hat\epsilon + \chi_{\SS +-} ) \,
  \exp \left(
             \left(- i m_{\SS L} - \frac{ \Gamma_{\SS L}}{2} \right)
             \frac{m }{ P} \, L
       \right)
  \right]
  \nonumber
\end{eqnarray}

Notons que si nous \'etions int\'eress\'es par le mode de d\'esint\'egration des
kaons neutres en $\pi^0\pi^0$, il suffirait de remplacer, dans les \'equations
ci-dessus, \mbox{${\cal M}(K_1 \!\to\! \pi^+\pi^- )$} par
\mbox{${\cal M}(K_1 \!\to\! \pi^0\pi^0 )$} et $\chi_{\SS +-}$ par
$\chi_{\SS 00}$, ce dernier param\`etre \'etant d\'efini par
$$
  \chi_{\SS 00} \equiv
  \frac{ {\cal M}(K_2 \!\to\! \pi^0\pi^0 ) }
       { {\cal M}(K_1 \!\to\! \pi^0\pi^0 ) } \, .
$$
A partir des expressions de ${\cal T}_{\SS +-}(L)$ et de
$\overline{ {\cal T}_{\SS +-} }(L)$, on extrait les rapports mesurables entre
l'amplitudes de d\'esint\'egration du $K_{\SS L}$ violant CP et l'amplitude
de d\'e\-sin\-t\'e\-gra\-tion du $K_{\SS S}$ conservant CP:
\begin{equation}
 \eta_{\SS +-} \equiv
  \frac{ {\cal M}(K_{\SS L} \!\to\! \pi^+\pi^- ) }
       { {\cal M}(K_{\SS S} \!\to\! \pi^+\pi^- ) }
  = \frac{ \hat\epsilon + \chi_{\SS +-} }
         {1+\chi_{\SS +-} \hat\epsilon }   \, ,
  \label{defetaCP1}
\end{equation}
ainsi que
\begin{equation}
 \eta_{\SS 00} \equiv
  \frac{ {\cal M}(K_{\SS L} \!\to\! \pi^0\pi^0 ) }
       { {\cal M}(K_{\SS S} \!\to\! \pi^0\pi^0 ) }
  = \frac{ \hat\epsilon + \chi_{\SS 00} }
         {1+\chi_{\SS 00} \hat\epsilon }   \, .
  \label{defetaCP2}
\end{equation}
Les param\`etres $\eta_{\SS +-}$ et $\eta_{\SS 00}$ sont couramment utilis\'es
pour param\'etriser la violation CP dans les d\'esint\'egrations \`a deux
pions du $K_{\SS L}$ (voir par exemple p. 107 de \cite{pdg}). Les relations
ci-dessus entre les quantit\'es mesurables et les param\`etres quantifiant la
violation CP directe et indirecte ont \'et\'e d\'eriv\'ees sans utiliser la
sym\'etrie d'isospin.
En outre, il est possible de montrer explicitement que les param\`etres
$\eta_{\SS +-}$ et $\eta_{\SS 00}$ sont ind\'ependants de la convention de
phase \'etrange choisie pour les kaons neutres, ce qui n'est pas le cas pour
les param\`etres $\hat\epsilon$ et $\chi_{\SS +-,00}$ (voir plus loin).

L'asym\'etrie pr\'ef\'er\'ee des exp\'erimentateurs est d\'efinie par
$$
  A_{\SS +-}(L) \equiv
  \frac{ |\overline{ {\cal T}_{\SS +-} }(L)|^2
         - |\sigma|^{-2}  \, |{\cal T}_{\SS +-}(L)|^2 }
       {|\overline{ {\cal T}_{\SS +-} }(L)|^2
         + |\sigma|^{-2} \, |{\cal T}_{\SS +-}(L)|^2 } \, ,
$$
o\`u $\sigma \equiv (1-\hat \epsilon)/ (1+\hat \epsilon)$ est le
param\`etre d\'ej\`a rencontr\'e lors de l'analyse des
d\'esint\'e\-gra\-tions semi-leptoniques.
A partir des \'equations pour ${\cal T}_{\SS +-}(L)$ et
$\overline{ {\cal T}_{\SS +-} }(L)$, on calcule que
\begin{equation}
  A_{\SS +-}(L) =
  -2 \, |\eta_{\SS +-}| \;
  \frac{ e^{(\Gamma_{\SS S} - \Gamma_{\SS L})mL/2P} \,
         \cos ( m\Delta m L/P - \phi_{\SS +-} ) }
       {1 + |\eta_{\SS +-}|^2 \, e^{(\Gamma_{\SS S} - \Gamma_{\SS L})mL/2P}} \, ,
  \label{asymcplear}
\end{equation}
o\`u $\eta_{\SS +-} \equiv |\eta_{\SS +-}| \, e^{ i\phi_{\SS +-} }$
et $\Delta m \equiv m_{\SS L} - m_{\SS S}$.\\
Cette \'equation se r\'eduit \`a l'expression d'\'evolution
temporelle utilis\'ee dans l'ana\-lyse de la collaboration CPLEAR \cite{adler+-},
si nous choisissons le rep\`ere centre de masse des deux pions produits par la
d\'esint\'egration du $K^0$ ou $\overline{K^0}$ et que nous substituons le
temps \`a l'espace par la relation $L=vT$.
Les quantit\'es $\sigma$, $|\eta_{\SS +-}|$ et $\phi_{\SS +-}$ s'obtiennent
simultan\'ement par un ajustement de la formule ci-dessus
aux donn\'ees exp\'erimentales de CPLEAR.
La quantit\'e $\Delta m$ provient d'autres mesures \cite{pdg}.
Les quantit\'es  $|\eta_{\SS 00}|$ et $\phi_{\SS 00}$ sont obtenues de
fa\c{c}on analogue \cite{adler00}.

\subsection{Transposition au syst\`eme $B^0\overline{B^0}$}

Tous ces calculs se transposent sans difficult\'e aux syst\`emes
$D^0\overline{D^0}$ et $B^0\overline{B^0}$.
R\'eglons tout de suite le compte du premier en notant que le mod\`ele standard
\cite{datta} pr\'edit que  m\'elange entre $D^0$ et $\overline{D^0}$ est
extr\^emement faible en raison du petit nombre de canaux communs de
d\'esint\'egration qui, en outre, sont fortement supprim\'es par de petits
angles de m\'elange.
Les $D^0$ et $\overline{D^0}$ se d\'esint\`egreront donc avant d'osciller
notablement. Il n'y a m\^eme pas lieu de mentionner la violation CP indirecte.
La violation CP directe est aussi pr\'edite d'un niveau n\'egligeable.

Passons au $B^0\overline{B^0}$.
En premier lieu, en ce qui concerne l'asym\'etrie semi-leptonique $A_T$ dans
ce syst\`eme,
les calculs th\'eoriques \cite{fleischer} montrent qu'elle est
proportionnelle \`a $m_c^2/m_t^2 = {\cal O}(10^{-4})$, ce qui est bien au del\`a
de la pr\'ecision exp\'erimentale actuelle. On s'attend par contre \`a une
violation CP importante dans les d\'esint\'egrations
(violation CP {\it directe}) \cite{carter}, ce qui est confirm\'e par les
premiers indices d'une mesure de la violation CP dans ce syst\`eme \cite{affolder}.
Par cons\'equent, la violation CP dans les oscillations
(violation CP {\it indirecte}) est en g\'en\'eral n\'eglig\'ee.
C'est la situation inverse du syst\`eme $K^0\overline{K^0}$.

En second lieu, en ce qui concerne les asym\'etries non leptoniques,
$A_{\SS +-}$ et $A_{\SS 00}$,
les formules pour les kaons ne leur sont pas appliqu\'ees telles quelles,
bien qu'il n'y ait pas d'obstacle \`a le faire.
Comme les \'etats finaux sont de toutes sortes,
les param\`etres $\eta_{\SS +-,00}$ et $\chi_{\SS +-,00}$ sont
not\'es $\eta_f$ et $\chi_f$.
Lors de l'\'etude de l'asym\'etrie $A_T$, nous avons vu que la
violation CP dans les oscillations est estim\'ee th\'eoriquement \`a
${\cal O}(10^{-4})$, tandis que la violation CP dans les d\'esint\'egrations
pourrait \^etre tr\`es importante puisque les trois g\'en\'erations de quarks
sont directement impliqu\'ees dans des processus comme $B_d \to J/\psi K_S$.
D\`es lors, la violation CP indirecte est habituellement n\'eglig\'ee et l'on
pose $|\sigma|=1$ (voir \'equation (\ref{asymlepto})).
Cette condition est facile \`a appliquer si l'on d\'efinit un param\`etre
$\mu_f$ associ\'e \`a $\chi_f$ par le m\^eme type de relation qui existe entre
$\hat \epsilon$ et $\sigma$:
\begin{equation}
  \mu_f \equiv \frac{1-\chi_f}{1+\chi_f}
  = \frac{ {\cal M}( \overline{B^0} \to f ) }
         { {\cal M}( B^0 \to f ) } \, .
  \label{defmu}
\end{equation}
Le param\`etre $\eta_f$ devient
\begin{equation}
  \eta_f = \frac{1-\sigma\mu_f}{1+\sigma\mu_f}
  \equiv \frac{1-\xi_f}{1+\xi_f} \, .
  \label{defxi}
\end{equation}
Le param\`etre $\xi_f$ est tr\`es int\'eressant car il peut s'exprimer
en bonne approximation, dans certains cas
($B^0_d \!\to\! J/\psi K_S$, $B^0_d \!\to\! \pi^+\pi^-$,
$B^0_s \!\to\! D^{*+}_s D^{*-}_s$, $B^0_s \!\to\! J/\psi \, \phi$ etc.),
uniquement en fonction des \'el\'ements de la matrice de m\'elange des quarks
(matrice CKM). Sa mesure permettra de v\'erifier si le mod\`ele standard
suffit \`a param\'etriser compl\`etement la violation CP.
Notons n\'eanmoins que les asym\'etries CP d\'ependent directement des
interactions dans l'\'etat final (qui fournissent les phases conservant CP des
amplitudes partielles, voir plus loin), dont les contributions \`a longue distance
sont difficiles \`a \'evaluer. La pr\'ediction de ces asym\'etries varie de
fa\c{c}on significative selon le mod\`ele choisi pour estimer ces interactions
dans l'\'etat final \cite{delepine}.

Comme pour l'exp\'erience CPLEAR, on s'int\'eresse \`a une asym\'etrie pour des
\'etats finaux \'etats propres de CP, d\'efinie par \cite{nir}
$$
  A_{CP}(B^0_d \to f,L) \equiv
  \frac{ \Gamma( B^0_d(L) \to f) - \Gamma( \overline{B^0_d}(L) \to f) }
       { \Gamma( B^0_d(L) \to f) + \Gamma( \overline{B^0_d}(L) \to f) } \, .
$$
Les \'etats propres de propagations ont des largeurs quasiment \'egales
\cite{fleischer} et sont diff\'erenci\'es par leurs masses. Supposons
que l'\'etat pair sous CP soit plus l\'eger
(not\'e $B_L$ pour {\sl Light})
que l'\'etat impair sous CP (not\'e $B_H$ pour {\sl Heavy}),
comme dans le cas des kaons et d\'efinissons
$\Delta m_d \equiv m_{\SS H} - m_{\SS L}$. 
Partant des expressions de ${\cal T}_{\SS +-}(L)$ et
$\overline{ {\cal T}_{\SS +-} }(L)$
(\'equations (\ref{ampli+-}) et (\ref{amplibar+-}))
et y substituant $B$ pour $K$ et
l'expression de $\xi_f$ (\'equation (\ref{defxi})),
on calcule l'asym\'etrie $A_{CP}$ avec les approximations
$|\sigma|=1$ et $\Gamma_H=\Gamma_L$:
$$
  A_{CP}(B^0_d \!\to\! f,L) =
  A_{CP}^{direct}(B^0_d \!\to\! f) \, \cos (\Delta m_d \frac{mL}{P})
  + A_{CP}^{induit}(B^0_d \!\to\! f) \, \sin (\Delta m_d \frac{mL}{P}) \, ,
$$
o\`u la violation CP directe,
\begin{equation}
  A_{CP}^{direct}(B^0_d \to f) \equiv
  \frac{1-|\xi_f|^2}{1+|\xi_f|^2} =
   \frac{1-|\mu_f|^2}{1+|\mu_f|^2} \, ,
  \label{violationdirecte}
\end{equation}
a \'et\'e s\'epar\'ee de la violation CP provenant de l'interf\'erence entre le
m\'elange et la d\'esint\'egration du B:
\begin{equation}
  A_{CP}^{induit}(B^0_d \to f) \equiv
  - \frac{2 \, {\cal I}\!m \,\xi_f }{1+|\xi_f|^2}
  =  - \frac{2 \, {\cal I}\!m \,(\sigma \mu_f) }{1+|\mu_f|^2} \, .
  \label{violationinduite}
\end{equation}
On voit ici l'importance de tenir compte du m\'elange pour la violation CP
m\^eme s'il n'y a pas de violation CP indirecte, puisque $A_{CP}^{induit}$
d\'epend de l'interf\'erence entre $\sigma$ et $\mu_f$.

Examinons de plus pr\`es le cas de l'\'etat final $f=J/\psi \, K_S$ dont
l'asym\'etrie vient d'\^etre mesur\'ee. Une \'etude th\'eorique
des diff\'erentes contributions au processus (voir par exemple \cite{fleischer})
montre que le facteur
$\xi_{\SS J/\psi \, K_S}$ est en tr\`es bonne approximation \'egal \`a
$\xi_{\SS J/\psi \, K_S} = e^{-2i\beta}$, o\`u $\beta$ est un angle du triangle
unitaire (pour la d\'efinition de ce triangle, voir la r\'ef\'erence \cite{pdg}, p. 105).
Ces angles sont ind\'ependants des conventions de
phase pour les quarks comme il se doit, puisque $\xi_f$ est aussi ind\'ependant
de phase.
Les asym\'etries directes (\ref{violationdirecte}) et induite
(\ref{violationinduite}) deviennent
$$
   A_{CP}^{direct}(B^0_d \to J/\psi \, K_S) = 0
  \quad \mbox{et} \quad
  A_{CP}^{induit}(B^0_d \to J/\psi \, K_S) = - \sin 2\beta \, .
$$
L'exp\'erience  \cite{affolder} donne
$$
  A_{CP}^{induit}(B^0_d \to J/\psi \, K_S) = - 0.79^{+0.41}_{-0.44} \,
  (\mbox{STAT+SYST}) \, ,
$$
et constitue le premier indice de l'existence d'une violation CP dans
le syst\`eme $B^0\overline{B^0}$, ainsi que la premi\`ere \'etape de la
v\'erification de l'unitarit\'e de la matrice de Cabibbo-Kobayashi-Maskawa.

\section{Production de kaons corr\'el\'es \`a DA$\Phi$NE}

Cette section est consacr\'ee aux oscillations d'une paire de
kaons neutres produits par l'annihilation \'electron-positron dans le
collisionneur DA$\Phi$NE \`a Frascati\cite{dafne}. Les r\'esultats obtenus dans notre
formalisme sont imm\'ediatement applicables \`a la description de la
production de paires de m\'esons $B$ dans la r\'egion de la r\'esonance
$\Upsilon$(4s) \cite{argus,kek}.

Les kaons neutres et charg\'es sont produits en quantit\'e
($\!\sim\! 10^9$ paires de $K^0\overline{K^0}$/an) dans des collisions
\'electron-positron \`a une \'energie de centre de masse dans la r\'egion de la
r\'esonance $\phi$(1020). Le m\'eson $\phi$ produit se d\'esint\`egre au point
$x$ en une paire $K^0\overline{K^0}$ avec un rapport de branchement de $34\%$.
Les kaons partent en sens oppos\'es avec une faible impulsion et ne
parcourent que 6 mm pendant le temps de vie moyen du $K_S$. Cette technique
permet de mesurer dans un m\^eme d\'etecteur les d\'esint\'egrations
des $K_S$ et des $K_L$ ainsi que leurs corr\'elations: si un $K_S$ est
identifi\'e par sa d\'esint\'egration en deux pions, la particule associ\'ee
est en principe un $K_L$.

Chaque kaon oscille entre ses composantes $K_L\!-\!K_S$ avant de se
d\'esint\'egrer en \'etats finaux $f_1(k_1)$ et $f_2(k_2)$ aux points $y_1$ et
$y_2$: $$
  \phi(q) \to K^0\overline{K^0} \to f_1(k_1) f_2(k_2) \, ,
$$
o\`u $q$, $k_1$ et $k_2$ sont les \'energies-impulsions correspondantes.

Chaque \'etat final pouvant \^etre produit soit par $K^0$, soit par
$\overline{K^0}$, il faut sommer les deux amplitudes provenant de l'\'echange
de $K^0$ et de $\overline{K^0}$ avant de prendre la norme au carr\'e.
Les nombres quantiques du $\phi$, $J^{PC}=1^{--}$, sont conserv\'es par les
interactions fortes provoquant la d\'esint\'egration. La paire
$K^0\overline{K^0}$ se trouve donc dans un \'etat antisym\'etrique sous $P$,
ainsi que sous $C$. Le signe relatif des deux contributions \`a
$\phi\!\to\!f_1f_2$ est donc n\'egatif \cite{lipkin1}.

Comme deux particules oscillent dans ce processus, notre formule d'oscillation
n'est pas directement applicable. Consid\'erons momentan\'ement le m\^eme
processus, mais pour des particules non m\'elang\'ees et non
antisym\'etris\'ees. L'amplitude du processus s'\'ecrit
selon la m\'ethode du chapitre \ref{production}, section \ref{amplitude}:
$$
  {\cal A} =
  \int [d{\bf q}] \, \Phi_P
  \int [d{\bf k}_1] \, \Phi_{D_1}^*
  \int [d{\bf k}_2] \, \Phi_{D_2}^* \;
  {\cal A}_{ondes \; planes}(q,k_1,k_2) \, ,
$$
avec
\begin{eqnarray*}
  {\cal A}_{ondes \; planes}(q,k_1,k_2) &\equiv&
  \int d^4x \, e^{ -i q \cdot x }
  \int d^4y_1 \, M_{D_1} \, e^{ i k_1 \cdot y_1 } \,
  \int d^4y_2 \, M_{D_2} \, e^{ i k_2 \cdot y_2 }
  \\ &\times& 
  \int \frac{d^4p_1}{(2\pi)^4} \, e^{ -ip_1 \cdot (y_1-x) } \, G_1(p_1^2) \,
  \int \frac{d^4p_2}{(2\pi)^4} \, e^{ -ip_2 \cdot (y_2-x) } \, G_2(p_2^2) \,
  M_P
\end{eqnarray*}
o\`u $G_i(p_i^2)$ symbolise le propagateur de la particule se d\'esint\'egrant
en $f_i(k_i)$ et les particules ext\'erieures sont sur leur couche de masse.
Les $M_P$ et $M_{D_i}$ sont les amplitudes des processus de production et de d\'etection.
Les paquets d'ondes $\Phi_P$ et $\Phi_{D_i}$
sont centr\'es dans l'espace des impulsions respectivement autour de
${\bf Q}$ et ${\bf K}_i$ tandis que leurs
transform\'ees de Fourier sont localis\'ees dans l'espace de configuration
respectivement autour de $x_{\SS P}$ et $y_{\SS D_i}$
o\`u $x_{\SS P}=(t_{\SS P},{\bf x}_{\SS P})$ et
$y_{\SS D_i}=(t_{\SS D_i},{\bf y}_{\SS D_i})$ avec $i=1,2$.
Dans la notation du chapitre \ref{production} (voir \'equation
(\ref{paquetonde})), ils s'\'ecrivent
\begin{eqnarray*}
  \Phi_P({\bf q},{\bf Q},{\bf x}_{\SS P},t_{\SS P})
  &=& \phi_P({\bf q},{\bf Q}) \, e^{ i q \cdot x_{\SS P} }
  \\
  \Phi_{D_i}({\bf k}_i,{\bf K}_i,{\bf y}_{\SS D_i},t_{\SS D_i})
  &=& \phi_{D_i}({\bf k}_i,{\bf K}_i) \, e^{ i k_i \cdot y_{\SS D_i} } \, .
\end{eqnarray*}
Apr\`es le changement de variables $x \!\to\! x + x_{\SS P}$,
$y_i \!\to\! y_i + y_{\SS D_i}$
et les int\'egrations sur les variables $x$ et $y_i$, l'amplitude devient
$$
  {\cal A} =  
  \int d^4p_1 \int d^4p_2 \, \varphi(p_1,p_2) \, G_1(p_1^2) \, G_2(p_2^2) \,
  e^{ -ip_1 \cdot (y_{\SS D_1}-x_{\SS P}) 
      -ip_2 \cdot (y_{\SS D_2}-x_{\SS P}) } \, ,
$$
o\`u la fonction-poids $\varphi(p_1,p_2)$ est une int\'egrale de recouvrement
des paquets d'ondes entrants et sortants. Elle est d\'efinie par
\begin{eqnarray*}
  \varphi(p_1,p_2) &\equiv&
  \int [d{\bf q}]  \, \phi_P({\bf q},{\bf Q})  
  \int [d{\bf k}_1] \, \phi_{D_1}^* ({\bf k}_1,{\bf K}_1)
  \int [d{\bf k}_2] \, \phi_{D_2}^* ({\bf k}_2,{\bf K}_2)
  \\
  &\times& (2\pi)^4 \, \delta^{(4)}(q-p_1-p_2) \, \delta^{(4)}(p_1-k_1) \,
  \delta^{(4)}(p_2-k_2) \, M_P \, M_{D_1} \, M_{D_2} \, .
\end{eqnarray*}
Les int\'egrations sur $p_1^0$ et $p_2^0$ se font par int\'egrale de contour,
selon la m\'ethode \'etablie au
chapitre \ref{production}, section \ref{analysetemp}. Il s'ensuit que
\begin{eqnarray*}
  {\cal A} &\approx&
  \int d^3p_1 \int d^3p_2 \, \psi(z_1,z_2,{\bf p_1},{\bf p_2})
  \\ &\times&
  \exp \left(
         -i \sqrt{ z_1 + {\bf p}_1^2 } \, T_1 + i {\bf p}_1 \cdot {\bf L}_1
         -i \sqrt{ z_2 + {\bf p}_2^2 } \, T_2 + i {\bf p}_2 \cdot {\bf L}_2
  \right)
\end{eqnarray*}
o\`u $T_i \equiv t_{\SS D_i}-t_{\SS P}$,
${\bf L}_i \equiv {\bf y}_{\SS D_i}-{\bf x}_{\SS P}$ et les $z_i$ sont les
p\^oles des $G_i(p_i^2)$.

La norme de la fonction $\psi$ est maximale pour
${\bf p}_i \cong {\bf K}_i$ {\sl et} ${\bf p}_1 + {\bf p}_2 \cong {\bf Q}$
{\sl et} les ${\bf L}_i$ align\'es avec ${\bf K}_i$.
L'\'evaluation des int\'egrales pose des difficult\'es \`a cause de la
corr\'elation entre ${\bf p}_1$ et ${\bf p}_2$, sauf si la contrainte
${\bf p}_1 + {\bf p}_2 \cong {\bf Q}$ est impos\'ee avec moins de pr\'ecision
que les contraintes ${\bf p}_i \cong {\bf K}_i$.
C'est justement le cas pour $\phi \!\to\! K^0\overline{K^0}$ puisque la
relation ${\bf p}_1 + {\bf p}_2 = {\bf Q}$ est exacte \`a
$\Gamma_\phi = 4.4$ MeV pr\`es, tandis que les relations
${\bf p}_i = {\bf K}_i$ sont exactes \`a l'incertitude sur la masse des kaons
pr\`es, bien inf\'erieure \`a $\Gamma_\phi$.
On peut donc effectuer ind\'ependamment les int\'egrales sur ${\bf p}_1$ et
${\bf p}_2$. Le reste du calcul s'op\`ere comme si les particules \'etaient
ind\'ependantes et notre prescription de calcul est applicable.

En fin de compte, si les directions de ${\bf L}_1$ et ${\bf L}_2$ sont fix\'ees,
l'amplitude antisym\'etris\'ee de d\'etection de $f_1$ \`a une distance $L_1$
et de $f_2$ \`a une distance $L_2$ est donn\'ee pour le syst\`eme m\'elang\'e
$K^0\overline{K^0}$ par
\begin{eqnarray*}
  {\cal T}_{f_1f_2}(L_1,L_2) &\sim& 
  \left( \quad
  \left[
  \left( 
         {\cal M}(K^0 \!\to\! f_1 ) \quad
         {\cal M}(\overline{K^0} \!\to\! f_1 ) 
  \right) \,
   V^{-1} \, \exp \left\{ -i M_1 \frac{L_1}{v_1} \right\} \, V \,
  \left(
  \begin{array}{c}
  1 \\ 0
  \end{array}
  \right)
  \right]
  \right.
  \\ & & \quad \times
  \left[
  \left( 
         {\cal M}(K^0 \!\to\! f_2) \quad
         {\cal M}(\overline{K^0} \!\to\! f_2 ) 
  \right) \,
   V^{-1} \, \exp \left\{ -i M_2 \frac{L_2}{v_2} \right\} \, V \,
  \left(
  \begin{array}{c}
  0 \\ 1
  \end{array}
  \right)
  \right]
  \\ & & \quad
  \left.
  - \left[
           \mbox{
           m\^eme expression avec
           $\left( \begin{array}{c}1 \\ 0 \end{array} \right)
            \leftrightarrow
            \left( \begin{array}{c}0 \\ 1 \end{array} \right)
           $}
   \right] \;
   \right) {\cal M}(\phi \to  K^0\overline{K^0})
\end{eqnarray*}
La notation $M_i$ signifie que la matrice de masse effective $M$ est \'evalu\'ee \`a
l'\'energie $E_i$ de l'\'etat final $f_i$. Les $v_i$ sont d\'efinis comme
d'habitude par la norme de \mbox{${\bf v}_i \equiv {\bf K}_i/E_i$}. 
L'amplitude se r\'e\'ecrit en fonction de la base propre de CP comme 
\begin{eqnarray*}
  & & {\cal T}_{f_1f_2}(L) \sim
  {\cal M}(\phi \to  K^0\overline{K^0}) \;
  \frac{1}{2(1-\hat\epsilon^2)}
  \\ & & \times
  \left( \;
  \left[
  \left( 
         {\cal M}(K_1 \!\!\to\!\! f_1 ) \quad
         {\cal M}(K_2 \!\!\to\!\! f_1 ) 
  \right) \,
  \left(
  \begin{array}{rr}
  1 & \hat \epsilon \\
  \hat \epsilon & 1
  \end{array}
  \right)
  \exp \left\{ -i M_1 \frac{L_1}{v_1} \right\}
  \left(
  \begin{array}{rr}
  1 & -\hat \epsilon \\
  -\hat \epsilon & 1
  \end{array}
  \right)
  \left(
  \begin{array}{c}
  1 \\ 1
  \end{array}
  \right)
  \right]
  \right.
  \\ & & \quad \, \times
  \left[
  \left( 
         {\cal M}(K_1 \!\!\to\!\! f_2) \quad
         {\cal M}(K_2 \!\!\to\!\! f_2 ) 
  \right)
  \left(
  \begin{array}{rr}
  1 & \hat \epsilon \\
  \hat \epsilon & 1
  \end{array}
  \right) \,
  \exp \left\{ -i M_2 \frac{L_2}{v_2} \right\}
  \left(
  \begin{array}{rr}
  1 & -\hat \epsilon \\
  -\hat \epsilon & 1
  \end{array}
  \right)
  \left(
  \begin{array}{r}
  1 \\ -1
  \end{array}
  \right)
  \right]
  \\ & & \quad
  \left. \,
  - \left[
           \mbox{
           m\^eme expression avec
           $\left( \begin{array}{r} 1 \\ 1 \end{array} \right)
            \leftrightarrow
            \left( \begin{array}{r} 1 \\ -1 \end{array} \right)
           $}
   \right]
   \right)
\end{eqnarray*}
Si l'on d\'efinit ${\cal M}_{ij} \equiv {\cal M}(K_i \!\to\! f_j)$,
l'amplitude devient
\begin{eqnarray*}
  & & {\cal T}_{f_1f_2}(L) \sim
  {\cal M}(\phi \to  K^0\overline{K^0}) \,
  \frac{1}{1-\hat\epsilon^2}
  \\ & & \times
  \Bigg[ \;
  - ( {\cal M}_{11} + \hat\epsilon {\cal M}_{21} ) \,
    ( \hat\epsilon {\cal M}_{12} + {\cal M}_{22} ) \,
  \exp \left(
             - \left( i m_{\SS S} + \frac{\Gamma_S}{2} \right) \frac{mL_1}{K_1}
             - \left( i m_{\SS L} + \frac{\Gamma_L}{2} \right) \frac{mL_2}{K_2}
       \right)
  \\ & & \quad \; \,
  + \, ( \hat\epsilon {\cal M}_{11} + {\cal M}_{21} ) \,
    ( {\cal M}_{12} + \hat\epsilon {\cal M}_{22} ) \,
  \exp \left(
              - \left( i m_{\SS S} + \frac{\Gamma_S}{2} \right) \frac{mL_2}{K_2}
              - \left( i m_{\SS L} + \frac{\Gamma_L}{2} \right) \frac{mL_1}{K_1}
       \right)
  \Bigg]
\end{eqnarray*}
L'expression obtenue est relativiste: aucun boost n'est n\'ecessaire pour
passer du rep\`ere au repos d'une particule oscillante au rep\`ere du
laboratoire. Observons que l'amplitude est nulle \cite{lipkin1} si
$f_1=f_2$ {\sl et} $K_1=K_2$ {\sl et} $L_1=L_2$. Pour l'application des
formules th\'eoriques aux exp\'eriences, nous r\'ef\'erons \`a la litt\'erature
\cite{dunietz}.

Une remarque \`a propos de la fr\'equence d'oscillation semble de mise.
Dans le rep\`ere centre de masse, $K_1=K_2\equiv K$ et le terme d'interf\'erence
oscille comme un cosinus de
$$
  \frac{m}{K} \, ( m_{\SS L} -m_{\SS S} ) \, (L_1-L_2) \, .
$$
Si l'on r\'e\'ecrit ce terme comme $\omega (L_1-L_2)$, la fr\'equence
d'oscillation $\omega$ vaut \mbox{$\omega=\frac{m\Delta m}{K}$}.
Le calcul de cette fr\'equence a
suscit\'e une controverse \`a la suite d'un article \cite{srivastava}
pr\'etendant qu'elle \'etait deux fois plus grande. Plusieurs
articles \cite{kayser1,lowe} ont \'et\'e \'ecrits pour r\'efuter cette
affirmation mais leurs arguments reviennent \`a contester le choix de
diff\'erents temps de d\'etection et de temps propres pour les diff\'erents
\'etats propres de masse. Tout cela est assez obscur. Seule une
d\'erivation en th\'eorie des champs donne sans ambigu\"{\i}t\'e la formule
d'oscillation. La valeur calcul\'ee est par ailleurs confirm\'ee \cite{kayser1}
par les exp\'eriences ARGUS et CLEO \cite{argus} dans le cas du processus
analogue $\Upsilon\!\to\! B^0\overline{B^0}$.

Un autre article controvers\'e \cite{widom} soutient que dans un processus tel
que $$
  \pi^- p \to \Lambda K^0 \, ,
$$
la particule $\Lambda$ associ\'ee \`a la production de la particule oscillante
$K^0$ oscillera aussi: elle serait une superposition de deux \'etats
d'\'energie-impulsion diff\'erentes par conservation de l'\'energie-impulsion
\`a la production.
Notre approche montre qu'il n'en est rien. Le processus peut \^etre d\'ecrit
compl\`etement comme ci-dessus, avec un propa\-gateur pour le kaon et un
propagateur pour le $\Lambda$. Le propagateur du kaon est une
superposition de deux propagateurs correspondant aux \'etats propres
de masse: \`a grande distance, deux p\^oles diff\'erents contribueront et
l'amplitude sera la somme de deux amplitudes correspondant \`a deux \'etats
propres de masse diff\'erents sur leur couche de masse. Ce n'est pas le cas du
$\Lambda$. On voit clairement dans cet exemple que la question ne peut \^etre
tranch\'ee sans le formalisme de th\'eorie des champs\footnote{Les r\'efutations
recourant \`a la  m\'ecanique quantique \cite{lowe} font de nouveau
appel \`a la notion de diff\'erents temps propres.}.
Comme nous l'avons vu dans l'\'equation (\ref{facteurB}),
les diff\'erences d'\'energie et d'impulsion se compensent dans le facteur
d'oscillation {\it sauf} la diff\'erence de masse entre les \'etats propres.
La m\^eme question s'est pos\'ee \cite{sassaroli,dolgov,shtanov,weiss,zralek}
pour le processus $\pi \!\to\! \mu\nu$. Le muon oscillerait-il? Par le m\^eme
raisonnement, on conclut que non.

\section{La param\'etrisation correcte de la violation CP}
\label{convphase}

Le d\'eveloppement du formalisme th\'eorique adapt\'e \`a la description de
l'ex\-p\'e\-rien\-ce CPLEAR nous a montr\'e que la violation CP peut \^etre
param\'etris\'ee dans le syst\`eme $K^0\overline{K^0}$ ou
$B^0\overline{B^0}$ par des param\`etres complexes $\hat \epsilon$ et
$\chi_{\SS f}$. 
Cette simplicit\'e cache certaines subtilit\'es dans l'extraction des
param\`etres \`a partir des donn\'ees.
Tout d'abord, il faut mentionner que ces param\`etres ne sont pas des
observables physiques car ils d\'ependent de certaines conventions de phase.
Ensuite, des approximations assez violentes sont souvent appliqu\'ees dans
le syst\`eme $K^0\overline{K^0}$, par exemple n\'egliger les termes en
$\hat \epsilon^2$, alors qu'ils sont du m\^eme ordre que les param\`etres
$\chi_{\SS f}$.
Les ambigu\"{\i}t\'es d'ordre ${\cal O}(\hat \epsilon^2)$ dues au
probl\`eme de la normalisation des \'etats $K_{S,L}$ sont aussi
oubli\'ees.
La plupart des analyses ne tiennent pas compte de ces points d\'elicats.
Il est vrai que les exp\'eriences ne sont pas encore assez pr\'ecises pour
discriminer les d\'emarches correctes des mauvaises mais il n'est plus l'heure
de passer \`a c\^ot\'e de telles incoh\'erences quand la violation CP
directe est confirm\'ee par deux exp\'eriences ind\'ependantes
\cite{burkhardt}.
Par ailleurs, les conventions de phase choisies pour le
$K^0\overline{K^0}$ ne sont pas n\'ecessairement pratiques pour le
$B^0\overline{B^0}$; il est d\`es lors utile de garder la li\-ber\-t\'e
de phase dans le formalisme.
Etant donn\'e que les d\'efinitions de nos param\`etres ne souffrent pas de
probl\`eme de normalisation, il est int\'eressant de montrer que l'a\-na\-ly\-se
peut \^etre continu\'ee rigoureusement jusqu'au bout dans l'extraction,
\`a partir des r\'esultats exp\'erimentaux, des param\`etres violant CP.
Nos r\'esultats finaux
ressemblent \`a ceux de la r\'ef\'erence \cite{chau} mais celle-ci utilise
des \'etats interm\'ediaires de kaons et ne discute pas explicitement de la
transformation de phase. Avant de d\'eterminer les valeurs des
param\`etres, on \'etudiera la question de la convention de phase.

\subsection{Phase \'etrange et param\`etres observables}

Une premi\`ere manifestation de la d\'ependance de phase de $\hat \epsilon$ est
apparue lors du calcul de l'asym\'etrie semi-leptonique $A_T$.
Une asym\'etrie non nulle est une mani\-festation de la violation CP et implique
que ${\cal R}\!e \, \hat\epsilon \neq 0$. Par contre, si
${\cal R}\!e \, \hat\epsilon = 0$ mais ${\cal I}\!m \, \hat\epsilon \neq 0$, cette asym\'etrie
s'annule et n'implique pas de violation CP.
La partie imaginaire de $\hat \epsilon$ n'a donc aucune
importance dans ce cas-ci.
Cet arbitraire dans la valeur de ${\cal I}\!m \, \hat \epsilon$ est en fait la
manifestation d'un ph\'enom\`ene plus g\'en\'eral.
La raison provient de la sym\'etrie $U(1)$ associ\'ee \`a chaque saveur, et qui
laisse le lagrangien total invariant, except\'e le terme de masse (ou la
matrice de m\'elange CKM apr\`es diagonalisation de la matrice de masse,
ce qui revient au m\^eme).
On s'int\'eresse ici \`a la transformation ${\cal S}$ de la phase
\'etrange \cite{wolfenstein,grimus1},
puisque les kaons poss\`edent une \'etranget\'e non nulle.
De nouveaux \'etats peuvent \^etre d\'efinis \`a l'aide de ${\cal S}$:
\begin{eqnarray*}
  |K^0_\alpha \rangle
  &=& e^{-i\alpha \hat S } \, |K^0 \rangle
  = e^{-i\alpha} \, |K^0 \rangle \, ,
  \\
  |\overline{K^0_\alpha} \rangle
  &=& e^{-i\alpha \hat S } \, |\overline{K^0} \rangle
  = e^{+i\alpha} \, |\overline{K^0} \rangle\, ,
\end{eqnarray*}
o\`u $\hat S$ est l'op\'erateur d'\'etranget\'e. Cette red\'efinition est
possible parce que $K^0$ et $\overline{K^0}$ sont produits par les interactions
fortes qui conservent l'\'etranget\'e.
Elle ne peut \'evidemment avoir d'effet observable: toute
quantit\'e observable doit \^etre ind\'epen\-dante de cette phase.
Dans le but de conserver la relation
\mbox{$|\overline{K^0} \rangle = CP \, |K^0 \rangle$}, 
la transformation CP est red\'efinie dans la nouvelle base:
$$
  (CP)_\alpha \equiv e^{-i\alpha \hat S } \, CP \, e^{i\alpha \hat S } \, ,
$$
de sorte que
$$
  |\overline{K^0_\alpha} \rangle = (CP)_\alpha \, |K^0_\alpha \rangle \, .
$$
La base propre sous $(CP)_\alpha$ est d\'efinie\footnote{Une
autre possibilit\'e consiste \`a ne pas red\'efinir l'op\'erateur CP. On a alors
$CP\, |K^0_\alpha \rangle = \eta^{\SS CP}_\alpha \, |\overline{K^0_\alpha} \rangle$,
o\`u $\eta^{\SS CP}_\alpha \equiv e^{-2i\alpha}$. Il faut d\'efinir diff\'eremment
les \'etats propres sous CP:
\mbox{$
|K_{1,2 \, \alpha} \rangle \equiv
  \frac{1}{\sqrt{2}} \,
  \left( |K^0_\alpha \rangle \pm CP \, |K^0_\alpha \rangle \right)
$} 
} par:
$$
  |K_{1,2 \, \alpha} \rangle \equiv
  \frac{1}{\sqrt{2}} \,
  \left( |K^0_\alpha \rangle \pm |\overline{K^0_\alpha} \rangle \right) \, .
$$
Comme la transformation ne laisse pas invariants les termes m\'elangeant des
particules d'\'etranget\'es diff\'erentes, les \'el\'ements non diagonaux du
propagateur complet du syst\`eme $K^0\overline{K^0}$ sont modifi\'es.
Se rappelant la param\'etrisation du propagateur (\'equation (\ref{propKK})),
$$
  i G^{-1}(p^2) =
  \left(
  \begin{array}{cc}
  \langle K^0 |\hat G^{-1}| K^0 \rangle &
  \langle K^0 |\hat G^{-1} |\overline{K^0} \rangle \\
  \langle \overline{K^0} |\hat G^{-1}| K^0 \rangle &
  \langle \overline{K^0} |\hat G^{-1}| \overline{K^0} \rangle
  \end{array}
  \right)
  \equiv
  \left(
  \begin{array}{cc}
  d   &  a+b  \\
  a-b &  d
  \end{array}
  \right) \; ,
$$
on voit que les termes non diagonaux sont transform\'es sous ${\cal S}$ en
\begin{eqnarray*}
  a_\alpha + b_\alpha &=& (a+b) \, e^{2i\alpha} \, ,
  \\
  a_\alpha - b_\alpha &=& (a-b) \, e^{-2i\alpha} \, .
\end{eqnarray*}

Quel est l'effet de ${\cal S}$ sur le param\`etre $\hat \epsilon$?
Ce param\`etre a \'et\'e d\'efini
(\'equation (\ref{defepsilon})) comme la solution de
$$
  \frac{\hat \epsilon}{1+\hat \epsilon^2} \equiv \frac{b}{2a} \, .
$$
Cette \'equation poss\`ede en fait deux solutions $\hat \epsilon_1$ et
$\hat \epsilon_2$ inverses l'une de l'autre:
$$
  \hat \epsilon_{1,2} = \frac{a}{b} \,
  \left( 1 \pm \sqrt{1 - \frac{b^2}{a^2} } \right) \, ,
$$
o\`u le signe de la racine est choisi de sorte que sa partie r\'eelle soit
positive.
Comme $\hat \epsilon_1 =1/\hat \epsilon_2$, une des deux solutions a
n\'ecessairement une norme inf\'erieure \`a 1. On choisit de travailler avec
cette solution, dans l'esprit de repr\'esenter la petite violation CP
observ\'ee par des param\`etres th\'eoriques petits. Nous choisissons donc
la solution avec le signe $-$:
$$
  \hat \epsilon \equiv \frac{a}{b} \,
 \left( 1 - \sqrt{1 - \frac{b^2}{a^2} } \right) \, .
$$
Lors du calcul de l'asym\'etrie $A_T$, un param\`etre $\sigma$ a \'et\'e
associ\'e \`a $\hat \epsilon$ (\'equation (\ref{defsigma})) par
$$
  \sigma \equiv \frac{1-\hat \epsilon}{1+\hat \epsilon}
  \quad \Leftrightarrow \quad
  \hat \epsilon \equiv \frac{1-\sigma}{1+\sigma} \, .
$$
Son expression en fonction de $a$ et $b$ s'\'ecrit
$$
  \sigma = \frac{ \sqrt{a^2 - b^2} }{ a+b } \, .
$$
Notons que
$$
  {\cal R}\!e \, \sigma =
  \frac{ 1-|\hat \epsilon|^2 }{1+ 2 {\cal R}\!e \, \hat \epsilon + |\hat \epsilon|^2}
  \ge 0
$$
puisque $|\hat \epsilon| \le 1$.

Sous la tranformation ${\cal S}$, le param\`etre $\sigma$ devient
$$
  \sigma_\alpha =
  \frac{ \sqrt{a^2_\alpha - b^2_\alpha} }{ a_\alpha + b_\alpha } =
  e^{-2i\alpha} \, \sigma \, ,
$$
Partant de l'\'equation liant $\hat \epsilon_\alpha$ et $\sigma_\alpha$,
$$
  \hat \epsilon_\alpha
  = \frac{ 1 - \sigma_\alpha }{ 1 + \sigma_\alpha } \, ,
$$
on \'etablit la loi de transformation de $\hat \epsilon$:
\begin{eqnarray}
  \hat \epsilon_\alpha
  &=& \frac{ \hat \epsilon + i \tan \alpha }{ 1 + i \hat \epsilon \tan \alpha }
  \nonumber \\
  &=& \frac{
             2 \, {\cal R}\!e \, \hat \epsilon
             + i \left[
                       (1 - |\hat \epsilon|^2) \, \sin 2\alpha
                       + 2 \, {\cal I}\!m \, \hat \epsilon \, \cos 2\alpha
                 \right] }
           { 1 + \cos 2\alpha - 2 \, {\cal I}\!m \, \hat \epsilon \, \sin 2\alpha
              + |\hat \epsilon|^2 \, ( 1 - \cos 2\alpha ) } \, .
  \label{transfoepsilon}
\end{eqnarray}
Remarquons d'abord que
${\cal R}\!e \, \hat \epsilon_\alpha \propto {\cal R}\!e \, \hat \epsilon$. La condition
${\cal R}\!e \, \hat \epsilon \neq 0$ est donc bien invariante sous ${\cal S}$ comme il
se doit.
Cependant, la transformation ${\cal S}$ ne peut \^etre tout \`a fait
arbitraire.  En effet, le choix de la valeur $\alpha=\pm \pi/2$ r\'esulte en la
transformation $\hat \epsilon_\alpha = 1/ \hat \epsilon$.
On doit donc restreindre le domaine de $\alpha$ si l'on veut respecter la
condition $|\hat \epsilon| \le 1$. Comme les expressions ne d\'ependent que de
$2\alpha$, il suffit de travailler sur l'intervalle
$-\pi/2 \!\le\! \alpha \!\le\! \pi/2$.
On calcule que $|\hat \epsilon_\alpha| \le 1$ pour
$$
  \frac{1}{2} \,
  \left(
       \arctan \frac{ 1 - |\hat \epsilon|^2 }{ 2 \, {\cal I}\!m \, \hat \epsilon } - \pi
  \right)
  \le \alpha \le
  \frac{1}{2} \,
  \arctan \frac{ 1 - |\hat \epsilon|^2 }{ 2 \, {\cal I}\!m \, \hat \epsilon } \, ,
$$
o\`u l'$\arctan$ est pris dans le premier ou deuxi\`eme quadrant.

\subsection{D\'etermination des param\`etres}
\label{determparam}

Examinons maintenant la premi\`ere donn\'ee exp\'erimentale dont nous
disposons pour les kaons.
La petitesse de l'asym\'etrie semi-leptonique, $A_T = (6.6 \pm 1.6) \times 10^{-3}$
\cite{angelopoulos2},
nous dit \`a travers l'\'equation (\ref{asymlepto})
que $|\sigma| = 1 + {\cal O}(10^{-3})$. Comme
$$
  {\cal R}\!e \, \hat \epsilon =
  \frac{ 1 - |\sigma|^2 }
  { 1 + 2 \, {\cal R}\!e \sigma + |\sigma|^2 } \, ,
$$
avec ${\cal R}\!e \, \sigma \ge 0$, on conclut que
${\cal R}\!e \, \hat \epsilon \sim (10^{-3})$.
La valeur de ${\cal R}\!e \, \hat \epsilon$ d\'epend bien s\^ur du choix de la
phase \'etrange mais on peut d\'ej\`a donner son ordre de grandeur qui sera
correct tant que la condition $|\hat \epsilon| \!<\! 1$ est maintenue.
Par contre, on ne peut rien dire sur ${\cal I}\!m \, \hat \epsilon$ sans faire un choix
de phase \'etrange, puisque
$$
  {\cal I}\!m \, \hat \epsilon =
  \frac{ -2 \, {\cal I}\!m \, \sigma }
  { 1 + 2 \, {\cal R}\!e \sigma + |\sigma|^2 } \, .
$$

Peut-on choisir la phase \'etrange telle que $|{\cal I}\!m \, \hat \epsilon| \ll 1$?
En utilisant l'\'equation (\ref{transfoepsilon}), on obtient
$$
  {\cal I}\!m \, \hat \epsilon_\alpha = 0
  \qquad \mbox{si} \qquad
  \tan 2\alpha = -\frac{ 2 \, {\cal I}\!m \hat \epsilon }{ 1 - |\hat \epsilon|^2 } \, .
$$
Si\footnote{On a choisi
des valeurs de $\alpha$ appartenant au domaine tel que
$| \hat \epsilon_\alpha | \le 1$.}
${\cal I}\!m \, \hat\epsilon \ge 0$, $-\pi/4 \le \alpha \le 0$.
Si ${\cal I}\!m \, \hat\epsilon \le 0$, $0 \le \alpha \le \pi/4$.
Par cons\'equent, il est toujours possible de choisir la phase \'etrange telle
que ${\cal I}\!m \, \hat \epsilon_\alpha =0$. Toute une s\'erie d'autres choix sont
possibles avec
$|{\cal I}\!m \, \hat \epsilon_\alpha| \ll 1$.

Poursuivons notre examen des donn\'ees exp\'erimentales.
Dans les d\'esint\'egrations des kaons neutres en pions, les observables li\'es
\`a la violation CP ont \'et\'e d\'efinis
(\'equations (\ref{defetaCP1}) et (\ref{defetaCP2})) par
$$
  \eta_f \equiv \frac{ {\cal M}( K_L\to f) }{ {\cal M}( K_S\to f) }
  =  \frac{ \hat \epsilon + \chi_{\SS f} }{ 1 + \chi_{\SS f} \hat \epsilon } \, ,
$$
o\`u $\chi_{\SS f} \equiv {\cal M}( K_2\to f)/{\cal M}( K_1\to f)$ et
$f$ est un \'etat pair sous CP.
Comme les observables doivent \^etre invariants sous la transformation de
phase ${\cal S}$,  les quantit\'es $\chi_{\SS f}$ compensent la variation
de $\hat \epsilon$ sous ${\cal S}$. On le v\'erifie le plus
ais\'ement \`a l'aide du param\`etre $\mu_{\SS f}$ associ\'e \`a $\chi_{\SS f}$
par l'\'equation (\ref{defmu}).
La transformation de $\mu_{\SS f}$ sous ${\cal S}$ est
tout simplement la transformation inverse de $\sigma$:
$$
  \mu_{f \, , \, \alpha} = e^{2i \alpha} \, \mu_{\SS f} \, .
$$
La quantit\'e $\xi_{\SS f} \equiv \sigma\mu_{\SS f}$ est donc invariante sous
${\cal S}$ et d\`es lors $\eta_f$ aussi (\'equation (\ref{defxi})).
On a v\'erifi\'e du m\^eme coup que le param\`etre $\xi_{\SS f}$, d\'ecrivant
la violation CP dans le syst\`eme $B^0\overline{B^0}$, est invariant de phase,
comme il se doit puisqu'il est mesurable.
Pour un usage futur, je donne la transformation explicite de $\chi_{\SS f}$:
\begin{equation}
  \chi_{{\SS f} \, , \, \alpha} =
  \frac{ \chi_{\SS f} - i \tan \alpha }{ 1 - i \chi_{\SS f} \, \tan \alpha} \, .
  \label{transfochi} 
\end{equation}
Les param\`etres $\eta_{\SS f} \equiv |\eta_{\SS f}| \, e^{i\phi_{\SS f}}$
ont \'et\'e mesur\'es par CPLEAR \cite{adler+-,adler00}:
\begin{eqnarray}
  |\eta_{\SS +-}| &=& (2.254 \pm 0.024_{\SS STAT} \pm 0.026_{\SS SYST} )
  \times 10^{-3} \, ,
  \nonumber \\
  \phi_{\SS +-} &=& (43.63 \pm 0.54_{\SS STAT} \pm 0.48_{\SS SYST} )^\circ \, ,
  \nonumber \\
  |\eta_{\SS 00}| &=& (2.47 \pm 0.31_{\SS STAT} \pm 0.24_{\SS SYST} )
  \times 10^{-3} \, ,
  \nonumber \\
  \phi_{\SS 00} &=& (42.0 \pm 5.6_{\SS STAT} \pm 1.9_{\SS SYST} )^\circ \, .
  \label{valeursetaphi}
\end{eqnarray}
D\`es lors, si la convention de phase est telle que
$|\hat \epsilon| \sim {\cal O}({\cal R}\!e \, \hat \epsilon) \sim {\cal O}(10^{-3})$,
les param\`etres $\chi_{\SS f}$ doivent \^etre au plus du m\^eme ordre:
$\chi_{\SS f}  \le {\cal O}(10^{-3})$.
La violation CP indirecte pour les kaons est donc au maximum
du m\^eme ordre que la violation CP directe.
Une diff\'erence entre $\eta_{\SS +-}$ et $\eta_{\SS 00}$ \'etablit l'existence
d'une violation CP directe. C'est ce qu'ont fait les exp\'eriences NA31, E731, NA48
et KTeV \cite{burkhardt} en mesurant
\begin{equation}
  |\eta_{\SS 00}|^2/|\eta_{\SS +-}|^2 =
  1 - 6 \times (21.2 \pm 4.7) \times 10^{-4} \, ,
  \label{valeurepsprime}
\end{equation}
o\`u l'on a fait la moyenne des r\'esultats \cite{fry}.

Dans le cas o\`u l'\'etat $\tilde f$ est impair sous CP, on d\'efinit le
param\`etre
$$
  \eta_{\tilde f} \equiv
  \frac{ {\cal M}( K_S \to \tilde f) }{ {\cal M}( K_L \to \tilde f) }
  =  \frac{ \hat \epsilon + \chi_{\SS \tilde f} }
          { 1 + \chi_{\SS \tilde f} \hat \epsilon } \, ,
$$
o\`u
$\chi_{\SS \tilde f} \equiv
{\cal M}( K_1\to \tilde f)/{\cal M}( K_2\to \tilde f)$.
Si la violation CP directe est petite par rapport \`a la violation CP indirecte,
il existe une convention de phase telle que
$\eta_{\tilde f} \cong \hat \epsilon \cong \eta_f$. On pr\'edit par exemple
$\eta_{000} = \eta_{00} + {\cal O}(10^{-6})$, o\`u l'indice ${\SS 000}$ 
symbolise l'\'etat \`a trois pions neutres. 

Pour aller plus loin et extraire les quantit\'es $\hat \epsilon$,
$\chi_{\SS +-}$ et $\chi_{\SS 00}$ (d\'ependant de 6 nombres r\'eels) de $A_T$,
$\eta_{\SS +-}$ et $\eta_{\SS 00}$ (d\'ependant de 5 nombres r\'eels), il faut
disposer d'une relation suppl\'ementaire fournie par la sym\'etrie d'isospin
des pions.

Cette technique bien connue (voir par exemple \cite{gerard}), permet de
d\'ecomposer les amplitudes de transition des kaons en pions en amplitudes
correspondant \`a des isospins d\'efinis:
\begin{eqnarray*}
  {\cal M}(K^0 \to \pi^+\pi^-) &=&
    \sqrt{ \frac{2}{3} } \, e^{ i\delta_{\SS 0} } \, A_0
  + \sqrt{ \frac{1}{3} } \, e^{ i\delta_{\SS 2} } \, A_2  \, ,
  \\
  {\cal M}(\overline{K^0 }\to \pi^+\pi^-) &=&
    \sqrt{ \frac{2}{3} } \, e^{ i\delta_{\SS 0} } \, A_0^*
  + \sqrt{ \frac{1}{3} } \, e^{ i\delta_{\SS 2} } \, A_2^*  \, ,
  \\
  {\cal M}(K^0 \to \pi^0\pi^0) &=&
  - \sqrt{ \frac{1}{3} } \, e^{ i\delta_{\SS 0} } \, A_0
  + \sqrt{ \frac{2}{3} } \, e^{ i\delta_{\SS 2} } \, A_2  \, ,
  \\
  {\cal M}(\overline{K^0 } \to \pi^0\pi^0) &=&
  - \sqrt{ \frac{1}{3} } \, e^{ i\delta_{\SS 0} } \, A_0^*
  + \sqrt{ \frac{2}{3} } \, e^{ i\delta_{\SS 2} } \, A_2^*  \, .
\end{eqnarray*}
L'\'etat $\pi^+\pi^-$ a \'et\'e sym\'etris\'e et l'on a introduit un
facteur $1/\sqrt{2}$ pour l'\'etat final $\pi^0\pi^0$ pour fournir le
facteur $1/2$ dans la probabilit\'e de d\'etection de deux parti\-cules
identiques.
La brisure d'isospin est prise en compte dans la param\'etrisation en
red\'efinissant les param\`etres $A_I$ par combinaison lin\'eaire. La
d\'ecomposition garde la m\^eme forme bien que la sym\'etrie d'isospin ne
soit plus respect\'ee.
Soit $A_I \equiv |A_I| \, e^{ i\zeta_{\SS I} }$.
Sous une transformation de phase ${\cal S}$,
$A_{I \, \alpha} = e^{-i\alpha} \, A_I$ et il est par exemple possible de
rendre $A_0$ r\'eel (convention de Wu-Yang). La diff\'erence de phase
$(\zeta_{\SS 0} - \zeta_{\SS 2})$ reste invariante.
L'asym\'etrie $A_{+-}$ d\'efinie par
$$
  A_{+-} \equiv
  \frac{  |{\cal M}( K^0 \to \pi^+\pi^- )|^2
        - |{\cal M}( \overline{K^0} \to \pi^+\pi^- )|^2  }
       { |{\cal M}( K^0 \to \pi^+\pi^- )|^2
        + |{\cal M}( \overline{K^0} \to \pi^+\pi^- )|^2  }
$$
est donn\'ee en terme des amplitudes d'isospin par
\begin{equation}
  A_{+-} =
 \frac{ - 2\sqrt{2} \,
        \sin ( \delta_{\SS 0} - \delta_{\SS 2} ) \,
        \sin ( \zeta_{\SS 0} - \zeta_{\SS 2} ) \, |A_0| \, |A_2| }
      {   2 \, |A_0|^2 + |A_2|^2
        + 2 \sqrt{2} \, |A_0| \, |A_2| \, 
          \cos ( \delta_{\SS 0} - \delta_{\SS 2} ) \,
          \cos ( \zeta_{\SS 0} - \zeta_{\SS 2} ) } \, .
  \label{asymdirecte}
\end{equation}
Elle est non nulle s'il y a une violation CP directe.
C'est le cas s'il y a au moins deux amplitudes partielles non nulles
($A_0$ et $A_2$) {\it avec} une phase relative
($\zeta_{\SS 0} - \zeta_{\SS 2}$) violant CP {\it et} une phase relative
($\delta_{\SS 0} - \delta_{\SS 2}$) conservant CP.
L'asym\'etrie $A_{00}$ correspondant aux pions neutres a une forme similaire.

Ces asym\'etries ne sont cependant pas observables telles quelles puisque les
kaons se propagent et oscillent avant de se d\'esint\'egrer, donnant lieu \`a
une violation CP indirecte en plus de la violation CP directe.
Comme on l'a vu ci-dessus, les quantit\'es mesur\'ees sont $\eta_{\SS +-}$
et $\eta_{\SS 00}$.
On voudrait param\'etriser ind\'ependamment les violations CP directe et
indirecte. Ce n'est pas possible avec la d\'efinition de $\eta_f$ en fonction
de $\hat \epsilon$ et $\chi_{\SS f}$ puisqu'ils ne sont pas s\'epar\'ement
invariants sous la transformation de phase ${\cal S}$: m\^eme en absence de
violation CP directe, le param\`etre $\chi_{\SS f}$ ne sera nul que pour un
choix particulier de phase.
Pour d\'eterminer l'expression de $\eta_{\SS +-}$ en l'absence de violation CP directe,
calculons-le d'abord en fonction des amplitudes d'isospin:
$$
  \eta_{\SS +-} =
  \frac{   \sqrt{2} \, {\cal R}\!e \, A_0 \, ( \hat \epsilon + i \kappa_{\SS 0} ) 
         + e^{i\delta} \, {\cal R}\!e \, A_2 \, ( \hat \epsilon + i \kappa_{\SS 2} ) }
       {   \sqrt{2} \, {\cal R}\!e \, A_0 \, ( 1 + i \hat \epsilon \, \kappa_{\SS 0} )
         + e^{i\delta} \, {\cal R}\!e \, A_2 \, ( 1 + i \hat \epsilon \, \kappa_{\SS 2} ) } \, ,
$$
o\`u $\kappa_{\SS I} \equiv {\cal I}\!m \, A_I/{\cal R}\!e \, A_I = \tan \zeta_{\SS I}$ et
$\delta \equiv \delta_{\SS 2} - \delta_{\SS 0}$.
La violation CP directe est par exemple absente si $A_2=0$. Dans ce cas,
$$
  \eta_{\SS +-} =
  \frac{ \hat \epsilon + i \kappa_{\SS 0} }
       { 1 + i \hat \epsilon \, \kappa_{\SS 0} }
  \equiv \epsilon \, ,
$$
que l'on prend comme d\'efinition d'un nouveau param\`etre, $\epsilon$,
invariant sous la transformation de phase ${\cal S}$.
On peut v\'erifier explicitement que $i\kappa_{\SS 0}$ se transforme comme
$\chi_{\SS f}$ (\'equation (\ref{transfochi})) et joue donc le r\^ole de
$\chi_{\SS f}$ dans la compensation de la variation de $\hat \epsilon$ sous
${\cal S}$:
$$
  i\kappa_{{\SS 0} \, , \, \alpha} =
  \frac{ i \kappa_{\SS 0} - i \tan \alpha}
       { 1 - i ( i \kappa_{\SS 0} ) \tan \alpha } \, .
$$
D'une part, la condition
${\cal R}\!e \, \epsilon \neq 0$ signale une violation CP dans l'asym\'etrie $A_{+-}$
car ${\cal R}\!e \, \epsilon \propto {\cal R}\!e \, \hat \epsilon$.
Comme elle rend l'asym\'etrie $A_T$ non nulle, on l'interpr\`ete comme
une violation CP dans les oscillations.
D'autre part, ${\cal I}\!m \, \epsilon \neq 0$ signale aussi une violation CP dans
l'asym\'etrie $A_{+-}(L)$,
qui r\'esulte de l'interf\'erence entre l'oscillation et la d\'esint\'egration.
On pourrait l'appeler {\it violation CP induite}, en analogie avec la
violation CP induite d\'efinie dans le syst\`eme $B^0\overline{B^0}$
(\'equation (\ref{violationinduite})).  
En utilisant la d\'efinition de $\epsilon$, on peut maintenant \'ecrire
l'expression de $\eta_{\SS +-}$ en pr\'esence de violation CP directe,
c'est-\`a-dire pour $A_2 \neq 0$:
\begin{equation}
  \eta_{\SS +-} = \epsilon + \frac{\epsilon'}{ 1+\omega/\sqrt{2} } \, ,
  \label{parameta+-}
\end{equation}
o\`u $\epsilon$ garde sa d\'efinition et o\`u $\omega$ et $\epsilon'$ sont
d\'efinis par
\begin{eqnarray*}
  \omega &\equiv&
  \frac{ 1 + i \hat \epsilon \, \kappa_{\SS 2} }
       { 1 + i \hat \epsilon \, \kappa_{\SS 0} } \,
  \frac{ {\cal R}\!e \, A_{\SS 2} }{ {\cal R}\!e \, A_{\SS 0} } \, e^{i\delta} \, ,
  \\
  \epsilon' &\equiv&
  \frac{ i \omega \, (1- \hat \epsilon^2) \,
         ( \kappa_{\SS 2} - \kappa_{\SS 0} ) }
       { \sqrt{2} \, (1 + i \hat \epsilon \, \kappa_{\SS 2}) \,
                     (1 + i \hat \epsilon \, \kappa_{\SS 0})     } \, . 
\end{eqnarray*}
Les param\`etres $\epsilon$, $\omega$ et $\epsilon'$ sont invariants sous la
transformation de phase ${\cal S}$.

On proc\`ede de m\^eme pour la d\'esint\'egration des kaons en pions et l'on
obtient
\begin{equation}
  \eta_{\SS 00} = \epsilon - \frac{ 2\epsilon'}{ 1-\sqrt{2}\omega } \, .
  \label{parameta00}
\end{equation}
Ces \'equations sont {\it exactes}, contrairement \`a la plupart des
expressions des $\eta_{\SS f}$ en fonction de $\epsilon$ et $\epsilon'$
trouv\'ees dans la litt\'erature, o\`u $\epsilon$ et $\epsilon'$ ne sont
m\^eme pas invariants de phase. Cette param\'etrisation exacte nous \'evite
de devoir n\'egliger des termes en $\hat \epsilon^2$ qui sont du m\^eme ordre
que $\epsilon'$.
Dans la litt\'erature (par exemple dans \cite{grimus1}), les d\'efinitions de
$\omega$ et $\epsilon'$ diff\`erent des n\^otres et ne sont pas invariantes de phase,
avec pour cons\'equence que les expressions pour $\eta_{\SS +-}$ et $\eta_{\SS 00}$,
bien qu'identiques \`a nos formules (\ref{parameta+-}) et (\ref{parameta00}),
ne peuvent \^etre obtenues qu'apr\`es des
approximations n\'egligeant $\epsilon^2$ par rapport \`a $\epsilon'$.

Que deviennent les diff\'erents cas (voir l'\'equation (\ref{asymdirecte}))
pour lesquels l'asy\-m\'e\-trie $A_{+-}$ s'annulait?
Entra\^{\i}nent-ils l'annulation de $\epsilon'$, le nouveau param\`etre
d\'ecrivant la violation CP directe?
\begin{enumerate}
  
  \item
  Si $\kappa_{\SS 2} = \kappa_{\SS 0}$
  (c'est-\`a-dire $\zeta_{\SS 2} = \zeta_{\SS 0}$),
  alors $\epsilon'=0$.
  
  \item
  Si $A_2=0$, alors $\epsilon'=0$.
  
  \item
  Si $A_0=0$, la d\'ecomposition ci-dessus est mal d\'efinie. On a plut\^ot
  $$
    \varrho = \frac{ \hat \epsilon + i \kappa_{\SS 2} }
                         { 1+i\hat \epsilon \kappa_{\SS 2} }
    \quad \mbox{et} \quad
    \eta_{\SS +-} = \varrho - \frac{ 2\varrho'}{ 1+\sqrt{2}\varpi } \, ,
  $$
  o\`u $\varrho'$ et $\varpi$ sont obtenus \`a partir de $\epsilon$ et
  $\omega$ en \'echangeant les indices $0$ et $2$
  (donc $\varrho'=-\epsilon/\omega^2$ et $\varpi=\omega^{-1}$).
  Si $A_0=0$, le param\`etre $\varrho'$ est nul.
  
  \item
  Si $\delta=0$ {\it et} ${\cal R}\!e \, \hat \epsilon =0$, alors ${\cal R}\!e \, \epsilon'=0$
  mais il est toujours possible que ${\cal I}\!m \, \epsilon' \neq 0$.
  Le choix de la phase tel que $\hat \epsilon =0$ ne change rien.
  On pourrait se demander comment il peut y avoir une violation CP alors qu'une des
  conditions de son apparition n'est pas satisfaite, c'est-\`a-dire l'existence
  d'une phase rela\-tive conservant CP entre les amplitudes partielles.
  Contrairement aux apparences, il n'en est rien! En effet, les conditions
  $\delta=0$ et $\hat \epsilon =0$ impliquent que
  ${\cal R}\!e \, \eta_{\SS +-} \sim {\cal R}\!e \, \epsilon' =0$,
  ou encore $\phi_{\SS +-} = \pm \pi/2$
  (rappelons que $\eta_{\SS +-} \equiv | \eta_{\SS +-}| \, e^{i\phi_{\SS +-}}$).
  L'examen de l'asym\'etrie $A_{+-}$ (\'equation  (\ref{asymcplear})) montre
  que cette asym\'etrie serait nulle sous ces conditions, si ce n'\'etait
  la pr\'esence de $\Delta m \neq 0$ dans l'argument du cosinus. La phase
  relative conservant CP est donc fournie par la diff\'erence de masse entre
  les \'etats se propageant.
\end{enumerate}
Observons aussi que m\^eme si l'asym\'etrie semi-leptonique $A_T$ est nulle
(${\cal R}\!e \, \hat \epsilon=0$), le param\`etre $\hat \epsilon$ joue toujours
un r\^ole dans l'asym\'etrie $A_{+-}$ \`a travers sa partie imaginaire
${\cal I}\!m \, \hat \epsilon$. Non seulement ${\cal I}\!m \, \hat \epsilon$ peut se combiner
avec $\kappa_{\SS 0}$ pour donner ${\cal I}\!m \, \epsilon \neq 0$ comme on l'a vu
ci-dessus, mais il interf\`ere aussi avec les param\`etres $A_I$, $\zeta_{\SS I}$
et $\delta$ \`a l'int\'erieur de $\epsilon'$.
En bref, le param\`etre de m\'elange $\hat \epsilon$ intervient dans tous les
types de violations CP.
Les expressions des $\eta_{\SS f}$ se simplifient si l'on utilise l'information
exp\'erimentale (r\`egle $\Delta I = 1/2$) que $|A_2|/|A_0| \cong1/22$
\cite{marshak}. Les termes en $\omega$ peuvent alors \^etre n\'eglig\'es
et les \'equations
(\ref{parameta+-}) et (\ref{parameta00}) prennent la forme
\begin{eqnarray*}
  \eta_{\SS +-} &\cong& \epsilon + \epsilon' \, ,
  \\
  \eta_{\SS 00} &\cong& \epsilon - 2 \epsilon' \, .
\end{eqnarray*}
Les valeurs trouv\'ees par CPLEAR (voir \'equations (\ref{valeursetaphi}))
\'etant \'egales aux erreurs exp\'erimentales pr\`es, on sait que $\epsilon'$
est beaucoup plus petit que $\epsilon$. Les exp\'eriences d\'edi\'ees \`a la
mesure de la violation CP directe mesurent plut\^ot
${\cal R}\!e \, (\epsilon'/\epsilon)$:
$$
  |\eta_{\SS 00}|^2/|\eta_{\SS +-}|^2 =
  1 - 6 {\cal R}\!e \, \frac{\epsilon'}{\epsilon}
  + {\cal O} \left( \frac{\epsilon'^2}{\epsilon^2} \right) \, .
$$
Cette \'equation est identique \`a celle trouv\'ee dans la litt\'erature
(voir par exemple \cite{fry}), mais a \'et\'e obtenue sans approximations
douteuses. Nous avons en effet r\'esolu, d'une part, le probl\`eme de la normalisation
\`a l'ordre $\epsilon^2$ des \'etats oscillants et, d'autre part, nous avons d\'efini
les param\`etres $\omega$ et $\epsilon'$ de sorte qu'il ne nous est pas n\'ecessaire de
recourir \`a des approximations incorrectes du type $\epsilon^2 \ll \epsilon'$.
La compilation des r\'esultats exp\'erimentaux (voir \'equation
(\ref{valeurepsprime})) fournit la valeur
$$
  {\cal R}\!e \, \frac{\epsilon'}{\epsilon} = (21.2 \pm 4.7) \times 10^{-4} \, .
$$
Avec une convention de phase telle que $|\hat \epsilon| \ll 1$,
les expressions de $\epsilon$, $\omega$ et
$\epsilon'$ prennent la forme approximative (et non invariante de phase):
\begin{eqnarray*}
  \epsilon &\cong& \hat \epsilon + i \kappa_{\SS 0} \, ,
  \\
  \omega &\cong& \frac{ {\cal R}\!e \, A_2 }{ {\cal R}\!e \, A_0 } \, e^{i\delta} \, ,
  \\
  \epsilon' &\cong&  \frac{i}{ \sqrt{2} } \, \omega \,
  ( \kappa_{\SS 2} - \kappa_{\SS 0}) \, ,
\end{eqnarray*}
Dans ce cas, ${\cal R}\!e \, \epsilon \cong {\cal R}\!e \, \hat \epsilon$ et la mesure des
$\eta_{\SS f}$ fournit la valeur de ${\cal R}\!e \, \hat \epsilon$.
Dans la m\^eme convention de phase, l'asym\'etrie  semi-leptonique $A_T$ devient
$A_T \cong 4 \, {\cal R}\!e \, \hat \epsilon$. On peut donc comparer les r\'esultats
exp\'erimentaux provenant de $A_T$ et de $A_{\SS +-}$ et $A_{\SS 00}$. Ils
co\"{\i}ncident aux erreurs exp\'erimentales pr\`es.
Dans la convention de phase de Kobayashi-Maskawa, ${\cal I}\!m \, A_2=0$ si la
sym\'etrie d'isospin est respect\'ee. Le terme $\kappa_2$ dans l'expression
de $\epsilon'$ param\'etrise donc la violation de l'isospin dans cette convention
\cite{buras}.

Sous les m\^emes conditions, les param\`etres $\chi_{\SS f}$ de violation CP
directe se r\'e\'ecrivent:
\begin{eqnarray*}
  \chi_{\SS +-} &\cong&  \epsilon' + i \kappa_{\SS 0} \, ,
  \\
  \chi_{\SS 00} &\cong&  - 2 \epsilon' + i \kappa_{\SS 0}\, ,
\end{eqnarray*}
Dans la convention de Wu-Yang (qui est une des nombreuses conventions
telles que $|\hat \epsilon| \ll1$), $\kappa_{\SS 0}=0$ donc
$\chi_{\SS +-} \cong  \epsilon'$ et $\chi_{\SS 00} \cong  - 2 \epsilon'$.

\addcontentsline{toc}{chapter}{Conclusion}

\chapter*{Conclusion}
\pagestyle{myheadings}\markboth{CONCLUSION}{CONCLUSION}

Les oscillations spatio-temporelles de la probabilit\'e de d\'etection dans la
propa\-gation de particules en superposition quantique ont une importance
capitale aussi bien du point de vue fondamental, dans l'\'etude de la
violation CP  dans les syst\`emes de kaons neutres et de m\'esons {\it B} et dans
la d\'etermination du spectre de masse des neutrinos, que du point de vue
ph\'enom\'enologique dans l'\'elucidation des anomalies observ\'ees dans le
flux des neutrinos atmosph\'eriques et solaires. Ces ph\'enom\`enes
en essence identiques n'ont pas re\c{c}u jusqu'\`a pr\'esent de
des\-cription th\'eorique unifi\'ee, couvrant \`a la fois le cas des particules
stables et instables, ainsi que les domaines relativistes et
non relativistes.

Dans le calcul traditionnel de la probabilit\'e d'oscillation en m\'ecanique
quantique, les \'etats oscillants sont consid\'er\'es comme une superposition
d'\'etats propres de masse et sont suppos\'es \^etre dot\'es d'une
\'energie-impulsion bien d\'efinie. Ce traitement simpliste m\`ene
directement \`a un paradoxe, puisque la connaissance exacte de l'\'energie-impulsion
force l'\'etat \`a se trouver dans un
\'etat propre de masse et supprime du m\^eme coup les oscillations. De plus,
comment une oscillation de la probabilit\'e de d\'etection de l'\'etat
pourrait-elle \^etre observable si la connaissance pr\'ecise de son
\'energie-impulsion implique une incertitude infinie sur sa position?

Il para\^{\i}t logique de r\'esoudre ce paradoxe en postulant l'existence d'une
incertitude sur l'\'energie-impulsion, c'est-\`a-dire en traitant l'\'etat
oscillant comme une superposition de paquets d'ondes. Cependant, des questions
de principe subsistent, que le formalisme soit relativiste ou
non: d'une part, l'interf\'erence entre \'etats est interdite en m\'ecanique
quantique non relativiste par la r\`egle de superposition de Bargmann,
d'autre part, il para\^{\i}t impossible de cons\-truire un espace de Fock pour
une particule qui n'est pas un \'etat propre de masse. Une formule
rela\-tiviste ne peut \^etre d\'eriv\'ee rigoureusement dans ces conditions.
Par ailleurs, la taille et la forme du paquet d'ondes sont ind\'etermin\'ees,
ce qui est insatisfaisant vu l'influence qu'elles ont sur l'observabilit\'e des
oscillations. L'effet des conditions de production et de d\'etection n'est pas
non plus pris en compte. Pour donner un exemple, une mesure pr\'ecise de
l'\'energie-impulsion des \'etats issus du lieu de d\'etection de la
particule oscillante, identifie l'\'etat propre de masse se propageant et
supprime les oscillations. Enfin, les particules oscillantes instables ne
peuvent pas \^etre d\'ecrites par un paquet d'ondes.

Il ne reste plus qu'\`a se tourner vers la th\'eorie des champs, o\`u les
particules oscillantes sont d\'ecrites comme des \'etats interm\'ediaires
virtuels, non observ\'es directement et se propageant entre une source et un
d\'etecteur. Dans le m\^eme esprit, la question de savoir comment un \'etat
quantique peut se signaler par une trajectoire bien d\'efinie dans un
d\'etecteur, a d\'ej\`a \'et\'e r\'esolue par Mott \cite{mott} en 1929, en ne
consid\'erant comme observables que les \'etats excit\'es le long de la
trajectoire par la particule se propageant dans le d\'etecteur (ces \'etats
excit\'es se manifestent par exemple par des bulles dans une chambre \`a
bulles). Pris deux \`a deux, les \'etats excit\'es forment une suite de
syst\`emes source-d\'etecteur. Contrairement \`a Mott qui utilise le
th\'eorie des perturbations non relativiste de la m\'ecanique quantique, nous
repr\'esentons les \'etats oscillants par un propagateur covariant sous les
transformations de Lorentz et qui, sous sa forme compl\`ete, contient la
description de l'instabilit\'e des particules \`a travers la localisation de
ses p\^oles complexes. Un syst\`eme de particules en m\'elange est
repr\'esent\'e par un propagateur matriciel non diagonal, qui rend possible la
propagation d'une particule entre deux points avec changement de saveur. Ce
formalisme permet d'\'eviter la d\'efinition d'\'etats de saveur,
c'est-\`a-dire d'\'etats de masse ind\'efinie. La matrice m\'elangeant les
\'etats en m\'ecanique quantique est remplac\'ee par la matrice diagonalisant
le propagateur. Les corrections non exponentielles \`a la propagation d'un
m\'elange peuvent \^etre facilement analys\'ees dans ce contexte et sont
n\'egligeables.

Dans ce formalisme, les oscillations sont mesur\'ees indirectement, comme dans
les exp\'eriences, par la d\'etection des particules issues de la source et du
d\'etecteur. Les notions de {\it source} et de {\it d\'etecteur} sont en
fait fictives, et symbolisent les processus de production et de
d\'etection de la particule oscillante. Les particules entrantes et
sortantes sont mod\'elis\'ees de mani\`ere r\'ealiste par des paquets
d'ondes. Par ce biais, en jouant sur la largeur des paquets d'ondes, il
est possible d'\'etudier l'influence des conditions de production et de
d\'etection sur l'observabilit\'e des oscillations, puisque celles-ci
disparaissent dans le cas o\`u les \'etats asymptotiques sont des ondes
planes.

Les probl\`emes r\'esultant du traitement traditionnel des oscillations
n'appa\-rais\-sent plus dans notre calcul. Ils \'etaient en effet li\'es
d'une part \`a l'attribution d'une s\'erie de propri\'et\'es (saveur,
\'energie-impulsion, etc.) \`a la particule oscillante, et d'autre
part \`a l'introduction de concepts classiques (temps moyen de propagation,
vitesse, etc.) par des hypoth\`eses ext\'erieures au formalisme. Notre
m\'ethode met du m\^eme coup un point final aux controverses sur la longueur
d'oscillation ainsi que sur l'oscillation des particules associ\'ees \`a la
production de l'\'etat oscillant. La question de l'\'energie-impulsion de la
particule oscillante est r\'esolue en d\'emontrant en m\^eme temps l'influence
des conditions exp\'erimentales sur l'\'energie-impulsion et l'ind\'ependance
du r\'esultat final (en tout cas au premier ordre en la diff\'erence de masse)
par rapport \`a elle. La formule d'oscillation obtenue contient diff\'erents
facteurs que l'on peut identifier comme la d\'ecroissance exponentielle en
fonction de la distance pour une particule instable, l'oscillation d\'ependant
de la diff\'erence de masse et de la distance, la d\'ecoh\'erence supprimant
les oscillation \`a grande distance pour les particules relativistes et
finalement l'influence des conditions exp\'erimentales sur
l'observabilit\'e des oscillations. Les conditions sous lesquelles la
formule obtenue co\"{\i}ncide avec la formule classique apparaissent donc
explicitement dans notre r\'esultat.

Les caract\'eristiques principales de la probabilit\'e d'oscillation sont
ensuite repri\-ses dans une prescription de calcul s'appliquant directement \`a
l'amplitude d'un processus. Les cas des exp\'eriences CPLEAR (oscillation
d'une seule particule) et de DA$\Phi$NE sont examin\'es. Un traitement
coh\'erent de la violation CP dans le syst\`eme des kaons neutres
s'ensuit, contrairement au traitement traditionnel o\`u des probl\`emes de
normalisation des \'etats entachent la pr\'ecision des pr\'edictions
th\'eoriques.

Bien que ayons consid\'er\'e un cas d'oscillation double (DA$\Phi$NE), les
hypoth\`eses que nous avons pos\'ees ne permettent pas une extension
imm\'ediate \`a d'autres situations. Il serait par exemple int\'eressant
d'\'etudier les oscillations en cascade
$B^0(\overline{B^0}) \!\rightarrow\! K^0(\overline{K^0}) \!\rightarrow\!
\pi\pi,\mu\pi\nu$ qui pourront \^etre \'etudi\'ees au LHC \cite{azimov,stodolsky}.
Une autre application int\'eressante est l'\'etude de l'influence de
l'instabilit\'e de la source de la particule oscillante, qui peut en
principe se faire en consid\'erant aussi cette source comme un \'etat
interm\'ediaire non observ\'e directement. Nos techniques d'int\'egration ne
conviennent pas \`a cette analyse. Enfin, notre formule s'applique \`a la
propagation de neutrinos instables, qu'ils soient relativistes ou non.
Cette instabilit\'e coupl\'ee \`a un m\'elange est un mod\`ele explicatif
des anomalies observ\'ees dans la mesure des neutrinos atmosph\'eriques et
solaires. Pour l'instant, une modification simple de la formule de
m\'ecanique quantique par une exponentielle d\'ecroissante suffit \`a
param\'etriser les mesures en raison du peu de donn\'ees disponibles, mais
l'astronomie au moyen des neutrinos ne fait que commencer!

\backmatter
\pagestyle{headings}
\addcontentsline{toc}{chapter}{Bibliographie}
\def\baselinestretch{1}


\small

\begin{thebibliography}{150}


\bibitem{fonda}
L. Fonda, G. C. Ghirardi et A. Rimini, {\sl Decay theory of unstable
quantum systems}, Rep.\ Prog.\ Phys.\ {\bf 41}, 587 (1978).

\bibitem{khalfin}
L. A. Khalfin, {\sl Contribution to the decay theory of a
quasi-stationary state}, Sov.\ Phys.\ JETP {\bf 6}, 1053 (1958);
une liste des travaux pionniers concernant les corrections non exponentielles se
trouve dans: R. G. Newton, {\sl Scattering theory of waves and particles}
(McGraw-Hill, New York, 1966), p. 608. 

\bibitem{sakurai}
Voir par exemple: J. J. Sakurai, {\sl Modern quantum mechanics}, rev.\ ed.
(Addison-Wesley, Reading, 1994), p.\ 481.

\bibitem{schwinger}
J. Schwinger, {\sl Field theory of unstable particles}, Ann.\ Phys.
{\bf 9}, 169 (1960).

\bibitem{jacob}
R. Jacob et R. G. Sachs, {\sl Mass and lifetime of unstable
particles}, Phys.\ Rev.\ {\bf 121}, 350 (1961).

\bibitem{brown}
L. S. Brown, {\sl Quantum Field Theory} (Cambridge University Press,
Cambridge, 1992) p.\ 293 et p.\ 339.

\bibitem{greenland}
P. T. Greenland, {\sl Seeking non-exponential decay}, Nature {\bf
335}, 298 (1988).

\bibitem{peierls}
R. E. Peierls,  {\sl  Proc. of Glasgow Conference on Nuclear and
Meson Physics} (Pergamon Press, 1954) p.\ 296.

\bibitem{pais}
M. Gell-Mann et A. Pais,
{\sl Behavior of neutral particles under charge conjugation},
Phys.\ Rev.\ {\bf 97}, 1387 (1955).

\bibitem{lande}
K. Land\'e, E. T. Booth, J. Impeduglia, L. M. Lederman et W. Chinowsky,
{\sl Observation of long-lived neutral V articles},
Phys.\ Rev.\ {\bf 103}, 1901(L) (1956).

\bibitem{christenson}
J. H. Christenson, J. W. Cronin, V. L. Fitch et R. Turlay,
{\sl Evidence for the $2\pi$ decay of the $K^0_2$ meson},
Phys.\ Rev.\ Lett.\ {\bf 13}, 138 (1964).

\bibitem{burkhardt}
H. Burkhardt {\sl et al}, NA31 Collaboration,
{\sl First evidence for direct CP violation},
Phys.\ Lett.\ {\bf B206}, 169 (1988);
G. D. Barr {\sl et al}, NA31 Collaboration,
{\sl A new measurement of direct CP violation in the neutral kaon system},
Phys.\ Lett.\ {\bf B317}, 233 (1993);
L. K. Gibbons {\sl et al}, E731 Collaboration,
{\sl Measurement of the CP-violation parameter $Re \, (\epsilon'/\epsilon)$},
Phys.\ Rev.\ Lett.\ {\bf 70}, 1203 (1993);
V. Fanti {\sl et al}, NA48 Collaboration,
{\sl A new measurement of direct CP violation in two pion decays of the
neutral kaon}, Phys.\ Lett.\ {\bf B465}, 335 (1999);
KTeV Collaboration,
{\sl Observation of Direct CP Violation in $K_{S,L} \to \pi \pi$ Decays},
Phys.\ Rev.\ Lett.\ {\bf 83}, 22 (1999).

\bibitem{wigner}
E. Wigner et V. F. Weisskopf, {\sl Berechnung der nat\"urlichen
Linienbreite auf Grund der Diracschen Lichttheorie}, Z.\ Phys. {\bf 63}, 54
(1930); {\sl ibid.}, {\sl \"Uber die nat\"urliche Linienbreite in der Strahlung
des harmonischen Oszillators}, Z.\ Phys. {\bf 65}, 18 (1930);
pour une pr\'esentation moderne, voir par exemple la r\'ef. \cite{nachtmann},
p.\ 509.

\bibitem{lipshutz}
N. R. Lipshutz, {\sl Invariance principles and the $K^0\!-\!\overline{K^0}$
propagator matrix}, Phys. Rev. {\bf 144}, 1300 (1966). 

\bibitem{albrecht1}
H. Albrecht {\sl et al}, ARGUS Collaboration, {\sl Observation of
$B^0\!-\!\overline{B^0}$ mixing}, Phys. Lett. {\bf B192}, 245 (1987);
T. Affolder {\sl et al}, CDF Collaboration, 
{\sl Measurement of the $B^0\overline{B^0}$ oscillation frequency using
$l^-D^{*+}$ pairs and lepton flavor tags}, Phys. Rev. {\bf D60}, 112004 (1999).

\bibitem{affolder}
T. Affolder {\sl et al}, CDF Collaboration, {\sl A measurement of $\sin 2 \beta$
from $B \!\to\! J/\psi K^0_S$ with the CDF detector},
Phys. Rev. {\bf D61}, 072005 (2000).
 
\bibitem{lipkin1}
T. B. Day, {\sl Demonstration of quantum mechanics in the large}, Phys.
Rev. {\bf 121}, 1204 (1961);
H. Lipkin, {\sl CP violation and coherent
decays of kaon pairs}, Phys. Rev. {\bf 176}, 1715 (1968).

\bibitem{davis}
R. Davis, D. S. Harmer et K. C. Hoffman, {\sl Search for neutrinos from the
sun}, Phys. Rev. Lett. {\bf 20}, 1205 (1968).

\bibitem{bahcall}
J. N. Bahcall, {\sl Neutrino astrophysics} (Cambridge University Press,
Cambridge, 1989).

\bibitem{bilenky}
S. M. Bilenky, C. Giunti et W. Grimus, {\sl Phenomenology of neutrino
oscillations}, Prog. Part. Nucl. Phys. {\bf 43}, 1 (1999)
et r\'ef\'erences incluses.

\bibitem{pontecorvo}
B. Pontecorvo, {\sl Neutrino experiments and the problem of conservation
of leptonic charge}, Sov. Phys. JETP {\bf 26}, 984 (1968); V. Gribov et B.
Pontecorvo, {\sl Neutrino astronomy and lepton charge}, Phys. Lett. {\bf
28B}, 493 (1969).

\bibitem{petcov}
J. N. Bahcall, N. Cabibbo et A. Yahil, {\sl Are neutrinos stable particles?},
Phys. Rev. Lett. {\bf 28}, 316 (1972);
J. N. Bahcall, S. T. Petcov, S. Toshev et J. W. F. Valle, {\sl Tests of
neutrino stability}, Phys. Lett. {\bf B181}, 369 (1986).

\bibitem{fukuda}
Y. Fukuda {\sl et al}, Super-Kamiokande Collaboration,
{\sl Evidence for oscillation of atmospheric neutrinos},
Phys. Rev. Lett. {\bf 81}, 1562 (1998);
K. S. Hirata {\sl et al}, Kamiokande-II Collaboration, {\sl Experimental study
of the atmospheric neutrino flux}, Phys. Lett. {\bf B205}, 416 (1988).

\bibitem{learned}
J. G. Learned, S. Pakvasa et T. J. Weiler, {\sl Neutrino mass and mixing implied
by underground deficit of low energy muon-neutrino events},
Phys. Lett. {\bf B207}, 79 (1988); 
V. Barger et K. Whisnant, {\sl The effects of neutrino oscillations with one
mass scale on the atmospheric neutrino flux}, Phys. Lett. {\bf B209}, 365
(1988);
K. Hidaka, M. Honda et S. Midorikawa, {\sl Neutrino oscillations and the
anomalous atmospheric neutrino flux}, Phys. Rev. Lett. {\bf 61}, 1537 (1988).

\bibitem{barger}
V. Barger, J. G. Learned, S. Pakvasa et T. J. Weiler,
{\sl Neutrino Decay as an Explanation of Atmospheric Neutrino Observations},
Phys. Rev. Lett. {\bf 82}, 2640 (1999);
V. Barger, J. G. Learned, P. Lipari, M. Lusignoli, S. Pakvasa et T. J. Weiler,
{\sl Neutrino Decay and Atmospheric Neutrinos},
Phys. Lett. {\bf B462}, 109 (1999).

\bibitem{lsnd}
C. Athanassopoulos {\sl et al}, LSND Collaboration, {\sl Evidence for
$\bar\nu_\mu\to\bar\nu_e$ oscillations from the LSND experiment at LAMPF},
Phys. Rev. Lett. {\bf 77}, 3082 (1996);
C. Athanassopoulos {\sl et al}, LSND Collaboration, {\sl Results on
$\nu_\mu \!\to\!\nu_e$ neutrino oscillations from the LSND experiment},
Phys. Rev. Lett. {\bf 81}, 1774 (1998).

\bibitem{msw}
L. Wolfenstein, {\sl Neutrino oscillations in matter}, Phys. Rev. {\bf D17},
2369 (1978);
S. P. Mikheyev et A. Yu. Smirnov, {\sl Resonance enhancement of
oscillations in matter and solar neutrino spectroscopy}, Sov. J. Nucl. Phys.
{\bf 42}, 913 (1985);
H. A. Bethe, {\sl Possible explanation of the solar-neutrino puzzle},
Phys. Rev. Lett. {\bf 56}, 1305 (1986).

\bibitem{bilenkygiunti}
S. M. Bilenky et C. Giunti, {\sl See-saw type mixing and
$\nu_\mu \!\to\! \nu_\tau$ oscillations}, Phys. Lett. {\bf B300}, 137 (1993).

\bibitem{maki}
Z. Maki, M. Nakagawa et S. Sakata, {\sl Remarks on the unified model of
elementary particles}, Prog. Theor. Phys. {\bf 28}, 870 (1962).

\bibitem{giunti1}
C. Giunti, C. W. Kim, U. W. Lee, {\sl Remarks on the weak states of
neutrinos}, Phys. Rev. {\bf D45}, 2414 (1992). 

\bibitem{nachtmann}
O. Nachtmann, {\sl Elementary particle physics} (Springer-Verlag, Berlin,
1990) p. 443.

\bibitem{sachs}
R. G. Sachs, {\sl Interference phenomena of neutral K mesons}, Ann.
Phys. {\bf 22}, 239 (1963).

\bibitem{enz}
C. P. Enz et R. R. Lewis, {\sl On the phenomenological description of
CP violation for K-mesons and its consequences}, Helv. Phys. Acta
{\bf 38}, 860 (1965).

\bibitem{alvarez}
L. Alvarez-Gaum\'e, C. Kounnas, S. Lola et P. Pavlopoulos, {\sl
Violation of time-reversal invariance and CPLEAR measurements},
Phys. Lett. {\bf B458}, 347 (1999).

\bibitem{jarlskog}
C. Jarlskog, {\sl Commutators of the quark mass matrices in the standard
electroweak model and a measure of maximal CP violation},
Phys. Rev. Lett. {\bf 55}, 1039 (1985).

\bibitem{gottfried}
K. Gottfried et V. F. Weisskopf, {\sl Concepts of Particle Physics}
(Clarendon Press, Oxford, 1984) vol. I p. 151.

\bibitem{srivastava}
Y. Srivastava, A. Widom et E. Sassaroli, {\sl Spatial correlations
in two neutral kaon decays}, Z. Phys. {\bf C66}, 601 (1995).

\bibitem{lipkin2}
H. J. Lipkin, {\sl Theories of non-experiments in coherent decays
of neutral mesons}, Phys. Lett. {\bf B348}, 604 (1995).

\bibitem{lowe}
J. Lowe, B. Bassaleck, H. Burkhardt, A. Rusek, G. J. Stephenson
Jr. et T. Goldman, {\sl No $\Lambda$ oscillations},
Phys. Lett. {\bf B384}, 288 (1996);
H. Burkhardt, J. Lowe, G. J. Stephenson Jr. et T. Goldman, {\sl
Oscillations of recoil particles against mixed states},
Phys. Rev. {\bf D59}, 054018 (1999).

\bibitem{kayser1}
B. Kayser, {\sl The frequency of neutral meson
and neutrino oscillation}, SLAC-PUB-7123;
B. Kayser, {\sl CP violation, mixing, and quantum mechanics},
Proc. of 28th HEP Conf., Varsovie 1996, p. 1135 (hep-ph/9702327). 

\bibitem{pal}
Voir par exemple: R. N. Mohapatra et P. B. Pal, {\sl Massive neutrinos in
physics and astrophysics}, (World Scientific, Singapore, 1991), p. 156.

\bibitem{winter}
R. G. Winter, {\sl Neutrino oscillation kinematics}, Lett. Nuovo Cimento
{\bf 30}, 101 (1981).

\bibitem{goldman}
T. Goldman, {\sl Source dependence of neutrino oscillations}, hep-ph/9604357.

\bibitem{dolgov}
A. D. Dolgov, A. Yu. Morozov, L. B. Okun et M. G. Schepkin,
{\sl Do muons oscillate?}, Nucl. Phys. {\bf B502}, 3 (1997).

\bibitem{lipkin3}
H. J. Lipkin, {\sl Quantum mechanics of neutrino oscillations - Hand waving
for pedestrians}, hep-ph/9901399.

\bibitem{bargmann}
V. Bargmann, {\sl On unitary ray representations of continuous
groups}, Ann. Math. {\bf 59}, 1 (1954);
A. Galindo et P. Pascual, {\sl Quantum Mechanics},
Springer-Verlag, 1990, p. 292;
F. A. Kaempffer, {\sl Concepts in quantum mechanics} (Academic
Press, 1965), p. 341; cette derni\`ere r\'ef\'erence est aussi
int\'eressante pour sa pr\'esentation des th\'eories de Yang-Mills
et de la gravitation comme th\'eorie de jauge.

\bibitem{blasone}
M. Blasone, P. A. Henning et G. Vitiello, {\sl The exact formula for
neutrino oscillations}, Phys. Lett. {\bf B451}, 140 (1999);
M. Blasone et G. Vitiello, {\sl Remarks on the neutrino oscillation
formula}, Phys. Rev. {\bf D60}, 111302 (1999).

\bibitem{beuthe}
M. Beuthe, G. L\'{o}pez Castro et J. Pestieau,
{\sl Field theory approach to $K^0 \!-\! \overline{K^0}$ and
$B^0 \!-\! \overline{B^0}$ systems}, Int. J. Mod. Phys. {\bf A13}, 3587
(1998).

\bibitem{veltman}
M. Veltman, {\sl Unitarity and causality in a renormalizable field
theory with unstable particles}, Physica {\bf 29}, 186 (1963);
{\sl ibid.}, {\sl Diagrammatica} (Cambridge University Press, Cambridge,
1994).

\bibitem{widom}
Y. Srivastava, A. Widom et E. Sassaroli, {\sl $\Lambda$ oscillations},
Phys. Lett. {\bf B344}, 436 (1995).

\bibitem{sassaroli}
Y. Srivastava, A. Widom et E. Sassaroli, {\sl Charged lepton
oscillations}, hep-ph/9509261;
{\sl ibid.}, {\sl Charged leptons and neutrino oscillations}, Eur. Phys.
J. {\bf C2}, 769 (1998);
Y. Srivastava et A. Widom, {\sl Of course muons can oscillate},
hep-ph/9707268.

\bibitem{kayser2}
B. Kayser, {\sl On the quantum mechanics of neutrino oscillation}, Phys.
Rev. {\bf D24}, 110 (1981).

\bibitem{giunti2}
C. Giunti, C. W. Kim, J. A. Lee et U. W. Lee,
{\sl Treatment of neutrino oscillations without resort to weak
eigenstates}, Phys. Rev. {\bf D48}, 4310 (1993);
C. Giunti, C. W. Kim et U. W. Lee, {\sl When do neutrinos cease to
oscillate?}, Phys. Lett. {\bf B421}, 237 (1998).

\bibitem{stuart}
R. G. Stuart, {\sl Gauge invariance, analyticity and physical observables at
the $Z^0$ resonance}, Phys. Lett. {\bf B262}, 113 (1991);
U. Baur et D. Zeppenfeld, {\sl Finite width effects and gauge invariance in
radiative W production and decay}, Phys. Rev. Lett. {\bf 75}, 1002 (1995);
J. Papavassiliou et A. Pilaftsis, {\sl Gauge-independent approach to resonant
transition amplitudes}, Phys. Rev. {\bf D53}, 2128 (1996);
M. Beuthe, R. Gonzalez Felipe, G. L\'{o}pez Castro et J. Pestieau,
{\sl Behaviour of the absorptive part of the $W^\pm$ electromagnetic vertex},
Nucl. Phys. {\bf B498}, 55 (1997) et r\'ef\'erences incluses.

\bibitem{kallen}
G. K\"all\'en, {\sl On the definitions of the renormalization constants},
Helv. Phys. Acta {\bf 25}, 417 (1957);
H. Lehmann, {\sl On the properties of propagation functions and
renormalization constants of quantized fields}, Nuovo
Cimento {\bf 11}, 342 (1954);
pour une pr\'esentation moderne, voir par exemple \cite{brown}, p. 282.

\bibitem{cutkosky}
R. E. Cutkosky, {\sl Singularities and discontinuities of Feynman
amplitudes}, J. Math. Phys. {\bf 1}, 429 (1960); voir aussi \cite{peskin},
p. 235.

\bibitem{buchalla}
G. Buchalla, A. J. Buras et M. E. Lautenbacher, {\sl Weak decays beyond leading
logarithms}, Rev. Mod. Phys. {\bf 68}, 1125 (1996).

\bibitem{okun}
L. Okun, {\sl Leptons and quarks} (North-Holland, Amsterdam, 1982), p. 85.

\bibitem{baulieu}
L. Baulieu et R. Coquereaux, {\sl Photon-Z mixing in the
Weinberg-Salam model}, Ann. Phys. {\bf 140}, 163 (1982).

\bibitem{harte}
S. Coleman et H. J. Schnitzer, {\sl Mixing of elementary particles}, Phys. Rev.
{\bf 134}, B863 (1964);
H. B. O'Connell, {\sl Recent developments in rho-omega mixing}, Aust. J. Phys.
{\bf 50}, 255 (1997), hep-ph/9604375.

\bibitem{rich}
J. Rich, {\sl Quantum mechanics of neutrino oscillations}, Phys.
Rev. {\bf D48}, 4318 (1993).

\bibitem{mohanty}
W. Grimus, P. Stockinger et S. Mohanty, {\sl The field-theoretical
approach to coherence in neutrino oscillations}, Phys. Rev. {\bf
D59}, 013011 (1999).

\bibitem{campagne}
J. E. Campagne, {\sl Neutrino oscillations from a pion decay in
flight}, Phys. Lett. {\bf B400}, 135 (1997).

\bibitem{grimus}
W. Grimus et P. Stockinger, {\sl Real oscillations of virtual
neutrinos}, Phys. Rev. {\bf D54}, 3414 (1996).

\bibitem{peskin}
M. E. Peskin et D. V. Schroeder, {\sl An introduction to quantum
field theory} (Addison-Wesley, Reading, 1995) p. 102.

\bibitem{sudarsky}
D. Sudarsky, E. Fischbach, C. Talmadge, S. H. Aronson et H.-Y.
Cheng, {\sl Effects of external fields on the neutral kaon system},
Ann. Phys. {\bf 207}, 103 (1991).

\bibitem{peres}
A. Peres, {\sl Nonexponential decay law}, Ann. Phys. {\bf 129}, 33 (1980).

\bibitem{fry}
J. R. Fry, {\sl CP violation and the standard model}, Rep. Prog. Phys. {\bf 63},
117 (2000).

\bibitem{pdg}
C. Caso {\sl et al}, Particle Data Group, {\sl Review of Particle Physics},
Eur. Phys. J. {\bf C3}, 1 (1998);
la nouvelle \'edition vient de para\^{\i}tre: D. E. Groom {\sl et al},
Particle Data Group, {\sl Review of Particle Physics},
Eur. Phys. J. {\bf C15}, 1 (2000); on peut aussi consulter le site http://pdg.lbl.gov/

\bibitem{fleischer}
R. Fleischer, {\sl CP Violation and the role of electroweak penguins in
nonleptonic B decays}, Int. J. Mod. Phys. {\bf A12}, 2459 (1997), et
r\'ef\'erences incluses.

\bibitem{chiu}
C. B. Chiu et E. C. G. Sudarshan, {\sl Decay and evolution of the neutral
kaon}, Phys. Rev. {\bf D42}, 3712 (1990).

\bibitem{wang}
Q. Wang et A. I. Sanda, {\sl Neutral kaon system reinvestigated}, Phys. Rev.
{\bf D55}, 3131 (1997).

\bibitem{giunti3}
C. Giunti, C. W. Kim et U. W. Lee, {\sl When do neutrinos really oscillate?
Quantum mechanics of neutrino oscillations}, Phys. Rev. {\bf D44}, 3635 (1991);
C. W. Kim et A. Pevsner, {\sl Neutrinos in physics and astrophysics}
(Harwood Academic Publishers, Chur, 1993), chap. 9;
C. Giunti et  C. W. Kim, {\sl Coherence of neutrino oscillations in the wave
packet approach}, Phys. Rev. {\bf D58}, 017301 (1998).

\bibitem{nussinov}
S. Nussinov, {\sl Solar neutrinos and neutrino mixing}, Phys. Rev. {\bf
63B}, 201 (1976);
S. M. Bilenky et B. Pontecorvo, {\sl Lepton mixing and neutrino
oscillation}, Phys. Rep. {\bf 41C}, 225 (1978).

\bibitem{kiers}
K. Kiers, S. Nussinov et N. Weiss,
{\sl Coherence effects in neutrino oscillations},
Phys. Rev. {\bf D53}, 537 (1996).

\bibitem{frautschi}
J. N. Bahcall et S. C. Frautschi, {\sl Lepton non-conservation and solar
neutrinos}, Phys. Lett. {\bf 29B}, 623 (1969).

\bibitem{glashow}
S. L. Glashow et L. M. Krauss, {\sl "Just so" neutrino oscillations}, Phys.
Lett. {\bf B190}, 199 (1987).

\bibitem{wilczek}
L. Krauss et F. Wilczek, {\sl Solar-neutrino oscillations}, Phys. Rev. Lett.
{\bf 55}, 122 (1985).

\bibitem{ahluwalia}
D. V. Ahluwalia et T. Goldman, {\sl Interplay of non-relativistic and
relativistic effects in neutrinos oscillations}, Phys. Rev. {\bf D56}, 1698
(1997).

\bibitem{reinartz}
P. Reinartz et L. Stodolsky, {\sl Neutrino masses and mixings in supernova
bursts}, Z. Phys. {\bf C27}, 507 (1985).

\bibitem{schechter}
J. Schechter et J. W. F. Valle, {\sl Neutrinos masses in $SU(2) \times U(1)$
theories}, Phys. Rev. {\bf D22}, 2227 (1980).

\bibitem{perrier}
B. Kayser, F. Gibrat-Debu et F. Perrier, {\sl The physics of massive neutrinos}
(World Scientific, Singapore, 1989), p. 59.

\bibitem{doi}
I. Yu. Kobzarev, B. V. Martem'yanov, L. B. Okun' et M. G. Shchepkin,
{\sl Phenomenology of neutrino oscillations}, Sov. J. Nucl. Phys. {\bf 32},
823 (1980);
S. M. Bilenky, J. Ho\v{s}ek et S. T. Petcov, {\sl On the oscillations of neutrinos
with Dirac and Majorana masses}, Phys. Lett. {\bf 94B}, 495 (1980).

\bibitem{cardall}
C. Y. Cardall, {\sl Coherence of neutrino flavor mixing in quantum
field theory}, Phys. Rev. {\bf D61}, 073006 (2000);
C. Y. Cardall et D. J. H. Chung,
{\sl The MSW effect in quantum field theory}, Phys. Rev. {\bf D60}, 073012
(1999).

\bibitem{shtanov} 
Yu. V. Shtanov, {\sl Space-time description of neutrino flavour
oscillations}, Phys. Rev. {\bf D57}, 4418 (1998).

\bibitem{weiss}
K. Kiers et N. Weiss, {\sl Neutrino oscillations in a model with
a source and detector}, Phys. Rev. {\bf D57}, 3091 (1998).

\bibitem{ioannisian}
A. Ioannisian et A. Pilaftsis, {\sl Neutrino oscillations in space
within a solvable model}, Phys. Rev. {\bf D59}, 053003 (1999).

\bibitem{dafne}
{\sl The DA$\Phi$NE Physics Handbook} Vol. I, eds. L.
Maiani, G. Pancheri et N. Paver (INFN-LNF Publications, 1992);
P. H. Eberhard, in: {\sl The Second DA$\Phi$NE Physics Handbook},
Vol. I, p. 99, eds. L. Maiani, G. Pancheri et N. Paver (INFN-LNF Publications,
1995).

\bibitem{argus}
H. Albrecht {\sl et al}, Collaboration ARGUS, {\sl A study
of $\overline{B^0}\to D^{*+}l^-\bar\nu$ and $B^0 \overline{B^0}$ mixing
using partial $D^{*+}$ reconstruction}, Phys. Lett. {\bf B324}, 249
(1994);
J. Bartelt {\sl et al}, Collaboration CLEO, {\sl Two measurements
of $B^0 \overline{B^0}$ mixing}, Phys. Rev. Lett. {\bf 71}, 1680 (1993);
H. Schr\"oder, {\sl $B\bar B$ mixing}, in: {\sl B decays}, Rev. 2nd ed.,
ed. S. Stone (World Scientific, Singapore, 1994) p. 449. 

\bibitem{kek}
{\sl KEK-B 1995 B factory Design Report KEK};
{\sl BABAR 1995 Technical Design Report} SLAC-R-95-0457.

\bibitem{dunietz}
I. Dunietz, J. Hauser et J. Rosner, {\sl Proposed experiment addressing
CP and CPT violation in the $K^0\!-\!\overline{K^0}$ system}, Phys. Rev. {\bf
D35}, 2166 (1987);
C. D. Buchanan, R. Cousins, D. Dib, R. D. Peccei et J. Quackenbush,
{\sl Testing CP and CPT violation in the neutral kaon system at a $\phi$
factory}, Phys. Rev. {\bf D45}, 4088 (1992);
M. Hayakawa et A. I. Sanda, {\sl Searching for T, CP, CPT, and
$\Delta S = \Delta Q$ rule violations in the neutral $K$ system: A guide},
Phys. Rev. {\bf D48}, 1150 (1993).

\bibitem{zralek}
M. Zra\makebox[0pt][l]{/}lek, {\sl From kaons to neutrinos:
quantum mechanics of particle oscillations}, Acta Phys. Polon.
{\bf B29}, 3925 (1998).

\bibitem{angelopoulos1}
A. Angelopoulos {\sl et al}, CPLEAR Collaboration, {\sl Measurement of the
$K_L\!-\!K_S$ mass difference using semileptonic decays of tagged neutral
kaons}, Phys. Lett. {\bf B444}, 38 (1998).

\bibitem{angelopoulos2}
A. Angelopoulos {\sl et al}, CPLEAR Collaboration, {\sl First direct observation
of time-reversal non-invariance in the neutral-kaon system},
Phys. Lett. {\bf B444}, 43 (1998).

\bibitem{adler+-}
R. Adler {\sl et al}, CPLEAR Collaboration, {\sl First determination of CP
violation parameters from $K^0\!-\!\overline{K^0}$ decay asymmetry},
Phys. Lett. {\bf B286}, 180 (1992);
A. Apostolakis {\sl et al}, CPLEAR Collaboration, {\sl A determination of the CP
violation parameter $\eta_{\SS +-}$ from the decay of strangeness-tagged neutral
kaons}, Phys. Lett. {\bf B458}, 545 (1999).

\bibitem{adler00}
R. Adler {\sl et al}, CPLEAR Collaboration, {\sl First observation of a
particle-antiparticle asymmetry in the decay of neutral kaons into
$\pi^0\pi^0$}, Z. Phys. {\bf C70}, 211 (1996);
A. Angelopoulos {\sl et al}, CPLEAR Collaboration, {\sl Measurement of the
CP-violation parameter $\eta_{\SS 00}$ using tagged $\overline{K^0}$ and $K^0$},
Phys. Lett. {\bf B420}, 191 (1998).

\bibitem{datta}
A. Datta et D. Kumbhakar, {\sl $D^0\!-\!\overline{D^0}$ mixing: a possible test
of physics beyond the standard model}, Z. Phys. {\bf C27}, 515 (1985).

\bibitem{carter}
A. B. Carter et A. I. Sanda, {\sl CP nonconservation in cascade decays of B
mesons}, Phys. Rev. Lett. {\bf 45}, 952 (1980);
I. I. Bigi et A. I. Sanda. {\sl Notes on the observability of CP violations in
B decays}, Nucl. Phys. {\bf B193}, 85 (1981);
{\sl ibid.}, {\sl CP violation in heavy flavor decays: predictions and search
strategies}, {\bf B281}, 41 (1987);
voir aussi la r\'ef\'erence \cite{fleischer}.

\bibitem{delepine}
D. Del\'epine, J.-M. G\'erard, J. Pestieau et J. Weyers, {\sl Final state interaction
phases in $(B\!\rightarrow\! K\pi)$ decay amplitudes}, Phys. Lett. {\bf B429}, 106 (1998).

\bibitem{nir}
Y. Nir et H. R. Quinn, {\sl Theory of CP violation in B decays}, in:
{\sl B decays}, Rev. 2nd ed., ed. S. Stone (World Scientific, Singapore, 1994) p. 520;
une synth\`ese figure dans la r\'ef\'erence \cite{pdg}, p. 555 \`a 562.

\bibitem{chau}
L.-L. Chau, {\sl Quark mixing in weak interactions},
Phys. Rep. {\bf 95}, 1 (1983).

\bibitem{grimus1}
W. Grimus, {\sl CP violating phenomena and theoretical results},
Fortschr. Phys. {\bf 36}, 201 (1988) et r\'ef\'erences incluses.

\bibitem{wolfenstein}
L. Wolfenstein, {\sl Models of CP violation}, in: {\sl Theory and phenomenology in particle
physics}, ed. A. Zichichi (Academic Press, New York, 1969), Part A, p. 218.

\bibitem{gerard}
J.-M. G\'erard, {\sl CP- and T- violations in the standard model}
in: {\sl Carg\`ese, 1992: Quantitative particle physics}, p. 149.

\bibitem{marshak}
R. E. Marshak, Riazuddin et C. P. Ryan, {\sl Theory of weak interactions in
particle physics}, (Wiley-Interscience, New York, 1969), p. 546.

\bibitem{buras}
A. J. Buras et J. M. G\'erard, {\sl Isospin breaking contributions to
$\epsilon'/\epsilon$}, Phys. Lett. {\bf B192}, 156 (1987).

\bibitem{mott}
N. F. Mott, {\sl The wave mechanics of $\alpha$-ray tracks}, Proc. R. Soc.
London, {\bf A126}, 79 (1929), reprinted in {\sl Quantum theory and
measurement}, ed. J. A. Wheeler et W. H. Zurek (Princeton University Press,
Princeton, 1983);
L. I. Schiff, {\sl Quantum mechanics}, 3rd ed. (McGraw-Hill Kogakusha, Tokyo,
1968), p. 335.

\bibitem{azimov}
Ya. I. Azimov, {\sl $K^0$ decays as analyzers of $B^0$ decays: How to measure
the sign of $\Delta m_B$}, Phys. Rev. {\bf D42}, 3705 (1990);
{\sl ibid.}, {\sl Phenomenology of neutral $D$-meson decays and double flavor
oscillations}, Eur. Phys. J. {\bf A4}, 21 (1999).

\bibitem{stodolsky}
B. Kayser et L. Stodolsky, {\sl Cascade mixing, a new kind of particle mixing
phenomenon}, hep-ph/9610522;
B. Kayser, {\sl Cascade mixing and the CP-violating angle beta}, hep-ph/9709382.



\end{thebibliography}
\end{document}